\documentclass[preprint,12pt]{elsarticle}
\usepackage[pdftex]{hyperref}
\usepackage{graphicx}

\usepackage{amssymb}
\usepackage{geometry}
\usepackage[fleqn]{amsmath}
\usepackage{bm}
\usepackage{amsbsy}
\usepackage{lineno}
\usepackage{physics}
\usepackage{mathtools}
\usepackage{xparse}
\usepackage{accents}

\usepackage{natbib}
\biboptions{comma,round, sort}
\numberwithin{equation}{section}

\newcommand{\extk}{
\mathrm{k}
}

\newcommand{\imagi}{
\mathrm{i}
}

\newcommand{\extket}[2]{
\ket{
\begin{array}{c}
#1 \\
#2
\end{array}}
}

\newcommand{\extbra}[2]{
\bra{
\begin{array}{c}
#1 \\
#2
\end{array}}
}

\newcommand{\extinnerprod}[4]{
\bra{
\begin{array}{c}
#1 \\
#2
\end{array}}\ket{
\begin{array}{c}
#3 \\
#4
\end{array}}
}

\newcommand{\classoperator}[2]{
\left[
\begin{array}{c}
#1\\ \\
#2
\end{array} \right]
}

\newcommand{\extscalar}[2]{
\left(
\begin{array}{c}
#1\; \\
#2
\end{array} \right)
}

\newcommand{\extscalarBra}[2]{
\left(
\begin{array}{c}
#1^\bullet \\
#2^\bullet
\end{array} \right)
}
\newcommand{\extscalarReal}[2]{
\left(
\begin{array}{c}
#1\; \\
#2
\end{array} \right)
}

\newcommand{\extcolumn}[2]{
\left(
\begin{array}{c}
#1 \\
#2
\end{array} \right)
}

\newcommand{\extoperator}[2]{
\left(
\begin{array}{c}
\bm{#1}^\dagger \\
\bm{#2\;}
\end{array} \right)
}

\newcommand{\extproduct}[4]{
\left [
\begin{array}{c}
#1 \quad #3\\
#2 \quad #4
\end{array} \right ]
}

\newcommand{\extaverage}[6]{
\bra{
\begin{array}{c}
#1 \\
#2
\end{array}}
\begin{array}{c}
\mathbf{#3}^\dagger \\
\mathbf{#4}\;
\end{array}
\ket{
\begin{array}{c}
#5 \\
#6
\end{array}}
}

\newcommand{\extaveragearg}[8]{
\bra{
\begin{array}{c}
#1 \\
#2
\end{array}}
\begin{array}{c}
\mathbf{#3}^\dagger #4 \\
\mathbf{#5\;} #6
\end{array}
\ket{
\begin{array}{c}
#7 \\
#8
\end{array}}
}
\newcommand{\biaverage}[2]{
\left\langle
\begin{array}{c}
\bm{#1\;} \\
\bm{#2}
\end{array} \right\rangle
}

\newcommand{\cbullet}{{\mathrlap{\bigcirc}\;\bullet}}

\newcommand{\rom}[1]{\mathrm{\romannumeral #1}}

\ExplSyntaxOn\makeatletter
\renewcommand*\dddot[1]{%
  \placeaccent{\acc@dot\mkern1.4mu\acc@dot\mkern1.4mu\acc@dot}{#1}%
  }
\renewcommand*\ddddot[1]{%

\placeaccent{\acc@dot\mkern1.4mu\acc@dot\mkern1.4mu\acc@dot\mkern1.4mu\acc@dot}{#1}%
  }
\NewDocumentCommand \vardot {O{1} m }
  {
    \int_compare:nNnTF
      {#1} = {1}
      {\dot #2}
      {\placeaccent{\prg_replicate:nn {#1-1} {\acc@dot\mkern1.4mu}\acc@dot}{#2}}
  }
\newcommand*\placeaccent[2]{%
  \begingroup
  \def\acc@dot{\kern-0.08em.\kern-0.08em}%
  \def\acc@skip{\ifx\macc@style\displaystyle0.32
           \else\ifx\macc@style\textstyle0.32
           \else\ifx\macc@style\scriptstyle0.22
           \else0.15\fi\fi\fi ex}%
  \def\mathaccent##1##2{%
    \setbox6\hbox{$\m@th\macc@style#1$}%
    \@tempdima\wd4
    \advance\@tempdima\macc@kerna
    \advance\@tempdima-\wd6
    \divide\@tempdima\tw@
    \@tempdimb\z@
    \ifdim\@tempdima<\z@ \@tempdimb-\@tempdima \@tempdima\z@ \fi
    \vbox{\offinterlineskip
          \moveright\@tempdima\box6
          \kern\acc@skip
          \moveright\@tempdimb\box4}%
  }%
  \macc@depth\@ne
  \let\math@bgroup\@empty \let\math@egroup\macc@set@skewchar
  \mathsurround\z@ \frozen@everymath{\mathgroup\macc@group\relax}%
  \macc@set@skewchar\relax
  \let\mathaccentV\macc@nested@a
  \macc@nested@a\relax111{#2}%
  \endgroup
}
\makeatother\ExplSyntaxOff

\journal{arXiv.org}

\begin{document}

\begin{frontmatter}


\title{A new proposal for a quantum theory  for isolated n-particle systems with variable masses connected by a field with variable form}



\author{Israel A. Gonz\'alez Medina}

\address{Instituto Superior de Tecnologias y Ciencias Aplicadas.\\ Universidad de la Habana. \\Cuba}
\ead{israelariel.gonzalezmedina@gmail.com}

\begin{abstract}
We propose a new quantum approach for describing a system of $n$ interacting particles with variable mass connected by an unknown field with variable form ($n$-VMVF systems). Instead of assuming any particular nature for variation of the masses and field, we propose to find them from the laws that rule the dynamic of the systems. In this context, we consider masses and field are functions depending only on the particle positions and velocities, excluding the presence of hidden variables. The quantum approach follows the modern method for constructing the quantum theory where quantum states are defined over a Hilbert space, and the operators are obtained from the canonical transformations of the corresponding classical theory. The classical theory results on two sets of second-order Hamilton constrained equations, while the Hilbert space is defined over the extension of the complex number, which is done from an unsolvable equation on the complex domain. This new approach does not contradict the existing theoretical theories in Particle Physics.

We add a new set of equations to the rectangular's in order to solve the classical problem and in agreement with the inclusion of new dynamical variables. The new and independent set of equations are the Lagrange equations using the $3$-D set of angular coordinates. The four-dimensional space-time naturally appears in the problem when the position of the particle is expressed as a function of angular coordinates. This transformation set the 3-D space of the angular coordinates as the stereographic projection of the 4-D sphere defined by the Lorentz condition in the space-time. The inclusion of mass as a variable quantity forces us to modify the D'Alembert's principle to ensure the compliance of the relativity principle under a Galilean transformation for the particle system. In consequence, we identify two sets of constraints, each one for every coordinate system. We construct the Lagrange function from the starting Lagrangian $L = \sum_n \frac{1}{2}m_n x^\nu_n x_{\nu,n} - A^\nu \dot{x}_{\nu,n}$ and forcing the system to satisfy linear and angular conservation laws. The obtained constraints are added to the initial Lagrangian by the Lagrange multiplier method and obtain not one, but two Lagrange functions and with them, two set of Lagrange equations for finding the final solution. Also, the $\ddot{x}$ dependency's of the constraints functions set the necessity of extending the classical theory up to second order of the Lagrange function $L(q,\dot{q},\ddot{q})$.

The extension of the classical Lagrange's theory up to the second order derivative results on a new set of second order Hamilton equations defining a new canonical variable $s = \frac{\partial L}{\partial \ddot{q}}$. The identity transformation introduces a new set of canonical variables $f_i$ as the poles a set of constraints which depend only on the velocities of all particles $\mathcal{F}_{2_i}(\bar{\dot{q}})$, removing the Ostrogradsky's instability. The new variable $s_i$ shows to be the generator for the negative displacement of the pole $f_i(\bar{\dot{q}})$ whose value is modified along the evolution of the system, being $f_i = 0$ the value for the identity transformation. The momentum $p_i$ remains as the generator of the displacement of coordinate $q_i$. We also evaluate the canonical transformations with the angular coordinates which have the angular momentum $l_i$ as the generator of the rotation of angular coordinate $\xi_i$ and the angular momentum $b_i = \frac{\partial L}{\partial \ddot{\xi}}$ as the generator of an negative displacement of the pole $f_i(\bar{\dot{\xi}})$.

We propose the extension of the complex numbers to be domain where the new quantum space should be developed. Beside it seem more appropriated to applied on the obtained classical mechanics, it also introduces new concepts, like negative and imaginary probabilities which may solve the problems of infinities. The new vector space named Extended Complex and represented as $\mathbb{E}$, has a new extended complex unit $k$, defined as the solution of the unsolvable equation in the complex domain: $|k|^2= k^* k = i$, so it does not violate the fundamental theorem of algebra. The new space definition also requests the inclusion of a new mapping operation represented as $k^\bullet$ so the product $k^{*\bullet} k^\bullet k^* k =i^*i=1$. We study the properties of the vector space like positive-definiteness, linearity, and conjugated symmetry. 

In the last, we put together all the obtained results and propose a new framework for the quantum treatment for $n$-VMVF systems. In that, any physical state is represented by what we define as the extended ket $\extket{\alpha}{\beta}$. We also define others two components mathematical entities like bras and operators. The action of a quantum operator acting over a space ket is represented now as: 
\begin{equation*}
\extoperator{A}{B} \extket{\alpha}{\beta} = \extket{\gamma}{\zeta}.
\end{equation*}
Based on the properties of the extended complex vector space, two new normalization conditions are settled for any new physical states. We study the eigenvalues equations, the series expansion, measurements and the quantum representation for momentum operator $p$ and $s$. Also, time evolution in the Schr\"odinger and Heisenberg pictures are introduced within this approach to obtain the equation for the eigenvalue of energy.

\end{abstract}


\end{frontmatter}

\newpage
\tableofcontents
\newpage
\section*{Dedication}
\thispagestyle{empty}
    \null\vspace{\stretch {1}}
        \begin{flushright}
        \vfill \vfill
			\textbf{To} my parents\\
			\vfill
			\textbf{To} my mentor Dr. Fernando Guzm\'an Mart\'inez
       \end{flushright}
\vspace{\stretch{2}}\null
\newpage
\section*{Acknowledgement}
I deeply thank my mentor Dr. Fernando Guzm\'an Mart\'inez and my personal friends Dr. Yoelvis Orosco Gonz\'alez, Dr. Juan Alejandro Garcia, Dr. Yansel Guerrero for informative discussions and correspondence.

\section*{Introduction}
Particle physics is the branch of physics that studies the constituents of matter and radiation knowns as particles. Specifically, it investigates the irreducibly smallest detectable particles and the fundamental interactions necessary to explain their behavior. Modern particle physics research is addressed mainly to subatomic particles, including the atomic constituents such as electrons, protons, neutrons. Those particles are produced by radioactive and scattering processes, such as photons, neutrinos, and muons, as well as a wide range of exotic particles. At the same time, there are others particles from what protons, neutrons are made and they are known as elementary or fundamental particles. An elementary particle is a subatomic particle with no substructure, thus not composed of other particles like the fundamental fermions (quarks, leptons, antiquarks, and antileptons) as well as the fundamental bosons (gauge bosons and the Higgs boson). Those last ones generally are the "force particles" that mediate interactions among the fermions.

The quantum Revolution took place in the first quarter of the twentieth century in the understanding of microscopic phenomena. Quantum mechanics not only replaced classical mechanics as the theory to explain microscopic experiments as Plank radiation laws, the Bohr atom, de Broglie's matter wave and so forth, but it also revised fundamental concepts of what most people know as reality. Quantum mechanic also governs the dynamics of particles; which exhibit wave-particle duality, displaying particle-like behavior under certain experimental conditions and wave-like behavior in others. In more technical terms, they are described by quantum state vectors in a Hilbert space. This approach is resumed in what some authors call ``quantum mechanical way of thinking''. This modern point treats quantum states as vectors in an abstract complex vector space whose dimension is determined by the degree of freedom of the physical system under study. The algebra of the complex space gives the algebra of the theory while physics equations are included from classical mechanics, specifically from Lagrangian/Hamiltonian formulation. The success of the quantum mechanic theory is undisputed, pointing to a trustworthy strategy to follow when we study any physical phenomena at the quantum length scale. Quantum mechanics also introduce fundamental concepts and materialistic conceptions from its results like the ladder operators, $a^\dagger, a^-$, which increases and decreases the eigenvalue of another operator, respectively.

Quantum mechanic works fine until it deals with a relativistic particle, starting with photons which have rest mass zero, and correspondingly travel in the vacuum at speed $c$. Once the conceptual framework of quantum mechanics was established, theoreticians like Born, Jordan, Heisenberg, Dirac, P. Jordan among others, direct their efforts to extend quantum methods to fields and particles starting precisely with photons and the quantization of electromagnetic field. The most successful outcome of these works was the inception of quantum field theory (QFT) which provide a suitable procedure for transferring the discreteness of physical quantities on the treatment of particles in quantum mechanics to an equal treatment of fields. 

Most of the problems that puzzle theoretical physics are related to the processes that take place at high energies and involve the variation of the properties of the particles. Examples of this process are the nuclear reactions, particle decay, fusion process, among others. One of the properties that change in a particle reaction is the mass of the particle. Several variable masses theories in modern physics have been applied to study this problem. However, most of them are based on spontaneous symmetry breaking which is rooted in quantum field theory. 

The theoretical bases of QFT are extracted from the quantum mechanic theory, which relies on classical mechanics as the theory from where it extracts the expressions for quantum operators. This fact can be troublesome since the Lagrange, and Hamilton's classical theory is developed considering that particle mass remains constant. The theoretical frame for including the appearance or vanishing of particles is the definition of the ``annihilation'' and ``creation'' operators, from the notions of the ladder operators on Quantum mechanic, for the destruction and creation of particles in the QFT theory.

The String theory has also evolved into a candidate for describing the dynamics of the elementary particle not without controversy. It proposes the unification of quantum mechanics and general relativity assuming that particles are replaced by small strings and branes. The theory claims that the particle with a single mass value and the force's charge are created because of a specific oscillatory pattern of the string. String theories usually require extra dimensions of spacetime for their mathematical consistency. In bosonic string theory, spacetime is 26-dimensional, while in superstring theory it is 10-dimensional, and in M-theory it is 11-dimensional.  This feature and the lack of experimental verification of some of its results are the main drawbacks of the theory.

Recently, some authors have the opinion that rest mass should not be treated as a fixed quantity. An interesting review of this topic is found in Journal of Physics: Conference Series 615 (2015) 012016. The author cites some anomalous nuclear reactions in condensed matter that can be explained assuming the variation of the rest mass. Among other approaches of variable mass theory, we can find the historical time models, introduced in the 1930s by Fock and Steuckelberg and developed later by Horwitz and others \cite{Davidson2014, Davidson2015}. Also, Greenberger develops a modern variant of the content of general relativity and the principle of equivalence. In the first approach, the proper time of particles is treated as the quantum operator with time coordinate describing the causal evolution of the system while Greenberger's approach addresses the time coordinate as an operator, with proper time (which is formally similar to historical time) being a real-valued parameter which parametrizes the causal evolution of the system \cite{Davidson2015}.  An important conclusion extracted from these works is that variable mass theories are compatible with the requirement of general relativity.

It is our understanding that any quantum mechanic based theory, like QFT, constructed to describe the system with variable masses should be built on a quantum theory which in turn is built over a classical theory that consider the the mass as a variable quantity.

Its also know the connexion between the occurrence of subatomic particles and fields.  In the Standard Model, the electromagnetic, strong, and weak interactions are associated with elementary particles, whose behaviors are modeled in quantum mechanics. For example, the strong interaction is carried by gluons, binding the quarks to form hadrons, such as protons and neutrons. The weak interaction is carried by the $W$ and $Z$ bosons and is responsible for the radioactive decay while photons carry the electromagnetic field interaction. In theories of quantum gravity, the graviton is the hypothetical elementary particle that mediates the force of gravity. The treatment of the gravitational field within quantum mechanics bring up one of the major unsolved problems in physics since there is no theory of gravitational interaction that reconciles with the currently popular Standard Model of particle physics. The main obstacle for combining the general relativity with quantum mechanics is a mathematical problem with the renormalization of the gravitational field.

Because of the relationship between fields and mass generation, it is also our understanding that any quantum mechanic based theory whose treat mass as a variable quantity, should also treat the field as a variable magnitude for a consistent approach.

\subsection*{Main Tasks, methodology, and assumptions}

The modern formalism of quantum mechanic success in the construction of the theory using elemental axioms. Thus, the intention to follow this formalism in the in the study of variable masses systems under quantum mechanics scope. While look at the idea of associate quantum generators to generators of  classical canonical transformations, the following questions come out: 

- What will be the quantum generator, which changes the state described by mass $m_0$ to $m_0 + dm$? In quantum operators terms what will be the operator $\mathcal{M}$ whose action over a state ket is
\begin{equation}
\mathcal{M}(dm) | m_0 \rangle = | m_0 + dm\rangle ?
\end{equation}
and, if such an operator exists,

- what would be its associated canonical transformation in classical mechanics?

Both questions are the primary motivation for this work. Although Davidson's approach \citep{Davidson2014} is radical and it must be acknowledged as visionary, we have the opinion that the analysis should begin by trying, first, to answer previous questions. As references  \cite{Davidson2014, Davidson2015} recall, there is no direct evidence that rest masses can change. However, it is our point of view that the inertial mass, as the measurement of an object's resistance to being accelerated by force, indeed changes. For example, if there is a neutron in the initial state which decays into an electron, a proton, and an electron anti-neutrino, we can assume that the mass of the particles of the system change if the mass of the ``appeared'' particles vary from the zero value. Our election of setting the inertial mass as a variable quantity ignores any nature of such variation or any particular law or principle which may imply any assumption about it, among then the relativistic. In that case, the concept of the rest mass can be no longer applied in here. Instead, the variation of the particle masses should be given only by the universal laws of motion.

We have to set out starting point for our approach. The modern construction of the quantum mechanics is done by passing over from the Hamiltonian theory to the quantum theory. The quantum field theories, which are based on the quantum theory, present a high degree of difficulty. As an alternative, many others theories have been developed. However, it is our thinking that the assumption of the variable character of the inertial mass is so fundamental that it can not be included any other way different from starting from the classical action. There are many others reasons for that choice. They are well discussed by Dirac at ``Lectures of Quantum mechanics'' \cite{dirac2001lectures}. In the text, Dirac concludes by saying: ``I don't think one can in any way shot-circuit the route of starting with an action integral, getting a Lagrangian, passing from the Lagrangian to the Hamiltonian, and then passing from the Hamiltonian to the quantum theory.''

Now there is the issue related to the dependency of the unknown variables of masses and field. We set the field and masses as variables. However, should they be treated in the same framework than the position of the particle? In the affirmative case, the classical theory defines a generalized momentum associated with a conservation law, which at the same time should be connected to a property of the space or matter. In the absence of a generalized coordinate, its generalized momentum must be conserved, and the alleged property of the space or matter must emerge. To the best of our knowledge, there is no known property from space or matter which lead to a conservation law for the momentum of the mass as a generalized coordinate. Because of that, the positions of the particles should be the only generalized coordinates of the system. We solve this issue by proposing the masses of the particle and the field as functions of the position of the particle and its derivatives.

We expect momentum depends not only on mass but also on its derivative. That could allow the description of particles with zero mass value and non-zero mass derivative. In that case, we will have a particle with zero mass value but with a well-defined momentum. We also presume that the variable mass approach will allow the study of physical systems including particle with zero mass and zero momentum value, which is impossible in modern quantum mechanic. In that case, we may be describing the ground state of a particle which means we may be in the presence of de description of a particle of vacuum. Also, the creation and annihilation of particles during field interaction could mean the variation of the mass of particles from zero to some real value and vice versa.

We study particle systems with a fixed number of particles which depends on the phenomena under analysis. We consider the deviation of masses for all particles forming part of the reaction: photon electrons, protons, deuterons, vacuum particles and others. The number of particles of the system is then defined as the maximum number of particles involved in the reaction during the reaction's time. For instance, in the electron-positron pair formation, we consider a two-particle system from the beginning: one being the energetic photon and other being a zero mass particle which may be defined, after the final result of the approach are obtained, as one vacuum particle. Same way, if there is a process which includes the creation or destruction of an intermediate particle, then that particle should also be included as part of the system. 

Following modern quantum mechanic approach, to accomplish the main objective, three tasks are defined:
\begin{enumerate}
\item to develop a classical theory that includes particle mass and field variation with the final goal of obtaining the infinitesimal canonical transformations of the system.
\item to create a vector space where the quantum mechanic for variable mass can be developed in
\item to construct a quantum theory that includes mass variation using previous results, including and adapting quantum axioms that from our point of view, must be taking into account. 
\end{enumerate}

Since the Lagrangian and Hamiltonian approaches, as we know them today, are constant mass-based theories, they are probably not sufficient to describe the physics needed for particles system with variable masses. However, the clear understanding and abstract concepts that Lagrangian or Hamiltonian formulation provided, make both of them an excellent start point for include mass variation within the classical theory.

In section \ref{QMsection}, we review the construction of the ordinary quantum mechanic as the path to follow for construct a quantum mechanic with variable masses and field. Among the main ideas, we reinterpret the normalization axiom as a relation between expectation values and space vectors properties. In section \ref{ClassicalLagSection} we develop the classical theory, starting for the D'Alembert principle, which must be modified because of the mass variation inclusion. In that section, relativistic effects are introduced, and with the previous principle modification, we construct two Lagrangians of second order needed for describing $n$-VMVF systems. In section \ref{ClassicalHamSection}, we develop the second order Hamilton theory for each obtained Lagrangian in the previous chapter, which differs from the Ostrogradsky's higher order Hamiltonian construction. We obtain the canonical transformations and their generators. In Section \ref{ExtendedNumSection} is developed a new space where the theory can be developed, in logical consistency with the classical results and also as a possible solution for the normalization issues of the actual quantum theories. Finally, in Section \ref{ExtendedNumSection} the obtained results from classical theory are added to a new Hilbert space taken over the extension of the complex numbers, following the same methodology than the modern construction of Quantum Mechanics. In this chapter, we introduce new axioms according to the outcome from previous chapters like two new normalization conditions for the state vector. We also cover the main aspects of conventional quantum mechanic theory as eigenvalues and eigenvectors, measurement, quantum dynamics and temporal evolution.

\newpage

\section{Modern Quantum mechanics}\label{QMsection}
We briefly describe this formalism and comment the main concepts that, from our point of view, are the cornerstone for developing a quantum approach for particle systems that include masses and field variations.

\subsection{Mathematical description of Quantum mechanics}
Several experiments, such the Stern-Gerlach's, show the complex space as the analytical framework to develop the quantum theory \citep{sakurai} pp.23, pp.40. On quantum theory, a physical state is represented by a vector in a complex based vector space called by Dirac as ``ket'' and denoted by $|\alpha \rangle$.

This vector space must adhere to a number of requirements called axioms. Let \textbf{u}, \textbf{v} and \textbf{w} are vectors in $V$ and $a$, $b$ are scalars in $F$, then the following relations are satisfied:
\begin{itemize} \label{vectorSpaceAxioms}
\item Associativity of addition: \textbf{u} + (\textbf{v} + \textbf{w}) = (\textbf{u} + \textbf{v}) + \textbf{w}
\item Commutativity of addition: \textbf{u} + \textbf{v} = \textbf{v} + \textbf{u}
\item Identity element of addition:	There exists an element \textbf{0} $\in V$, called the zero vector, such that \textbf{v} + \textbf{0} = \textbf{v} for all \textbf{v} $\in V$.
\item Inverse elements of addition:	For every $v \in V$, there exists an element −\textbf{v} $\in V$, called the additive inverse of \textbf{v}, such that \textbf{v} + ($-$ \textbf{v}) = \textbf{0}.
\item Compatibility of scalar multiplication with field multiplication:	\\ a(b\textbf{v}) = (ab)\textbf{v}
\item Identity element of scalar multiplication: 1\textbf{v} = \textbf{v}, where 1 denotes the multiplicative identity in F
\item Distributivity of scalar multiplication with respect to vector addition:	a(\textbf{u} + \textbf{v}) = a\textbf{u} + a\textbf{v}
\item Distributivity of scalar multiplication with respect to field addition:	\\ (a + b)\textbf{v} = a\textbf{v} + b\textbf{v}
\end{itemize}

These axioms allows states to satisfy the superposition principle in Quantum mechanics meaning that if a state is described by two kets $|\alpha \rangle$ and $|\beta \rangle$ then the ket vector
\begin{equation}
|\gamma \rangle = a |\alpha \rangle + b |\beta \rangle
\end{equation}
also describe the system by itself being $a$, $b$ complex numbers.

Operators X are the mathematical entities that modify the states and use to represent physical magnitudes like position, momentum, etc. Operators act over the kets from the left side, and the result is a new ket like:
\begin{equation}
X|\alpha \rangle =|\gamma \rangle.
\end{equation}

If a product of one operator with a vector state results in a same vector state times a complex number:
\begin{equation}
A|a'\rangle = a'|a'\rangle, \;A|a''\rangle = a''|a''\rangle, \;....
\end{equation}
then the set of complex numbers ${a',\;a''...}$ are known as ``eigenvalues, '' and the set of vectors state ${|a'\rangle, \; |a''\rangle\;...}$ as ``eigenvectors'' of that operator. 

To introduce intuitive geometrical notions such as the length of a vector, the angle between two vectors or the orthogonality between vectors it must be defined an operation, $mapping$,  that associates each pair of vectors in the space with a scalar quantity. That operation is called \textit{inner product}. Formally the inner product space is a vector space with an additional structure $V$ over the field $F$ of complex numbers $\mathbb{C}$ together with an inner product:

\begin{equation}
\langle \cdot,\cdot \rangle : V \times V \to F. \label{innerProdRepr}
\end{equation}

The inner product satisfied three axioms for all vectors  $x,y,z \in V$ and all scalar $a \in F$:
\begin{itemize}
\item Conjugate symmetry:
\begin{equation}
\langle x,y \rangle = \overline{\langle y,x \rangle }
\end{equation}
\item Linearity in the first argument:
\begin{align}
\langle x,ay \rangle &= a \langle x, y \rangle \nonumber \\
\langle ax,y \rangle &= \bar{a} \langle x, y \rangle \nonumber \\
\langle x,y + z \rangle &= \langle x,y\rangle + \langle x,z \rangle \nonumber \\
\langle x + y,z \rangle &= \langle x,z\rangle + \langle y,z \rangle
\end{align}
\item Positive-definiteness:
\begin{align}
\langle x, x \rangle &\geq 0 \\
\langle x, x \rangle &= 0 \Rightarrow x = 0
\end{align}
\end{itemize}
where $\bar{z}$ denotes the complex conjugate of $z$.

The necessity of introducing inner product lead to the concept of a 1-1 mapping space called ``bra'' space. If ket vector exists then bra vector must also exist and also the dual vector of its spanned eigenkets which are named eigenbras. The dual correspondence between ket and bra space is:
\begin{align}
c|\alpha\rangle \quad &\overset{\text{DC}}{\longleftrightarrow} \quad c^*\langle \alpha|
\nonumber \\
a'|a'\rangle, a''|a''\rangle,... \quad &\overset{\text{DC}}{\longleftrightarrow} \quad \langle a'|a'^*,  \langle a''| a''^*,...
\nonumber \\
a |\alpha \rangle + b |\beta \rangle \quad &\overset{\text{DC}}{\longleftrightarrow} \quad \langle \alpha|c_\alpha^* +  \langle \beta| c_\beta^*
\end{align}
Operators act on a bra from the right side, and the result is also another bra vector. The corresponding mapping of the ket $X|\alpha \rangle$ is $ \langle \alpha|X^{\dagger}$ being $X^{\dagger}$ the Hermitian adjoint operator of X. An operator X is to say to be Hermitian if $X^{\dagger}=X$.

Inner product is defined as
\begin{equation}
\langle \alpha|\beta \rangle = (\langle \alpha|) \cdot (|\beta \rangle).
\end{equation}

Its postulate, following complex vector space axioms, that inner product satisfy
\begin{align}
\langle \alpha|\alpha \rangle &\geq 0
\nonumber \\
\langle \alpha|\beta \rangle &= \langle \beta|\alpha \rangle^*,
\end{align}
from where can deduce $\langle \alpha|\alpha \rangle$ is a real nonnegative number. Experiments, observations, and optical phenomena probe the probabilistic nature of quantum mechanics, included in the formalism through the axiom:
\begin{equation}
\langle \alpha|\alpha \rangle = 1. \label{normCond}
\end{equation}

An observable is a real magnitude represented by a hermitian operator. It can include that the set of eigenvectors of observable forms an orthonormal basis and its eigenvalues are real.

Previous axiom eq. \ref{normCond} together with the composition axiom derive the probabilistic interpretation of quantum states in modern quantum theory. Indeed, an arbitrary ket can be spanned by eigenkets $|a'\rangle$:
\begin{equation}
|\alpha\rangle =  \sum_{a'} c_{a'} |a' \rangle.
\end{equation}
Expansion coefficients $c_{a'}$ are found using the unitary operator representation over an orthonormal basis
\begin{equation}
\bm{1} =  \sum_{a'} |a' \rangle  \langle a'|.
\end{equation} 
Indeed, any state vector can be spanned using unitary operator
\begin{align}
|\alpha\rangle &=  \bm{1} |\alpha\rangle = \sum_{a'}|a' \rangle  \langle a'| |\alpha\rangle
\nonumber \\
&= \sum_{a'} ( \langle a'|\alpha\rangle )|a' \rangle \qquad c_{a'} \equiv  \langle a'|\alpha\rangle.
\end{align}
On the other side, normalization condition \ref{normCond} impose
\begin{align}
\langle \alpha|\alpha \rangle &=  \sum_{a' a''}  c_{a''}^*   c_{a'} \langle a'' |a' \rangle
\nonumber \\
&=\sum_{a'}  |c_{a'}|^2 = 1.
\end{align}
If the sum of all real and nonnegative coefficients is equal to the unity then the probabilistic association can be made. This idea is reinforced when studying the expectation value for an operator.

The expectation value of an observable for system at the state vector $| \alpha \rangle$,  is defined as:
\begin{equation}
\langle A \rangle = \langle \alpha|A|\alpha \rangle
\end{equation} 
The measured value of an observable A can be computed by expanding the state vector as a linear combination of its eigenstates and using the orthonormal condition
\begin{equation}
\langle \alpha|A|\alpha \rangle = \sum_{a' a''} \langle \alpha|a'\rangle \langle a' |A|a''\rangle \langle a''|\alpha \rangle = \sum_{a'} a'|\langle a' |\alpha \rangle|^2 =  \sum_{a'} a' |c_{a'}|^2 \equiv  \sum_{a'} a'\mathcal{P}_{a'},
\end{equation}
where $\mathcal{P}_{a'}$ is the weight of the state $|a'\rangle$ on the general state $|\alpha \rangle$, pointing out the probabilistic nature of the theory.

Observables A and B are said to be compatible if $[A,B]=0$, otherwise they are classified as non compatible observables. Operator
\begin{equation}
[\; ,\; ](X,Y)\Rightarrow [X,Y]
\end{equation}
Is called commutator operator of X and Y and it plays a crucial role in quantum mechanics. An outstanding property inherent to quantum systems with no analogue in the classical mechanic theory is the uncertainty relation of Heisenberg. Following the complex vector space algebra it is shown, that the two operators $\bm{A}$ and $\bm{B}$ satisfy the referred relation
\begin{equation}
\langle (\Delta A)^2 \rangle \langle \Delta B)^2 \rangle \geq \frac{1}{4} | \langle [A,B] \rangle|^2.
\end{equation}
The uncertain relation asserts a fundamental limit to the precision with which two mean value of operators can be measured. There is no limit for compatible observables. In fact, two compatible observables with non-degenerate spectra share the same set of eigenvalues and eigenvectors. The uncertainty relation has no analogue in classical mechanics and is a mathematical inequality obtained from the Schwarz's lemma, and the pure real and pure complex character of the expectation value of Hermitian and anti-Hermitian operators respectively, in a complex based Hilbert's space \cite{sakurai}.

\subsection{The relation between expectation value and normalization} \label{SectNorm}
Quantum Mechanics states that $|\alpha \rangle$ and $c |\alpha \rangle$ represent the same state. This axiom leads to the normalization as a requirement for physical states: 
\begin{equation}
\langle \alpha|\alpha \rangle =1. \label{QM_NormCond0}
\end{equation} 
This relation is the cornerstone for the probabilistic interpretation of quantum mechanic. However, the last axiom may have a different point of view that we hope it can be useful for obtaining equivalent relations on others vector spaces. Normalization condition can be seen as a relationship more than an axiom if it is written such as:
\begin{quote}
``The expectation value of the operator identity defined on a vector space \textbf{1} must be 1 for any physical vector of such space''. \label{expectNormRelExpres}
\end{quote}

In others words, for complex vector space and expectation value defined as $\langle A \rangle = \langle \alpha | A | \alpha \rangle$ we have
\begin{equation}
\langle \alpha|\textbf{1}|\alpha \rangle =1. \label{QM_NormCond}
\end{equation} 

For this statement take place, expectation value must be already defined. This fact is kind of contradictory since both concepts: normalization condition and expectation value are Quantum mechanic's axioms. However, it is our thought, both concepts are related, and they must be consistently defined. The definition of the expectation value of an operator has to take into account the form of how an operator must act over a state; in this case $\textbf{A}|\alpha \rangle = |\beta \rangle$; and what happens when the resulting state remain the same: $\textbf{A}|\alpha \rangle = c_{\alpha}|\alpha \rangle$, being $c_{\alpha} \in \mathbb{C}$. 

The logic to follow is:
\begin{enumerate}
\item the expectation value of an operator laying on an arbitrary state must be an operation involving the algebraic entities: kets, bras, operators and scalars and the result of such operation is a scalar.
\item the expectation value must be a function, named $\mathcal{F}$, which depend on the inner product of space vectors modified by the operator, since this is the only defined operation whose result is a scalar:
\begin{equation}
\langle \bm{A}\rangle = \mathcal{F}(V \times (\bm{A} V)) = \mathcal{F}((V \bm{A})\times  V) \to F,
\end{equation}
where we use the notation of eq. \ref{innerProdRepr} and $\bm{A}$ represents the operator.
\item the expectation value of operator identity is a function depending only inner products and must be equal to scalar unity $1$ for any physical state as: 
\begin{equation}
\langle \bm{I}\rangle = \mathcal{F}(V \times (\bm{I} V) ) = \mathcal{F}(V \times  V ) \to 1,
\end{equation}
which is a real nonnegative number.
\end{enumerate}

Resuming, we need to determine a compatible expectation value definition that will lead to a basic vector space operation depending on the inner product that permits us to write a relation like expression \ref{expectNormRelExpres}. Following these ideas, and taking into account the properties of the inner product in complex space, the simplest proposal for the function which describes the expectation value of the unitary operator is:
\begin{equation}
\langle \bm{I}\rangle = \mathcal{F}(V \times  V ) \equiv \langle \alpha | \alpha \rangle,
\end{equation}
whose result, according to complex number domain, is always a real and nonnegative number. 

The new restriction for space vector representing a physical state is:
\begin{equation}
\langle \alpha | \alpha \rangle=1
\end{equation}
and the expectation value for any operator $\bm{A}$ in the physical state $\alpha$ is be defined as:
\begin{equation}
\langle \bm{A} \rangle \equiv \langle \alpha |\bm{A}| \alpha \rangle
\end{equation}

In general, space vectors are constrained only to the axioms of complex numbers domain; however, once operators, as the analytic representation of measurable quantities, are included, then the previous restriction should also be added for normalizing the results.

\subsection{Translation operator}
The dynamic equations to describe the quantum state evolution are obtained by introducing the concept of translation or spatial displacement. The infinitesimal translation operator $\mathcal{T}(d\textbf{x})$ translates a well-localized state around $\textbf{x}$ into another well localized state around $\textbf{x} + d\textbf{x}$:
\begin{equation}
\mathcal{T}(d\textbf{x})|\textbf{x}\rangle = |\textbf{x} + d\textbf{x} \rangle,
\end{equation}
where a possible arbitrary phase factor is set to the unit by convention.

Translation operator must fulfill some properties. To preserve the norm of the translated ket, operator $\mathcal{T}(d\textbf{x})$ needs to be unitary. Also, two successive infinitesimal translation must be equal to a single translation operation by the sum of the two displacement vectors \textit{i.e}
\begin{equation}
\mathcal{T}(d\textbf{x}')\mathcal{T}(d\textbf{x}'')= \mathcal{T}(d\textbf{x}' + d\textbf{x}'')
\end{equation}
Also, translation in the opposite-direction are expected to be the same as the inverse of the original translation:
\begin{equation}
\mathcal{T}(-d\textbf{x}) = \mathcal{T}^{-1}(d\textbf{x})
\end{equation}
The fourth property demands that the translation operator reduces to identity operator when $d\textbf{x} \to 0$.
The infinitesimal translation operator has the form:
\begin{equation}
\mathcal{T}(d\textbf{x}) = 1 - \imagi\textbf{K}\cdot d\textbf{x}, \label{infOper}
\end{equation}
being the components of vector operator $\textbf{K}$ Hermitian operators. In fact, most of the infinitesimal operators that perform a variable displacement can be written as the sum of the identity operator plus the generator operator times the differential value of the displaced quantity.

The action of the position operator $\textbf{x}$ over a position ket is
\begin{equation}
\textbf{x}|\textbf{x}'\rangle = x'|\textbf{x}'\rangle,
\end{equation}
and it can be proved, ignoring second order terms in $d\textbf{x}$, that position and infinitesimal translation operators satisfy the relation
\begin{equation}
[\textbf{x},\mathcal{T}(d\textbf{x}')]=d\textbf{x}',
\end{equation}
or
\begin{equation}
[\textbf{x}_i,\textbf{K}_j]=i\delta_{ij}\textbf{1}. \label{xKcommutator}
\end{equation}
Now it only rests to find the form of the action of the physical operators like momentum over a state ket. That is where classical mechanic enters in the construction of the theory.

\subsection{Classical mechanics in Quantum mechanics}
So far we show quantum mechanics space properties can be established using the properties of complex vector space. Also, transformations of quantum states, like translation operator, was defined using common sense and also the algebraical properties the operators and the resulting state must have. Until this point, nothing but complex number algebra and logic was used. To complete the theory, physics need to be applied concepts like momentum, mass, velocity, and laws like momentum and energy conservation need also be incorporated. This is masterly done by borrowing the notion of infinitesimal canonical transformation from classical mechanics as generators of the same transformations in quantum mechanics. For example, linear momentum is the classical generator for the infinitesimal translation \citep{goldstein}
\begin{equation}
\textbf{x}_{new} \equiv \textbf{X}= \textbf{x} + d\textbf{x}, \;\;\;\;\;\; \textbf{p}_{new} \equiv \textbf{P}=\textbf{p}.
\end{equation}
The generating function $\textit{F}_2$ of the canonical transformation depending on generalized momentum and coordinates have the form:
\begin{equation}
\textit{F}(\textbf{x}, \textbf{P})  = \textbf{x} \cdot \textbf{P} + \textbf{p} \cdot d\textbf{x}, \label{GenFunc}
\end{equation}
where generating function
\begin{equation}
\textit{F}(\textbf{x}, \textbf{P})  = \textbf{x} \cdot \textbf{P}
\end{equation}
is proven to be the identical canonical transformation.

The expression of eq. \ref{GenFunc} has a remarkable similarity to infinitesimal translation operator of eq.   \ref{infOper}. The exact relation between the \textbf{K} and \textbf{p} can be settled from a dimensional analysis and the expression is:
\begin{equation}
\textbf{K}=\frac{\textbf{p}}{\text{universal constant with dimension of action}}.
\end{equation}
This constant is critical since it connects quantum mechanics expectation values with macroscopic quantities that were defined before quantum mechanics was developed. The universal constant is the well-known $\hbar$ constant. The infinitesimal translation operator is then written as:
\begin{equation}
\mathcal{T}(d\textbf{x}) = \mathbf{1} - \imagi\frac{\textbf{p}\cdot d\textbf{x}}{\hbar },
\end{equation}
being $\mathbf{1}$ the identity operator defined in the quantum space.

The finite translation operator is defined assuming that finite spatial displacement $\Delta x'\mathbf{\hat{x}}$ value is compounded of $N$-infinitesimal translations of $\Delta x'\mathbf{\hat{x}}/N$ letting $N \to \infty$. Using the successive infinitesimal translations property of infinitesimal translation operator, the finite operator has the form:
\begin{align}
\mathcal{T}(\Delta x'\mathbf{\hat{x}}) &= \lim_{N\to \infty} \Big( 1 - \frac{\imagi\textbf{p}\cdot \mathbf{\hat{x}}}{\hbar}   \frac{\Delta x'}{N} \Big)^N \nonumber \\
&= \exp \Big( - \frac{\imagi\textbf{p}\cdot \mathbf{\hat{x}} \Delta x'}{\hbar} \Big) \label{finiteOperator}.
\end{align}

The relation between  $\textbf{K}$ and $\textbf{p}$ can be included in relations \ref{xKcommutator} which coerce operators\textbf{x} and  \textbf{p} to satisfy:
\begin{equation}
[\mathbf{x} _i,\mathbf{p}_j]=\imagi\delta_{ij}\hbar,
\end{equation}
and set, according Heisenberg's uncertain relation, the limit:
\begin{equation}
\langle (\Delta \mathbf{x})^2 \rangle \langle \Delta \mathbf{p})^2 \rangle \geq \frac{\hbar^2}{4}.
\end{equation}

The commutators between momentum and position operators are
\begin{equation}
[\mathbf{x}_i,\mathbf{x}_j]=0, \;\;\;\;\; [\mathbf{p}_i,\mathbf{p}_j]=0 \text{ and } \;\;\;\;\; [\mathbf{x}_i,\mathbf{p}_j]=\mathrm{i}\delta_{ij}\hbar,
\end{equation}
and set one of the cornerstones of quantum mechanics.

Dirac observes that quantum and classical relations are related just by replaced classical Poisson brackets by commutators as:
\begin{equation}
[\;,\;]_{cassical} \to \frac{[\;,\;]}{\imagi\hbar} \label{DiracRule}
\end{equation}
where
\begin{equation}
[\;,\;]_{cassical} \equiv \sum_s \Big( \frac{\partial A}{\partial q_s} \frac{\partial B}{\partial p_s} - \frac{\partial A}{\partial p_s} \frac{\partial B}{\partial q_s}\Big)
\end{equation}

The reason is that quantum commutators and classical Poisson brackets satisfied the same algebraic properties \cite{sakurai}. There are significant differences between them, however, despite the classical or quantum nature of the operator, the following properties can be proved:
\begin{align}
[\mathbf{A},\mathbf{A}]&=0, \nonumber \\
[\mathbf{A},\mathbf{B}]&=[\mathbf{B},\bm{A}], \nonumber \\
[\mathbf{A},c]&=0 \; \; \text{being $c$ an scalar},  \nonumber \\
[\mathbf{A} + \mathbf{B},\mathbf{C}]&=[\mathbf{A},\mathbf{C}] + [\mathbf{B},\mathbf{C}], \nonumber \\
[\mathbf{A},\mathbf{B} \mathbf{C}]&=[\mathbf{A},\mathbf{B}]\mathbf{C} + \mathbf{B}[\mathbf{A},\mathbf{C}], \nonumber \\
[\mathbf{A},[\mathbf{B}, \mathbf{C}]]+[\mathbf{B}[\mathbf{C},\mathbf{A}]] + [\mathbf{C}[,\mathbf{B}]]&=0.
\end{align}
The last relation is known as Jacobi identity is a measures of how the order of the placement of parentheses in a multiple product affects the result of the operation.

Modern quantum mechanics formalism set the quantum dynamics equations first examining momentum operator representation in $x$-basis. The action of infinitesimal translation operator acting over an arbitrary ket can be studied expanding the ket into a $x$-basis vectors:

\begin{align}
\Big( 1 - \frac{\imagi \mathbf{p}\Delta \mathbf{x}'}{\hbar}\Big) | \alpha \rangle &= \int dx' \mathcal{T}(d\mathbf{x}')|x'\rangle \langle x'|\alpha \rangle \nonumber \\
&= \int dx' |x' + dx'\rangle \langle x'|\alpha \rangle \nonumber \\
&= \int dx' |x'\rangle \langle x' - dx'|\alpha \rangle \nonumber \\
&= \int dx' |x'\rangle \Big( \langle x' |\alpha \rangle  - dx'\frac{\partial}{\partial x'} \langle x' |\alpha \rangle \Big).
\end{align}

The comparison of both sides leads to:
\begin{equation}
\langle x'|\mathbf{p}|\alpha \rangle = -\imagi\hbar\frac{\partial}{\partial x'} \langle x'|\alpha \rangle.
\end{equation}
In general
\begin{equation}
\langle x'|\textbf{p}^n|\alpha \rangle = (-\imagi\hbar)^n \frac{\partial^n}{\partial x^{'n}} \langle x'|\alpha \rangle. \label{pOperatorXrepr}
\end{equation}

Similar results are obtained for the rotation operator acting over a quantum state described with the angular coordinate $\ket{\theta}$.

\subsection{Quantum dynamics}

The study of the evolution of the quantum system follows a similar procedure of modifying the quantum state by a transformation operator extracted from the classical mechanic. However, in this case, time is considered as a continuous parameter instead of a degree of freedom. There are several approaches for the time evolution of quantum systems.  Two of the most important is the Schr\"odinger picture where the operators are constant with time, and the vector states vary with the evolution of system; on the contrary of  Heisenberg picture, where the vector state remains fixed, and the operators change with time. Nevertheless, their different point of view, the probability amplitude, known as the transition amplitude, is the same for both approaches.

Time evolution operator $\mathcal{U}(t,t_0)$ describes the system evolution with time, changing a state that stays at some reference time $t_0$ to time instant $t$. Time evolution is studied in a similar way than translation, and that reinforces the accuracy of the formalism.

In Schr\"odinger picture time-dependent state is denoted by:
\begin{equation}
|\alpha ,t_0;t \rangle, \;\;\; (t>t_0),
\end{equation}
that change with time as:
\begin{equation}
|\alpha ,t_0;t \rangle = \mathcal{U}(t,t_0) |\alpha ,t_0 \rangle.
\end{equation}
It can be showed that infinitesimal time evolution operator have the form:
\begin{equation}
\mathcal{U}(t_0 + d t,t_0) = 1 - \imagi\frac{Hdt}{\hbar },
\end{equation}
where the generator of the time evolution is the Hamiltonian operator $H$  times a universal constant. Again, the similarity is extracted from systems time evolution in classical theory. 

The eigenvalues of the Hamiltonian operator
\begin{equation}
H|E' \rangle = E'|E' \rangle
\end{equation}
are known as the energy eigenvalues of the system. The constant character of energy eigenvalues also has a remarkable similarity to the constant character of Hamiltonian in the classical theory. The total time derivative of Lagrange functions in the classical theory is
\begin{equation}
\frac{dL}{dt} = \sum_j \frac{\partial L}{\partial q_j} \frac{d q_j}{d t} + 
\frac{\partial L}{\partial \dot{q}_j} \frac{d \dot{q}_j}{d t} + \frac{\partial L}{\partial t} 
\end{equation}
from where its defined the energy function:
\begin{equation}
h =  \sum_j \dot{q}_j \frac{\partial L}{\partial \dot{q}_j }  -L,
\end{equation}
that remains constant if the Lagrange function does not explicitly involve time. On the other side, Hamilton function, $H(q,p)$, corresponds to a different mathematical formalism and provide a more abstract understanding of the classical theory, and with it, a set of equations to solve a classical problem. In the classical theory, the Hamilton function is proven to be the generator for the time evolution of the system. If energy function $h$ and Hamiltonian $H(q,p)$ are the same function, then the energy of the system remains constant with time. However, it is essential, for future developments, to differentiate both concepts.

The Shr\"odinguer equation for time evolution is obtained using the composition property and have the expression
\begin{equation}
\imagi\hbar \frac{\partial}{\partial t} \mathcal{U}(t,t_0) = H \mathcal{U}(t,t_0),
\end{equation}
or
\begin{equation}
\imagi\hbar \frac{\partial}{\partial t} |\alpha,t_0;t \rangle = H |\alpha,t_0;t \rangle.
\end{equation}

In the Shr\"odinguer picture, nevertheless the state ket changes, eigenkets remain constant with time. By left multiplying bra $\langle \mathbf{x'}|$, we find that the \textbf{x}-basis expansion coefficients satisfy,
\begin{equation}
\imagi\hbar \frac{\partial}{\partial t} \langle \mathbf{x'}|\alpha,t_0;t \rangle = \langle \mathbf{x'} |H |\alpha,t_0;t \rangle. \label{Schrod1}
\end{equation}
If Hamiltonian operator $H$ is expressed in the \textbf{x} representation like
\begin{equation}
H=\frac{\mathbf{p}}{2m} + V(\mathbf{x})
\end{equation}
where momentum operator representation is given by equation \ref{pOperatorXrepr} and the potential energy $V(\mathbf{x})$ is a local hermitian operator $i.e$ $ \langle \mathbf{x}' |V(\mathbf{x}) | \mathbf{x}''  \rangle = V(\mathbf{x}') \delta ( \mathbf{x}' -  \mathbf{x}'' )$, then eq. \ref{Schrod1} have the form
\begin{equation}
\imagi\hbar \frac{\partial}{\partial t} \langle \mathbf{x'}|\alpha,t_0;t \rangle = -\frac{\hbar^2}{2m} {\nabla'}^2 \langle \mathbf{x'} |\alpha,t_0;t \rangle + V(\mathbf{x}') \langle \mathbf{x'}  |\alpha,t_0;t \rangle, \label{Schrod2}
\end{equation}
which is the well-known Schr\"odinger equation in the \textbf{x}-basis representation.

The Sch\"rodinger and Heisenberg approaches appear as the result of the associative property of vector states. Because of this property, inner product can be written as:
\begin{align}
\langle \beta |\mathbf{X}|\alpha \rangle &\to  (\langle \beta | \mathcal{U}^\dagger)(t,t_0) \mathbf{X} (\mathcal{U}(t,t_0) |\alpha \rangle) = \langle \beta,t_0;t | \mathbf{X} |\alpha,t_0;t \rangle  \label{assocSchro} \\
&\text{or}  \nonumber \\
&\to \langle \beta | (\mathcal{U}^\dagger(t,t_0) X \mathcal{U}(t,t_0)) |\alpha \rangle = \langle \beta | X(t_0;t) |\alpha \rangle. \label{assocHeis}
\end{align}
Time evolution on Heisenberg's picture now takes place only for operators. According to the above equation \ref{assocHeis},  operators change like
\begin{equation}
\mathbf{A}^{(H)}(t) = \mathcal{U}^\dagger (t) \mathbf{A}^{(S)}\mathcal{U}(t), \label{Heise1}
\end{equation}
where $H$ and $S$ superscripts stand for Heisenberg and Schr\"odinger respectively. Assuming that $A^{(S)}$ do not explicitly depend on time, the total time derivative is obtained by differentiating equation \ref{Heise1}:
\begin{align}
\frac{d \mathbf{A}^{(H)}}{dt} &= \frac{\partial \mathcal{U}^\dagger}{\partial t}  \mathbf{A}^{(S)}\mathcal{U} + \mathcal{U}^\dagger  \mathbf{A}^{(S)} \frac{\partial\mathcal{U} }{\partial t}
\nonumber \\
&= -\frac{1}{\imagi\hbar} \mathcal{U}^\dagger \mathbf{H} \mathcal{U} \; \mathcal{U}^\dagger \mathbf{A}^{(S)}\mathcal{U} + \frac{1}{\imagi\hbar} \mathcal{U}^\dagger \mathbf{A}^{(S)} \mathcal{U} \; \mathcal{U}^\dagger \mathbf{H} \mathcal{U}.
\end{align} 
Finally, the equation of motion in the Heisenberg picture is
\begin{equation}
\frac{d \mathbf{A}^{(H)}}{dt}=\frac{1}{\hbar} [\mathbf{A}^{(H)},\mathbf{H}].
\end{equation}
This equation is another example of Dirac's rule eq. \ref{DiracRule} and shows the classical time derivative of classical observable $A$:
\begin{equation}
\mathbf{A}|a'\rangle = a' |a'\rangle
\end{equation}
Finally, it is important to check how the eigenvalue equation 
\begin{equation}
\mathbf{A}|a'\rangle = a' |a'\rangle
\end{equation}
evolves with time under Heisenberg picture. Multiplying by $\mathcal{U}^\dagger (t)$ operator and using unitary condition, eigenvalue equation can transform as
\begin{equation}
\mathcal{U}^\dagger\mathbf{A} \;\mathcal{U} \;\mathcal{U}^\dagger|a'\rangle = a' \mathcal{U}^\dagger|a'\rangle.
\end{equation}
The eigenvalue equation under Heisenberg picture, where operators change as equation \ref{Heise1}, can be written as:
\begin{equation}
\mathbf{A}^{(H)} (\mathcal{U}^\dagger|a'\rangle) = a' (\mathcal{U}^\dagger|a'\rangle).
\end{equation}
As time goes on, the base ket change as
\begin{equation}
|a'\rangle_{H} \equiv \mathcal{U}^\dagger|a'\rangle_{S}.
\end{equation}

For both approaches, the transition amplitude and expansion coefficients are the same: in the  Shr\"odinguer picture we have
\begin{equation}
c_{a'}^{Sch}(t) = \langle a'|\cdot (\mathcal{U}\;| \alpha,t_0=0 \rangle )
\end{equation}
while in the Heisenberg picture, the coefficients are
\begin{equation}
c_{a'}^{Heis}(t) = (\langle a' | \mathcal{U}) \cdot| \alpha,t_0=0 \rangle.
\end{equation}

We have exposed the main topics that, from our point of view, should be taken into account in developing a quantum approach that includes the variable feature of the particle masses and field. Once the normalization relation is related to the mean value of the Identity operator, we think that the method shows no forced idea in the construction of the theory, following the ``quantum mechanical way of thinking''. We have to define then a right vector space which provides the analytical properties of the state vector, and the classical generator of the classical transformations which reveals the form of the operators when acting over such states.

\newpage
\section{Lagrange theory for $n$-VMVF systems.}\label{ClassicalLagSection}
This work proposes a new quantum theory for $n$-VMVF systems. It follows the modern quantum mechanic formalism where states are vectors and defined quantum generators from canonical transformations in classical theory \cite{sakurai}. In this section, we propose the classical Lagrange approach for $n$-VMVF systems with the objective of finding the Lagrange function for those kinds of systems.

The main difficulty constructing a theory that includes variable particle masses is that we do not know how masses or field vary; so how can we assure for an isolated system that relevant laws like linear or angular momentum conservation are satisfied? In other words, we need to know the solution of the equation to solve the equation. Is that possible? The answer is yes because we do have the solution for isolated systems which, according to D'Alembert Principle, is the sum of applied and inertial forces acting on every particle of the system. In this case, we can solve the problem by finding those forces and constrain the system, so the solution matches the referred forces.

\subsection{Equation of motions for single isolated particle with variable mass}
The construction of a classical theory that takes into account masses and field variation must start from the very beginning. Variable masses in time, mostly depending on velocity, has been studied on several problems of the classical mechanic using Newton's second law. It is well-known that the mass variation of a single particle violates the relativity principle under a Galilean transformation. The second Newton law state that the force $\mathbf{F}$ acting on a particle with mass $m$ and velocity $\mathbf{v} $ is equal to the time derivative of the momentum:
\begin{equation}
\mathbf{F} = \frac{d(m\mathbf{v})}{dt} = m\frac{d\mathbf{v}}{dt} + \mathbf{v} \frac{dm}{dt} \label{Newtown2Law}
\end{equation} 
A contradiction appears in the presence of null force $\mathbf{F}=\mathbf{0}$ if mass varies with time ($\frac{dm}{dt}\neq 0$). In effect, Newton's first law says particle will remain at rest in a system where it is originally at rest, but, according to equation \ref{Newtown2Law}, it is also accelerated by a force of value  $-\dot{m}\mathbf{v}$ in a system where particle moves with velocity $\mathbf{v}$ \cite{Plastino1992}. 

One way to solve this contradiction is to set constraints on the variation of mass with time like: to consider the mass lost as isotropic, so the last term of equation \ref{Newtown2Law} vanish \cite{Plastino1992}. However, in this work, we plan to study particle mass variation, and for that, we consider it as a degree of freedom, same as position. The space properties and conservation laws should give the only acceptable constraints equations for the system.

We can conclude then, based on the previous discussion that isolated particle whose mass varies with no restriction can no longer exist, and this is the first axiom of this approach. 

The relativity principle under Galilean transformation will be satisfied for the particle with variable mass if there is some external ``action'' to suppress the violation. On an isolated particle system, the external action must proceed from the particles the system is composed of. Then, it should exist a coordinated action of all particles of the system to preserve the equilibrium of the system. We assume that the response time of the system of those variations is zero or what is the same, we are proposing a system whose components interact via instantaneous interactions. We have no knowledge of any existing classical phenomena involving isolated $n$-particles with masses changing in a coordinated way so the system, as a single object and also that the variation of the mass of one particle instantaneously causes a change in the system. However, the reason to develop a classical mechanic in the construction of quantum theory for $n$-particle system with variable masses is for obtaining the canonical transformations and use it as a mold for the quantum transformations to find the form of the quantum observable operators. Also, a correct classical theory will be the behavior of systems described by the new quantum approach the limit of large quantum numbers as stated by the correspondence principle. 

\begin{figure}[h!]
\centering
\includegraphics[width=0.5\textwidth]{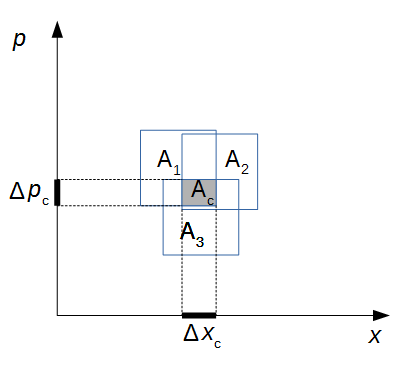}
\caption{Phase diagram for 3 particle at quantum scale.} \label{phaseDig}
\end{figure}

An important issue is the instantaneous interactions approach. We hypothesize that this behavior exists when length scale is in the order of system's de Broglie wavelength $e.i$ at the quantum scale. In quantum mechanics, because of the Heisenberg's Uncertainty Principle, if the distance between particles is small enough, there will exist a spatial volume where the probability to find all the particle at the same point is no null. The phase diagram of figure \ref{phaseDig} 
represents the phase volume of tree particles at the quantum scale. Each particle corresponding phase volume $A_1$, $A_2$ and $A_3$ is displayed as equal squares of volume  $\hbar$. At close enough distance there will exist a phase volume shared by all particles, represented by the gray area $A_c$. In that subspace, each point has a not null probability of finding all particle together at the same time. We suppose that the existence of this commonplace volume allows the particle system to behaves as a single object, so its components properties, such as every particle position, masses, became intrinsic properties of a unique object. Under this point of view, masses and field derivatives will instantaneously vary in a ``harmonic way'', so the system can satisfy conservation laws as one single physical object.

\subsection{D'Alembert's Principle}

From the classic theory, D'Alembert's principle states \textit{``The total virtual work of the impressed forces plus the inertial forces vanishes for reversible displacements''}. Indeed, the virtual work for particles in equilibrium condition is
\begin{equation}
\sum_n (\mathbf{F}^{(a)}_n - \dot{\mathbf{p}}_n + \mathbf{f}_n)\cdot\delta \mathbf{r}_n = 0,
\end{equation}
where $\mathbf{F}^{(a)}_n$ are the applied forces, $\dot{\mathbf{p}}_n$ are the constraint forces and the dynamical effects are included by a ``reversed effective force'' $- \dot{\mathbf{p}}_n$. The method of transforming the dynamical problem into a static phenomenon by the inclusion of the ``reversed effective force'' is known in the literature as the Bernoulli and D'Alembert's method.

The restriction of set the net virtual work of the constraint forces to zero,
\begin{equation}
\sum_n \mathbf{f}_n\cdot\delta \mathbf{r}_n = 0 \qquad \text{or}
\qquad
\sum_n (\mathbf{F}^{(a)}_n - \dot{\mathbf{p}}_n)\cdot\delta \mathbf{r}_n = 0, \label{DAlembertOrig}
\end{equation}
is known in the literature as the D'Alembert's principle. The existence of constraint forces implies that $\delta \mathbf{r}_n$ are not entirely independent but connected by constraint equations. This fact means coefficients $\delta \mathbf{r}_n $  can be no longer zero in equation \ref{DAlembertOrig} $e.i$ $\mathbf{F}^{(a)}_n - \dot{\mathbf{p}}_n \neq 0$.

The D'Alembert's principle as stated, however, can no longer be applied to an isolated particle with variable mass since it led to a contradiction in the application of Newton's second law. In effect, for particle number equal to one and in the absence of constraint forces $e.i$, for the free particle case, each component of $\delta \mathbf{r}_n \equiv \delta \mathbf{r}$ is considered independent. Then, either $\delta \mathbf{r}$ or $(\mathbf{F}^{(a)} - \dot{\mathbf{p}})$ must be zero, which means that particle is not allowed to move or the second Newton law must be satisfied. While $\delta \mathbf{r}=0$ provide no physical interest for us, last option was proven unsatisfactory for a free particle with a variable mass.

A particle system with variable masses, however, must satisfy the equilibrium condition as an isolate physical object, which means the net force on the particle must vanish as
\begin{equation}
\sum_n \mathbf{F}_n=0 \label{sysEquCond}.
\end{equation}
The equilibrium condition is related to physical laws such as the linear and the angular momentum conservation laws which reflect space properties like homogeneity and isotropy. The fact that second Newton's law can be no longer applied to each particle of the system, $\mathbf{F}_n\neq0$, means that there must exist a ``coordinated action'' between particles so that the system can remain in equilibrium. This actions must take place through an internal field, and it will depend on all the degrees of freedom of the system.

The force acting over every particle $\mathbf{F}_n$ can be divided into applied and constraint forces
\begin{align}
&\mathbf{F}_n(\ddot{\mathbf{r}}_1, \; \ddot{\mathbf{r}}_2 ...\ddot{\mathbf{r}}_n,\dot{\mathbf{r}}_1, \; \dot{\mathbf{r}}_2 ...\dot{\mathbf{r}}_n, \mathbf{r}_1, \; \mathbf{r}_2 ...\mathbf{r}_n) \equiv \mathbf{F}^{(a)}_n  - \dot{\mathbf{p}}_n + \mathbf{f}_n\neq 0.
\end{align}
In general, these forces depend on the position and velocities of all the particle of the system, so the virtual work of the force acting on particle $n$ by displacing any other particle of the system 
\begin{equation}
(\mathbf{F}_n^{(a)}  - \dot{\mathbf{p}}_n+ \mathbf{f}_n)\cdot \delta \mathbf{r}_{n'}  \qquad n\neq n'
\end{equation}
have no reason to be null. The virtual displacement of particle $n$ does a virtual work on the position $\mathbf{r}_{n}$ due to the $\delta \mathbf{r}_{n}$ term, but also on the other particles because the forces that act over the others particles also depend on the particle $n$ position and its derivatives with time. 

We can divide the constrained forces into two types: the constrained forces related to geometric restrictions and the constraints related to the coordinated action between the particles of the system for the particle system with variable mass satisfy the equilibrium condition. The first type of constraints groups those we are used to known, and they are usually connected to geometrical restrictions. For example, those that: keeps the distance between particles constant in a rigid body, or maintain the gas molecules moving inside a container, or force particles to move on a particular surface. The constraints of the first type can be removed from the physical system, and they set relations between the geometric coordinates of the system which originates the generalized coordinates. There are several classifications for this type of constraints. One of the most important, of such type of constraints are the holonomic constraints which relate the particle coordinates in the form
\begin{equation}
f(\mathbf{r}_1, \mathbf{r}_2,...\mathbf{r}_n, t)=0 \qquad \text{or} \qquad f(\{\mathbf{r}_n\}, t)=0.
\end{equation}
On the other side, the constraints related to the ``coordinate action'' between particles when masses vary taking into account are intrinsic of the system, and they can not be removed, otherwise the system won't satisfy the equilibrium condition. The constraint of the second type set relations to all the variables of the system, including masses and field derivatives. Such type of constraints relate all the variables of the system like
\begin{equation}
f(\{\mathbf{r}_n\}, \{\dot{\mathbf{r}}_n\}, \{m_n\}, \mathbf{A},t)=0.
\end{equation}
Note that is not possible to define independent generalized coordinates from intrinsic constraints such the particle's positions can be expressed in the form
\begin{equation}
\mathbf{r}_n = \mathbf{r}_n(q_1,q_2...q_{3N-k},t),
\end{equation}
where $k$ is the number of constraint equations.

We can separate then the constraints, $\mathbf{f}_n$, into the geometrical's $\mathbf{f}_n^{(g)}$ and the intrinsic constraints $\mathbf{f}_n^{(i)}$ like
\begin{equation}
\mathbf{f}_n = \mathbf{f}_n^{(g)} + \mathbf{f}_n^{(i)}.
\end{equation}

Starting from the equilibrium condition for the system, and decomposing all forces on the applied forces, the geometrical and the intrinsic constraint forces and substituting them on equilibrium condition for the $n$-VMVF systems, the eq. \ref{sysEquCond} becomes
\begin{equation}
\sum_n (\mathbf{F}_n^{(a)} + \mathbf{f}_n^{(i)} - \dot{\mathbf{p}}_n)\cdot\delta \mathbf{r}_{n'} + \sum_n \mathbf{f}_n^{(g)}\cdot \delta \mathbf{r}_{n'} = 0.
\end{equation}

We propose then a modification of the D'Alembert Principle for $n$-VMVF systems such as: 

\textit{``The total virtual work of the sum of the impressed, the intrinsic constraint and the inertial forces vanishes for the reversible displacements of any particle of the system''}. 

Note that we have included the terms ``intrinsic constraint'', ``system'' and ``any particle'' which covers all the above discussion for particle systems with variable masses. Also, if the masses of the particle are constants, all the constraints are geometrical and the second law of Newton is satisfied for every single particle. In that case, the modified principle reduces to the original D'Alembert principle.

The modified principle can be written as
\begin{equation}
\sum_n \mathbf{f}_n^{(g)}\cdot \delta \mathbf{r}_{n'} = 0 \label{DAlembertExt}.
\end{equation}
which led to the equation
\begin{equation}
\sum_n (\mathbf{F}_n^{(a)}  + \mathbf{f}_n^{(i)} - \dot{\mathbf{p}}_n )\cdot \delta \mathbf{r}_{n'} = 0.  \label{DAlembertExt1}
\end{equation}
The term constraint forces exclude those effects related to intern strengths that keep the equilibrium condition of the system and also that principle does not apply for irreversible displacements, such as sliding friction.

Equation \ref{DAlembertExt1} can be separated on contributions of constant and variable mass terms. Indeed, eq.  \ref{DAlembertExt1}
\begin{equation}
\sum_n \{\big(\mathbf{F}_n^{(a)}  + \mathbf{f}_n^{(i)}- \dot{\mathbf{p}}_n\big)_{[\dot{m}_n= 0]} 
+ \big(\mathbf{F}_n^{(a)}  + \mathbf{f}_n^{(i)} - \dot{\mathbf{p}}_n\big)_{[\dot{m}_n\neq 0]} \}\cdot \delta \mathbf{r}_{n'} = 0. \label{DAlembertExtSpl}
\end{equation}
The last term represents the dynamical, the intrinsic constraints and the applied force including mass variations while the first one includes only mass function dependency but not its derivatives. If we neglect mass changes, then there are no intrinsic constraints and the second Newton's law is satisfied:
\begin{equation}
(\mathbf{F}_n^{(a)} - \dot{\mathbf{p}}_n\big)_{[\dot{m}_n= 0]}=0,
\end{equation}
then the solution can be found by solving the set of equations:
\begin{equation}
\sum_n \big(\mathbf{F}_n^{(a)}  + \mathbf{f}_n^{(i)} - \dot{\mathbf{p}}_n\big)_{[\dot{m}_n\neq 0]} \cdot \delta \mathbf{r}_{n'} = 0.
\end{equation}

From Lagrange's point of view, equation \ref{DAlembertExt1} means that the application of Lagrange operator over the Lagrangian of the system, $L_{sys}$, using the degree of freedom $q$ of the particle $n$, will result in the equation of motion of the system in equilibrium.
\begin{align}
\mathcal{L}_{q_{n}} L_{sys} &\equiv \Big[ \frac{d}{dt}\Big(\frac{\partial}{\partial \dot{q_n}} \Big) - \frac{\partial}{\partial q_n}\Big]L_{sys} \nonumber \\ 
&= \sum_{n'=1}^N \dot{P}_{q_{n'}} - Q_{q_{n'}}= 0 \;\;\;\; \forall\; n=1,2...
\label{LagrangeForces}
\end{align}
where $\dot{P}_{q_{n'}}$, $ Q_{q_{n'}}$ are the generalized dynamical and applied force on particle $n$, respectively, and where the last term also includes the intrinsic constraint since such constraints are also a type an applied force. For example, the equation of motion for an isolated 2-particle system in equilibrium on the  $x$-direction is:
\begin{equation}
\Big[ \frac{d}{dt}\Big(\frac{\partial}{\partial \dot{x}_{1}} \Big) - \frac{\partial}{\partial x_{1}}\Big]L_{sys}= \Big[ \frac{d}{dt}\Big(\frac{\partial}{\partial \dot{x}_{2}} \Big) - \frac{\partial}{\partial x_{2}}\Big]L_{sys} = \dot{P}_{x_{1}} + \dot{P}_{x_{2}} - Q_{x_{1}}  - Q_{x_{2}}=0 \label{MotionEqTrans}
\end{equation}
The Lagrangian operator depends on both $x_1$ and $x_2$ acting over the function $L$ result in the same equation of motion, meaning equation \ref{MotionEqTrans} is $n$-degenerated. This degeneration feature is the base idea to obtain the final Lagrangian function.


We can divide the terms on those who contain the mass variation and those who do not, as shown equation \ref{DAlembertExtSpl}, and it suggests that the final Lagrangian can be split into two terms: one related to particle mass variations and other to the static masses:
\begin{equation}
L_{sys} = L_{[\dot{m}_n= 0]} + L_{[\dot{m}_n\neq 0]} = \sum_{n'=1}^N L_{sp_{n'}} + L_{[\dot{m}_n\neq 0]}. \label{splittedLagEq}
\end{equation}
The last term is the well known Lagrangian for interacting particles with constant masses system while $L_{[\dot{m}_n\neq 0]}$ vanishes for constant particle mass.

\subsection{Constructing Lagrangian for isolated n-VMVF systems. Initial assumptions.}

The present work is restricted to study isolated $n$-VMVF systems. The main difference of this classical approach of $n$-particle systems with the ordinary classical theory is that the particle masses and field are considered variables or degree of freedom of the system to be found in the final solution. 
The critical issue for this problem is that the Lagrange function for $n$-VMVF systems is unknown, mostly because there are no conservation laws related to these variables. The present section has the primary goal of constructing this function. Once the Lagrange function is fully determined, then externals field can be added in the standard way as an external field with a defined form and geometric constraint can be included as they commonly do.

Let us start by defining the generalized coordinates. For isolated $n$-VMVF systems there are no applied forces $\mathbf{F}_n^{(a)}$. Also, we can consider that there are no geometric constraint $\mathbf{f}_n^{(g)}$ or equations connecting coordinates. In that case, Cartesian coordinates $x,y,z$ describe the system as the generalized coordinates and masses and field will be considered as functions of all particle positions and velocities:
\begin{equation}
U(\mathbf{r}_1, \mathbf{r}_2, ..\mathbf{r}_n,\mathbf{\dot{r}}_1, \mathbf{\dot{r}}_2,...\mathbf{\dot{r}}_n),  \qquad
m_i(\mathbf{r}_1, \mathbf{r}_2, ..\mathbf{r}_n,\mathbf{\dot{r}}_1, \mathbf{\dot{r}}_2,...\mathbf{\dot{r}}_n).
\end{equation}
That means that the D'Alembert principle of equation \ref{DAlembertExt1} take the form
\begin{equation}
\sum_n (\mathbf{f}_n^{(i)}({\{\dot{\mathbf{r}}_l}\}, \{\mathbf{r}_l\}) - \dot{\mathbf{p}}_n ({\{\dot{\mathbf{r}}_l}\}, \{\mathbf{r}_l\}))\cdot \delta \mathbf{r}_{n'} = 0, \label{DAlembertExt2}
\end{equation}
where $\mathbf{r}_l \equiv \mathbf{r}_l(x,y,z)$. This equation means that the generalized forces $Q_{q_{n'}}$ in equation \ref{LagrangeForces} includes only the intrinsic constraint forces.
 
As the next step, we can make use of the classification for the Lagrangian terms relative to mass variation as shown in equation \ref{splittedLagEq} and associate the term $ L_{[\dot{m}_n= 0]}$, with no loss of generality, to the last term in \ref{DAlembertExt2}. Note that, until this point, we can assume $\mathbf{p} \equiv m \dot{\mathbf{r}}$ with no relation with the Hamiltonian momentum $p$ that we will extract once the Lagrangian is fully determined. 

According the classical mechanics theory \cite{goldstein}, the Lagrange function for isolated non-interacting particles considering constant mass, $e.i.$ $\mathbf{f}_n^{(i)}=0$, is the well-known expression of linear momentum $\dot{p} = m\dot{\mathbf{r}}$ and after few analytic operations:
\begin{equation}
\sum_i \dot{\mathbf{p}}_i \delta \mathbf{r}_i = \sum_{i,j} m_i \ddot{\mathbf{r}}_i \frac{\partial \mathbf{r}_i}{\partial q_j} \delta q_j
\end{equation}
where $q_j$ are the generalized coordinates. Using some relations between derivatives is obtained:
\begin{equation}
\sum_i m_i \ddot{\mathbf{r}}_i \frac{\partial \mathbf{r}_i}{\partial q_j} = \sum_i \Big\{ \frac{d}{dt} \Big[ \frac{\partial}{\partial \dot{q}_j}\Big(  \sum_i \frac{1}{2} m_i \dot{\mathbf{r}_i}^2 \Big) \Big]    - \frac{\partial}{\partial q_j} \Big (  \sum_i \frac{1}{2} m_i \dot{\mathbf{r}_i}^2 \Big)  \Big\},
\end{equation}
from where is defined the well-known Lagrange function for isolated noninteracting particle systems
\begin{equation}
L_0 = \sum_i \frac{1}{2} m_i \dot{\mathbf{r}_i}^2,
\qquad \text{such as} \qquad
 \frac{d}{dt} \Big[ \frac{\partial L_0}{\partial \dot{q}_j} - \frac{\partial L_0}{\partial q_j}\Big] = 0.
\end{equation}

The Lagrangian term in equation \ref{splittedLagEq} related to the mass variation incorporate then only the intrinsic constraints. It must be identified with a unique field for connecting particles and ``transport'' the information between particles as the coordinated action needed for the system satisfy the equilibrium condition. Let represent this action with the potential energy $U$. In this case, the Lagrangian will have the form:
\begin{equation}
L= \sum_n \frac{1}{2}m_n\dot{\textbf{r}}^2_n -U,
\end{equation}
where $U$ must tend to zero when the mass of the particle is constant. As mentioned before, the inclusion of particle mass variations comes with a big problem: both, Lagrangian, particularly the $U$ term, nor generalized momentum are unknown at this point. However, by making some considerations, it is possible to construct the Lagrangian for isolated $n$-VMVF systems. 

From the extension of D'Alembert principle, we know that the result of Lagrange equation is the sum of the applied, intrinsic and inertial forces over all particles. Lets name the sum of the applied and intrinsic forces as the applied net force for abbreviation. We can find those forces if we assume that
\begin{quote}\label{genForceApproach}
`The net applied and inertial forces acting on every particle of the system are the same forces acting on each particle measured by an external observer on an inertial frame. In that case, each particle of the system is considered as isolated and the external forces acting over it, as the action of the other particles through the field.''
\end{quote}

Even we show that particles with variable mass cannot be isolated without violating universal laws, the system can be described by an observer at rest in an inertial frame by computing every particle motion. From the observer point of view, there is no way to say whether the particle is or isn't part of an isolated system. Only the observation of the phenomena along the passing of time can tell if the set of particles behave as the components of such a system. At any time, the inertial observer can describe the motion of every particle assuming particle as isolated and influenced by the external action of a field. This field depends on several variables, in this case, the position and velocities of the other particles, which from an isolated one particle system, must be considered parameters. At any time then, the observer can compute the inertial and dynamical forces of every particle $\dot{P}_{n,i}$ and $Q_{n,i}$ respectively in any direction $i$. It is the zero summation of all measured forces during the evolution of the phenomenon, which indicates the presence of an isolated system of $n$ particles.

Under the previous statement, a particle system can be considered as a set of isolated particle subsystems where the forces acting on each one of them resume the action of the others particles. Each particle of the system feels the influence of the other particles through the action of the field. The Lagrangian $L_{{sp}_n} $ for a single particle under the action of external forces derivable from potential function $U$ is:
\begin{equation}
L_{{sp}_n} = \frac{1}{2}m_n\dot{\textbf{r}}^2_n -U.
\end{equation}

The net applied and inertial forces are obtained from the Lagrange equations
\begin{equation}
\dot{P}_{n,x} - Q_{n,x}  = \mathcal{L}_{n,x} L_{{sp}_n} 
 \equiv \Big[ \frac{d}{dt}\frac{\partial}{\partial \dot{x}_n} - \frac{\partial}{\partial x_n}\Big](\frac{1}{2}m_n\dot{\textbf{r}}^2_n -U ),
\end{equation}
where coordinates of particle $n$ are the variable while the coordinate of the others particle are considered parameters. Under these condition, the net forces are
\begin{align}
U(\mathbf{r}_1, \mathbf{r}_2, ..,\mathbf{\dot{r}}_1, \mathbf{\dot{r}}_2,...) &\equiv U(\mathbf{r}_n, \alpha_1, ..,\mathbf{\dot{r}}_n, \beta_2,...)   
\nonumber \\
m_n(\mathbf{r}_1, \mathbf{r}_2, ..,\mathbf{\dot{r}}_1, \mathbf{\dot{r}}_2,...) &\equiv m_n(\mathbf{r}_n, \alpha_1, ..,\mathbf{\dot{r}}_n, \beta_2,...).
\end{align} 
being $\alpha_i$, $\beta_i$ the before introduced parameters. 

Before we go further, we need to take a look at the form of the proposed Lagrangian. From the algebraic point of view, beyond the approach of particle system being considered as a set of isolated one particle systems, first the system must be solvable, which means the numbers of variables must be the same as the number of independent equations. To accomplish this objective and also to reduce the complexity of the solution, we need to make some assumptions for the mass and field functions(\textit{ansatz}):
\begin{description}\label{classicalAssump1}
\item Mass

We consider mass variation generated by internal changes in the structure of the particle, and it depends only on particle position, $e.i$
\begin{equation}
m_n \equiv m_n(\mathbf{r}_n)
\end{equation}
which means that mass derivatives satisfy
\begin{align}
&\frac{\partial^{(l)} m_n}{\partial x_{n}^{(l)}},\; \frac{\partial^{(l)} m_n}{\partial y_{n}^{(l)}}, \; \frac{\partial^{(l)} m_n}{\partial z_{n}^{(l)}}\neq 0, \qquad 
\frac{\partial^{(l)} m_n}{\partial x_{n'}^{(l)}},\; \frac{\partial^{(l)} m_n}{\partial y_{n'}^{(l)}}, \; \frac{\partial^{(l)} m_n}{\partial z_{n'}^{(l)}} = 0 \qquad \forall \; n \neq n'  
\nonumber \\
&\frac{\partial^{(l)} m_n}{\partial \dot{x}^{(l)}_{n'}},\; \frac{\partial^{(l)} m_n}{\partial \dot{y}^{(l)}_{n'}}, \; \frac{\partial^{(l)} m_n}{\partial \dot{z}^{(l)}_{n'}} = 0 \qquad \forall \; n , n' \qquad l=1,2...
\end{align}

Davidson reference \citep{Davidson2014} also presumes mass densities depend only on particle position

The mass variation we proposed, should be related to intrinsic properties. It differs from the change related to the inertial coordinate system that appeared at relativistic energies depending on particle velocity. The present classical approach with particle mass and field variation is developed, first, at non-relativistic speeds. Later, we expect to include relativistic concepts writing a suitable Lagrangian in its covariant form.

\item Field

In physics, the fundamental fields describe what are known as fundamental interactions. They are the interactions that do not appear to be reducible to more basic interactions in any processes in nature. Most of the fundamental theories for forces are developed using this characterization.  There are four fundamental interactions known to exist: the gravitational, electromagnetic, the strong and the weak interactions. All four are believed to be related, and to unite into a single force all energies and at any length scales, which is currently one of the major unsolved problems in physics. Over the past few centuries, all modern physics rests at two main theoretical frameworks: the general relativity and the quantum field theory. The main difficulty of the unification problem lies on include quantum mechanical effects into gravity.

One of the main consequences of treating field and mass as variables of the system to be found is precisely the no classification of the field. The final solution of that variable will include then, in one single function, all the mentioned fundamental interactions and their coupling, if they exist. 

Another consequence of the unique field approach is that the nature of the interaction, and with it, the physical properties of the matter that are the cause of the existence of those interactions like electric charge, spin, baryon charge among others, are all within a single field. It should not be an obstacle since this properties contribution are included in the interaction as constant value; for example, the potential function of the electromagnetic field has the form $U=q_n\mathbf{A}\cdot \mathbf{\dot{r}}$, being $q_n$ the electric charge of the particle.

Lagrange and Hamilton's classical theory include fields in the form of potential energy. In our case, the field is the medium to connect particle, and nothing is said about its nature which we assume undefined. However, the undoubted success and abstract concepts introduced in the electromagnetic theory led us to include some of its basics ideas. 
We can presume the existence of two fields generated from all the particles of the system: one scalar $\phi$ and other vectorial $\mathbf{A}$. We propose the form of the potential energy for the field connecting particles in the system as
\begin{equation}
U_n = q\mathbf{A}\cdot \dot{\mathbf{r}}_n -q\phi,  \label{PotentialEnergyForm}
\end{equation}
where fields $\mathbf{A}$ and $\phi$  are considered to be time independent. In this case, the proposition for the Lagrange function for particle $n$ under the action of an external field, $L_{{sp}_n}$, have the form:
\begin{equation}
L_{{sp}_n} = \frac{1}{2}m_n\dot{\textbf{r}}^2_n -\phi + \mathbf{A}\cdot \dot{\mathbf{r}}_n \label{singlePartLagran}
\end{equation}

From electromagnetic theory, we borrow the ideas of set scalar field $\phi$ depend only on fixed source contributions $e.i$
\begin{equation}
\phi \equiv \phi(\mathbf{r}_1, \mathbf{r}_2, ..)
\end{equation}

We also presume the Gauge invariance of these fields $e.i$, and they can be chosen to satisfy, the Lorentz condition:
\begin{equation}
\nabla \mathbf{A} + \frac{1}{c^2} \frac{\partial \phi}{\partial t} =0
\end{equation}

The preference for the form of potential energy like \ref{PotentialEnergyForm} is based on the following facts:
\begin{itemize}
\item It has a well tested covariant form for $U=A^\nu \dot{x}_\nu$ and $\partial_\nu A^\nu=0$ \cite{jackson.ElectroDynamics}.
\item Its Lagrange invariant equations combined with the Maxwell equation provide a complete description of classical dynamics of interacting particles and electromagnetic field.
\item The Lagrange equation of motions obtained using the electromagnetic potential energy, using $x$-coordinate as example
\begin{equation}
m \ddot{x} = \dot{x}\frac{\partial A_x}{\partial x} + \dot{y}\frac{\partial A_y}{\partial x} + \dot{z}\frac{\partial A_z}{\partial x} - \frac{\partial \phi}{\partial x } - \frac{d A_x}{dt}
\end{equation}
permit to define two vector fields $\mathbf{E}$ and $\mathbf{B}$, still with undefined nature, transforming motion equation like
\begin{equation}
m \ddot{x} = E_x + (\dot{\mathbf{x}} \times \mathbf{B})_x
\end{equation}
where 
\begin{equation}
\mathbf{E} = -\mathbf{\nabla} \phi - \frac{\partial \mathbf{A}}{\partial t}
\end{equation}
and 
\begin{equation}
\mathbf{B} = \mathbf{\nabla}\times A.
\end{equation}
Note that while the last equations resemble the homogeneous Maxwell's equations. However,  different from the Classic Electrodynamics theory, the dynamic behavior of fields $\mathbf{A}$ and $\phi$ is given by the Lagrange equations of motions and not by the set of inhomogeneous Maxwell's equations. This last set of equations includes what the electromagnetic theory acknowledge as the source of the electromagnetic field: charge density and current which are related to the dynamical properties of particles position and velocities respectively. In our case, the sources of the proposed field are the dynamic properties of the system, which means that they include the particle positions, velocities and masses, and field derivatives.
\item Being invariant under Gauge transformation make that vector fields $\mathbf{E}$ and $\mathbf{B}$ remain unchanged if fields $\mathbf{A}$ and $\phi$ transforms as  
\begin{equation}
\mathbf{A}' \to  \mathbf{A} + \mathbf{\nabla}\Lambda \qquad \text{and} \qquad
\phi' \to \phi = \frac{\partial \Lambda }{\partial t}.
\end{equation}
\end{itemize}
\end{description}

Once the Lagrangian for isolated particle systems using Cartesian coordinates is established, the analysis can be extended to others generalized coordinates and the action of external forces. 
\subsection{Single particle motion equation in Cartesian coordinates} \label{SinglePartEqMotionCart}

The extension of D'Alembert principle for $n$-VMVF systems states that the Lagrange equation of motion is the sum of the generalized inertial and net applied forces acting on every particle. This fact is the key in the construction of the final Lagrangian since we suppose that they are the same forces acting on each particle assuming every particle is isolated and affected by the others through the external action, as mentioned on section \ref{genForceApproach}. We are in the position now, after established previous assumptions for mass and field functions, to define the expression for the force acting on every particle of the isolated system in Cartesian coordinates.

For example, the inertial and net applied force in Cartesian coordinate on the $x$ direction for $n$ particle under the action of external field is found from the Lagrange equation
\begin{equation}
 \dot{P}_{n,x} - Q_{n,x} = \mathcal{L}_{n,x} L_{{sp}_n} \label{genP1},
\end{equation}
where the external influence of the particles over the particle $n$ is on fields  $\mathbf{A}$ and $\phi$. Using Lagrangian \ref{singlePartLagran} the inertial and net applied force in the $x$-direction have the form:
\begin{align}
 \dot{P}_{n,x} - Q_{n,x} &= \mathcal{L}_{n,x} L_{{sp}_n} \equiv \Big[ \frac{d}{dt}\frac{\partial}{\partial \dot{x}_n} - \frac{\partial}{\partial x_n}\Big](\frac{1}{2}m_n\dot{\textbf{r}}^2_n -\phi + \mathbf{A}\cdot \dot{\mathbf{r}}_n ) 
\nonumber \\
& =\frac{d}{dt}\Big(m_n \dot{x}_n + \dot{\mathbf{r}}_n\frac{\partial \mathbf{A}}{\partial \dot{x}_n} + A_x \Big) - \frac{1}{2}\frac{\partial m_n}{\partial x_n}\dot{\textbf{r}}^2_n +  \frac{\partial \phi}{\partial x_n}  -  \dot{\mathbf{r}}_n\frac{\partial \mathbf{A}}{\partial x_n}  
\nonumber \\ 
& =m_n\ddot{x}_n + (\vec{\nabla}_n m_n\cdot \dot{\mathbf{r}}_n) \dot{x}_n + \ddot{\mathbf{r}}_n\frac{\partial \mathbf{A}}{\partial \dot{x}} + \dot{\mathbf{r}}_n \frac{d}{dt}\Big(\frac{\partial \mathbf{A}}{\partial \dot{x}_n}\Big)
\nonumber \\
&\;\;\;+ (\vec{\nabla}_{r_n}  A_x) \dot{\textbf{r}}_{n} + (\vec{\nabla}_{\dot{r}_n}  A_x) \ddot{\textbf{r}}_{n} - \frac{1}{2}\frac{\partial m_n}{\partial x}\dot{\textbf{r}}^2_n +  \frac{\partial \phi}{\partial x_n}  -  \dot{\mathbf{r}}_n\frac{\partial \mathbf{A}}{\partial x_n} \label{genP} 
\\ 
\end{align}

The obtained inertial and net applied forces show its dependency with mass and potential derivatives. We treat these derivatives functions as a new set of variables of the system.
 
\subsection{The theory of Special Relativity in $n$-VMVF systems.}

The number of degrees of freedom of the system has increased. We must include then a new set of equations, so the equation system remains solvable. So far the set of equation of motion in eq. \ref{genP1} was settled using  Cartesian coordinates $r(x,y,z)$. This set of coordinates constitute a plane space, and they describe the translation of the system in a three-dimensional space. The Lagrange equations using this coordinates reproduce the conservation of linear momentum at all directions,  in correspondence with the homogeneity of the space. We should look for another set of coordinates which are related to other conservation laws and other properties of the space. Others coordinates system like cylindrical, spherical, elliptical, etc., won't increase the number of the independent equation because they are related to Cartesian coordinates through the transformation relations. The set of independent Lagrange equations must come from a set of coordinates also independent from Cartesian's.

We propose the set of rotation angles  $r(\theta,\phi,\chi)$, used in the description of the rotation of physical systems as the new independent set of coordinates needed for increasing the number of equations of the system. The Lagrange equations depending on the rotation angles led to the conservation of angular momentum law in the absence of external torques which is related to the isotropy of space. The set of angular coordinates forms a curved space and is independent of the Cartesian set of coordinates. In fact, it is not possible transforming the position vector expressed in Cartesian coordinates $r(x,y,z)$ into a vector expressed by the set of angles $r(\theta,\phi,\chi)$. There is a mathematical obstruction in that transformation. Also, it is well-known that the most general motion of a particle can is divided into one rotation plus one translation.

The situation does not get better if the angles are also added as three more variables to the problem. This issue can be overcome since the particle position can be represented in both of set of coordinate. We need to find the relation between both sets, however as already mentioned there is no coordinate transformation between them. Nevertheless, the relation can be set with an extra parameter. Indeed, a vector position can be expressed as functions of the angles of three independent rotations by describing the Cartesian space with an extra dimension $r(x,y,z) \to r(x,y,z,w)$ and imposing the condition:
\begin{equation}
x^2+y^2+z^2+w^2=R^2. \label{relatRelation}
\end{equation}
where $R=cte$ is the new parameter. The constant value for $R$ makes both sets of coordinates $\{x,y,z\}$ and $\{\theta,\phi,\chi\}$ independent. 

The equation \ref{relatRelation} is the well known Lorentz condition. Unexpectedly, we start developing the theory under Newtonian approach planning to extend the final result to its invariant form as every ``good'' theory should be. However, the need of describing the system using the 3-D angular coordinates to increase the number of independent equations, force the inclusion of the relativistic approach. 

Under this point of view, the Lorentz condition in the space-time defines the 4-D surface which sets the 3-D space for the angular variables for all particles. We can say that the 3-D space of the angular coordinates is the stereographic projection of the 4-D sphere given by the Lorentz condition in the space-time.

It well knows the relation of the 4-component with time. By this approximation, we should treat the fourth coordinate as a dependent variable to make space Euclidean, but based on all the knowledge acquired from physics community; we assume the modern approach of treating all coordinates as part of a non-Euclidean space. So, from now on, we develop the theory under Lorentz approach and apply the well know Minkowski algebra. We expand the theory increasing the Cartesian position vector $r(x,y,z)$ to a 4-vector, represented in its contravariant form by $r(x^0,x^1,x^2,x^3)$, whose inner product with its covariant form (\textit{1-form}) remains unchanged in all inertial frames of reference.

One of the most studied sets of angular coordinates is the Euler angles set, which are defined to describe the transformation from given Cartesian coordinates system to another by a set of three consecutive rotations. The position as a function of Euler angles includes parameter $R$ as the length of vector position. On the other side, a Lagrangian depending on Euler angles have a non-quadratic form, so they are not the best candidates in our study.  

We propose the use of the spherical coordinates for four-dimensional Euclidean space. The transformation equations between the two set of variables are:
\begin{align}
x\equiv x^1&=R \sin \theta \sin \phi \sin \chi, \nonumber \\
y\equiv x^2&=R \sin \theta \sin \phi \cos \chi, \nonumber \\
z\equiv x^3&=R \sin \theta \cos \phi, \nonumber \\
w\equiv x^0&=R \cos \theta. \label{PosAngleRelation}
\end{align}

The theory must now be extended to its relativistic formulation. We consider some aspects in this formulation:
\begin{itemize}
\item 
In the present theory, the four components of vector position $x^\nu$ are not independent due to  Lorentz constraint. However, we follow Dirac point of view, from where such constraint is treated as how he described: ``a weak condition'' \cite{goldstein321}, what means that the constraint should be imposed after all derivation processes have been carrying through. 
The scalar and vector fields,$\phi$ and $\mathbf{A}$ are now replaced by the four-dimensional field $A^\nu$. The proposed relativistic Lagrangian, $L_n$, for a single particle under the action of an external field, now has the form:
\begin{equation}
L_n=\frac{1}{2}m_n \dot{r}^\nu_n \dot{r}_{n,\nu} -A^\nu \dot{r}_{n,\nu} \label{RelLagrangian}
\end{equation}
\item All the angular and Lorentzian coordinates of the particles of the $n$-VMVF systems are related by equation like equation \ref{PosAngleRelation} like
\begin{align}
x^1_n&=R \sin \theta_n \sin \phi_n \sin \chi_n, \nonumber \\
x^2_n&=R \sin \theta_n \sin \phi_n \cos \chi_n, \nonumber \\
x^3_n&=R \sin \theta_n \cos \phi_n, \nonumber \\
x^0_n&=R \cos \theta_n. \label{PosAngleRelationN}
\end{align}
As the relation is a particular transformation between coordinates of each particle in the same space, the parameter $R$ should be considered not related to each particle itself but the entire system. The definition of this value must be settle on future works.

\item
The choice of writing the Lagrangian on its ``invariant'' form like equation \ref{RelLagrangian}, explicitly assume the first postulate of the Theory of Relativity known as the principle of relativity which states that 
\begin{quote}
``The laws of physics are the same for all observers in uniform motion relative to one another''. 
\end{quote}
However, the second postulate, which states that 
\begin{quote}
``The speed of light in a vacuum is the same for all observers, regardless of their relative motion or of the motion of the light source''
\end{quote}
cannot be applied in this approach, not without set restrictions to the motion of the particle system. Indeed, because of the chosen transformation from equation \ref{PosAngleRelationN}, the square value of the velocity of a particle is
\begin{equation}
\dot{r}^\nu_n \dot{r}_{n,\nu} = R^2 (\dot{\theta}^2_n  + \dot{\phi}^2_n \sin^2 \theta_n  +  \dot{\chi}^2_n \sin^2 \theta_n \sin^2 \phi_n),
\end{equation}
while from the second postulate of the relativistic theory, we have
\begin{equation}
\dot{r}^\nu_n \dot{r}_{n,\nu} = c^2,
\end{equation}
being $c$ the speed of the light. This relation implies that the angular velocity of the particle are restricted to the equation:
\begin{equation}
\dot{\theta}^2_n  + \dot{\phi}^2_n \sin^2 \theta_n  +  \dot{\chi}^2_n \sin^2 \theta_n \sin^2 \phi_n = \frac{c^2}{R^2}.
\end{equation}

Also, the kinetic energy of the particle system, neglecting all mass and field variations using Lorentzian and angular coordinates can be written as
\begin{equation}
E = \sum_{n = 1}^N \frac{1}{2}m_n \dot{r}^\nu_n \dot{r}_{n,\nu} = \sum_{n = 1}^N  \frac{1}{2}m_n R^2 (\dot{\theta}^2_n  + \dot{\phi}^2_n \sin^2 \theta_n  +  \dot{\chi}^2_n \sin^2 \theta_n \sin^2 \phi_n).
\end{equation}
In particular, the rotation energy will have a constant value equal $E_R =  \sum_{n = 1}^N \frac{1}{2}m_n c^2$. It is a well-known fact that the second postulate of the Theory of Relativity is based on the invariance of Maxwell's equations. However, as one of the main assumptions of this work is precisely the unknown form of the field, and because of that, the second postulate of the theory of relativity should be analyzed and also it relation with the parameter $R$. Also, because the above exposed, the first postulate should not be applied to laws of physics which include fields with a defined form.

\item According to the exposed ideas, we must discuss the relativity principle under this approach. As shown, the expansion of our 3-D to a 4-D space for relating the rectangular coordinates from vector position with its angular coordinates given by equation \ref{PosAngleRelationN}
\begin{align}
x^1_n&=R \sin \theta_n \sin \phi_n \sin \chi_n, \nonumber \\
x^2_n&=R \sin \theta_n \sin \phi_n \cos \chi_n, \nonumber \\
x^3_n&=R \sin \theta_n \cos \phi_n, \nonumber \\
x^0_n&=R \cos \theta_n,
\end{align}
results in the inclusion of the Lorentz condition of equation \ref{relatRelation}
\begin{equation*}
x_0^2+x_1^2+x_2^2+x_3^2=R^2
\end{equation*}
to the rectangular coordinates. It is well known the relation of the Lorentz condition with the relativity principle. We adopt in here such principle; however with some issues according to the assumptions of this work. From the accepted universal knowledge about this topic, we know the physical meaning of the new coordinate and its relation with time in the form $x_4 = \imagi c t$. The combination of the three-dimensional Euclidean space and time satisfying the Lorentz condition \ref{relatRelation} into a four-dimensional manifold is known as the  Minkowski space.

According to the second Einstein's postulate, $c$ is the speed of the light, and it is constant at all inertial frames. As explained in the last item, we should not make such an assumption because of the undefined form of the field feature of this work. However, $c$ should be a constant with the dimension of velocity, independent of the coordinates at that point. The conservation of the relation \ref{relatRelation} imply that the coordinates of two point of the space-time satisfy
\begin{equation}
\sum_{i=0}^3 x_i^2 = \sum_{k=0}^3 x_k^2,
\end{equation}
moreover, that they are connected by a linear coordinate transformations known as the Lorentz transformation which in general connect two coordinate frames that move at constant velocity relative to each other. The general form of the Lorentz transformation for a general boost is
\begin{equation}
\begin{bmatrix}
c\,t' \\ x' \\ y' \\ z'
\end{bmatrix}
=
\begin{bmatrix}
\gamma&-\beta_x\,\gamma&-\beta_y\,\gamma&-\beta_z\,\gamma\\
-\beta_x\,\gamma&1+(\gamma-1)\frac{\beta_{x}^{2}}{\beta^{2}}&(\gamma-1)\frac{\beta_{x}\beta_{y}}{\beta^{2}}&(\gamma-1)\frac{\beta_{x}\beta_{z}}{\beta^{2}}\\
-\beta_y\,\gamma&(\gamma-1)\frac{\beta_{y}\beta_{x}}{\beta^{2}}&1+(\gamma-1)\frac{\beta_{y}^{2}}{\beta^{2}}&(\gamma-1)\frac{\beta_{y}\beta_{z}}{\beta^{2}}\\
-\beta_z\,\gamma&(\gamma-1)\frac{\beta_{z}\beta_{x}}{\beta^{2}}&(\gamma-1)\frac{\beta_{z}\beta_{y}}{\beta^{2}}&1+(\gamma-1)\frac{\beta_{z}^{2}}{\beta^{2}}\\
\end{bmatrix}
\begin{bmatrix}
c\,t \\ x \\ y \\ z
\end{bmatrix}\ 
\end{equation}
where 
$\beta = \frac{v}{c}=\frac{\|\vec{v}\|}{c}$ and $\gamma = \frac{1}{\sqrt{1-\beta^2}}$.
\item One of the invariants against Lorentz transformation is the infinitesimal square of the distance in the Minkowski space,
\begin{equation}
ds^2 = ds^{'2} = dx_0^2+dx_1^2+dx_2^2+dx_3^2.
\end{equation}
For establishing clear relations for the differentiation of any vector in the spacetime, was defined as the invariant quantity known as the proper time of the system. By the inclusion of this concept to our problem, the covariant Lagrange equations evolve now, not on time, but on the proper time or $\tau$. The covariant Lagrange operator depending on the Lorentzian coordinates is:
\begin{equation}
\mathcal{L}_{n,\nu}  \equiv \Big[ \frac{d}{d\tau}\frac{\partial}{\partial \dot{x}^\nu_n} - \frac{\partial}{\partial x_n^\nu}\Big], \;\; x^\nu=\{x^0,x^1,x^2,x^3\} \label{lagrangetranslation}
\end{equation}
while the same operator depending on the angular coordinates is
\begin{equation}
\mathcal{L}_{n,\xi}  \equiv \Big[ \frac{d}{d\tau}\frac{\partial}{\partial \dot{\xi}_n} - \frac{\partial}{\partial \xi_n}\Big], \;\; \xi=\{ \theta,\phi,\chi\}. \label{lagrangeRotation}
\end{equation}
\item The relativistic formulation of the present approach, along with the extra dimension, it also includes new variables to the description of the problem related with this new coordinate. They are 
\begin{equation}
\frac{\partial m_n}{\partial x_{n}^0}, \;\;\; \frac{\partial m_n}{\partial \dot{x}_{n}^0}, \;\;\; \frac{\partial A^0}{\partial x^\nu_{n}} , \;\;\; \frac{\partial A^0}{\partial \dot{x}^\nu_{n}}, \;\;\; \frac{\partial A^i}{\partial x^0_n}, \;\;\; \frac{\partial A^i}{\partial \dot{x}^0_{n}}.\label{relConstraint}
\end{equation}
The new variables which appear because the inclusion of the relativistic framework, however, they must not increase internal degrees of freedom of the system. These quantities must be related trough constraints in somehow with the already existing variables in the non-relativistic picture. In that case, to keep variables number equal to the number of equations, some approximations need to be made. Our approximations star from the initial assumptions for particle masses and field functions on section \ref{classicalAssump1}. These conjectures include particle masses depending only on its particle position, the scalar field depending only on fixed source contributions also and the Gauge invariant condition to connect the scalar and the vectorial field. Now, in the relativistic frame, we use the simplified approach of considering the scalar field as the four component of a four-component field set the gauge condition as the connection to relate its components. The approaches of section  \ref{classicalAssump1} in the relativistic frame, have the form:
\begin{equation}
\frac{\partial m_n}{x^\nu_{n'}} \equiv \frac{\partial m_n}{x^\nu_{n'}}\delta_{nn'}, \qquad  \frac{\partial m_n}{\dot{x}^\nu_{n'}} =0 \qquad  \frac{\partial A^0}{\partial \dot{x}^\nu_{n}}=0 , \qquad \partial_\nu A^\nu=0\;\; \forall\; n. \label{relConstraint1}
\end{equation}
\item Mass variations can be classified into two types: the ``structural'' variations and the inertial's. The first type includes the ones introduced in the non-relativistic approach, and that entitle this work. Their first derivatives are $\frac{\partial m_n}{\partial x^i_n}$ with $\{x^1_n,x^2_n,x^3_n\} \equiv \{x,y,z\}$. The inertial variation, $\frac{\partial m_n}{\partial x^0_n}$,  is related to the inertial frames of references and introduced within the Lorentz coordinate system.

The constraint equation for $\frac{\partial m_n}{\partial x^0_n}$ can be obtained from the relativistic motion equation for an isolated particle. The mass variation for an isolated particle in the Minkowski space, different from the Euclidean space, satisfies the relativistic principle under a Lorentzian transformation. It is the fourth mass derivative component, the quantity that allows the conservation of relativistic momentum. If structural mass variations are neglected, $e.i$ mass variations will depend only on $x_{n}^0$ coordinate. Using the relativistic Lagrangian of equation \ref{RelLagrangian} is applied in the isolated free particle case, the equation of motion has the form:
\begin{align}
&\mathcal{L}_{\mu} L \equiv \Big[ \frac{d}{d\tau}\frac{\partial}{\partial \dot{x}^\mu} - \frac{\partial}{\partial x^\mu}\Big] \Big( \frac{1}{2}m\dot{x}^\nu\dot{x}_\nu \Big)=0 
\nonumber \\
&\frac{d}{d\tau} (m\dot{x}_\mu)- \frac{\dot{x}^\nu\dot{x}_\nu}{2}\frac{\partial m}{\partial x^\mu}\delta_\mu^0 = \dot{m}\dot{x}_\mu + m\ddot{x}_\mu - \frac{\dot{x}^\nu\dot{x}_\nu}{2} \frac{\partial m}{\partial x^\mu}\delta_\mu^0=0. \nonumber
\end{align}
Multiplied both sides and sum by $\dot{x}^\mu$, we have:
\begin{align}
&\dot{m}\dot{x}_\mu \dot{x}^\mu + m\ddot{x}_\mu\dot{x}^\mu- \frac{\dot{x}^\nu\dot{x}_\nu}{2}\frac{\partial m}{\partial x^\mu}\delta_\mu^0 \dot{x}^\mu =0 
\nonumber \\
&\dot{m}c^2+m\ddot{x}_\mu\dot{x}^\mu - \frac{\dot{x}_\mu \dot{x}^\mu}{2} \frac{\partial m}{\partial x^0}  \dot{x}^0=0,
\end{align}
where we use the Einstein summation notation. We recall in here that in our approach $\dot{x}_\mu \dot{x}^\mu \neq c^2$. As we stated before, the only existed mass variation is related to inertial frames of reference, $i.e$ it depends only on the $x^0$ component, so the mass variation with time is:
\begin{equation}
\dot{m} = \frac{\partial m}{\partial x^0}\dot{x}^0.
\end{equation}
Constraint equation for the inertial variation of mass property have the final form:
\begin{equation}
\frac{1}{2}\frac{\partial m}{\partial x^0}\dot{x}^0 \dot{x}_\mu \dot{x}^\mu + m\ddot{x}_\mu\dot{x}^\mu=0. \label{mass0compConstraint}
\end{equation}
\end{itemize}

It is useful to obtain some relations for the Lorentzian and angular's coordinates. 

From Goldstein \citep{goldstein}, we get the equations:
\begin{equation}
 \frac{d}{dt}\Big(\frac{\partial x_l}{\partial q_i}\Big)= \frac{\partial \dot{x}_l}{\partial q_i} \;\; \text{ and } \;\; \frac{\partial \dot{x}_l}{\partial \dot{q}_i} = \frac{\partial x_l}{\partial q_i}, \label{relations0}
\end{equation}
However, using the compound derivative law, we obtain can other relations up to the second order derivatives. Indeed,
\begin{align}
\dot{x_l} &= \sum_i \frac{\partial x_l}{\partial q_i} \dot{q}_i + \frac{\partial x_l}{\partial t}
\nonumber\\
\ddot{x_l} &= \frac{d}{dt}\Big( \sum_i \frac{\partial x_l}{\partial q_i} \dot{q}_i + \frac{\partial x_l}{\partial t} \Big)
\nonumber\\
&= \sum_{i}\Big[ \sum_j \frac{\partial^2 x_l}{\partial q_i \partial q_j} \dot{q}_i \dot{q}_j + \frac{\partial x_l}{\partial q_i} \ddot{q}_i + 2\frac{\partial^2 x_l}{\partial q_i \partial t} \dot{q}_i \Big]+ \frac{\partial^2 x_l}{\partial t^2}. \label{relations1}
\end{align}
On the other side, deriving $\dot{x_l}$ and  $\ddot{x_l}$ we obtain
\begin{align}
\frac{\partial \dot{x_l}}{\partial q_k} &= \sum_i \frac{\partial^2 x_l}{\partial q_i \partial q_k} \dot{q}_i + \frac{\partial^2 x_l}{\partial t \partial q_k}
\nonumber\\
\frac{\partial \ddot{x_l}}{\partial q_k}&= \sum_{i}\Big[ \sum_j \frac{\partial^3 x_l}{\partial q_i \partial q_j \partial q_k} \dot{q}_i \dot{q}_j + \frac{\partial^2 x_l}{\partial q_i \partial q_k} \ddot{q}_i + 2\frac{\partial^3 x_l}{\partial q_i \partial q_k \partial t} \dot{q}_i \Big] + \frac{\partial^3 x_l}{ \partial t^2 \partial q_k}  
\nonumber\\
\frac{\partial \ddot{x_l}}{\partial \dot{q}_k} &= \sum_{i}\Big[ \sum_j \frac{\partial^2 x_l}{\partial q_i \partial q_j} (\dot{q}_i \delta_{jk} + \dot{q}_j \delta_{ik}) +  2\frac{\partial^2 x_l}{\partial q_i \partial t} \delta_{ik}
\Big] 
\nonumber\\
&=   \sum_i 2\frac{\partial^2 x_l}{\partial q_i \partial q_k} \dot{q}_i + 2\frac{\partial^2 x_l}{\partial t \partial q_k} = 2 \frac{\partial \dot{x_l}}{\partial q_k}
\nonumber\\
\frac{\partial \ddot{x_l}}{\partial \ddot{q}_k} &= \sum_i \frac{\partial x_l}{\partial q_i} \delta_{ik} \;\;=\frac{\partial x_l}{\partial q_k}. \label{relations2}
\end{align}
Also
\begin{align}
\frac{d}{dt}\Big( \frac{\partial \dot{x_l}}{\partial q_k} \Big) &= \frac{d}{dt}\Big(  \sum_i \frac{\partial^2 x_l}{\partial q_i \partial q_k} \dot{q}_i + \frac{\partial^2 x_l}{\partial t \partial q_k} \Big)
\nonumber\\
&= \sum_{i}\Big[ \sum_j \frac{\partial^3 x_l}{\partial q_i \partial q_j \partial q_k} \dot{q}_i \dot{q}_j + \frac{\partial^2 x_l}{\partial q_i \partial q_k} \ddot{q}_i + 2\frac{\partial^3 x_l}{\partial q_i \partial q_k \partial t} \dot{q}_i \Big] + \frac{\partial^3 x_l}{ \partial t^2 \partial q_k} \label{relations3}
\end{align}
comparing \ref{relations2} and \ref{relations3} we get the relation
\begin{equation}
\frac{d}{dt}\Big( \frac{\partial \dot{x_l}}{\partial q_k} \Big)= \frac{\partial \ddot{x_l}}{\partial q_k} \;\;\;\; \text{or} \;\;\;\; \frac{d^2}{dt^2}\Big( \frac{\partial x_l}{\partial q_k} \Big) = \frac{\partial \ddot{x_l}}{\partial q_k} \label{relations4}.
\end{equation}
Applying previous results to our coordinates systems and using compound derivative law up to the second order, the derivative on angular variables can be replaced as:
\begin{align}
\frac{\partial}{\partial \xi_i}&= \frac{\partial x^\nu}{\partial \xi_i} \frac{\partial}{\partial x^\nu} + \frac{\partial \dot{x}^\nu}{\partial \xi_i} \frac{\partial}{\partial \dot{x}^\nu} + \frac{\partial \ddot{x}^\nu}{\partial \xi_i} \frac{\partial}{\partial \ddot{x}^\nu}
\nonumber\\
&=\frac{\partial x^\nu}{\partial \xi_i} \frac{\partial}{\partial x^\nu} + \frac{d}{d\tau}\Big(\frac{\partial x^\nu}{\partial \xi_i}\Big) \frac{\partial}{\partial \dot{x}^\nu} + \frac{d^2}{d\tau^2}\Big(\frac{\partial x^\nu}{\partial \xi_i}\Big) \frac{\partial}{\partial \ddot{x}^\nu} 
\nonumber\\
\frac{\partial}{\partial \dot{\xi}_i}&= \frac{\partial \dot{x}^\nu}{\partial \dot{\xi}_i} \frac{\partial}{\partial \dot{x}^\nu} + \frac{\partial \ddot{x}^\nu}{\partial \dot{\xi}_i} \frac{\partial}{\partial \ddot{x}^\nu} 
\nonumber\\
&= \frac{\partial x^\nu}{\partial \xi_i} \frac{\partial}{\partial \dot{x}^\nu} + 2\frac{d}{d\tau}\Big(\frac{\partial x^\nu}{\partial \xi_i}\Big) \frac{\partial}{\partial \ddot{x}^\nu}
\nonumber\\
\frac{\partial}{\partial \ddot{\xi}_i}&= \frac{\partial \ddot{x}^\nu}{\partial \ddot{\xi}_i} \frac{\partial}{\partial \ddot{x}^\nu}  \;\; = \frac{\partial x^\nu}{\partial \xi_i} \frac{\partial}{\partial \ddot{x}^\nu} \label{relations5}
\end{align}
In matrix notation we have:
\begin{align}
\vec{\nabla}_\xi&= D_{\;\xi}^\nu \square_\nu + \frac{d }{d\tau}\Big(D_{\;\xi}^\nu \Big) \square_{\dot{\nu}} + \frac{d^2}{d\tau^2}\Big(D_{\;\xi}^\nu \Big) \square_{\ddot{\nu}} 
\nonumber\\
\vec{\nabla}_{\dot{\xi}}&= D_{\;\xi}^\nu \square_{\dot{\nu}} + 2\frac{d }{d\tau}\Big(D_{\;\xi}^\nu \Big) \square_{\ddot{\nu}}
\nonumber\\
\vec{\nabla}_{\ddot{\xi}}&= D_{\;\xi}^\nu \square_{\ddot{\nu}} \label{relationsMatrix}
\end{align}
where elements $D_{\;\xi}^\nu$ correspond to a $3\times 4$ matrix of the $\frac{\partial x^\nu}{\partial \xi_i}$ derivatives. The resulting relations was computed up to second order derivative for later uses. $D_{\;\xi}^\nu$ expressions are obtained from transformation relations between sets $D_{\;\xi}^\nu$ expressions are obtained from transformation relations between sets $\{x^0,x^1,x^2,x^3\}$ and $\{\theta,\phi,\chi\}$ showed in eq. \ref{PosAngleRelation}. From this definition, we get:
\begin{align}
&\cos \theta = x^0/R \\
&\cos \phi = x^3/(R\sin \theta)=	x^3/\sqrt{R^2-(x^0)^2}\\
&\cos \chi = x^2/(R\sin \theta \sin \phi)=	x^2/\sqrt{R^2-(x^0)^2 - (x^3)^2}\\
&\tan \chi =x^1/x^2.
\end{align}
After some straightforward derivation steps we obtain matrix $D_{\;\xi}^\nu$ :
\begin{equation}
D_{\;\xi}^\nu = 
\begin{bmatrix}
\frac{x^1x^0}{\sqrt{R^2-({x^0})^2}} & \frac{x^2 x^0}{\sqrt{R^2-({x^0})^2}} & \frac{x^1x^3}{\sqrt{R^2-({x^0})^2}} & -\frac{\sqrt{R^2-({x^0})^2}}{R}\\
\frac{x^1x^3}{\sqrt{R^2-({x^0})^2-({x^3})^2}} & \frac{x^2x^3}{\sqrt{R^2-({x^0})^2-({x^3})^2}} & - {\scriptstyle \sqrt{R^2-({x^0})^2-({x^3})^2}} & 0 \\
x^2 & -x^1 & 0 & 0
\end{bmatrix}.
\end{equation}

\subsection{Single particle motion equation in Lorentzian and angular coordinates}
We update now the inertial and net applied forces acting on an isolated particle under the action of an external field of section \ref{SinglePartEqMotionCart} and use the new sets of coordinates: the Lorentzian and the angular's.

The net applied, and inertial forces in Lorentzian coordinates are computed from the single particle relativistic Lagrangian assuming system depend only on the $n$-particle coordinates while other particles coordinates are considered as parameters. In that case, the $\mu$ component of the inertial and applied forces acting on particle $n$ is
\begin{align}
&\dot{P}_{n,\mu} - Q_{n,\mu} = \mathcal{L}_{n,\mu}(\frac{1}{2}m_n \dot{x}^\nu_n \dot{x}_{n;\nu} -A^\nu \dot{x}_{n;\nu} ) \nonumber \\
&\equiv \Big[ \frac{d}{d\tau}\frac{\partial}{\partial \dot{x}_n^\mu} 
- \frac{\partial}{\partial x_n^\mu}\Big](\frac{1}{2}m_n \dot{x}^\nu_n \dot{x}_{n;\nu} 
-A^\nu \dot{x}_{n;\nu} ) \nonumber \\
& =\frac{d}{d\tau}\Big[m_n \dot{x}_{n;\mu} 
- \Big(\frac{\partial A^\nu}{\partial \dot{x}_n^\mu}\Big) \dot{x}_n^\nu 
- A_\mu \Big]
 - \Big[ \frac{1}{2}\frac{\partial m_n}{\partial x_n^\mu}\dot{x}^\nu_n \dot{x}_{n;\nu} 
 -  \Big(\frac{\partial A^\nu}{\partial x_n^\mu}\Big)\dot{x}_{n;\nu}  \Big] \label{genPRel} \\
\end{align}
or
\begin{align}
&\dot{P}_{n,\mu} - Q_{n,\mu} =m_n \ddot{x}_{n;\mu} 
+ \Big[\square_n^\nu m_n \dot{x}_{n;\nu}\Big] \dot{x}_{n;\mu} 
- \Big(\frac{\partial A^\nu}{\partial \dot{x}_n^\mu}\Big)\ddot{x}_{n;\nu} 
- \frac{d}{d\tau}\Big(\frac{\partial A^\nu}{\partial \dot{x}_n^\mu} \Big) \dot{x}_{n;\nu}
\nonumber\\ 
& \;\;\; - \frac{\partial A_\mu}{\partial x_n^\nu}\dot{x}_n^\nu 
- \frac{\partial A_\mu}{\partial \dot{x}_n^\nu}\ddot{x}_n^\nu  
- \frac{1}{2}\frac{\partial m_n}{\partial x_n^\mu}\dot{x}^\nu_n \dot{x}_{n;\nu} 
+ \Big(\frac{\partial A^\nu}{\partial x_n^\mu}\Big)\dot{x}_{n;\nu} 
\label{genP1Rel},
\end{align}

On the other side, is we replace $\xi_{n,i}...\; i=(1,2,3)$ by the already introduced components $\theta,\phi,\chi$ for particle $n$, respectively, the $\xi_i$ component of the angular forces acting on particle $n$, known in literature as ``torque'' and represented as $\dot{L}_{\;\xi_{n,i}} - T_{\;\xi_{n,i}}$, is:
\begin{align}
&\dot{L}_{\;\xi_{n,i}} - T_{\;\xi_{n,i}} = \mathcal{L}_{\xi_{n,i}}(\frac{1}{2}m_n \dot{x}^\nu_n \dot{x}_{n;\nu} -A^\nu \dot{x}_{n;\nu} ) \nonumber \\
&\equiv \Big[ \frac{d}{d\tau}\frac{\partial}{\partial \dot{\xi}_{n,i}} - \frac{\partial}{\partial \xi_{n,i}}\Big](\frac{1}{2}m_n \dot{x}^\nu_n \dot{x}_{n;\nu} -A^\nu \dot{x}_{n;\nu} ) \nonumber \\
&= \Big[ \frac{d}{d\tau}\Big(  D_{\;\xi_{n,i}}^\mu \square_{n,\dot{\mu}} \Big) - D_{\;\xi_{n,i}}^\mu \square_{n,\mu} - \frac{d }{d\tau}\Big(D_{\;\xi_{n,i}}^\mu \Big) \square_{n,\dot{\mu}}\Big](\frac{1}{2}m_n \dot{x}^\nu_n \dot{x}_{n;\nu} -A^\nu \dot{x}_{n;\nu} ) \nonumber \\
&= D_{\;\xi_{n,i}}^\mu \Big[  \frac{d}{d\tau}\Big(  \square_{n,\dot{\mu}} \Big) - \square_{n,\mu} \Big](\frac{1}{2}m_n \dot{x}^\nu_n \dot{x}_{n;\nu} -A^\nu \dot{x}_{n;\nu} ) \nonumber \\
&=  D_{\;\xi_{n,i}}^\mu \Big( \dot{P}_{n,\mu} - Q_{n,\mu} \Big),  \label{genLRel}
\end{align}
where the Lagrange operator is transformed into the $x^\nu_n$'s representation using the relations \ref{relationsMatrix} and taking into account that because there is no $\ddot{x}^\nu_n$ dependency in the relativistic Lagrangian, $\square_{\ddot{\nu}}=0$. An example of the previous result, the angular component $\xi_{n,3}\equiv \chi$ of the resultant force has the expression:
\begin{align}
&\mathcal{L}_{\xi_{n,3}}(\frac{1}{2}m_n \dot{x}^\nu_n \dot{x}_{n;\nu} -A^\nu \dot{x}_{n;\nu} ) =  \dot{L}_{\;\xi_{n,3}} - T_{\;\xi_{n,3}} 
\nonumber \\
&= D_{\;\xi_{n,3}}^\mu \Big( \dot{P}_{n,\mu} - Q_{n,\mu} \Big) = x^2 \Big( \dot{P}_{n,1} - Q_{n,1} \Big)-x^1 \Big( \dot{P}_{n,2} - Q_{n,2} \Big),
\end{align}
which is the well-known expression in tree dimensions $\mathbf{T}=\mathbf{r} \times \mathbf{F}$.
\subsection{Equations of motion for n-VMVF systems}

The Lagrangian of $n$-VMVF is unknown at this point. However, from the extension of the D'Alembert principle, we know that the solution of Lagrange equation is the sum of the inertial and the net applied forces of all particles. On the other case, under the assumption of section \ref{genForceApproach}, we have that those forces are the net forces for isolated particles under the action of the external field. Starting from an initial Lagrange function, if we compare the forces from both approaches, we can obtain then a set of relations that will constraint the motion of the particles of the system.
We need to perform such comparison for both set of independent equations. In general, for isolated systems, the net applied and inertial forces in the Lorentzian and angular coordinates are just the results we obtain on the previous section. We have then the solution for the Lagrange equation for both sets of coordinates.

In summary, our set of equations is:
\begin{align}
&\mathcal{L}_{n,\mu}L_{sys} =0  \label{partSysEq1} \\ 
&\mathcal{L}_{n,\mu}L_{sys}= \sum_{n'} \dot{P}_{{n'},\mu} - Q_{{n'},\mu}= 0 \label{partSysEq2}\\
&\mathcal{L}_{\xi_{n,i}}L_{sys} =0 \label{partSysEq3} \\
&\mathcal{L}_{\xi_{n,i}}L_{sys}= \sum_{n'}\dot{L}_{\;\xi_{{n'},i}} - T_{\;\xi_{{n'},i}}= \sum_{n'} D_{\;\xi_{{n'},i}}^\mu \Big( \dot{P}_{{n'},\mu} - Q_{{n'},\mu} \Big)=0. \label{partSysEq4} 
\end{align}

Equations \ref{partSysEq1} and \ref{partSysEq3} correspond to the least action principle for any isolated system while equations \ref{partSysEq2} and \ref{partSysEq4} are related to the conservation law of linear and angular momentum respectively for isolated systems, which, at the same time, is connected with space's homogeneity and isotropy and also with the conservation the of the total energy. The masses and the field derivatives are degrees of freedom of the whole system; however, they are not generalized coordinates of the variational problem. This solution, which involves this kind of treatment and includes two sets of constrained Lagrange equations which at the same time outcome in two different Lagrangian, is new as a solution of a classical problem, at least to the best of our knowledge.

Recalling equation \ref{splittedLagEq}, the total Lagrangian for particle $n$-VMVF systems can be divided into two terms, 

\begin{equation*}
L_{sys} = L_{[\dot{m}_n= 0]} + L_{[\dot{m}_n\neq 0]} = \sum_{n'=1}^N L_{sp_{n'}} + L_{[\dot{m}_n\neq 0]},
\end{equation*}
where the first term considers constant masses and is equal to the well-known Lagrangian for particle system under the action of a field $L_{sp}$:
\begin{equation}
L_{sp} = \sum_{n'} L_{sp_{n'}} = \sum_{n'} \frac{1}{2}m_{n'} \dot{x}^\nu_{n'} \dot{x}_{n';\nu} -A^\nu \dot{x}_{n';\nu}. \label{nonIntPartSys}
\end{equation}

The strategy to follow for obtaining the final Lagrangian for $n$-VMVF systems is to propose an initial Lagrangian and obtain its Lagrange equations. The $n$-degeneracy characteristics of the Lagrange equations led to $n$ independent equations by comparing the derived equations with the equations \ref{partSysEq1} and \ref{partSysEq3}. Those constraints should be included in the final Lagrangian using the Lagrange multiplier method.

According to the classification for the Lagrange function in \ref{splittedLagEq}, we can start assuming the Lagrange function $L_{sys}= L_{sp}$. From there we obtain the Lagrange equations, this time assuming all the position of the particles as variables of the system.

\subsubsection*{Motion and constraint equation for $n$-VMVF systems using Lorentzian coordinates}
The motion equation of particle $n$ in the $\mu$ direction as a constituent of an isolated $n$-VMVF system using the proposed Lagrangian is
\begin{align}
&\mathcal{L}_{n,\mu}( L_{sp}) = \mathcal{L}_{n,\mu}( \sum_{n'} \frac{1}{2}m_{n'} \dot{x}^\nu_{n'} \dot{x}_{n';\nu} -A^\nu \dot{x}_{n';\nu} ) =0 \nonumber \\
&\equiv \Big[ \frac{d}{d\tau}\frac{\partial}{\partial \dot{x}_n^\mu} - \frac{\partial}{\partial x_n^\mu}\Big]( \sum_{n'} \frac{1}{2}m_{n'} \dot{x}^\nu_{n'} \dot{x}_{{n'};\nu} -A^\nu \dot{x}_{{n'};\nu}. )=0 \nonumber 
\end{align}
We divide the Lagrange's function into on term related to the particle $n$ and another grouping the terms of the rest of the particles
\begin{equation}
\mathcal{L}_{n,\mu}( \sum_{n'} L_{n'})= \mathcal{L}_{n,\mu}( L_n + \sum_{{n'} \neq n} L_{n'}).\label{lagranDivision1}
\end{equation}

The first term, using equation \ref{genP1Rel}, have the form:
\begin{align}
& \mathcal{L}_{n,\mu} L_n=\Big[ \frac{d}{d\tau}\frac{\partial}{\partial \dot{x}_n^\mu} - \frac{\partial}{\partial x_n^\mu}\Big](\frac{1}{2}m_n \dot{x}^\nu_n \dot{x}_{n;\nu} -A^\nu \dot{x}_{n;\nu} )
\nonumber \\
&=\frac{d}{d\tau}\Big[m_n \dot{x}_{n;\mu} - \Big(\frac{\partial A^\nu}{\partial \dot{x}_n^\mu}\Big) \dot{x}_{n;\mu} - A_\mu \Big] - \Big[ \frac{1}{2}\frac{\partial m_n}{\partial x_n^\mu}\dot{x}^\nu_n \dot{x}_{n;\nu} -  \Big(\frac{\partial A^\nu}{\partial x_n^\mu}\Big)\dot{x}_{n;\mu}  \Big] 
\nonumber \\
&= \dot{P}_{n,\mu} - Q_{n,\mu} - \sum_{l \neq n} \Big( \frac{\partial A_\mu}{\partial x_{l}^\nu}\dot{x}_l^\nu + \frac{\partial A_\mu}{\partial \dot{x}_l^\nu}\ddot{x}_l^\nu \Big), \label{TransConstraintEq0} 
\end{align}
where all particle potential dependency is now taking into account, since field, we are now considering include all particles contributions as constituent parts of the system.

The only nonnull terms in the last part of motion equation \ref{lagranDivision1} is related to potential since $\frac{1}{2}m_n \dot{x}^\nu_n \dot{x}_{n;\nu}$ depends only on particle $n$ coordinates. In that case, we obtain:
\begin{align}
&\mathcal{L}_{n,\mu}(\sum_{{n'} \neq n} L_{n'}) = \mathcal{L}_{n,\mu}(\sum_{{n'} \neq n} -A^\nu \dot{x}_{{n'};\nu})
\nonumber \\
&= - \sum_{{n'} \neq n} \Big\{ \frac{d}{d\tau}\Big[ \Big(\frac{\partial A^\nu}{\partial \dot{x}_{n}^\mu}\Big) \dot{x}_{n';\nu} \Big] - \Big(\frac{\partial A^\nu}{\partial x_{n}^\mu}\Big)\dot{x}_{n';\nu} \Big\}. \label{TransConstraintEq1}
\end{align}
Putting all together, the motion equation of particle $n$ in the $\mu$ direction is
\begin{align}
\mathcal{L}_{n,\mu}L_{sp} &= \dot{P}_{n,\mu} - Q_{n,\mu} 
- \sum_{{n'} \neq n} \Big\{ \frac{d}{d\tau}\Big[ \Big(\frac{\partial A^\nu}{\partial \dot{x}_{n}^\mu}\Big) \dot{x}_{n';\nu} \Big] - \Big(\frac{\partial A^\nu}{\partial x_{n}^\mu}\Big)\dot{x}_{n';\nu}  
\nonumber \\ 
& + \frac{\partial A_\mu}{\partial x_{n'}^\nu}\dot{x}_{n'}^\nu + \frac{\partial A_\mu}{\partial \dot{x}_{n'}^\nu}\ddot{x}_{n'}^\nu \Big\}=0.
\end{align}

Comparing the obtained motion equation and the result of the extension of D'Alembert principle
\begin{equation}
\mathcal{L}_{n,\mu}( L_{sp}) =  \sum_{n'} \dot{P}_{{n'},\mu} - Q_{{n'},\mu},
\end{equation}
we obtain we obtained 3-$n$ independent equation:
\begin{align}
\Phi_{\mu_n} &= \sum_{{n'} \neq n} \Big\{ (\dot{P}_{{n'},\mu} - Q_{{n'},\mu})  + \frac{d}{d\tau} \Big(\frac{\partial A^\nu}{\partial \dot{x}_{n}^\mu}\Big) \dot{x}_{n';\nu} + \frac{\partial A^\nu}{\partial \dot{x}_{n}^\mu} \ddot{x}_{n';\nu}
 \nonumber  \\
& - \frac{\partial A^\nu}{\partial x_{n}^\mu}\dot{x}_{n';\nu} 
+ \frac{\partial A_\mu}{\partial x_{n'}^\nu}\dot{x}_{n'}^\nu 
+ \frac{\partial A_\mu}{\partial \dot{x}_{n'}^\nu}\ddot{x}_{n'}^\nu \Big\}   =0.
\end{align}
After rearrange the last equation and put it together to the least action principle equation we have the 6-$n$ independent equations needed to solve the problem: 
\begin{equation}
\mathcal{L}_{n,\mu}( L_{sp}) = \mathcal{L}_{n,\mu}( \sum_{n'} \frac{1}{2}m_{n'} \dot{x}^\nu_{n'} \dot{x}_{n';\nu} -A^\nu \dot{x}_{n';\nu} ) =0 \label{TransLeastActionEq} 
\end{equation}
and 
\begin{align}
\Phi_{\mu_n} = \sum_{{n'} \neq n}  &\Big[
\Big(\square_{n'}^\alpha m_n' \dot{x}_{n';\alpha}\Big)g^\nu_\mu
- \frac{1}{2}\frac{\partial m_{n'}}{\partial x_{n'}^\mu}\dot{x}^\nu_{n'}
+ \frac{d}{d\tau} \Big(\frac{\partial A^\nu}{\partial \dot{x}_{n}^\mu}
- \frac{\partial A^\nu}{\partial \dot{x}_{n'}^\mu}
\Big) 
+ \frac{\partial A^\nu}{\partial x_{n'}^\mu}
- \frac{\partial A^\nu}{\partial x_{n}^\mu}
\Big]\dot{x}_{n';\nu}
\nonumber \\
&+ \Big[ 
m_{n'}g^\nu_\mu + \frac{\partial A^\nu}{\partial x_{n}^\mu}
- \frac{\partial A^\nu}{\partial x_{n'}^\mu}
+ \frac{\partial A_\mu}{\partial x_{n'}^\nu}
\Big] \ddot{x}_{n';\nu}
\label{TransConstraintEq}
\end{align}
Actually, there are 8-$n$ equations, however, Minkowski's constraint reduce to 6-$n$ independent equations as mentioned before. 

Equations depend on variables
\begin{align}
&\{x_n^\nu\},\{\dot{x}_n^\nu\},\{\ddot{x}_n^\nu\},\{\frac{\partial m_n}{\partial x_n^\mu }\},\{\frac{d}{dt} \big( \frac{\partial m_n}{\partial x_n^\mu }\big)\},\{ \frac{d^2}{dt^2}\big(\frac{\partial m_n}{\partial x_n^\mu }\big)\}, 
\nonumber \\
&\{\frac{\partial A^\nu}{\partial x_n^\mu }\},\{ \frac{d}{dt} \big(\frac{\partial A^\nu}{\partial x_n^\mu }\}\big),\{\frac{d^2}{dt^2} \big(\frac{\partial A^\nu}{\partial x_n^\mu }\}\big),\{\frac{\partial A^\nu}{\partial \dot{x}_n^\mu }\},\{ \frac{d}{dt} \big(\frac{\partial A^\nu}{\partial \dot{x}_n^\mu }\big)\},\{ \frac{d^2}{dt^2} \big(\frac{\partial A^\nu}{\partial \dot{x}_n^\mu} \big)\}.
\end{align}
which means we need more equations for the system being solvable. Thus, the inclusion of angular variables.

\subsubsection*{Motion and constraint equation for $n$-VMUF systems using angular coordinates}
We now compute the inertial and the net applied forces using angular coordinates to obtain the other set of independent equations needed to solve the problem. The motion equation of particle $n$ rotating around axis in the $i$ direction for isolated $n$-VMVF systems using the starting Lagrangian is
\begin{align}
&\mathcal{L}_{\xi_{n,i}}( L_{sp}) = \mathcal{L}_{\xi_{n,i}}( \sum_{n'} \frac{1}{2}m_{n'} \dot{x}^\nu_{n'} \dot{x}_{n';\nu} -A^\nu \dot{x}_{n';\nu} ) =0 \nonumber \\
&=  D_{\;\xi_{n,i}}^\mu \Big[  \frac{d}{d\tau}\Big(  \square_{n,\dot{\mu}} \Big) - \square_{n,\mu} \Big]( \sum_{n'} \frac{1}{2}m_{n'} \dot{x}^\nu_{n'} \dot{x}_{{n'};\nu} -A^\nu \dot{x}_{{n'};\nu} )=0 \nonumber \\
&\equiv  D_{\;\xi_{n,i}}^\mu \Big[  \frac{d}{d\tau}\Big(  \square_{n,\dot{\mu}} \Big) - \square_{n,\mu} \Big] (L_n + \sum_{{n'} \neq n} L_{n'}).
\end{align}
Using previous results in eq. \ref{TransConstraintEq0}, the first term in the sum is:
\begin{align}
\mathcal{L}_{\xi_{n,i}}( L_n) &= D_{\;\xi_{n,i}}^\mu \Big[  \frac{d}{d\tau}\Big(  \square_{n,\dot{\mu}} \Big) - \square_{n,\mu} \Big]L_n \nonumber \\ 
& = D_{\;\xi_{n,i}}^\mu \Big[ \dot{P}_{n,\mu} - Q_{n,\mu} - \sum_{n' \neq n} \Big( \frac{\partial A_\mu}{\partial x_{n'}^\nu}\dot{x}_{n'}^\nu + \frac{\partial A_\mu}{\partial \dot{x}_{n'}^\nu}\ddot{x}_{n'}^\nu\Big) \Big].
\end{align}
The only nonnull terms of the last part are related to the field's addends as eq. \ref{TransConstraintEq1}. In that case, we obtain:
\begin{align}
\mathcal{L}_{\xi_{n,i}}(\sum_{{n'} \neq n} L_{n'})&=  D_{\;\xi_{n,i}}^\mu \Big[  \frac{d}{d\tau}\Big(  \square_{n,\dot{\mu}} \Big) - \square_{n,\mu} \Big] (\sum_{{n'} \neq n} L_{n'})\nonumber \\
&= - D_{\;\xi_{n,i}}^\mu \sum_{{n'} \neq n} \Big\{ \frac{d}{d\tau}\Big[ \Big(\frac{\partial A^\nu}{\partial \dot{x}_{n}^\mu}\Big) \dot{x}_{n';\nu} \Big] - \Big(\frac{\partial A^\nu}{\partial x_{n}^\mu}\Big)\dot{x}_{n';\nu} \Big\}. \label{ConstraintDep}
\end{align}
Putting all together we have
\begin{align}
\mathcal{L}_{\xi_{n,i}}L_{sp} &=   D_{\;\xi_{n,i}}^\mu \Big\{ \dot{P}_{n,\mu} - Q_{n,\mu} - \sum_{{n'} \neq n}  \frac{d}{d\tau}\Big[ \Big(\frac{\partial A^\nu}{\partial \dot{x}_{n}^\mu}\Big) \dot{x}_{n';\nu} \Big] - \Big(\frac{\partial A^\nu}{\partial x_{n}^\mu}\Big)\dot{x}_{n';\nu} \nonumber \\ 
& +\frac{\partial A_\mu}{\partial x_{n'}^\nu}\dot{x}_{n'}^\nu + \frac{\partial A_\mu}{\partial \dot{x}_{n'}^\nu}\ddot{x}_{n'}^\nu  \Big\} 
\nonumber \\
&=\dot{L}_{\;\xi_{{n},i}} - T_{\;\xi_{{n},i}} - D_{\;\xi_{n,i}}^\mu \sum_{{n'} \neq n} \Big\{ \frac{d}{d\tau} \Big(\frac{\partial A^\nu}{\partial \dot{x}_{n}^\mu}\Big) \dot{x}_{n';\nu} + \frac{\partial A^\nu}{\partial \dot{x}_{n}^\mu} \ddot{x}_{n';\nu}  
\nonumber \\
& - \Big(\frac{\partial A^\nu}{\partial x_{n}^\mu}\Big)\dot{x}_{n';\nu}
+\frac{\partial A_\mu}{\partial x_{n'}^\nu}\dot{x}_{n'}^\nu + \frac{\partial A_\mu}{\partial \dot{x}_{n'}^\nu}\ddot{x}_{n'}^\nu \Big\}. \label{RotConstraintEq0}
\end{align}
Comparing with D'Alembert extended principle for angular coordinates:
\begin{equation}
\mathcal{L}_{\xi_{n,i}}L_{sp} = \sum_{n'} \dot{L}_{\;\xi_{{n'},i}} - T_{\;\xi_{{n'},i}} 
= \sum_{n'} D_{\;\xi_{n',i}}^\mu ( \dot{P}_{n',\mu} - Q_{n',\mu}  ),
\end{equation}
we obtain we other set of 3-$n$ independent equations:
\begin{align}
\Psi_{i_n}  = &\sum_{{n'} \neq n}  \dot{L}_{\;\xi_{{n'},i}} - T_{\;\xi_{{n'},i}}  
+ D_{\;\xi_{n,i}}^\mu \Big[ \frac{d}{d\tau} \Big(\frac{\partial A^\nu}{\partial \dot{x}_{n}^\mu}\Big) \dot{x}_{n';\nu} 
+ \frac{\partial A^\nu}{\partial \dot{x}_{n}^\mu} \ddot{x}_{n';\nu}   
 \nonumber \\
& - \Big(\frac{\partial A^\nu}{\partial x_{n}^\mu}\Big)\dot{x}_{n';\nu} 
+ \frac{\partial A_\mu}{\partial x_{n'}^\nu}\dot{x}_{n'}^\nu 
+ \frac{\partial A_\mu}{\partial \dot{x}_{n'}^\nu}\ddot{x}_{n'}^\nu  \Big] = 0, 
\end{align}

By rearranging the last equation and put it together with the least action principle in angular coordinates, we another 6-$n$ independent equations:
\begin{equation}
\mathcal{L}_{\xi_{n,i}}L_{sp}( L_{sp}) = \mathcal{L}_{\xi_{n,i}}L_{sp}( \sum_{n'} \frac{1}{2}m_{n'} \dot{x}^\nu_{n'} \dot{x}_{n';\nu} -A^\nu \dot{x}_{n';\nu} ) =0  \label{RotLeastActionEq}
\end{equation}
and 
\begin{align}
\Psi_{i_n}  = \sum_{{n'} \neq n} &\Big\{
 D_{\;\xi_{n',i}}^\mu \Big[ 
\Big(\square_{n'}^\alpha m_n' \dot{x}_{n';\alpha}\Big)g^\nu_\mu
- \frac{1}{2}\frac{\partial m_{n'}}{\partial x_{n'}^\mu}\dot{x}^\nu_{n'}
-\frac{d}{d\tau} \Big(
\frac{\partial A^\nu}{\partial \dot{x}_{n'}^\mu}
\Big) 
- \frac{\partial A_\mu}{\partial x_{n';\nu}}
+ \frac{\partial A^\nu}{\partial x_{n'}^\mu}
 \Big]
 \nonumber \\
&+D_{\;\xi_{n,i}}^\mu \Big[
\frac{d}{d\tau} \Big(
\frac{\partial A^\nu}{\partial \dot{x}_{n}^\mu}
\Big) 
+ \frac{\partial A_\mu}{\partial x_{n';\nu}}
- \frac{\partial A^\nu}{\partial x_{n}^\mu} 
  \Big]
\Big\} \dot{x}_{n';\nu}
\nonumber \\
+& \Big\{
 D_{\;\xi_{n',i}}^\mu \Big[ 
 m_{n'}g^\nu_\mu - \frac{\partial A^\nu}{\partial x_{n'}^\mu}
  \Big]
 +D_{\;\xi_{n,i}}^\mu \Big[ 
\frac{\partial A^\nu}{\partial x_{n}^\mu}
+\frac{\partial A_\mu}{\partial x_{n';\nu}}
\Big]
\Big\} \ddot{x}_{n';\nu}
\label{RotConstraintEq}
\end{align}

We also note the same dependency of variables as Lorentzian case as shown in equation \ref{ConstraintDep}.

We are then, in the presence of a non-holonomic constraint problem. Equations connecting the generalized coordinates of a system are written as 
\begin{equation}
f(\ddot{\mathbf{r}}_1, \; \ddot{\mathbf{r}}_2 ...,\dot{\mathbf{r}}_1, \; \dot{\mathbf{r}}_2 ..., \mathbf{r}_1, \; \mathbf{r}_2 ...) = 0.
\end{equation}
We have, in total, a set of 12-$n$ independent equations depending on variables \ref{ConstraintDep}.

There are some comments on the method we use. The first one is the fact that coordinates are now connected by the constraint equations \ref{TransConstraintEq} and \ref{RotConstraintEq}. This may look contradictory since position coordinates were set as independent from the beginning. However, we lost no generality if we set the new degree of freedom of the system $e.i$ masses and field derivatives as the dependent variables. The second issue is the one-to-one correspondence between constraint equations and the initial proposed Lagrangian. In fact, if another initial Lagrangian is proposed, then others constraint equations will be obtained. In other words, the construction of the final Lagrangian depends only on 
\begin{itemize}
\item the proposition for the inertial and the net applied forces acting on all particles of the system
\item the degeneracy of the extension of the D'Alembert Principle
\item our proposal for considering the inertial the net applied forces as the same forces acting on isolated particles as mentioned in section \ref{genForceApproach}.
\end{itemize}

\subsection{Second order Lagrangian.}
On the last sections, we obtain 2 set of 6-$n$ independent equations of motions for $n$-VMVF systems eqs. \ref{TransConstraintEq}, \ref{TransLeastActionEq}, \ref{RotConstraintEq} and \ref{RotLeastActionEq}, each for every independent set of coordinates. The constraint relations from equation ref{RotConstraintEq} and \ref{TransConstraintEq} have the dependency
\begin{equation}
\varphi_i(x,y_1,y_2,...y_n,\dot{y}_1,\dot{y}_2,...\dot{y}_n,\ddot{y}_1,\ddot{y}_2,...\ddot{y}_n)=0 \;\; (i=1,2,....m;m<n),
\end{equation}
so, the integrand also should have the form
\begin{equation}
L_{sys}(x,y_1,y_2,...y_n,\dot{y}_1,\dot{y}_2,...\dot{y}_n,\ddot{y}_1,\ddot{y}_2,...\ddot{y}_n)=0.
\end{equation}
This dependency implies the necessity to expand Euler-Lagrange equations. We follow the same method shown on Goldstein's \citep{goldstein} or Elsgoltz's \citep{elsgoltz1977} textbooks for generating these equations from least action principle. 

Let us consider Lagrangian as part of the family of functions  $F$ that satisfied
\begin{equation}
\delta J(y_1,y_2,...y_n) = \int_{x_0}^{x_1} F(x,y_1,y_2,...y_n,\dot{y}_1,\dot{y}_2,...\dot{y}_n,\ddot{y}_1,\ddot{y}_2,...\ddot{y}_n)dx = 0,
\end{equation}
Let us also consider Lagrangian dependent of parameter $\alpha$ which labels possibles solutions functions $y_i(x,\alpha)$. These functions are expressed as
\begin{align}
&y_1(x,\alpha) = y_1(x,0) + \alpha \eta_1(x) \nonumber \\
&y_2(x,\alpha) = y_2(x,0) + \alpha \eta_2(x) \nonumber \\
&\ldots  \ldots  \ldots  \ldots  \ldots  \ldots  \ldots  \ldots  \nonumber \\
&y_n(x,\alpha) = y_n(x,0) + \alpha \eta_n(x),
\end{align}
being $y_i(x,0)$,  the solutions for the extremal problem and $\eta_i(x)$ are arbitrary and well-behaved functions. $\eta_i(x)$ functions are continuous and nonsingular between $x_0$ and $x_1$ with also continuous first, second and third derivatives. Functions $\eta_i(x)$ and their derivatives vanish at $x_0$ and $x_1$.  Variational principle states $J$ depending on parameter $\alpha$
\begin{equation}
J(y_1,y_2,...y_n) = \int_{x_0}^{x_1} F(x, \{y_i(x,\alpha)\},\{\dot{y}_i(x,\alpha)\},\{\ddot{y}_i(x,\alpha)\})dx,
\end{equation}
will have a stationary point if the variation
\begin{equation}
\Big( \frac{d J}{d \alpha} \Big)_{\alpha = 0}=0.
\end{equation}
Differentiating variational $J$ by standard methods we obtain
\begin{equation}
 \frac{d J}{d \alpha} =  \int_{x_0}^{x_1}  \sum_i \Big( \frac{\partial F}{\partial y_i} \frac{\partial y_i}{\partial \alpha} +  \frac{\partial F}{\partial \dot{y}_i} \frac{ \partial\dot{y}_i}{\partial \alpha} +  \frac{\partial F}{\partial \ddot{y}_i} \frac{ \partial \ddot{y}_i}{\partial \alpha} \Big)dx.
\end{equation}
Integrating by parts, the second integrand becomes:
\begin{align}
&\int_{x_0}^{x_1}  \sum_i  \frac{\partial F}{\partial \dot{y}_i} \frac{\partial\dot{y}_i}{\partial \alpha} dx =  \int_{x_0}^{x_1}  \sum_i  \frac{\partial F}{\partial \dot{y}_i} \frac{d}{dx}\Big(\frac{\partial y_i}{\partial \alpha}\Big) dx \nonumber \\ 
&=  \frac{\partial F}{\partial \dot{y}_i} \frac{\partial y_i}{\partial \alpha}\bigg \vert_{x_0}^{x_1} - \int_{x_0}^{x_1}  \sum_i  \frac{d}{dx}\Big(\frac{\partial F}{\partial \dot{y}_i} \Big) \frac{\partial y_i}{\partial \alpha} dx \nonumber \\ 
&=  \frac{\partial F}{\partial \dot{y}_i} \eta_i(x)\bigg \vert_{x_0}^{x_1} - \int_{x_0}^{x_1}  \sum_i  \frac{d}{dx}\Big(\frac{\partial F}{\partial \dot{y}_i} \Big) \frac{\partial y_i}{\partial \alpha} dx= - \int_{x_0}^{x_1}  \sum_i  \frac{d}{dx}\Big(\frac{\partial F}{\partial \dot{y}_i} \Big) \frac{\partial y_i}{\partial \alpha} dx.
\end{align}
Integrating by parts twice, the third integrand becomes:
\begin{align}
&\int_{x_0}^{x_1}  \sum_i  \frac{\partial F}{\partial \ddot{y}_i} \frac{\partial\ddot{y}_i}{\partial \alpha} dx =  \int_{x_0}^{x_1}  \sum_i  \frac{\partial F}{\partial \ddot{y}_i} \frac{d}{dx} \Big(\frac{\partial \dot{y}_i}{\partial \alpha}\Big) dx \nonumber \\ 
&=  \frac{\partial F}{\partial \ddot{y}_i} \frac{\partial \dot{y}_i}{\partial \alpha}\bigg \vert_{x_0}^{x_1} - \int_{x_0}^{x_1}  \sum_i  \frac{d}{dx}\Big(\frac{\partial F}{\partial \ddot{y}_i} \Big) \frac{\partial \dot{y}_i}{\partial \alpha} dx \nonumber \\
&=  \frac{\partial F}{\partial \ddot{y}_i} \dot{\eta}_i(x)\bigg \vert_{x_0}^{x_1} - \int_{x_0}^{x_1}  \sum_i  \frac{d}{dx}\Big(\frac{\partial F}{\partial \ddot{y}_i} \Big) \frac{\partial \dot{y}_i}{\partial \alpha} dx = -\int_{x_0}^{x_1}  \sum_i  \frac{d}{dx}\Big(\frac{\partial F}{\partial \ddot{y}_i} \Big) \frac{\partial \dot{y}_i}{\partial \alpha} dx \nonumber \\
&=  - \frac{d}{dt} \Big( \frac{\partial F}{\partial \ddot{y}_i}\Big) \frac{\partial y_i}{\partial \alpha}\bigg \vert_{x_0}^{x_1} + \int_{x_0}^{x_1}  \sum_i  \frac{d^2}{dx^2}\Big(\frac{\partial F}{\partial \ddot{y}_i} \Big) \frac{\partial y_i}{\partial \alpha} dx = \int_{x_0}^{x_1}  \sum_i  \frac{d^2}{dx^2}\Big(\frac{\partial F}{\partial \ddot{y}_i} \Big) \frac{\partial y_i}{\partial \alpha} dx.
\end{align}
Substituting in the differentiation of the variational $J$
\begin{equation}
\frac{dJ}{d \alpha}  = \int_{x_0}^{x_1}  \sum_i \Big[ \frac{d^2}{dx^2}\Big(\frac{\partial F}{\partial \ddot{y}_i} \Big) - \frac{d}{dx}\Big(\frac{\partial F}{\partial \dot{y}_i} \Big) + \frac{\partial F}{\partial y_i}\Big] \frac{\partial y_i}{\partial \alpha} dx.
\end{equation}
Multiplying by $\delta \alpha$ and using:
\begin{equation}
\Big( \frac{d J}{d \alpha} \Big)_{\alpha = 0} \delta \alpha = \delta J\;\; \text{and} \;\; \Big(\frac{\partial y_i}{d \alpha} \Big)_{\alpha = 0}\delta \alpha = \delta y_i
\end{equation}
we obtain
\begin{equation}
\delta J = \int_{x_0}^{x_1}  \sum_i \Big[ \frac{d^2}{dx^2}\Big(\frac{\partial F}{\partial \ddot{y}_i} \Big) - \frac{d}{dx}\Big(\frac{\partial F}{\partial \dot{y}_i} \Big) + \frac{\partial F}{\partial y_i}\Big]  \delta y_i dx 
\end{equation}
this relation all variations $\delta y_i$ are considered independents. We can apply the so-called fundamental lemma of the calculus of variations and obtain the extended Euler-Lagrange equations:
\begin{equation}
\frac{d^2}{dx^2}\Big( \frac{\partial F}{\partial \ddot{y}_i}\Big) - \frac{d}{dx}\Big( \frac{\partial F}{\partial \dot{y}_i}\Big) + \frac{\partial F}{\partial y_i} =0\label{extLagrangeEq2Order}
\end{equation}
Following the same procedure, it can be proved that for higher $n$-order variational 
\begin{equation}
\int F(x, y,y^{(1)},... y^{(n)})dx,
\end{equation}
Euler-Lagrange equations have the form:
\begin{equation}
\sum_{i=1}^n (-1)^n \frac{d^n}{dx^n}\Big( \frac{\partial F}{\partial y_i^{(n)}}\Big) + \frac{\partial F}{\partial y_i} = 0.
\end{equation}

These results are summarize on the textbook of R. Courant and D. Hilbert ``Methods of Mathe\-matical Physics'' \cite{Courant53physics}.

Note that the extension of Euler-Lagrange equations do not affect the way the constraint equation were found since the initial non-interacting particles Lagrangian have no $\ddot{\mathbf{r}}$ dependency. 
\subsection{Second order constrained Lagrangian}
We defined a set of constraint that connects the coordinates of the particle up the second-order and also the masses, and field derivatives showed in equations (eq. \ref{TransConstraintEq}) (\ref{RotConstraintEq}). The existence of these relations means that some virtual displacements of the generalized coordinates from the second-order Lagrangian
\begin{equation}
L(x, y_1,y_2...y_n,\dot{y}_1,\dot{y}_2...\dot{y}_n,\ddot{y}_1,\ddot{y}_2...\ddot{y}_n) 
\end{equation}
where
\begin{equation}
\{y_{n}\} \equiv \{\mathbf{r}_n,\frac{\partial m_n}{\partial x_{n,i}},\frac{\partial \mathbf{A}}{\partial x_{n,i}}, \frac{\partial \mathbf{A}}{\partial \dot{x}_{n,i}}\},
\end{equation}
are not independent. In this case, the fundamental lemma of calculus of variations to obtain Euler-Lagrange equations can no longer be applied, and the extended Lagrange equations \ref{extLagrangeEq2Order} are no longer valid.

One of the most efficient procedures to treat these displacements is the well-known method of the undetermined multipliers \citep{elsgoltz1977} developed by Lagrange. This method is a strategy to solve the extremal problem for functions or functionals subject to equality constraint relations. Thus, given the functional
\begin{equation}
J = \int_{x_0}^{x_1} F(x, y_1,y_2...y_n,\dot{y}_1,\dot{y}_2...\dot{y}_n,\ddot{y}_1,\ddot{y}_2...\ddot{y}_n)dx,
\end{equation}
and the equality constraint relations
\begin{equation}
\varphi_i(x, y_1,y_2...y_n,\dot{y}_1,\dot{y}_2...\dot{y}_n,\ddot{y}_1,\ddot{y}_2...\ddot{y}_n)=0 \;\; (i=1,2,....m;m<n),
\end{equation}
under a proper selection of constants $\lambda_i$, it is proved that if functions $y_j\;\; j=(1,2,..n)$ are extremes of the problem then, they also are extremes for the generalized functional:
\begin{equation}
J^* = \int_{x_0}^{x_1} \Big( F + \sum_{i=1}^m \lambda_i(x)\varphi_i \Big)dx = \int_{x_0}^{x_1} F^*dx.
\end{equation}
The system is then solved considering functions $y_1$, $y_2$..., $y_n$ and $\lambda_1$, $\lambda_2$ ..., $\lambda_m$ as arguments of variational $J^*$.

The proof is similar to the method exposed on reference \citep{elsgoltz1977} for $\varphi (y_i,\dot{y}_i)$ dependence. The constraint equations $\varphi_i$ are independent, so they satisfy
\begin{equation}
\frac{D(\varphi_1, \varphi_2,....\varphi_m) }{D( \ddot{y}_1,\;\ddot{y}_2\;....\ddot{y}_m)}\neq 0.
\end{equation}
In that case, $\ddot{y}_1$, $\ddot{y}_2$ ... $\ddot{y}_m$ variables can be determined as
\begin{equation}
\ddot{y}_i = \Psi_i (x, y_1,y_2...y_n,\dot{y}_1,\dot{y}_2...\dot{y}_n,\ddot{y}_1,\ddot{y}_2...\ddot{y}_n) \quad i = 1,2...,m,
\end{equation}
being $y_i \;(i=m+1,m+2...n)$ functions whose variations $\delta y_i$ are arbitrary and, also with the variations of its derivatives $\delta \dot{y}_i$ and $\delta \ddot{y}_i$, vanish at $x_0$ and $x_1$ points. If $y_1...y_n$ are arbitrary functions that satisfies  $m$ equations $\varphi_i$, the variation of these equations are
\begin{equation}
\delta \varphi_i=\sum_{j=1}^n \frac{\partial \varphi_i}{\partial y_j} \delta y_j + \frac{\partial \varphi_i}{\partial \dot{y}_j} \delta \dot{y}_j + \frac{\partial \varphi_i}{\partial \ddot{y}_j} \delta \ddot{y}_j =0,
\end{equation}
where higher arbitrary variation order of $\delta y_j $, $\delta \dot{y}_j$ and $\delta \ddot{y}_j$ are omitted since their influence are minimized when computing the variation of the variational function where only first order rearranging of $\delta y_j $, $\delta \dot{y}_j$ and $\delta \ddot{y}_j$ matters \citep{elsgoltz1977}.  

Multiplying by undetermined factor $\lambda_i(x)$ and integrating over $dx$:
\begin{equation}
\int_{x_0}^{x_1} \lambda_i(x)\delta \varphi_i dx= \int_{x_0}^{x_1} \Big\{ \sum_{j=1}^n \lambda_i(x)\frac{\partial \varphi_i}{\partial y_j} \delta y_j + \lambda_i(x)\frac{\partial \varphi_i}{\partial \dot{y}_j} \delta \dot{y}_j + \lambda_i(x)\frac{\partial \varphi_i}{\partial \ddot{y}_j} \delta \ddot{y}_j\Big\}dx =0.
\end{equation}
Integrating by parts, the second integrand become
\begin{align}
&\int_{x_0}^{x_1} \sum_{j=1}^n \lambda_i(x)\frac{\partial \varphi_i}{\partial \dot{y}_j} \delta \dot{y}_j dx=  \sum_{j=1}^n \lambda_i(x)\frac{\partial \varphi_i}{\partial \dot{y}_j} \delta {y}_j\Big\vert_{x_0}^{x_1} - \int_{x_0}^{x_1} \sum_{j=1}^n \frac{d}{dx}\Big( \lambda_i(x)\frac{\partial \varphi_i}{\partial \dot{y}_j}\Big) \delta y_j dx \nonumber \\
&= - \int_{x_0}^{x_1} \sum_{j=1}^n \frac{d}{dx}\Big( \lambda_i(x)\frac{\partial \varphi_i}{\partial \dot{y}_j}\Big) \delta y_j dx
\end{align}
and the third integrand become
\begin{align}
&\int_{x_0}^{x_1} \sum_{j=1}^n \lambda_i(x)\frac{\partial \varphi_i}{\partial \ddot{y}_j} \delta \ddot{y}_j dx =  \sum_{j=1}^n \lambda_i(x)\frac{\partial \varphi_i}{\partial \ddot{y}_j} \delta \dot{y}_j\Big\vert_{x_0}^{x_1} - \int_{x_0}^{x_1} \sum_{j=1}^n \frac{d}{dx}\Big( \lambda_i(x)\frac{\partial \varphi_i}{\partial \ddot{y}_j}\Big) \delta \dot{y}_j dx \nonumber \\
&= \sum_{j=1}^n -\frac{d}{dx}\Big( \lambda_i(x)\frac{\partial \varphi_i}{\partial \ddot{y}_j}\Big) \delta y_j\Big\vert_{x_0}^{x_1} + \int_{x_0}^{x_1} \sum_{j=1}^n \frac{d^2}{dx^2}\Big( \lambda_i(x)\frac{\partial \varphi_i}{\partial \ddot{y}_j}\Big) \delta y_j dx \nonumber \\
&= \int_{x_0}^{x_1} \sum_{j=1}^n \frac{d^2}{dx^2}\Big( \lambda_i(x)\frac{\partial \varphi_i}{\partial \ddot{y}_j}\Big) \delta y_j dx,
\end{align}
because variations $\delta y_i$ and its derivatives $\dot{y}_i$ at $x_0$ and $x_1$ points are null. Putting all together, we have:
\begin{align}
\int_{x_0}^{x_1} \lambda_i(x)\delta \varphi_i dx &= \int_{x_0}^{x_1} \sum_{j=1}^n \Big[ \lambda_i(x)\frac{\partial \varphi_i}{\partial y_j} - \frac{d}{dx}\Big( \lambda_i(x)\frac{\partial \varphi_i}{\partial \dot{y}_j}\Big)  + \frac{d^2}{dx^2}\Big( \lambda_i(x)\frac{\partial \varphi_i}{\partial \ddot{y}_j}\Big) \Big] \delta y_j dx.
\end{align}
Adding these $m$ equations to $\delta J$ variation
\begin{equation}
\delta J = \int_{x_0}^{x_1}  \sum_{j=1}^n \Big[ \frac{d^2}{dx^2}\Big(\frac{\partial F}{\partial \ddot{y}_j} \Big) - \frac{d}{dx}\Big(\frac{\partial F}{\partial \dot{y}_j} \Big) + \frac{\partial F}{\partial y_j}\Big]  \delta y_j dx 
\end{equation}
we obtain
\begin{equation}
\delta J = \int_{x_0}^{x_1}  \sum_{j=1}^n \Big[ \frac{d^2}{dx^2}\Big(\frac{\partial F^*}{\partial \ddot{y}_j} \Big) - \frac{d}{dx}\Big(\frac{\partial F^*}{\partial \dot{y}_j} \Big) + \frac{\partial F^*}{\partial y_j}\Big]  \delta y_j dx \label{EulerLagrangeEq1}
\end{equation}
being
\begin{equation}
F^*= F+\sum_{i=1}^m \lambda_i(x)\varphi_i.
\end{equation}
Variations $\delta y_i$ are not arbitrary since they are restricted with constraints $\varphi_i$. However, factors $\lambda_i(x)$ can be chosen to satisfy
\begin{equation}
\frac{d^2}{dx^2}\Big(\frac{\partial F^*}{\partial \ddot{y}_j} \Big) - \frac{d}{dx}\Big(\frac{\partial F^*}{\partial \dot{y}_j} \Big) + \frac{\partial F^*}{\partial y_j}=0 \;\;\; (j=1,2,...m),
\end{equation}
We have then a set of linear equations depending on
\begin{equation}
\lambda_i,\;\;\;\; \frac{\partial \lambda_i}{\partial x} \;\; \text{and}\;\; \frac{\partial^2 \lambda_i}{\partial x^2}.
\end{equation}
If $\delta y_j$, $(j=1,2,...m)$, are chosen, without any loss of generality, as the nonarbitrary variations then equations \ref{EulerLagrangeEq1} reduce to
\begin{equation}
\delta J = \int_{x_0}^{x_1}  \sum_{j=m+1}^n \Big[ \frac{d^2}{dx^2}\Big(\frac{\partial F^*}{\partial \ddot{y}_j} \Big) - \frac{d}{dx}\Big(\frac{\partial F^*}{\partial \dot{y}_j} \Big) + \frac{\partial F^*}{\partial y_j}\Big]  \delta y_j dx, \label{IndepVariations}
\end{equation}
where $\delta y_j$ $(j=m+1,m+2,...n)$ are now independent and allow to apply fundamental lemma of the calculus of variations and obtain:
\begin{equation}
\frac{d^2}{dx^2}\Big(\frac{\partial F^*}{\partial \ddot{y}_j} \Big) - \frac{d}{dx}\Big(\frac{\partial F^*}{\partial \dot{y}_j} \Big) + \frac{\partial F^*}{\partial y_j}=0 \;\;\; (j=m+1,m+2,...n).
\end{equation}
Thereby, functions $y_1(x), y_2(x)...y_n(x)$ that extreme the variational $J(y_1,y_2,...y_n)$ and constants $\lambda_1(x), \lambda_2(x)...\lambda_m(x)$ must satisfy the set of $n+m$ equations
\begin{equation}
\frac{d^2}{dx^2}\Big(\frac{\partial F^*}{\partial \ddot{y}_j} \Big) - \frac{d}{dx}\Big(\frac{\partial F^*}{\partial \dot{y}_j} \Big) + \frac{\partial F^*}{\partial y_j}=0 \;\;\; (j=1,2,...n)
\end{equation}
and
\begin{equation}
\varphi_i(x, y_1,y_2...y_n,\dot{y}_1,\dot{y}_2...\dot{y}_n,\ddot{y}_1,\ddot{y}_2...\ddot{y}_n)=0 \;\; (i=1,2,....m).
\end{equation}

\subsection{Lagrangian for n-VMVF systems}
So far, we extend the Lagrange equations up to the second order because of the $\mathbf{\ddot{r}}$ dependency in the constraints in Lagrange equations for $n$-VMVF systems. Also, the Lagrange method of the undetermined multipliers was extended to treat the second order constraints. We are in a position now to apply all previous results to the $n$-VMVF isolated systems and find the finals Lagrangians.

The second order Lagrange operators in Lorentzian and angular coordinates are:
\begin{equation}
\mathcal{L}_{n,\nu}  \equiv \Big[ -  \frac{d^2}{d\tau^2}\frac{\partial}{\partial \ddot{x}^\nu_n}  + \frac{d}{d\tau}\frac{\partial}{\partial \dot{x}^\nu_n} - \frac{\partial}{\partial x_n^\nu}\Big], \;\; x^\nu=\{x^1,x^2,x^3,x^4\} \label{lagrangetranslationExt}
\end{equation}
and
\begin{equation}
\mathcal{L}_{n,\xi}  \equiv \Big[ - \frac{d^2}{d\tau^2}\frac{\partial}{\partial \ddot{\xi}_n} + \frac{d}{d\tau}\frac{\partial}{\partial \dot{\xi}_n} - \frac{\partial}{\partial \xi_n}\Big], \;\; \xi=\{ \theta,\phi,\chi\}. \label{lagrangeRotationExt}
\end{equation}

Using relations \ref{relationsMatrix}, the angular extended Lagrangian eq \ref{lagrangeRotationExt}  takes the form:
\begin{align}
\mathcal{L}_{n,\xi_i} &=  - \frac{d^2}{d\tau^2} \Big[  D_{\;\xi_{n,i}}^\mu \square_{n,\ddot{\mu}} \Big] + \frac{d}{d\tau} \Big[  D_{\;\xi_{n,i}}^\mu \square_{n,\dot{\mu}}  +  2\frac{d}{d\tau}\Big(  D_{\;\xi_{n,i}}^\mu \Big) \square_{n,\ddot{\mu}} \Big]
\nonumber \\
&- \Big[ D_{\;\xi_{n,i}}^\mu \square_{n,\mu}  +  \frac{d}{d\tau}\Big(  D_{\;\xi_{n,i}}^\mu \Big)\square_{n,\dot{\mu}}  + \frac{d^2}{d\tau^2}\Big(  D_{\;\xi_{n,i}}^\mu \Big)\square_{n,\ddot{\mu}} \Big]
\nonumber \\
&= - \frac{d^2}{d\tau^2} \Big(  D_{\;\xi_{n,i}}^\mu \Big) \square_{n,\ddot{\mu}} - 2\frac{d}{d\tau} \Big(  D_{\;\xi_{n,i}}^\mu \Big) \frac{d}{d\tau} \Big( \square_{n,\ddot{\mu}} \Big) - D_{\;\xi_{n,i}}^\mu  \frac{d^2}{d\tau^2} \Big(\square_{n,\ddot{\mu}} \Big)
\nonumber \\
&+ \frac{d}{d\tau} \Big(  D_{\;\xi_{n,i}}^\mu \Big) \square_{n,\dot{\mu}}  +  D_{\;\xi_{n,i}}^\mu  \frac{d}{d\tau} \Big( \square_{n,\dot{\mu}} \Big) +  2\frac{d^2}{d\tau^2}\Big(  D_{\;\xi_{n,i}}^\mu \Big) \square_{n,\ddot{\mu}} 
\nonumber \\
&+  2\frac{d}{d\tau}\Big(  D_{\;\xi_{n,i}}^\mu \Big) \frac{d}{d\tau} \Big( \square_{n,\ddot{\mu}} \Big) -  D_{\;\xi_{n,i}}^\mu \square_{n,\mu}  -  \frac{d}{d\tau}\Big(  D_{\;\xi_{n,i}}^\mu \Big)\square_{n,\dot{\mu}}
\nonumber \\
&  - \frac{d^2}{d\tau^2}\Big(  D_{\;\xi_{n,i}}^\mu \Big)\square_{n,\ddot{\mu}}.
\nonumber \\
& = - D_{\;\xi_{n,i}}^\mu  \frac{d^2}{d\tau^2} \Big(\square_{n,\ddot{\mu}} \Big) +  D_{\;\xi_{n,i}}^\mu  \frac{d}{d\tau} \Big( \square_{n,\dot{\mu}} \Big) -  D_{\;\xi_{n,i}}^\mu \square_{n,\mu}.
\end{align}
Finally,
\begin{equation}
\mathcal{L}_{n,\xi_i} = D_{\;\xi_{n,i}}^\mu \Big[ -\frac{d^2}{d\tau^2} \Big(\square_{n,\ddot{\mu}} \Big) +  \frac{d}{d\tau} \Big( \square_{n,\dot{\mu}} \Big) - \square_{n,\mu} \Big]. \label{lagrangeRotationExt1}
\end{equation}

We can apply the undetermined multipliers method of Lagrange for rearranging the obtained equations of motions for $n$-VMVF systems resumed in equations \ref{TransLeastActionEq},\ref{RotLeastActionEq}, \ref{TransConstraintEq} and \ref{RotConstraintEq}. According to the previous section, solutions that minimize the former Lagrangian and satisfy two independent set of constraint equations for each set of coordinates, also minimize two Lagrangian:
\begin{equation}
L_1 = L_{sp} + \sum_{n} \lambda^\nu_n \Phi_{\nu_n} \qquad
\text{and}
\qquad
L_2 = L_{sp} + \sum_{i,n} \beta_{i_n} \Psi_{i_n}.
\end{equation}
Let's name the constructed Lagrangian with the Lorentzian coordinates as $L_T$ and the Lagrangian constructed using the angular as $L_R$. $T$ and $R$'s index stands for the translation and rotation transformations since they are the operation described with Lorentzian and angular coordinates, respectively. Also as both Lagrangians describe the physical system, we can write them as a single two-components Lagrangian like
\begin{equation}
L_{sys} \equiv 
\classoperator{L_T}{ L_R}= 
\classoperator{L_{sp} + \sum_{n} \lambda^\nu_n \Phi_{\nu_n}}
{L_{sp} + \sum_{i,n} \beta_{i_n} \Psi_{i_n}},\label{extLagran}
\end{equation}
where we introduce 7-$n$ independent constants $\lambda^\nu_n$ and $\beta_{i_n}$. Note that, while $\lambda^\nu_n$ constants need to be included in the covariant form, $\beta_{i_n}$ constant is invariant under Lorentz transformation. We use the matrix notation for expressing the general Lagrangian. The matrix representation remark the necessity of solving the two Lagrangians to find the solution for all the new degrees of freedom of the system
\begin{equation}
\{x_n^\nu\},\{\frac{\partial m_n}{\partial x_n^\mu }\},\{\frac{\partial A^\nu}{\partial x_n^\mu }\},\{\frac{\partial A^\nu}{\partial \dot{x}_n^\mu }\}.
\end{equation}
Thereby, the solutions that extreme Hamilton's least action principle and constants $\lambda^\nu_n$ and $\beta_{i_n}$ for $n$-VMVF systems must satisfy the equations, using the extended Lagrangian expressions eq. \ref{lagrangetranslationExt} and \ref{lagrangeRotationExt1},
\begin{equation}
\mathcal{L}_{n,\mu,\xi_i}L_{sys} \equiv \classoperator{
\mathcal{L}_{n,\mu} L_T}{
\mathcal{L}_{n,\xi_i} L_R} =
\classoperator{
\Big[ - \frac{d^2}{d\tau^2}\frac{\partial}{\partial \ddot{x}_n^\mu} 
+ \frac{d}{d\tau}\frac{\partial}{\partial \dot{x}_n^\mu} 
- \frac{\partial}{\partial x_n^\mu}\Big] L_T}
{
D_{\;\xi_{n,i}}^\mu \Big[ -  
\frac{d^2}{d\tau^2}\Big(  \square_{n,\ddot{\mu}} \Big) + 
\frac{d}{d\tau}\Big(  \square_{n,\dot{\mu}} \Big) - 
\square_{n,\mu} \Big] L_R}=0 \label{extEulerLagranEq}
\end{equation}
and
\begin{align}
&\Omega_{n,\mu,\xi_i} \equiv 
\classoperator
{\Phi_{\nu_n}}
{\Psi_{i_n}} 
= 0. \label{extEulerLagranConst}
\end{align}

$L_T$ and $L_R$ functions can be obtained substituting \ref{extEulerLagranConst} in \ref{extLagran} as
\begin{align}
L_T &= L_{sp} + \sum_{n} \lambda^\mu_n \Phi_{\mu_n}
\nonumber \\
&=  \sum_{n}   \frac{1}{2}m_{n} \dot{x}^\nu_{n} \dot{x}_{n;\nu} -A^\nu \dot{x}_{n;\nu} 
+ \sum_{n'\neq n} \lambda^\mu_{n} \Big[ (\dot{P}_{{n'},\mu} - Q_{{n'},\mu}) 
+ \frac{d}{d\tau} \Big(\frac{\partial A^\nu}{\partial \dot{x}_{n}^\mu}\Big) \dot{x}_{n';\nu} 
\nonumber \\
&+ \frac{\partial A^\nu}{\partial \dot{x}_{n}^\mu} \ddot{x}_{n';\nu}  
- \frac{\partial A^\nu}{\partial x_{n}^\mu}\dot{x}_{n';\nu}
+ \frac{\partial A_\mu}{\partial x_{n'}^\nu}\dot{x}_{n'}^\nu 
+ \frac{\partial A_\mu}{\partial \dot{x}_{n'}^\nu}\ddot{x}_{n'}^\nu  \Big] 
\nonumber \\
&=  \sum_{n}  \frac{1}{2}m_{n} \dot{x}^\nu_{n} \dot{x}_{n;\nu} -A^\nu \dot{x}_{n;\nu} 
+ \sum_{n'\neq n} \lambda^\mu_{n} \Big[  m_{n'} \ddot{x}_{{n'};\mu} + (\square_{n'}^\nu m_{n'} \dot{x}_{{n'};\nu}) \dot{x}_{{n'};\mu} 
\nonumber \\ 
- &\frac{\partial A^\nu}{\partial \dot{x}_{n'}^\mu}\ddot{x}_{{n'};\nu} 
- \frac{d}{d\tau}\Big(\frac{\partial A^\nu}{\partial \dot{x}_{n'}^\mu} \Big) \dot{x}_{{n'};\nu}
- \frac{\partial A_\mu}{\partial x_{n'}^\nu}\dot{x}_{n'}^\nu 
- \frac{\partial A_\mu}{\partial \dot{x}_{n'}^\nu}\ddot{x}_{n'}^\nu  
- \frac{1}{2}\frac{\partial m_{n'}}{\partial x_{n'}^\mu}\dot{x}^\nu_{n'} \dot{x}_{{n'};\nu} 
\nonumber \\
&+ \frac{\partial A^\nu}{\partial x_{n'}^\mu}\dot{x}_{n';\nu}
+ \frac{d}{d\tau} \Big(\frac{\partial A^\nu}{\partial \dot{x}_{n}^\mu}\Big) \dot{x}_{n';\nu} 
+ \frac{\partial A^\nu}{\partial \dot{x}_{n}^\mu} \ddot{x}_{n';\nu}  
- \frac{\partial A^\nu}{\partial x_{n}^\mu}\dot{x}_{n';\nu}
+ \frac{\partial A_\mu}{\partial x_{n'}^\nu}\dot{x}_{n'}^\nu 
+ \frac{\partial A_\mu}{\partial \dot{x}_{n'}^\nu}\ddot{x}_{n'}^\nu \Big],
\end{align}
to finally obtain 
\begin{align}
L_T &=  \sum_{n}  \frac{1}{2}m_{n} \dot{x}^\nu_{n} \dot{x}_{n;\nu} -A^\nu \dot{x}_{n;\nu} 
+ \sum_{n'\neq n} \lambda^\mu_{n} \Big[  m_{n'} \ddot{x}_{{n'};\mu} + (\square_{n'}^\nu m_{n'} \dot{x}_{{n'};\nu}) \dot{x}_{{n'};\mu} 
\nonumber \\ 
- &\frac{\partial A^\nu}{\partial \dot{x}_{n'}^\mu}\ddot{x}_{{n'};\nu} 
- \frac{d}{d\tau}\Big(\frac{\partial A^\nu}{\partial \dot{x}_{n'}^\mu} \Big) \dot{x}_{{n'};\nu}
- \frac{1}{2}\frac{\partial m_{n'}}{\partial x_{n'}^\mu}\dot{x}^\nu_{n'} \dot{x}_{{n'};\nu} 
+ \frac{\partial A^\nu}{\partial x_{n'}^\mu}\dot{x}_{n';\nu}
+ \frac{d}{d\tau} \Big(\frac{\partial A^\nu}{\partial \dot{x}_{n}^\mu}\Big) \dot{x}_{n';\nu} 
\nonumber \\
&+ \frac{\partial A^\nu}{\partial \dot{x}_{n}^\mu} \ddot{x}_{n';\nu}  
- \frac{\partial A^\nu}{\partial x_{n}^\mu}\dot{x}_{n';\nu}
\Big]  \label{extTransLagrangianAppx}
\end{align}
The rotation Lagrangian for $n$-VMVF systems is
\begin{align}
L_R &= L_{sp} + \sum_{i,n} \beta_{i_n} \Psi_{i_n}
\nonumber \\
&=  \sum_{n}  \frac{1}{2}m_{n} \dot{x}^\nu_{n} \dot{x}_{n;\nu} -A^\nu \dot{x}_{n;\nu} 
+ \sum_{i,n'\neq n} \beta_{i_n} \Big[ \dot{L}_{\;\xi_{{n'},i}} - T_{\;\xi_{{n'},i}}  + D_{\;\xi_{n,i}}^\mu \Big[ \frac{d}{d\tau} \Big(\frac{\partial A^\nu}{\partial \dot{x}_{n}^\mu}\Big) \dot{x}_{n';\nu} 
\nonumber \\  
&+ \frac{\partial A^\nu}{\partial \dot{x}_{n}^\mu} \ddot{x}_{n';\nu}
- \frac{\partial A^\nu}{\partial x_{n}^\mu}\dot{x}_{n';\nu} 
+ \frac{\partial A_\mu}{\partial x_{n'}^\nu}\dot{x}_{n'}^\nu 
+ \frac{\partial A_\mu}{\partial \dot{x}_{n'}^\nu}\ddot{x}_{n'}^\nu  \Big] 
\end{align}
or
\begin{align}
L_R &=  \sum_{n} \frac{1}{2}m_{n} \dot{x}^\nu_{n} \dot{x}_{n;\nu} - A^\nu \dot{x}_{n;\nu} + \sum_{i,n'\neq n} \beta_{i_n}   D_{\;\xi_{n',i}}^\mu \Big[ m_{n'} \ddot{x}_{{n'};\mu} + (\square_{n'}^\nu m_{n'} \dot{x}_{{n'};\nu}) \dot{x}_{{n'};\mu} 
\nonumber \\ 
- &\frac{\partial A^\nu}{\partial \dot{x}_{n'}^\mu}\ddot{x}_{{n'};\nu} 
- \frac{d}{d\tau}\Big(\frac{\partial A^\nu}{\partial \dot{x}_{n'}^\mu} \Big) \dot{x}_{{n'};\nu}
- \frac{\partial A_\mu}{\partial x_{n';\nu}}\dot{x}_{n';\nu} 
- \frac{\partial A_\mu}{\partial \dot{x}_{n';\nu}}\ddot{x}_{n';\nu}  
- \frac{1}{2}\frac{\partial m_{n'}}{\partial x_{n'}^\mu}\dot{x}^\nu_{n'} \dot{x}_{{n'};\nu} 
+ \frac{\partial A^\nu}{\partial x_{n'}^\mu}\dot{x}_{n';\nu} \Big] 
 \nonumber \\  
&+ D_{\;\xi_{n,i}}^\mu \Big[ \frac{d}{d\tau} \Big(\frac{\partial A^\nu}{\partial \dot{x}_{n}^\mu}\Big) \dot{x}_{n';\nu} 
+ \frac{\partial A^\nu}{\partial \dot{x}_{n}^\mu} \ddot{x}_{n';\nu}
- \frac{\partial A^\nu}{\partial x_{n}^\mu}\dot{x}_{n';\nu} 
+ \frac{\partial A_\mu}{\partial x_{n';\nu}}\dot{x}_{n';\nu} 
+ \frac{\partial A_\mu}{\partial \dot{x}_{n';\nu}}\ddot{x}_{n';\nu}  \Big] 
\label{extRotLagrangianAppx}
\end{align}

So far, we obtain the extended rotation and Translation Lagrangians. However, both Lagrangians depend on the defined degrees of freedom of this methodology, means the particle coordinates and masses and field derivative but also on the particle mass $m_n$ and the vector potential $A^\nu$. Because of that, we need to express both functions with expressions depending on their derivatives. Using Taylor series expansion, mass, and vector potential can be written as:
\begin{align}
m_n(x_n^\nu) &= m_n(0) + \frac{\partial m_n}{\partial x^\mu_n} \Big |_0 x_n^\mu + \frac{1}{2!} \frac{\partial^2 m_n}{\partial x^\mu_n x^\alpha_n}\Big |_0{x_n^\mu}^2 {x_n^\alpha}^2 ....
\nonumber \\
A^\nu(\{x^\mu\}, \{\dot{x}^\mu\}) &= A^\nu(0) + \sum_n \frac{\partial A^\nu}{\partial x^\mu_n} \Big |_0 x_n^\mu 
+ \frac{\partial A^\nu}{\partial \dot{x}^\mu_n} \Big |_0 \dot{x}_n^\mu 
+ \sum_{n'} \frac{1}{2!} \frac{\partial^2 A^\nu}{\partial x^\mu_n x^\alpha_{n'}}\Big |_0 {x_n^\mu}^2 {x_{n'}^\alpha}^2 
\nonumber \\
&+ \frac{1}{2!} \frac{\partial^2 A^\nu}{\partial \dot{x}^\mu_n \dot{x}^\alpha_{n'}}\Big |_0 (\dot{x}_n^\mu)^2 (\dot{x}_{n'}^\alpha)^2 
+ 2\frac{1}{2!} \frac{\partial^2 A^\nu}{\partial x^\mu_n \dot{x}^\alpha_{n'}}\Big |_0 (x_n^\mu)^2 (\dot{x}_{n'}^\alpha)^2 ....
\end{align}

If we consider series expansion up to linear terms, the final extended Lagrangians have the form:
\begin{align}
L_T &=  \sum_{n}  \frac{1}{2}(m_n(0) + \frac{\partial m_n}{\partial x^\mu_n}  x_n^\mu) \dot{x}^\nu_{n} \dot{x}_{n;\nu} -(A^\nu(0) + \frac{\partial A^\nu}{\partial x^\mu_n} x_n^\mu 
+ \frac{\partial A^\nu}{\partial \dot{x}^\mu_n} \dot{x}_n^\mu ) \dot{x}_{n;\nu} 
+ \sum_{n'\neq n} \lambda^\mu_{n} \Big[  (m_{n'}(0) 
\nonumber \\ 
&+ \frac{\partial m_{n'}}{\partial x^\mu_{n'}} x_{n'}^\mu) \ddot{x}_{{n'};\mu} 
+ (\square_{n'}^\nu m_{n'} \dot{x}_{{n'};\nu}) \dot{x}_{{n'};\mu} 
- \frac{\partial A^\nu}{\partial \dot{x}_{n'}^\mu}\ddot{x}_{{n'};\nu} 
- \frac{d}{d\tau}\Big(\frac{\partial A^\nu}{\partial \dot{x}_{n'}^\mu} \Big) \dot{x}_{{n'};\nu}
- \frac{1}{2}\frac{\partial m_{n'}}{\partial x_{n'}^\mu}\dot{x}^\nu_{n'} \dot{x}_{{n'};\nu} 
\nonumber \\
&+ \frac{\partial A^\nu}{\partial x_{n'}^\mu}\dot{x}_{n';\nu}
+ \frac{d}{d\tau} \Big(\frac{\partial A^\nu}{\partial \dot{x}_{n}^\mu}\Big) \dot{x}_{n';\nu} 
+ \frac{\partial A^\nu}{\partial \dot{x}_{n}^\mu} \ddot{x}_{n';\nu}  
- \frac{\partial A^\nu}{\partial x_{n}^\mu}\dot{x}_{n';\nu}
\Big]   \label{extTransLagrangianAppx1}
\end{align}
and 
\begin{align}
L_R &=  \sum_{n} \frac{1}{2}(m_n(0) + \frac{\partial m_n}{\partial x^\mu_n}  x_n^\mu) \dot{x}^\nu_{n} \dot{x}_{n;\nu} -(A^\nu(0) + \frac{\partial A^\nu}{\partial x^\mu_n} x_n^\mu 
+ \frac{\partial A^\nu}{\partial \dot{x}^\mu_n} \dot{x}_n^\mu ) \dot{x}_{n;\nu} + 
\nonumber \\ 
&\sum_{i,n'\neq n} \beta_{i_n}   D_{\;\xi_{n',i}}^\mu \Big[ (m_{n'}(0) 
+ \frac{\partial m_{n'}}{\partial x^\nu_{n'}} x_{n'}^\nu) \ddot{x}_{{n'};\mu} + (\square_{n'}^\nu m_{n'} \dot{x}_{{n'};\nu}) \dot{x}_{{n'};\mu} 
- \frac{\partial A^\nu}{\partial \dot{x}_{n'}^\mu}\ddot{x}_{{n'};\nu} 
- \frac{d}{d\tau}\Big(\frac{\partial A^\nu}{\partial \dot{x}_{n'}^\mu} \Big) \dot{x}_{{n'};\nu}
\nonumber \\
&- \frac{\partial A_\mu}{\partial x_{n';\nu}}\dot{x}_{n';\nu} 
- \frac{\partial A_\mu}{\partial \dot{x}_{n';\nu}}\ddot{x}_{n';\nu}  
- \frac{1}{2}\frac{\partial m_{n'}}{\partial x_{n'}^\mu}\dot{x}^\nu_{n'} \dot{x}_{{n'};\nu} 
+ \frac{\partial A^\nu}{\partial x_{n'}^\mu}\dot{x}_{n';\nu} \Big] 
 \nonumber \\  
&+ D_{\;\xi_{n,i}}^\mu \Big[ \frac{d}{d\tau} \Big(\frac{\partial A^\nu}{\partial \dot{x}_{n}^\mu}\Big) \dot{x}_{n';\nu} 
+ \frac{\partial A^\nu}{\partial \dot{x}_{n}^\mu} \ddot{x}_{n';\nu}
- \frac{\partial A^\nu}{\partial x_{n}^\mu}\dot{x}_{n';\nu} 
+ \frac{\partial A_\mu}{\partial x_{n';\nu}}\dot{x}_{n';\nu} 
+ \frac{\partial A_\mu}{\partial \dot{x}_{n';\nu}}\ddot{x}_{n';\nu}  \Big] .
 \label{extRotLagrangianAppx1}
\end{align}

Also the constraint equations \ref{TransConstraintEq} and \ref{RotConstraintEq} are modified substituting the approximation for particle mass and field functions like
\begin{align}
\Phi_{\mu_n}^{(L)} = \sum_{{n'} \neq n}  &\Big[
\Big(\square_{n'}^\alpha m_n' \dot{x}_{n';\alpha}\Big)g^\nu_\mu
- \frac{1}{2}\frac{\partial m_{n'}}{\partial x_{n'}^\mu}\dot{x}^\nu_{n'}
+ \frac{d}{d\tau} \Big(\frac{\partial A^\nu}{\partial \dot{x}_{n}^\mu}
- \frac{\partial A^\nu}{\partial \dot{x}_{n'}^\mu}
\Big) 
+ \frac{\partial A^\nu}{\partial x_{n'}^\mu}
- \frac{\partial A^\nu}{\partial x_{n}^\mu}
\Big]\dot{x}_{n';\nu}
\nonumber \\
&+ \Big[ 
\Big(m_{n'}(0) + \frac{\partial m_{n'}}{\partial x^\mu_{n'}}\Big) g^\nu_\mu + \frac{\partial A^\nu}{\partial x_{n}^\mu}
- \frac{\partial A^\nu}{\partial x_{n'}^\mu}
+ \frac{\partial A_\mu}{\partial x_{n'}^\nu}
\Big] \ddot{x}_{n';\nu}
\label{TransConstraintEqApprx}
\end{align}
and
\begin{align}
\Psi_{i_n}^{(L)}  = \sum_{{n'} \neq n} &\Big\{
 D_{\;\xi_{n',i}}^\mu \Big[ 
\Big(\square_{n'}^\alpha m_n' \dot{x}_{n';\alpha}\Big)g^\nu_\mu
- \frac{1}{2}\frac{\partial m_{n'}}{\partial x_{n'}^\mu}\dot{x}^\nu_{n'}
-\frac{d}{d\tau} \Big(
\frac{\partial A^\nu}{\partial \dot{x}_{n'}^\mu}
\Big) 
- \frac{\partial A_\mu}{\partial x_{n';\nu}}
+ \frac{\partial A^\nu}{\partial x_{n'}^\mu}
 \Big]
 \nonumber \\
&+D_{\;\xi_{n,i}}^\mu \Big[
\frac{d}{d\tau} \Big(
\frac{\partial A^\nu}{\partial \dot{x}_{n}^\mu}
\Big) 
+ \frac{\partial A_\mu}{\partial x_{n';\nu}}
- \frac{\partial A^\nu}{\partial x_{n}^\mu} 
  \Big]
\Big\} \dot{x}_{n';\nu}
\nonumber \\
+& \Big\{
 D_{\;\xi_{n',i}}^\mu \Big[ 
\Big(m_{n'}(0) + \frac{\partial m_{n'}}{\partial x^\mu_{n'}} \Big) g^\nu_\mu - \frac{\partial A^\nu}{\partial x_{n'}^\mu}
  \Big]
 +D_{\;\xi_{n,i}}^\mu \Big[ 
\frac{\partial A^\nu}{\partial x_{n}^\mu}
+\frac{\partial A_\mu}{\partial x_{n';\nu}}
\Big]
\Big\} \ddot{x}_{n';\nu}
\label{RotConstraintEqApprx}
\end{align}

We refer these constraints like $\Phi_{\mu_n}^{(L)} $ $\Psi_{i_n}^{(L)}$, indicating the Lagrange picture, so we won't confuse it with the Hamilton picture that we study in the next chapter.

Under an external potential depending on particle positions, the extended Lagrangian equation should have have the form:
\begin{equation}
\mathcal{L}_{n,\mu,\xi_i}L \equiv 
\classoperator{
\mathcal{L}_{n,\mu} (L_T - V(x^\nu))}
{\mathcal{L}_{n,\xi_i} (L_R - V(\xi_i))} =0
\;\;\;\;\;\text{and } \;\;\;\;
\Omega_{n,\mu,\xi_i} \equiv 
\classoperator
{\Phi_{\nu_n}}{ 
\Psi_{i_n}} = 0
\label{extEulerLagranEqV}
\end{equation}

Constraint equation has been included on Lagrangian using Lagrange multiplier method increasing the new variables $\lambda^\mu_{n}$ and $\beta_{i_n}$. Some constraints were used in the generation of previous results like some equations of \ref{relConstraint1}
\begin{equation*}
\frac{\partial m_n}{x^\nu_{n'}} \equiv \frac{\partial m_n}{x^\nu_{n'}}\delta_{nn'}, \qquad  \frac{\partial m_n}{\dot{x}^\nu_{n'}} =0 \qquad  \frac{\partial A^0}{\partial \dot{x}^\nu_{n}}=0.
\end{equation*}
However, not all constraint equations were added. Indeed, relativistic constraints equations \ref{relatRelation}, the last equations of \ref{relConstraint1} and \ref{mass0compConstraint}:
\begin{align}
&x^\nu_n x_{\nu;n} =R^2_n , \qquad \partial_\nu A^\nu=0\;\; \forall\; n 
\nonumber \\
&\frac{1}{2}\frac{\partial m_n}{\partial x^0_n}\dot{x}^0_n \dot{x}_\mu \dot{x}^\mu + m_n\ddot{x}_{\mu;n}\dot{x}^\mu_n=0. \label{relConstraintFinal}
\end{align}
have a different nature and are related to the inertial frame of references. In this case, we chose to adhere to the Dirac's idea of treating them as weak constraints and include these relations after all derivative process have been carrying out.
\newpage

\section{Hamilton theory for $n$-VMVF systems.}\label{ClassicalHamSection}
In the previous section, we obtained a set of equations needed to solve the dynamics of $n$-VMVF systems. They are the result of applying the extension of the Lagrange operator up to second order for two independent set of coordinates over a staring Lagrangian for particle systems with fixed masses, which outcome in a set of constraints. Using those constraints, we construct two finals extended Lagrangians with a $\ddot{x}^\nu$ dependency with the Lagrange multiplier method. After some approximations, we were able to obtain a set of equations and the same numbers of variables including the particle masses and the field derivatives.

On the other hand, while Lagrangian has no physical meaning, Hamiltonian provide more in-depth knowledge of the classical mechanic structure and sets equals status for coordinates and momenta as independent variables. Hamiltonian is related to essential system features such as energy and also provide significant relations between symmetry and conservation laws. From Hamilton theory, we can define infinitesimal canonical transformations and their generators. Finding the classical generators is the primary objective of our classical approach, so we can use them in a quantum theory that includes particle masses and field variations.

\subsection{Ostrogradsky's Hamiltonian construction for Two Derivative Lagrangian}
Theories with higher order Lagrangians have been explored along the evolution of physics. Individually, theories of second-order Lagrangian rise a remarkable interest because they are renormalizable \cite{PhysRevD.16.953} in four dimensions. For the treatment of such higher derivative systems, Ostrogradsky generalizes the construction of the Hamilton function \cite{Ostrogradsky:1850fid}. We briefly expose his main ideas. 

Let us consider the extended Lagrange equation for a single particle \ref{extLagrangeEq2Order}
\begin{equation*}
\frac{d^2}{d t^2}\Big( \frac{\partial L}{\partial \ddot{q}}\Big) - \frac{d}{dt}\Big( \frac{\partial L}{\partial \dot{q}}\Big) + \frac{\partial L}{\partial q} =0
\end{equation*}
describing a system whose Lagrange function $L(q,\dot{q},\ddot{q})$ depends nondegenerately upon $\ddot{q}$, which means imply that the Hessian $\frac{\partial^2 L}{\partial \ddot{q}^2}\neq 0$. In this case, the four derivative term can be expressed as function of the others as
\begin{equation}
\ddddot{q} = \ddddot{q}( \dddot{q}, \ddot{q}, \dot{q},q), \label{ortroNonDeg0}
\end{equation}
or what is the same, the solution depend on four quantities of initial data
\begin{equation}
q = q(\dddot{q}_0, \ddot{q}_0, \dot{q}_0,q_0,t). \label{ortroNonDeg}
\end{equation}
This solution indicates the existences of four canonical variables in the phase space. Ostrogradsky \cite{Ostrogradsky:1850fid} propose to define these variables as
\begin{align}
Q_1 \equiv q, \qquad 
Q_2 \equiv \dot{q}, \qquad 
P_1 \equiv \frac{\partial L}{\partial \dot{q}} - \frac{d}{dt} \frac{\partial L}{\partial \ddot{q}}, \qquad
P_2 \equiv \frac{\partial L}{\partial \ddot{q}}\;\;.
\end{align}
The nondegeneracy of the Lagrange function, implies that, $\ddot{q}$ can be solve in terms of $\ddot{q} = A(Q_1, Q_2, P_2)$ excluding momentum $P_1$ which is only needed for the third derivative.

Ostrogradsky’s Hamiltonian is obtained using the Legendre transformation
\begin{equation}
H(Q_1,Q_2, P_1, P_2) = P_1 Q_2 + P_2 A(Q_1,Q_2, P_2) − L(Q_1,Q_2,A(Q_1,Q_2, P_2). \label{OrtoHamiltonian}
\end{equation}
The Hamilton equations are
\begin{align}
\dot{Q}_1 &= \frac{\partial H}{\partial P_1} = Q_2
\\
\dot{Q}_2 &= \frac{\partial H}{\partial P_2} = A + P_2 \frac{\partial A}{\partial P_2} -
\frac{\partial L}{\partial \ddot{q}} \frac{\partial A}{\partial P_2} = A
\\
\dot{P}_2 &= - \frac{\partial H}{\partial Q_2} = -P_1 -P_2 \frac{\partial A}{\partial Q_2}
+ \frac{\partial L}{\partial \dot{q}} 
+ \frac{\partial L}{\partial \ddot{q}} \frac{\partial A}{\partial Q_2}
= -P_1 + \frac{\partial L}{\partial \dot{q}}
\\
\dot{P}_1 &= - \frac{\partial H}{\partial Q_1} =  -P_2 \frac{\partial A}{\partial Q_1} 
+ \frac{\partial L}{\partial q} 
+ + \frac{\partial L}{\partial \ddot{q}} \frac{\partial A}{\partial Q_1}
= \frac{\partial L}{\partial q} 
\end{align}
The first two equations reproduce the phase space transformation $\dot{q} = Q_2$ and $\ddot{q} = \dot{Q}_2$ while the others show that the evolution of momentum $P_1$ depends on the evolution of $P_2$. The equations exhibit the time evolution of the system generated by Ostrogradsky’s Hamiltonian, which is also the conserved Noether current if Lagrangian have no explicit dependence of time \cite{Woodard:2015zca}.

The Ostrogradsky’s Hamiltonian \ref{OrtoHamiltonian} is linear in the canonical momentum $P_1$, which means that there is an instability related to the fact that the Lagrangian depends on fewer coordinates than the defined canonical coordinates. Besides, from our point of view, this method also mixture the derivatives $\frac{\partial^{(n)} L}{\partial q^{(n)}}$, what means that the new canonical variables involve combinations of values $\dddot{q}_0, \ddot{q}_0, \dot{q}_0,q_0$ in equation \ref{ortroNonDeg}. This feature may unnecessarily obscure the physical meaning of the future canonical variables in systems where each addend depend only on one of the variable's derivative ($\ddot{q}, \dot{q},q$) like the one we are studying.

\subsection{Extended Hamilton equations}
Our proposal for obtaining the canonical variables for an extended Lagrangian start by writing the total time derivative of a general extended Lagrangian as:
\begin{equation}
\frac{dL}{dt} = \sum_i \frac{\partial L}{\partial q_i} \frac{d q_i}{dt} +  \frac{\partial L}{\partial \dot{q}_i} \frac{d\dot{q}_i}{dt} +  \frac{\partial L}{\partial \ddot{q}_i} \frac{d \ddot{q}_i}{dt} +  \frac{\partial L}{\partial t} 
\end{equation} 
or, using the extended Lagrange equations \ref{extLagrangeEq2Order}
\begin{equation*}
\frac{d^2}{d t^2}\Big( \frac{\partial L}{\partial \ddot{q}_i}\Big) - \frac{d}{dt}\Big( \frac{\partial L}{\partial \dot{q}_i}\Big) + \frac{\partial L}{\partial q_i} =0
\end{equation*}
\begin{equation}
\frac{dL}{dt} = \sum_i \Big[ \frac{d}{dt} \Big( \frac{\partial L}{\partial \dot{q}_i}\Big) - \frac{d^2}{dt^2}\Big( \frac{\partial L}{\partial \ddot{q}_i} \Big) \Big] \dot{q}_i+ \frac{\partial L}{\partial \dot{q}_i} \frac{d\dot{q}_i}{dt} +  \frac{\partial L}{\partial \ddot{q}_i} \frac{d \ddot{q}_i}{dt} +  \frac{\partial L}{\partial t}.
\end{equation}
After some derivative steps we obtain:
\begin{equation}
\frac{d L}{dt} = \sum_i \frac{d}{dt} \Big[ \frac{\partial L}{\partial \dot{q}_i} \dot{q}_i - \frac{d}{dt}  \Big( \frac{\partial L}{\partial \ddot{q}_i}\Big) \dot{q}_i + \frac{\partial L}{\partial \ddot{q}_i}\ddot{q}_i \Big]+\frac{\partial L}{\partial t}
\end{equation}
or
\begin{equation}
\sum_i \frac{d}{dt} \Big[ \frac{\partial L}{\partial \dot{q}_i} \dot{q}_i - \frac{d}{dt}  \Big( \frac{\partial L}{\partial \ddot{q}_i}\Big) \dot{q}_i + \frac{\partial L}{\partial \ddot{q}_i}\ddot{q}_i  - L \Big] +\frac{\partial L}{\partial t} = 0.
\end{equation}
We can define an extended energy function \\ $h(t, q_1,q_2...q_n,\dot{q}_1,\dot{q}_2...\dot{q}_n,\ddot{q}_1,\ddot{q}_2...\ddot{q}_n)$ as:
\begin{equation}
h= \frac{\partial L}{\partial \dot{q}_i} \dot{q}_i - \frac{d}{dt}  \Big( \frac{\partial L}{\partial \ddot{q}_i}\Big) \dot{q}_i + \frac{\partial L}{\partial \ddot{q}_i}\ddot{q}_i  - L
\end{equation}
where
\begin{equation}
\frac{dh}{dt}=-\frac{\partial L}{\partial t}.
\end{equation}
If Lagrangian doesn't explicit depend on time, then $h$ will remain constant in time. We can define the first and second order momentums
\begin{equation}
p_i=\frac{\partial L}{\partial \dot{q}_i} \;\;\; \text{and} \;\;\;\; s_i=\frac{\partial L}{\partial \ddot{q}_i}
\end{equation}
so the energy function is written as
\begin{equation}
h= \sum_i p_i \dot{q}_i - \dot{s}_i\dot{q}_i + s_i\ddot{q}_i  - L. \label{extEnergyFunction}
\end{equation}
If we compute the total differential for the energy function
\begin{align}
dh = &\sum_i p_i d\dot{q}_i +  \dot{q}_i  dp_i - \dot{s}_id\dot{q}_i - \dot{q}_id\dot{s}_i  + s_id\ddot{q}_i + \ddot{q}_ids_i  
\nonumber \\
&- \Big[\frac{\partial L}{\partial q_i} dq_i + \frac{\partial L}{\partial \dot{q}_i} d\dot{q}_i + \frac{\partial L}{\partial \ddot{q}_i} d\ddot{q}_i + \frac{\partial L}{\partial t} dt\Big]
\end{align}
and substitute the momentums $p$ and $s$ definitions, we obtain
\begin{equation}
dh + \sum_i d(\dot{s}_i\dot{q}_i) = \sum_i -(\dot{p}_i-\ddot{s}_i) dq_i +  \dot{q}_i  dp_i  + \ddot{q}_i ds_i -  \frac{\partial L}{\partial t} dt,
\end{equation}
were we use the relation
\begin{equation}
\frac{\partial L}{\partial q_i}= \frac{d}{dt} \Big( \frac{\partial L}{\partial \dot{q}_i}\Big) - \frac{d^2}{dt^2}\Big( \frac{\partial L}{\partial \ddot{q}_i} \Big)=\dot{p}_i-\ddot{s}_i.
\end{equation}
We can define the function
\begin{equation}
H = h + \sum_i \dot{s}_i\dot{q}_i = \sum_i p_i \dot{q}_i + s_i\ddot{q}_i  - L. \label{ExtHamiltonDef}
\end{equation}
as a function depending only on variables $q,p,s$, whose differential is
\begin{equation}
d H = \sum_i -(\dot{p}_i-\ddot{s}_i)dq_i +   \dot{q}_i  dp_i + \ddot{q}_ids_i -  \frac{\partial L}{\partial t} dt.
\end{equation}
The differential of function $H$ can be also written as
\begin{equation}
dH = \sum_i \frac{\partial H}{\partial q_i} dq_i + \frac{\partial H}{\partial p_i} dp_i + \frac{\partial H}{\partial s_i} ds_i + \frac{\partial H}{\partial t} dt
\end{equation}
from we obtain $3n+1$ relations:
\begin{align}
&\frac{\partial H}{\partial q_i} =  -(\dot{p}_i-\ddot{s}_i)
\nonumber \\
&\frac{\partial H}{\partial p_i} = \dot{q}_i
\nonumber \\
&\frac{\partial H}{\partial s_i} = \ddot{q}_i
\nonumber \\
&\frac{\partial H}{\partial t} = -  \frac{\partial L}{\partial t} \label{ExtHamiltonEq}
\end{align}
Mathematically speaking, the set of $3n+1$ second-order extended Hamilton equations replace the set of $n+1$ four-order extended Lagrange equations.

We are in the presence of a problem where the energy function is different from Hamiltonian. This issue is in contrast to the classical mechanic where both functions are the same if there is no explicit time dependence on Hamiltonian. If such dependency is missing, the energy function is related to the energy of the system as the quantity that remains invariant with time, while Hamiltonian is the function from we can study the evolution of every degree of freedom in the system.

Since index summation $n$ stands for particle iteration on both extended Lagrangian and Hamiltonian, we can also define the particle energy function as:
\begin{equation}
h_n = H_n - \dot{s}_{n}\dot{q}_{n}. \label{extPartEnergyFunction}
\end{equation}
In this case, particle energies are no longer constant with time. Only its summation over all particles remains invariant.

Let us summarise the so far obtained results:
\begin{table}[h]
\caption{Summary of the number of equation, variables and degree of freedom for the extended classical theory for Lagrange and Hamilton} \label{extSummaryTable}
\begin{center}
\begin{tabular}{@{} p{15mm}  p{7cm} p{4cm}  p{4cm} @{}}
\hline \hline
 & Equations  & Variables & Degree of freedom	\\ 
\hline
Lagrange
&
$n$-four order equations:
\begin{equation*}
\frac{d^2}{dx^2}\Big( \frac{\partial L}{\partial \ddot{q}_i}\Big) - \frac{d}{dx}\Big( \frac{\partial L}{\partial \dot{q}_i}\Big) + \frac{\partial L}{\partial q_i} =0
\end{equation*}
& 
5-$n$ +1 initial values:
\newline
$\ddddot{q}_{i_0},\dddot{q}_{i_0}, \ddot{q}_{i_0}, \dot{q}_{i_0},q_{i_0},t$
& 5$n$+1 - 1$n$ = 4$n$+1\\ 
\hline
Hamilton
&
$3n$-second order equations:
$
\begin{array} {r@{} l@{}}
\dfrac{\partial H}{\partial q_i} {}&=  -(\dot{p}_i-\ddot{s}_i)
\\ 
\dfrac{\partial H}{\partial p_i} {}&= \dot{q}_i
\\ 
\dfrac{\partial H}{\partial s_i} {}&= \ddot{q}_i
\end{array}$
& 
8-$n$ +1 initial values:
\newline
$\ddot{q}_{i_0}, \dot{q}_{i_0},q_{i_0}, \dot{p}_{i_0},p_{i_0}, 
\newline
\ddot{s}_{i_0}, \dot{s}_{i_0},s_{i_0},t$
& 8$n$+1 - 3$n$ = 5$n$+1\\ 
\hline
\end{tabular} 
\end{center}
\end{table}

Note that the proposal extended Hamiltonian have $n$ more degrees of freedom that the Lagrange approach, which resembles the Ostrogradsky's instability. This fact means that another set of equations is needed for the correct description of the system. The requested equations are obtained in the next sections.

\subsection{Hamilton equation for n-VMVF systems}
In this section, we obtain the extended Hamiltonian for isolated systems contained $n$-VMVF systems. Similar to the classical Hamiltonian, we construct the extended Hamiltonian by the following the next steps:
\begin{itemize}
\item Construct the Lagrangian with the appropriate set of coordinates.
\item Define the generalized first order momentum $p_i$ and second order momentum $s_i$.
\item Substitute $p_i$ and $s_i$ in equation \ref{ExtHamiltonDef}. The extended Hamiltonian is now a mixed function of variable $q_i$, $\dot{q}_i$, $\ddot{q}_i$ $p_i$, $s_i$ and $t$. 
\item Substitute $\dot{q}_i$ and $\ddot{q}_i$ from the inversion of the definition of $p_i$ and $s_i$ so the extended Hamiltonian function depends only on variables $q_i$,$p_i$, $s_i$ and $t$.
\end{itemize}
Following the previous methodology, we obtain two Hamiltonians for solving the $n$-VMVF isolated systems, according to the two constructed Lagrangians in the previous chapter. We also obtain two extended set of Hamilton equations in both set of coordinates for describing the position vector. Let's name these Hamiltonians as $H_T$ and $H_R$ related to the translation Lagrangian $L_T$ and the rotation Lagrangian $L_R$ respectively.

\subsubsection*{Hamiltonian for n-VMVF systems using Lorentzian coordinates}
Recalling equation \ref{extTransLagrangianAppx1}, the translation Lagrangian is,
\begin{align*}
L_T &=  \sum_{n}  \frac{1}{2}(m_n(0) 
+ \frac{\partial m_n}{\partial x^\mu_n}  x_n^\mu) \dot{x}^\nu_{n} \dot{x}_{n;\nu} 
-(A^\nu(0) + \frac{\partial A^\nu}{\partial x^\mu_n} x_n^\mu 
+ \frac{\partial A^\nu}{\partial \dot{x}^\mu_n} \dot{x}_n^\mu ) \dot{x}_{n;\nu} 
+ \sum_{n'\neq n} \lambda^\mu_{n} \Big[  (m_{n'}(0) 
\nonumber \\ 
&+ \frac{\partial m_{n'}}{\partial x^\mu_{n'}} x_{n'}^\mu) \ddot{x}_{{n'};\mu} 
+ (\square_{n'}^\nu m_{n'} \dot{x}_{{n'};\nu}) \dot{x}_{{n'};\mu} 
- \frac{\partial A^\nu}{\partial \dot{x}_{n'}^\mu}\ddot{x}_{{n'};\nu} 
- \frac{d}{d\tau}\Big(\frac{\partial A^\nu}{\partial \dot{x}_{n'}^\mu} \Big) \dot{x}_{{n'};\nu}
- \frac{1}{2}\frac{\partial m_{n'}}{\partial x_{n'}^\mu}\dot{x}^\nu_{n'} \dot{x}_{{n'};\nu} 
\nonumber \\
&+ \frac{\partial A^\nu}{\partial x_{n'}^\mu}\dot{x}_{n';\nu}
+ \frac{d}{d\tau} \Big(\frac{\partial A^\nu}{\partial \dot{x}_{n}^\mu}\Big) \dot{x}_{n';\nu} 
+ \frac{\partial A^\nu}{\partial \dot{x}_{n}^\mu} \ddot{x}_{n';\nu}  
- \frac{\partial A^\nu}{\partial x_{n}^\mu}\dot{x}_{n';\nu}
\Big].   
\end{align*}
We can group the $\dot{x}_{n;\nu} $ and $\ddot{x}_{n;\nu} $ terms like
\begin{align}
\mathcal{L}_T &= \sum_n \Big\{ \Big[ \frac{1}{2}\Big(m_n(0) + \frac{\partial m_n}{\partial x^\mu_n}  x_n^\mu\Big )
 + \sum_{n'\neq n} \lambda^\mu_{n'} \Big( 
 \square_{n}^\gamma m_{n} g_{\gamma \mu}
- \frac{1}{2}\frac{\partial m_{n}}{\partial x_{n}^\mu} \Big) \Big] \dot{x}^\nu_{n}
\nonumber \\  
&+ \sum_{n'\neq n} \lambda^\mu_{n'} \Big[
\frac{d}{d\tau} \Big(\frac{\partial A^\nu}{\partial \dot{x}_{n}^\mu}
- \frac{\partial A^\nu}{\partial \dot{x}_{n'}^\mu} \Big)
+ \frac{\partial A^\nu}{\partial x_{n'}^\mu} - \frac{\partial A^\nu}{\partial x_{n}^\mu}
\Big] 
-\Big[A^\nu(0) + \frac{\partial A^\nu}{\partial x^\mu_n} x_n^\mu 
+ \frac{\partial A^\nu}{\partial \dot{x}^\mu_n} \dot{x}_n^\mu\Big]
\Big\} \dot{x}_{n';\nu}
\nonumber \\  
&+ \sum_{\substack{n\\n'\neq n}}
\lambda^\mu_{n'} \Big[ 
\Big( m_n(0) + \frac{\partial m_n}{\partial x^\mu_n}  x_n^\mu \Big) g_\mu^\nu
+ \frac{\partial A^\nu}{\partial \dot{x}_{n'}^\mu}
- \frac{\partial A^\nu}{\partial \dot{x}_{n}^\mu}
\Big]  \ddot{x}_{n';\nu}
\end{align}

After an straight derivation we obtain the $\alpha$ component of the momentum $p$ for particle $l$:
\begin{align}
{{p_T}_l}^\alpha  =& \sum_n \Big \{ \Big [ 
\Big( m_n(0) 
+ \frac{\partial m_n}{\partial x^\mu_n}x_n^\mu 
- \sum_{n'\neq n} {\lambda_n'}^\mu \frac{\partial m_n}{\partial{x_n}^\mu} 
\Big )g^\alpha_\nu 
+  \sum_{n'\neq n} ({\lambda_{n'}}^\alpha {\square_n}_\nu m_n 
+ {\lambda_{n'}}_\nu \frac{\partial m_n}{\partial {x_n}_\alpha} ) 
\Big] \delta_{n,l}
\nonumber \\
&- \Big( \frac{\partial A^\alpha}{\partial \dot{x}^\nu_n} 
+ \frac{\partial A_\nu}{\partial \dot{x}_{n;\alpha}}
\Big) 
\Big \} {\dot{x}_n}^\nu
- \frac{\partial A^\alpha}{\partial x^\mu_n} x_n^\mu 
+ \Big \{ \sum_{n'\neq n} {\lambda_{n'}}^\mu \Big[ \frac{d}{d \tau}\Big( \frac{\partial A^\alpha}{\partial {\dot{x}_{n'}}^\mu} 
- \frac{\partial A^\alpha}{\partial {\dot{x}_n}^\mu}\Big)  
+ \frac{\partial A^\alpha}{\partial {x_n}^\mu} 
- \frac{\partial A^\alpha}{\partial {x_{n'}}^\mu} \Big] 
\nonumber \\
&- A^\alpha(0) \Big \}\delta_{nl}
\label{TrasLinearP}
\end{align}
while the $\alpha$ component of the variable $s$ for particle $l$ is:
\begin{align}
{{s_T}_l}^\alpha & \equiv \frac{\partial L_T}{\partial \ddot{x_l}_\alpha} = \sum_{n\neq l} 
\Big \{
{\lambda_n}^\alpha \Big[ 
m_l(0) 
+ \frac{\partial m_{l}}{\partial x^\nu_{l}} x^\nu_{l}
\Big]
+
{\lambda_n}^\mu 
\Big[ 
\frac{\partial A^\alpha}{\partial {\dot{x}_n}^\mu} 
- \frac{\partial A^\alpha}{\partial {\dot{x}_l}^\mu} 
\Big]
\Big \}. 
\end{align}

Using $p_n^\nu$ and $s_n^\nu$ definitions we have
\begin{align}
\mathcal{L}_T &= \sum_n p_n^\nu \dot{x}_{n}^\nu  +  s_n^\nu \ddot{x}_{n}^\nu
- \Big[ \frac{1}{2} \Big( m_n(0) + \frac{\partial m_n}{\partial x^\mu_n}  x_n^\mu
 - \sum_{n'\neq n} \lambda^\mu_{n'} 
\frac{\partial m_{n}}{\partial x_{n}^\mu} \Big) \dot{x}_{n}^\nu  \dot{x}_{n;\nu}
-  \sum_{n'\neq n} \lambda_{n';\alpha} 
\frac{\partial m_{n}}{\partial x_{n;\nu}}  \dot{x}_{n}^\alpha  \dot{x}_{n;\nu} 
\nonumber \\
&+ \sum_{n'}\frac{\partial A^\nu}{\partial \dot{x}^\mu_{n'}} \dot{x}_{n'}^\mu \dot{x}_{n;\nu}
\Big]
\end{align}

Momentum ${{p_T}_l}^\alpha$ can be rewritten in the matrix form as:
\begin{equation}
{{P_T}_l}^\alpha = \sum_n \big[ {{\mathcal{A}_T}_{ln}}^\alpha_\nu  
+  {{\mathcal{B}_T}_{ln}}^\alpha_{\nu \mu } X^\mu_n \big] \dot{X}^\nu_n 
+ {{\mathcal{C}_T}_{ln}}^\alpha
- {{\mathcal{D}_T}_{ln}}^\alpha_{\mu } X^\mu_n 
\end{equation}
where the matrix's element are
\begin{align}
&{[{\mathcal{A}_T]}_{ln}}^\alpha_\nu  \;=\Big [ 
\Big( m_n(0) 
- \sum_{n'\neq n} {\lambda_n'}^\mu \frac{\partial m_n}{\partial{x_n}^\mu} 
\Big )g^\alpha_\nu 
+  \sum_{n'\neq n} ({\lambda_{n'}}^\alpha {\square_n}_\nu m_n 
+ {\lambda_{n'}}_\nu \frac{\partial m_n}{\partial {x_n}_\alpha} ) 
\Big] \delta_{n,l}
- \Big( \frac{\partial A^\alpha}{\partial \dot{x}^\nu_n} 
+ \frac{\partial A_\nu}{\partial \dot{x}_{n;\alpha}}
\Big)
\nonumber \\
&{[{\mathcal{B}_T}]_{ln}}^\alpha_{\nu \mu } = \frac{\partial m_n}{\partial x^\mu_n} g^\alpha_\nu  \delta_{n,l}
;\qquad 
{[{\mathcal{D}_T}]_{ln}}^\alpha_{\mu } \;\;=  \frac{\partial A^\alpha}{\partial x^\mu_n}
\nonumber \\
&{[{\mathcal{C}_T}]_{ln}}^\alpha \;\;\;=  \Big \{ \sum_{n'\neq n} {\lambda_{n'}}^\mu \Big[ \frac{d}{d \tau}\Big( \frac{\partial A^\alpha}{\partial {\dot{x}_{n'}}^\mu} 
- \frac{\partial A^\alpha}{\partial {\dot{x}_n}^\mu}\Big)  
+ \frac{\partial A^\alpha}{\partial {x_n}^\mu} 
- \frac{\partial A^\alpha}{\partial {x_{n'}}^\mu} \Big] 
- A^\alpha(0) \Big \}\delta_{nl}.
\end{align}

The velocity of the particle can be isolated as 
\begin{equation}
\dot{X}^\nu_l = \big[ 
{{P_T}_l}^\alpha 
- \sum_n {{\mathcal{C}_T}_{ln}}^\alpha 
+  {{\mathcal{D}_T}_{ln}}^\alpha_{\mu } X^\mu_n 
\big] 
\big[ \sum_n {{\mathcal{A}_T}_{ln}}^\alpha_\nu  
+  {{\mathcal{B}_T}_{ln}}^\alpha_{\nu \mu } X^\mu_n \big]^-
\label{extHamiltVelocityT}
\end{equation}
from where we extract the Lorentz condition for momentum ${{P_L}_n}^\alpha$:
\begin{align}
\dot{X}^\nu_l \dot{X}_{l;\nu} =&
\big[ 
{{P_T}_l}^\alpha  
- \sum_n {{\mathcal{C}_T}_{ln}}^\alpha 
+  {{\mathcal{D}_T}_{ln}}^\alpha_{\mu } X^\mu_n 
\big] 
\big[ \sum_n {{\mathcal{A}_T}_{ln}}^\alpha_\nu  
+  {{\mathcal{B}_T}_{ln}}^\alpha_{\nu \mu } X^\mu_n \big]^- \times
\nonumber \\
g_{\nu \gamma} &\big[ 
{{P_T}_l}^\beta 
- \sum_n {{\mathcal{C}_T}_{ln}}^\beta 
+  {{\mathcal{D}_T}_{ln}}^\beta_{\mu } X^\mu_n 
\big] 
\big[ \sum_n {{\mathcal{A}_T}_{ln}}^\beta_\gamma  
+  {{\mathcal{B}_T}_{ln}}^\beta_{\gamma \mu } X^\mu_n \big]^- = \mathrm{c}^2.
\end{align}

The translation Hamiltonian is obtained then by replacing the previous result on the extended Hamiltonian definition:
\begin{equation}
H_T = \sum_n {{p_T}_n}^\nu \dot{x}_{n;\nu} + {{s_T}_n}^\nu \ddot{x}_{n;\nu}  - L_T
\end{equation}
\begin{align}
H_T &=  \sum_{n}  \Big \{ \frac{1}{2}\Big( m_n(0) 
+ \frac{\partial m_n}{\partial x^\mu_n}  x_n^\mu 
- \sum_{n'\neq n} {\lambda_{n'}}^\mu \frac{\partial m_n}{\partial{x_n}^\mu}
\Big) \dot{x}^\nu_{n} \dot{x}_{n;\nu} 
+ \sum_{n'\neq n} {\lambda_{n'}}_\mu \frac{\partial m_{n}}{\partial{x_{n}}^\nu} {\dot{x}_n}^\mu \dot{x}_{n;\nu}
\nonumber \\
&+ \sum_{n'}\frac{\partial A^\nu}{\partial \dot{x}^\mu_{n'}} \dot{x}_{n'}^\mu \dot{x}_{n;\nu }
\Big \}
\end{align}
In matrix representation we have
\begin{equation}
H_T = \sum_n\Big\{ 
{H_{T_1}}_n \dot{X}_n^\nu \dot{X}_{n;\nu}
+ {{H_{T_2}}_n}_\mu X_n^\mu \dot{X}_n^\nu \dot{X}_{n;\nu}
+ {{H_{T_3}}_n}_\mu^\nu \dot{X}_n^\mu \dot{X}_{n;\nu}
+ \sum_{n'} {{H_{T_4}}_{n n'}}_\mu^\nu \dot{X}_{n'}^\mu \dot{X}_{n;\nu}
\Big\}
\end{equation}
or, after rearranging index
\begin{equation}
H_T = \sum_{nn'}\Big[ 
\Big ( {H_{T_1}}_n g_{\mu \nu}
+ {{H_{T_2}}_n}_\gamma X_n^\gamma g_{\mu \nu}
+ {{H_{T_3}}_n}_{\mu \nu}\Big )\delta_{nn'}
+ {{H_{T_4}}_{n n'}}_{\mu \nu}
\Big] \dot{X}_{n'}^\mu \dot{X}_n^\nu
\end{equation}
where the matrix components of the second order Hamiltonian have the form:
\begin{align}
[{H_{T_1}}_n]\quad &= \frac{1}{2}\Big( m_n(0) 
- \sum_{n'\neq n} {\lambda_{n'}}^\mu \frac{\partial m_n}{\partial{x_n}^\mu} \Big)
&[{{H_{T_2}}_n}_\mu] \;\; &= \frac{\partial m_n}{\partial x^\mu_n} 
\nonumber \\
[{{H_{T_3}}_n}_\mu^\nu]\; \; &= \sum_{n'\neq n} {\lambda_{n'}}_\mu \frac{\partial m_{n}}{\partial{x_{n}}^\nu}
&[{{H_{T_4}}_{nn'}}_\mu^\nu] &= \frac{\partial A^\nu}{\partial \dot{x}^\mu_{n'}} .
\end{align}
Substituting the velocity of the particle from equation \ref{extHamiltVelocityT}, we obtain the final form for translation Hamiltonian:
\begin{align}
H_T = \sum_{nn'}&\Big[ 
\Big ( {H_{T_1}}_n g_{\mu \nu}
+ {{H_{T_2}}_n}_\gamma X_n^\gamma g_{\mu \nu}
+ {{H_{T_3}}_n}_{\mu \nu}\Big )\delta_{nn'}
+ {{H_{T_4}}_{n n'}}_{\mu \nu}
\Big] \times
\nonumber \\ 
&\Big[ 
{{P_T}_{n}}^\alpha  
- \sum_l {{\mathcal{C}_T}_{nl}}^\alpha 
+  {{\mathcal{D}_T}_{nl}}^\alpha_{\delta } X^\delta_l
\big] 
\big[ \sum_l {{\mathcal{A}_T}_{nl}}^\alpha_\mu  
+  {{\mathcal{B}_T}_{nl}}^\alpha_{\mu \delta } X^\delta_l \Big]^- \times
\nonumber \\
&\Big[ 
{{P_T}_{n'}}^\beta 
- \sum_l {{\mathcal{C}_T}_{n'l}}^\beta 
+  {{\mathcal{D}_T}_{n'l}}^\beta_{\delta } X^\delta_l
\big] 
\big[ \sum_l {{\mathcal{A}_T}_{n'l}}^\beta_\nu  
+  {{\mathcal{B}_T}_{n'l}}^\beta_{\nu \delta } X^\delta_l \Big]^- \label{extLinearHamiltonian}
\end{align}

\subsubsection*{Hamiltonian for $n$-VMVF systems using angular coordinates}

The rotation Hamiltonian, $H_R$, is defined depending on the rotation angles $\theta$,$\phi$ and $\chi$ coordinates. However, it is useful to express it as a function of the lorentzian coordinates, using the transformation equations \ref{PosAngleRelation}, so both equation systems depend on the same variables.  Using the relations \ref{relations5}, the angular momentum $l$ and the angular $s$-momentum, $b$, have the form
\begin{align}
l_{n_i} & \equiv \frac{\partial L_R}{\partial \dot{\xi}_{n_i}} = D_{n;i}^{\;\;\;\nu} \frac{\partial L_R}{\partial {\dot{x}_n}^\nu} + 2\frac{d}{d\tau}\Big(D_{n;i}^{\;\;\;\nu} \Big) \frac{\partial L_R}{\partial {\ddot{x}_n}^\nu} 
\nonumber \\
&=  D_{n;i}^{\;\;\;\nu} {{p_R}_n}_\nu + 2\frac{d}{d\tau}\Big(D_{n;i}^{\;\;\;\nu} \Big) {{s_R}_n}_\nu
\nonumber \\
b_{n_i}& \equiv \frac{\partial L_R}{\partial  \ddot{\xi}_{n_i}} =  D_{n;i}^{\;\;\;\nu} \frac{\partial L_R}{\partial {\ddot{x}_n}^\nu}
\nonumber \\
&= D_{n;i}^{\;\;\;\nu} {{s_R}_n}_\nu
\end{align}
where 
\begin{equation}
{p_R}_n^\alpha \equiv \frac{\partial L_R}{\partial \dot{x}_{n;\alpha}} \;\;\; \text{and } \;\;\; 
{s_R}_n^\alpha \equiv \frac{\partial L_R}{\partial \ddot{x}_{n;\alpha}}.
\end{equation}
On the other hand, angular velocity and acceleration have the expression:
\begin{align}
\dot{\xi}_{n_i} &= \frac{\partial \xi_{n_i}}{\partial x^\nu_n}\dot{x}^\nu_n \equiv  D_{n;i\nu}^{-} \dot{x}^\nu_n 
\nonumber \\
\ddot{\xi}_{n_i} &= \frac{d}{d \tau} \Big(\frac{\partial \xi_{n_i}}{\partial x^\nu_n} \Big)\dot{x}^\nu_n  + \frac{\partial \xi_{n_i}}{\partial x^\nu_n}\ddot{x}^\nu_n 
\equiv   \frac{d}{d \tau} \Big( D_{n;i\nu}^{-} \Big) \dot{x}^\nu_n + D_{n;i\nu}^{-} \ddot{x}^\nu_n \label{InvPosAngRelation}
\end{align} 
where
\begin{equation}
D_{n;i\nu}^{-}\equiv \frac{\partial \xi_{n_i}}{\partial x^\nu_n}.
\end{equation}
are the components of the inverted coordinates transformation. The relation between $D_{n;i\nu}^{-}$ and $D_{n;i}^{\;\;\;\nu}$ are obtained since both set of transformation 
\begin{equation}
\dot{\xi}_{n_i} = \frac{\partial \xi_{n_i}}{\partial x^\nu_n}\dot{x}^\nu_n \qquad \text{and} \qquad 
\dot{x}^\nu_n = \sum_i \frac{\partial x^\nu_n}{\partial \xi_{n_i}}\dot{\xi}_{n_i} 
\end{equation}
must be consistent. The substitution of one into the other, result in
\begin{equation}
\dot{\xi}_{n_i} = \frac{\partial \xi_{n_i}}{\partial x^\nu_n}\dot{x}^\nu_n 
= \sum_\nu \frac{\partial \xi_{n_i}}{\partial x^\nu_n} \frac{\partial x^\nu_n}{\partial \xi_{n_j}} \dot{\xi}_{n_j} = 
\sum_\nu D_{n;i\nu}^{-} D_{n;j}^{\;\;\;\nu} \; \dot{\xi}_{n_j}.
\end{equation}
As coordinates are independent, the components of both matrix must satisfy
\begin{equation}
\sum_\nu D_{n;i\nu}^{-} D_{n;j}^{\;\;\;\nu} = \delta_{ij} \label{dMatrixRel1}
\end{equation}
Same analysis is applied inverting the substitution order, where
\begin{align}
\dot{x}^\nu_n &= \sum_i \frac{\partial x^\nu_n}{\partial \xi_{n_i}}\dot{\xi}_{n_i} 
= \sum_i \frac{\partial x^\nu_n}{\partial \xi_{n_i}} \frac{\partial \xi_{n_i}}{\partial x^\mu_n} \dot{x}^\mu_n = 
\sum_i D_{n;i}^{\;\;\;\nu} D_{n;i\mu}^{-}  \; \dot{x}^\mu_n.
\end{align}
Then,  components of both matrix also satisfy the relation
\begin{align}
&\sum_i D_{n;i}^{\;\;\;\nu} D_{n;i\mu}^{-} = \delta^\nu_\mu. \label{dMatrixRel2}
\end{align}

Rotation Hamiltonian is computed from its definition using the approximated Rotation Lagrangian \ref{extRotLagrangianAppx1}:
\begin{align}
H_R  &=  \sum_{n,i} ( b_{n_i} \ddot{\xi}_{n_i}+  l_{n_i} \dot{\xi}_{n_i}) - L_R
\nonumber \\
&=  \sum_{n,i\nu\mu} \Big\{ D_{n;i}^{\;\;\;\nu} \frac{\partial L_R}{\partial \ddot{x}^\nu_n} \Big[  
\frac{d}{d \tau} \Big( D_{n;i\mu}^{-} \Big) \dot{x}^\mu_n + D_{n;i\mu}^{-} \ddot{x}^\mu_n
\Big] 
+ \Big[  D_{n;i}^{\;\;\;\nu}  \frac{\partial L_R}{\partial \dot{x}^\nu_n}  +  2\frac{d}{d\tau}\Big(  D_{n;i}^{\;\;\;\nu} \Big)  \frac{\partial L_R}{\partial \ddot{x}^\nu_n} \Big] D_{n;i\mu}^{-} \dot{x}^\mu_n  \Big\}
\nonumber \\
&\qquad - L_R
\nonumber \\
&=  \sum_{n,i\nu\mu} \Big\{  \Big[  \frac{d}{d \tau} \Big( D_{n;i}^{\;\;\;\nu}  D_{n;i\mu}^{-}\Big) + \frac{d}{d\tau}\Big(  D_{n;i}^{\;\;\;\nu} \Big) D_{n;i\mu}^{-} \Big] \frac{\partial L_R}{\partial \ddot{x}^\nu_n} \dot{x}^\mu_n  
+ D_{n;i}^{\;\;\;\nu}  D_{n;i\mu}^{-} \frac{\partial L_R}{\partial \ddot{x}^\nu_n} \ddot{x}^\mu_n 
\nonumber \\
&\qquad + D_{n;i}^{\;\;\;\nu}  D_{n;i\mu}^{-} \frac{\partial L_R}{\partial \dot{x}^\nu_n} \dot{x}^\mu_n \Big\} -L_R
\end{align}
Summing over  summation index $i$ on first, third and fourth terms and applying relations \ref{dMatrixRel1} and \ref{dMatrixRel2}, Rotation Hamiltonian results in
\begin{align}
H_R &=  \sum_{n,i\nu\mu}  \Big[ \frac{d}{d\tau}\Big(  D_{n;i}^{\;\;\;\nu} \Big) D_{n;i\mu}^{-} \Big] \frac{\partial L_R}{\partial \ddot{x}^\nu_n} \dot{x}^\mu_n  
+ \sum_{n,\nu} \frac{\partial L_R}{\partial \ddot{x}^\nu_n} \ddot{x}^\nu_n 
+ \sum_{n,\nu} \frac{\partial L_R}{\partial \dot{x}^\nu_n} \dot{x}^\nu_n  -L_R
\nonumber \\
&\text{or}
\nonumber \\
H_R &=  \sum_{n,i\nu\mu}  G_{n;i\gamma}^{\;\;\;\;\;\nu} D_{n;i\mu}^{-} s_{R n;\nu} \;\dot{x}^\gamma_n  \dot{x}^\mu_n
+ \sum_{n,\nu} s_{R n;\nu}\;\ddot{x}^\nu_n 
+ \sum_{n,\nu} p_{R n;\mu} \;\dot{x}^\nu_n  -L_R \label{ExtRotationHamiltDef}
\end{align}
where 
\begin{equation}
\frac{d}{d\tau}\Big(  D_{n;i}^{\;\;\;\nu} \Big) = \sum_\gamma G_{n;i\gamma}^{\;\;\;\;\;\nu}  \dot{x}^\gamma_n \;\;\;\;\;\text{being} \;\; G_{n;i\gamma}^{\;\;\;\;\;\nu}  \equiv \frac{\partial  D_{n;i}^{\;\;\;\nu} }{\partial x^\gamma_n}.
\end{equation}
The extended classical Lagrangian of Rotation of equation \ref{extRotLagrangianAppx1}
\begin{align*}
L_R &=  \sum_{n} \frac{1}{2}(m_n(0) + \frac{\partial m_n}{\partial x^\mu_n}  x_n^\mu) \dot{x}^\nu_{n} \dot{x}_{n;\nu} -(A^\nu(0) + \frac{\partial A^\nu}{\partial x^\mu_n} x_n^\mu 
+ \frac{\partial A^\nu}{\partial \dot{x}^\mu_n} \dot{x}_n^\mu ) \dot{x}_{n;\nu} + 
\nonumber \\ 
&\sum_{i,n'\neq n} \beta_{i_n}   D_{\;\xi_{n',i}}^\mu \Big[ (m_{n'}(0) 
+ \frac{\partial m_{n'}}{\partial x^\nu_{n'}} x_{n'}^\nu) \ddot{x}_{{n'};\mu} + (\square_{n'}^\nu m_{n'} \dot{x}_{{n'};\nu}) \dot{x}_{{n'};\mu} 
- \frac{\partial A^\nu}{\partial \dot{x}_{n'}^\mu}\ddot{x}_{{n'};\nu} 
- \frac{d}{d\tau}\Big(\frac{\partial A^\nu}{\partial \dot{x}_{n'}^\mu} \Big) \dot{x}_{{n'};\nu}
\nonumber \\
&- \frac{\partial A_\mu}{\partial x_{n';\nu}}\dot{x}_{n';\nu} 
- \frac{\partial A_\mu}{\partial \dot{x}_{n';\nu}}\ddot{x}_{n';\nu}  
- \frac{1}{2}\frac{\partial m_{n'}}{\partial x_{n'}^\mu}\dot{x}^\nu_{n'} \dot{x}_{{n'};\nu} 
+ \frac{\partial A^\nu}{\partial x_{n'}^\mu}\dot{x}_{n';\nu} \Big] 
 \nonumber \\  
&+ D_{\;\xi_{n,i}}^\mu \Big[ \frac{d}{d\tau} \Big(\frac{\partial A^\nu}{\partial \dot{x}_{n}^\mu}\Big) \dot{x}_{n';\nu} 
+ \frac{\partial A^\nu}{\partial \dot{x}_{n}^\mu} \ddot{x}_{n';\nu}
- \frac{\partial A^\nu}{\partial x_{n}^\mu}\dot{x}_{n';\nu} 
+ \frac{\partial A_\mu}{\partial x_{n';\nu}}\dot{x}_{n';\nu} 
+ \frac{\partial A_\mu}{\partial \dot{x}_{n';\nu}}\ddot{x}_{n';\nu}  \Big]
\end{align*}
can be rewritten grouping in $\dot{x}_{n;\nu}$ and $\ddot{x}_{n;\nu}$ terms like
\begin{align}
L_R &=  \sum_n \Big \{ \Big [ 
\Big( m_n(0) 
+ \frac{\partial m_n}{\partial x^\mu_n}x_n^\mu 
- \sum_{i;n'\neq n}  \beta_{n'_i} D_{n;i}^\mu  \frac{\partial m_n}{\partial{x_n}^\mu} 
\Big )g_\alpha^\nu 
+  \sum_{i;n'\neq n}  \beta_{n'_i} (D_{n;i}^\nu {\square_n}_\alpha m_n 
+ D_{n;i \alpha}   \frac{\partial m_n}{\partial {x_n}_\nu} ) 
\nonumber \\
&- \Big( \frac{\partial A^\nu}{\partial \dot{x}^\alpha_n} 
+ \frac{\partial A_\alpha}{\partial \dot{x}_{n;\nu}}
\Big) 
\Big ] \dot{x}_n^\alpha
- \frac{\partial A^\nu}{\partial x^\mu_n} x_n^\mu 
+  \sum_{i,n'\neq n} \beta_{i_{n'}} 
\Big \{
D_{\;\xi_{n,i}}^\mu  
\Big[
\frac{\partial A^\nu}{\partial x_n^\mu} 
- \frac{\partial A_\mu}{\partial x_{n;\nu}} 
- \frac{d}{d \tau}\Big( \frac{\partial A^\nu}{\partial {\dot{x}_{n'}}^\mu} 
\Big)
\Big]
\nonumber \\
&- D_{\;\xi_{{n'},i}}^\mu  
\Big[
\frac{\partial A^\nu}{\partial x_{n'}^\mu} 
- \frac{\partial A_\mu}{\partial x_{n;\nu}} 
- \frac{d}{d \tau}\Big( \frac{\partial A^\nu}{\partial {\dot{x}_{n'}}^\mu} 
\Big)
\Big] - A^\nu(0) \Big \}\dot{x}_{n;\nu}
\nonumber \\
& +  \sum_{i,n'\neq n} \Big \{ \beta_{i_{n'}} 
\Big[  
D_{\;\xi_{{n},i}}^\nu \Big(
m_{n}(0) 
+ \frac{\partial m_{n}}{\partial x^\nu_{n}} x^\nu_{n}
-\big( \frac{\partial A^\nu}{\partial \dot{x}_n^\mu} 
+ \frac{\partial A_\mu}{\partial \dot{x}_{n;\nu}} \big)
\Big)
+ D_{\;\xi_{n',i}}^\mu  
\Big(
\frac{\partial A^\nu}{\partial \dot{x}_{n'}^\mu} 
+ \frac{\partial A_\mu}{\partial \dot{x}_{n;\nu}} 
\Big)
\Big]
\Big \}
 \ddot{x}_{n;\nu}
\nonumber \\
&-\sum_i \Big[ \frac{1}{2}\Big( m_n(0) 
+ \frac{\partial m_n}{\partial x^\mu_n}  x_n^\mu 
- \sum_{i;n'\neq n} \beta_{n'_i} D_{n;i}^\mu  \frac{\partial m_n}{\partial{x_n}^\mu} \Big) {\dot{x}_n}^\nu \dot{x}_{n;\nu} 
\nonumber \\
&+ \sum_{n'\neq n} \beta_{n'_i} D_{n;i\mu}  \frac{\partial m_{n}}{\partial{x_{n}}^\nu} {\dot{x}_n}^\mu \dot{x}_{n;\nu}
+ \sum_{n'}\frac{\partial A^\nu}{\partial \dot{x}^\mu_{n'}} \dot{x}_{n'}^\mu \dot{x}_{n;\nu }
\Big]
\end{align}

We find ${p_R}_l^\alpha$ and ${s_R}_l^\alpha$ by straightforward derivation, and retaking duplex index Einstein sum notation:
 Momentum ${{p_R}_l}^\alpha$ have the form
\begin{align}
{{p_R}_l}^\alpha  =& \sum_n \Big \{ \Big [ 
\Big( m_n(0) 
+ \frac{\partial m_n}{\partial x^\mu_n}x_n^\mu 
- \sum_{i;n'\neq n}  \beta_{n'_i} D_{n;i}^\mu  \frac{\partial m_n}{\partial{x_n}^\mu} 
\Big )g^\alpha_\nu 
+  \sum_{i;n'\neq n}  \beta_{n'_i} (D_{n;i}^\alpha {\square_n}_\nu m_n 
+ D_{n;i \nu}   \frac{\partial m_n}{\partial {x_n}_\alpha} ) 
\Big] \delta_{n,l}
\nonumber \\
&- \Big( \frac{\partial A^\alpha}{\partial \dot{x}^\nu_n} 
+ \frac{\partial A_\nu}{\partial \dot{x}_{n;\alpha}}
\Big) 
\Big \} {\dot{x}_n}^\nu
- \frac{\partial A^\alpha}{\partial x^\mu_n} x_n^\mu 
+  \sum_{i,n\neq n} \beta_{i_{n'}} 
\Big \{
D_{\;\xi_{n,i}}^\mu  
\Big[
\frac{\partial A^\alpha}{\partial x_n^\mu} 
+ \frac{\partial A_\mu}{\partial x_{n;\alpha}} 
- \frac{d}{d \tau}\Big( \frac{\partial A^\alpha}{\partial {\dot{x}_{n'}}^\mu} 
\Big)
\Big]
\nonumber \\
&- D_{\;\xi_{{n'},i}}^\mu  
\Big[
\frac{\partial A^\alpha}{\partial x_{n'}^\mu} 
- \frac{\partial A_\mu}{\partial x_{n;\alpha}} 
- \frac{d}{d \tau}\Big( \frac{\partial A^\alpha}{\partial {\dot{x}_{n'}}^\mu} 
\Big)
\Big] - A^\alpha(0) \Big \}\delta_{nl}
\label{TrasAngularP}
\end{align}
and 
\begin{align}
{{s_R}_l}^\alpha & = \sum_{i,n\neq l} \Big \{ \beta_{i_n} 
\Big[  
D_{\;\xi_{l,i}}^\alpha \Big(
m_l(0) 
+ \frac{\partial m_{l}}{\partial x^\nu_{l}} x^\nu_{l}
\Big)
- D_{\;\xi_{l,i}}^\mu  
\Big(
\frac{\partial A^\alpha}{\partial \dot{x}_l^\mu} 
+ \frac{\partial A_\mu}{\partial \dot{x}_{l;\alpha}} 
\Big)
+ D_{\;\xi_{n,i}}^\mu  
\Big(
\frac{\partial A^\alpha}{\partial \dot{x}_n^\mu} 
+ \frac{\partial A_\mu}{\partial \dot{x}_{l;\alpha}} 
\Big)
\Big]
\Big \}.
\end{align}

We can express momentum ${{p_R}_l}^\alpha$ in its matrix form
\begin{equation}
{{P_R}_l}^\alpha = \sum_n \big[ {{\mathcal{A}_R}_{ln}}^\alpha_\nu  
+  {{\mathcal{B}_R}_{ln}}^\alpha_{\nu \mu } X^\mu_n \big] \dot{X}^\nu_n 
+ {{\mathcal{C}_R}_{ln}}^\alpha
- {{\mathcal{D}_R}_{ln}}^\alpha_{\mu } X^\mu_n, 
\end{equation}
whose matrix elements are
\begin{align}
&{[{\mathcal{A}_R]}_{ln}}^\alpha_\nu  \;=\Big [ 
\Big( m_n(0) 
- \sum_{i;n'\neq n}  \beta_{n'_i} D_{n;i}^\mu  \frac{\partial m_n}{\partial{x_n}^\mu} 
\Big )g^\alpha_\nu 
+  \sum_{i;n'\neq n}  \beta_{n'_i} (D_{n;i}^\alpha {\square_n}_\nu m_n 
+ D_{n;i \nu}   \frac{\partial m_n}{\partial {x_n}_\alpha} ) 
\Big] \delta_{n,l}
\nonumber \\
& \qquad \qquad \;\; - \Big( \frac{\partial A^\alpha}{\partial \dot{x}^\nu_n} 
+ \frac{\partial A^\nu}{\partial \dot{x}_{n;\alpha}}
\Big)
\nonumber \\
&{[{\mathcal{B}_R}]_{ln}}^\alpha_{\nu \mu } = \frac{\partial m_n}{\partial x^\mu_n} g^\alpha_\nu  \delta_{n,l}
;\qquad 
{[{\mathcal{D}_R}]_{ln}}^\alpha_{\mu } \;\;=  \frac{\partial A^\alpha}{\partial x^\mu_n}
\nonumber \\
&{[{\mathcal{C}_R}]_{ln}}^\alpha \;\;\;= \sum_{i,n\neq n} \beta_{i_{n'}} 
\Big \{
D_{\;\xi_{n,i}}^\mu  
\Big[
\frac{\partial A^\alpha}{\partial x_n^\mu} 
+ \frac{\partial A_\mu}{\partial x_{n;\alpha}} 
- \frac{d}{d \tau}\Big( \frac{\partial A^\alpha}{\partial {\dot{x}_{n'}}^\mu} 
\Big)
\Big]
- D_{\;\xi_{{n'},i}}^\mu  
\Big[
\frac{\partial A^\alpha}{\partial x_{n'}^\mu} 
- \frac{\partial A_\mu}{\partial x_{n;\alpha}} 
- \frac{d}{d \tau}\Big( \frac{\partial A^\alpha}{\partial {\dot{x}_{n'}}^\mu} 
\Big)
\Big] 
\nonumber \\
&\qquad \qquad \;\;- A^\alpha(0) \Big \}\delta_{nl}.
\end{align}

Particle velocity is isolated as
\begin{equation}
\dot{X}^\nu_n = \big[ 
{{P_R}_l}^\alpha 
- \sum_n {{\mathcal{C}_R}_{ln}}^\alpha 
+  {{\mathcal{D}_R}_{ln}}^\alpha_{\mu } X^\mu_n 
\big] 
\big[ \sum_n {{\mathcal{A}_R}_{ln}}^\alpha_\nu  
+  {{\mathcal{B}_R}_{ln}}^\alpha_{\nu \mu } X^\mu_n \big]^-
\label{extHamiltVelocityR}
\end{equation}
from where can be obtained the Lorentz condition for momentum ${{P_R}_n}^\alpha$:
\begin{align}
\dot{X}^\nu_l \dot{X}_{l;\nu} =&
\big[ 
{{P_R}_l}^\alpha
- \sum_n {{\mathcal{C}_R}_{ln}}^\alpha 
+  {{\mathcal{D}_R}_{ln}}^\alpha_{\mu } X^\mu_n 
\big] 
\big[ \sum_n {{\mathcal{A}_R}_{ln}}^\alpha_\nu  
+  {{\mathcal{B}_R}_{ln}}^\alpha_{\nu \mu } X^\mu_n \big]^- \times
\nonumber \\
g_{\nu \gamma}&\big[ 
{{P_R}_l}^\beta 
- \sum_n {{\mathcal{C}_R}_{ln}}^\beta 
+  {{\mathcal{D}_R}_{ln}}^\beta_{\mu } X^\mu_n 
\big] 
\big[ \sum_n {{\mathcal{A}_R}_{ln}}^\beta_\gamma  
+  {{\mathcal{B}_R}_{ln}}^\beta_{\gamma \mu } X^\mu_n \big]^- = \mathrm{c}^2.
\end{align}

The extended Rotation Hamiltonian eq. \ref{ExtRotationHamiltDef} 
\begin{align*}
H_R  &=  \sum_{n} \big[  \sum_i G_{n;i\gamma}^{\;\;\;\;\;\nu} D_{n;i\mu}^{-} s_{R n;\nu} \;\dot{x}^\gamma_n  \dot{x}^\mu_n
+ s_{R n;\nu}\;\ddot{x}^\nu_n 
+ p_{R n;\mu} \;\dot{x}^\nu_n \big]  -L_R, 
\end{align*}
can be rewritten as
\begin{align}
H_R &=  \sum_{n}  \Big[ \sum_{i}G_{n;i\gamma}^{\;\;\;\;\;\nu} D_{n;i\mu}^{-} s_{R n;\nu} \;\dot{x}^\gamma_n  \dot{x}^\mu_n + \frac{1}{2}\Big( m_n(0) 
+ \frac{\partial m_n}{\partial x^\mu_n}  x_n^\mu 
- \sum_{i;n'\neq n} \beta_{n'_i} D_{n;i}^\mu  \frac{\partial m_n}{\partial{x_n}^\mu} \Big) {\dot{x}_n}^\nu \dot{x}_{n;\nu} 
\nonumber \\
&+ \sum_{i;n'\neq n} \beta_{n'_i} D_{n;i\mu}  \frac{\partial m_{n}}{\partial{x_{n}}^\nu} {\dot{x}_n}^\mu \dot{x}_{n;\nu}
+ \sum_{n'}\frac{\partial A^\nu}{\partial \dot{x}^\mu_{n'}} \dot{x}_{n'}^\mu \dot{x}_{n;\nu } \Big]
\end{align}
or, using matrix representation as
\begin{align}
H_R &= \sum_n\Big\{ 
\sum_{i}G_{n;i\gamma}^{\;\;\;\;\;\nu} D_{n;i\mu}^{-} S_{R n;\nu} \;\dot{X}^\gamma_n  \dot{X}^\mu_n 
+ {H_{R_1}}_n \dot{X}_n^\nu \dot{X}_{n;\nu}
+ {{H_{R_2}}_n}_\mu X_n^\mu \dot{X}_n^\nu \dot{X}_{n;\nu}
\nonumber \\
&+ {{H_{R_3}}_n}_\mu^\nu \dot{X}_n^\mu \dot{X}_{n;\nu}
+ \sum_{n'} {{H_{R_4}}_{n n'}}_\mu^\nu \dot{X}_{n'}^\mu \dot{X}_{n;\nu}
\Big\}
\end{align}
Rearranging index we have
\begin{align}
H_R = \sum_{nn'}\Big[ 
&\Big ( \sum_{i}G_{n;i\mu}^{\;\;\;\;\;\gamma} D_{n;i\gamma}^{-} S_{R n;\nu}
+{H_{R_1}}_n g_{\mu \nu}
+ {{H_{R_2}}_n}_\gamma X_n^\gamma g_{\mu \nu}
+ {{H_{R_3}}_n}_{\mu \nu}\Big )\delta_{nn'}
\nonumber \\
&+ {{H_{R_4}}_{n n'}}_{\mu  \nu}
\Big] \dot{X}_{n'}^\mu \dot{X}_n^\nu
\end{align}
where the matrix components are:
\begin{align}
[{H_{R_1}}_n]\quad &= \frac{1}{2}\Big( m_n(0) 
- \sum_{i;n'\neq n} \beta_{n'_i} D_{n;i}^\mu  \frac{\partial m_n}{\partial{x_n}^\mu} \Big) 
&[{{H_{R_2}}_n}_\mu] \;\;&= \frac{\partial m_n}{\partial x^\mu_n} 
\nonumber \\
[{{H_{R_3}}_n}_\mu^\nu]\;\; &= \sum_{i;n'\neq n} \beta_{n'_i} D_{n;i\mu}  \frac{\partial m_{n}}{\partial{x_{n}}^\nu} 
&[{{H_{R_4}}_{nn'}}_\mu^\nu] &= \frac{\partial A^\nu}{\partial \dot{x}^\mu_{n'}}.
\end{align}
Substituting particle velocity, we obtain the final expression for the extended Rotation Hamiltonian
\begin{align}
H_R = \sum_{nn'}\Big[ 
&\Big ( \sum_{i}G_{n;i\mu}^{\;\;\;\;\;\gamma} D_{n;i\gamma}^{-} S_{R n;\nu}
+{H_{R_1}}_n g_{\mu \nu}
+ {{H_{R_2}}_n}_\gamma X_n^\gamma g_{\mu \nu}
+ {{H_{R_3}}_n}_{\mu \nu}\Big )\delta_{nn'}
\nonumber \\
&+ {{H_{R_4}}_{n n'}}_{\mu  \nu}
\Big] \Big[ 
{{P_R}_{n}}^\alpha  
- \sum_l {{\mathcal{C}_R}_{nl}}^\alpha 
+  {{\mathcal{D}_R}_{nl}}^\alpha_{\delta } X^\delta_l
\big] 
\big[ \sum_l {{\mathcal{A}_R}_{nl}}^\alpha_\mu  
+  {{\mathcal{B}_R}_{nl}}^\alpha_{\mu \delta } X^\delta_l \Big]^- \times
\nonumber \\
&\Big[ 
{{P_R}_{n'}}^\beta 
- \sum_l {{\mathcal{C}_R}_{n'l}}^\beta 
+  {{\mathcal{D}_R}_{n'l}}^\beta_{\delta } X^\delta_l
\big] 
\big[ \sum_l {{\mathcal{A}_R}_{n'l}}^\beta_\nu  
+  {{\mathcal{B}_R}_{n'l}}^\beta_{\nu \delta } X^\delta_l \Big]^- \label{extAngularHamiltonian}
\end{align}
\subsubsection{General Hamiltonian function and constraint equations for $n$-VMVF systems}
After obtaining the extended Translation and Rotation Hamiltonians, we can put they together as one single function, needed to solve a $n$-VMVF system:
\begin{equation}
H_{sys} \equiv 
\classoperator
{H_T}
{H_R} 
\end{equation}
and the $3n+1$ extended Hamilton equations, \ref{ExtHamiltonEqNew}, as
\begin{align}
&\frac{\partial H_{sys}}{\partial q_i} \equiv 
\begin{bmatrix}
\frac{\partial }{\partial x^\nu_n} & 0 \\
\\
0 & \frac{\partial }{\partial \xi_{n_i}}
\end{bmatrix} 
\classoperator
{H_T}
{H_R}
=-
\classoperator
{(\dot{p}_{T_{n_\nu}}-\ddot{s}_{T_{n_\nu}})}
{(\dot{l}_{R_{n_i}}-\ddot{b}_{R_{n_i}})}
\nonumber \\
&\frac{\partial H_{sys}}{\partial p_i} \equiv
\begin{bmatrix}
\frac{\partial }{\partial p^\nu_{T_n}} & 0\\ 
\\
0 & \frac{\partial }{\partial l_{R_i}}
\end{bmatrix}
\classoperator
{H_T}
{H_R}
=\classoperator
{ \dot{x}^\nu_n}
{ \dot{\xi}_{n_i}}
\nonumber \\
&\frac{\partial H_{sys}}{\partial s_i} \equiv
\begin{bmatrix}
\frac{\partial }{\partial s^\nu_{T_n}} & 0 \\
\\
0 & \frac{\partial }{\partial b_{R_i}}
\end{bmatrix}
\classoperator
{H_T}
{H_R}
=
\classoperator
{ \ddot{x}^\nu_n}
{ \ddot{\xi}_{n_i}}
\nonumber \\
&\frac{\partial H_{sys}}{\partial \tau} \equiv
\begin{bmatrix}
\frac{\partial }{\partial \tau} & 0 \\
\\
0 & \frac{\partial }{\partial \tau}
\end{bmatrix}
\classoperator
{H_T}
{H_R}
= -
\begin{bmatrix}
\frac{\partial }{\partial \tau} & 0 \\
\\
0 & \frac{\partial }{\partial \tau}
\end{bmatrix}
\classoperator
{L_T}
{L_R}
\end{align}
plus last relations in equations \ref{extEulerLagranEqV}
\begin{equation}
\Omega_{n,\mu,\xi_i} \equiv \left\{
\begin{array}{c}
\Phi_{\nu_n}\\ 
\Psi_{i_n}
\end{array} \right\} = 0 \label{extClassHamConst}
\end{equation}

\subsection{Mass and fields derivative approach for n-VMVF systems.} \label{massFieldApprox}
Along the development of this work, we extent the classical theory considering particle masses and the field as new degrees of freedom of the system, or more specifically, their first order derivatives:
\begin{align}
\{\frac{\partial m_n}{\partial x_n^\mu }\}, \quad
\{\frac{\partial m_n}{\partial \dot{x}_n^\mu }\} , \quad
\{\frac{\partial A^\nu}{\partial x_n^\mu }\},\quad
\{\frac{\partial A^\nu}{\partial \dot{x}_n^\mu }\}. \label{newDegreeFreedom}
\end{align}
We obtain a set of classical equations, using either the extension of the Lagrange or the Hamilton theory, whose functions depend on particle positions and the first-order derivative of masses and field with the particle coordinates as equation \ref{newDegreeFreedom}. 

We briefly discussed the nature of these new degrees of freedom; however nothing has been formally said about the dependencies and the constraint for masses and field derivatives. For example, we assume from the beginning of this work, for the sake of simplicity, that the mass functions should not depend on the particle velocity.

In general, the only requirement these derivatives must have is that the system of equations must be solvable. It means that the total number of variables must be equal to the total number of equations.

From the extended Lagrangian formulation, we have a total of 14 $n$ equations: the extended Lagrange equation for the Lorentzian $x_n^\nu$ and angular coordinates $\xi_n$ plus the constraint equation in both coordinate systems. Nevertheless, 2$n$ equations are restricted to the Lorentz constraints, letting the total number of equations equal to 12 $n$. On the other hand, mass and field derivative's dependencies are unknown. They are the variables to be determined, but if we sum all possibilities in equation \ref{newDegreeFreedom}, we will have a total of $40n$ independent variables. That is the reason why we need to set some restrictions on the degrees of freedom, so the equation system becomes solvable.

We have 4$n$ independent variables from particle position set of coordinates $x_n^\nu$, restricted by Lorentz constraint, which leads to 3$n$ independent variables. We recall that the equations depending on $\xi_n$ are also expressed as functions of Lorentzian coordinates, so we perform the analysis using only the Lorentzian set of coordinates. To solve the equation system, we need then to attribute the left 9$n$ degrees of freedom divided into the mass and field derivatives. This characteristic reflects the constructive character of this methodology since some approach needs to be included to keep the system solvable. The particle masses and fields derivatives can be settled in different ways as long the system remains soluble. We choose to divide the $9n$ number of variables as:
\begin{enumerate}
\item $3n $ for $ \{ \frac{\partial m}{\partial x_n^\nu}\}  $ and for $ \{ \frac{\partial m}{\partial \dot{x}_n^\nu}\}  $
\item $3n $ for $ \{ \frac{\partial A^\mu}{\partial \dot{x}_n^\nu}\}  $
\item $3n $ for $ \{ \frac{\partial A^\mu}{\partial x_n^\nu}\}  $
\end{enumerate}
According to our understanding, we analyze different approaches according to the previous division. Other variables distribution and others approximations can be chosen as long the number of independent variables remains equal to the number of equations.

\begin{itemize}
\item Masses derivatives $\frac{\partial m_n}{x^\nu_{n'}} $ and $\frac{\partial m_n}{\dot{x}^\nu_{n'}} $.

From the beginning of this work, we restrict particle masses depending only on it particle's position which means
\begin{equation}
\frac{\partial m_n}{x^\nu_{n'}} \equiv \frac{\partial m_n}{x^\nu_{n'}}\delta_{nn'} \;\;\; \text{and} \;\;\; \frac{\partial m_n}{\dot{x}^\nu_{n'}} =0.
\end{equation}
Also, as showed in the previous section, the four derivative component $\frac{\partial m_n}{\partial x^0_n}$ is restricted to the constraint \ref{mass0compConstraint}:
\begin{equation*}
\frac{1}{2}\frac{\partial m}{\partial x^0}\dot{x}^0 \dot{x}_\mu \dot{x}^\mu + m\ddot{x}_\mu\dot{x}^\mu=0.
\end{equation*}

We have defined then 3$n$ more variables for the system, remaining 6$n$ more to obtain. 

\item Field derivatives on particle velocities $\frac{\partial A^\nu}{\partial \dot{x}^\mu_{n}}$.

We defined the field as the medium for connecting particles and transporting information between them. It is through the field that particles feel the influence of others particles. In general, every field component should depend on the positions and the velocities of all particles. In that case, we have $4\times 4n = 16n$ variables related to the derivative with position, plus $4\times 4n = 16n$ degrees of freedom associated with the derivative with velocity. We may choose several approaches for treating this issue. From the analytic point of view, there will be no loss of generality in choosing an approach over another, as long as the system remains solvable. However, the results and their physical interpretation will be according to the chosen approach. We base some of our approximation same as the Electromagnetic field, because of the general concepts and laws it reflexes.

For clarity purposes, let us explicitly separate the $4\times 4n = 16n$ $\frac{\partial A^\nu}{\partial \dot{x}^\mu_{n}}$ variables:
\begin{align}
\Big \{\frac{\partial A^\nu}{\partial \dot{x}^\mu_{n}} \Big \} &= 
\Big \{\frac{\partial A^0}{\partial \dot{x}^0_{n}} \Big \}
+ \Big \{\frac{\partial A^i}{\partial \dot{x}^0_{n}} \Big \}
+ \Big \{\frac{\partial A^0}{\partial \dot{x}^i_{n}} \Big \}
+ \Big \{\frac{\partial A^i}{\partial \dot{x}^j_{n}} \Big \} \quad \forall \; i,j =1,2,3 
\nonumber \\
\{16n\} &= \;\;\; \{n\} \quad +\;\;  \{3n\} \; + \;\; \{3n\} \;\;\; + \;\;\{9n\},
 \label{fieldVarDiv}
\end{align}
where the first set of variables sums $n$ equations, the second $3n$, the third $3n$ and the last $3\times 3n = 9n$ for a total number of $16n$ variables.

As discussed in section \ref{classicalAssump1} and resumed in equations \ref{relConstraint1}, we suppose that the $A^0$ field component has the same behavior as the scalar field $\phi(\mathbf{x})$ on the Electromagnetic field, which is related to the field's fixed sources contribution. We have then \begin{equation*}
\frac{\partial A^0}{\partial \dot{x}^\nu_{n}} = 0.
\end{equation*}
This condition removes the first and the third set from the independent group of variables in \ref{fieldVarDiv} letting the total variables number equal to $12n$. On the other hand, the field derivative $\frac{\partial A^\nu}{\partial \dot{x}^\mu_{n}} $ may be inferred by also analyzing Electromagnetic field were the field component depends on the same velocity component of the particle. We extrapolate this dependency to our universal field $e.i$,  $A^\nu$ depends only on the $\dot{x}^{\nu}$ component of every particle, then
\begin{equation}
\frac{\partial A^\nu}{\partial \dot{x}^\mu_{n}} \equiv \frac{\partial A^\nu}{\partial \dot{x}^\mu_{n}} \delta^\mu_\nu.\label{fieldVelDep}
\end{equation}
This approach set eliminates the second group of variables in \ref{fieldVarDiv} and decreases the number of variables of the last group to $3n$ as needed. We have defined then $3n$ more degree of freedom to the equations system remaining $3n$ more to establish.

\item Field derivative on particle positions $\frac{\partial A^\nu}{\partial x^\mu_{n}}$.

Given the different natures and forms proposed for the field, this is the more complex degree of freedom and various approaches can be made for different physical systems. In this case, our propositions depend on the numbers of particles of the system. The field derivative with $x_n^0$ is chosen to be constrained to the Gauge invariance, as shown in equation \ref{relConstraint1} and discussed in section \ref{classicalAssump1}. We have then that for every particle:
\begin{equation}
\partial_{n,\nu} A^\nu \equiv \sum_\nu \frac{\partial A^\nu}{\partial x_n^\nu} = 0.
\end{equation}

Initially, for the position derivative, we have $4\times 4n = 16n$ independent variables, explicitly we can divided them as
\begin{align}
\Big \{\frac{\partial A^\nu}{\partial x^\mu_{n}} \Big \} &= 
\Big \{\frac{\partial A^0}{\partial x^0_{n}} \Big \}
+ \Big \{\frac{\partial A^i}{\partial x^0_{n}} \Big \}
+ \Big \{\frac{\partial A^0}{\partial x^i_{n}} \Big \}
+ \Big \{\frac{\partial A^i}{\partial x^j_{n}} \Big \} \quad \forall \; i,j =1,2,3
\nonumber \\
\{16n\} &= \;\;\; \{n\} \quad +\;\;  \{3n\} \; + \;\; \{3n\} \;\;\; + \;\;\{9n\}
 \label{fieldVarDiv1}
\end{align}

Same as the field derivative depending on velocity, some assumptions must be made to reduce the number of degree of freedom to 3$n$. We find essential 3 different cases:
\begin{enumerate}
\item Field component derivative depending on particle position same component.

Just as the approach for field derivative with respect the to velocity, we can also set field derivative concerning to the particle's position depending on the same component of the position of the particle as:
 \begin{equation}
\frac{\partial A^\nu}{\partial x^\mu_{n}} \equiv \frac{\partial A^\nu}{\partial x^\mu_{n}} \delta^\mu_\nu.
\end{equation}

By using this approach, the second and third set of variables in \ref{fieldVarDiv1} are set to zero, while the number of the fourth group of variables is reduced to $3n$ for a total of $4n$ independent. Finally, taking into account the Gauge condition \ref{relConstraint1} equations, we have defined 3-$n$ more independent variables, completing the $12n$ variables of the system described with $12n$ equations.

This approach may have a weak physical meaning, but it can be used to study isolated $n$-VMVF systems with any number of particles. Indeed, the field will not have the ordinary physical meaning depending on distances between particles, but it shall play its role as the mathematical entity that ``connect'' particles and ``transport information''  between them, whose only restriction is that equations system are solvable.

\item Field component derivative depending on the distance between particles. 

The most accepted idea on physics for any field connecting particles is that it depends on the distance between particles. This type of fields is known in the literature as central fields. In this case, the degree of freedom of the system is $\frac{\partial A^\nu}{\partial s_{ij}}$ where the Lorentz invariant $s_{ij}$ is the distance between particle $i$ and $j$, defined as:
\begin{equation}
s_{ij}=\sqrt{(x^0_i-x^0_j)^2 + (x^1_i-x^1_j)^2+ (x^2_i-x^2_j)^2+ (x^3_i-x^3_j)^2}.
\end{equation}

The number of the field derivatives with distances needs to be equal to 3$n$. The number of distances is the number of 2-combinations of $n$ particles:
\begin{equation}
N(s_{ij}) = \binom{n}{2} = \frac{n(n-1)}{2}
\end{equation}
Under this approach, if the number of particle increase, the number of field derivative depending on the distance between particle will surpass the 3$n$ variables limit at some point.

The Gauge $n$-conditions now have the form:
\begin{equation}
\partial_{n,\nu} A^\nu 
\equiv \sum_\nu \frac{\partial A^\nu}{\partial x_n^\nu} = 
\sum_{n'\neq n} \frac{\partial A^\nu}{s_{nn'}} \frac{s_{nn'}} {\partial x_n^\nu}
=0.
\end{equation}
Note that the gauge conditions are independent for each particle except for the number of particles being two.

We study two different cases:
\begin{enumerate}
\item The most general case is when all derivatives $\frac{\partial A^\nu}{\partial s_{ij}}$ are different: 
\begin{equation}
\frac{\partial A^0}{\partial s_{ij}} 
\neq \frac{\partial A^1}{\partial s_{ij}} 
\neq \frac{\partial A^2}{\partial s_{ij}}
\neq \frac{\partial A^3}{\partial s_{ij}}.
\end{equation}
Using the distance between particles, the number of independent variables is the number of field component (4) times the number of distances $n$ minus $n$ corresponding to the number of Gauge conditions for every particle. If we equate this number of variables to our limit $3n$ we have:
\begin{equation}
4 N(s_{ij}) -n = 4 \frac{n(n-1)}{2} -n = 3n \qquad  \therefore n= 3.
\end{equation}
This result means that we can successfully describe an $3$-VMF systems assuming previous restrictions and field depending on the distances between particles. If the number of particles is greater than 3, then central field approach cannot be used. 

If the number of particles is two, it is possible to describe the field using the $s_{ij}$'s dependency; however, this time the number of equations is greater than the number of variables, and because of that, others degree of freedom must be added. In the case of two particles, the number of distance is
$N(s_{ij}) = n(n-1)/ 2 = 1$ and the gauge $2$-conditions reduce to 1 since they are not independent:
\begin{equation}
0=\partial_{1,\nu} A^\nu
=\frac{\partial A^\nu}{s_{12}} \frac{s_{12}} {\partial x_1^\nu}
=- \frac{\partial A^\nu}{s_{12}} \frac{s_{12}} {\partial x_2^\nu} 
= -\partial_{2,\nu} A^\nu.
\end{equation}
The number of independent variables is $4 N(s_{ij}) -n_{Gauge} = 3$. Now we are short on the number of independent variables. We must then, add another dependency. If  we modify velocity field dependence on equation \ref{fieldVelDep} to
\begin{equation}
\frac{\partial A^\nu}{\partial \dot{x}^\mu_{n}} = \frac{\partial A^i}{\partial \dot{x}^j_{n}} \delta^i_j \quad \text{where }\; i,j=1,2,3,\label{fieldVelDep1}
\end{equation}
then, the second group of variables of expression \ref{fieldVarDiv} is no longer zero. Three more independent variables are added to obtain the finals six independent variables for the system.

\item Other case is when the three degrees of freedom related to the field derivative component $\frac{\partial A^\nu}{\partial s_{ij}}$ have the same dependency, $e.i$
\begin{equation}
\frac{\partial A^0}{\partial s_{ij}} 
= \frac{\partial A^1}{\partial s_{ij}} 
= \frac{\partial A^2}{\partial s_{ij}}
= \frac{\partial A^3}{\partial s_{ij}}.
\end{equation}
This approach resembles Electromagnetic vector field where all its component depends on distance as $s_{ij}^{-2}$. In this case, the number of distances cannot be greater that $3n$
\begin{equation}
\frac{n(n-1)}{2} - n = 3n \quad n(n-9) \therefore n= 9.
\end{equation}
We can, then, describe $n$-VMVF system with this conditions and previous restrictions, up to 9 particles. In the case of a lesser number of particles, others mass or field derivative need to be added as variables of the system, same as the previous dependency.
\end{enumerate}
\end{enumerate}
\end{itemize}

We expect the finals results reveal the different natures for interaction between particles. The mathematical solubility condition shall prevail over the nature of the dependency of the magnitudes. In fact, under that point of view, systems can be successfully studied using different approaches. Only the experiment and observation will tell which of them are more fundamental or illuminating. For example, two particle system problem can be solved by assuming two different mass and field dependency:
\begin{align}
&\frac{\partial m_n}{x^\nu_{n'}} \equiv \frac{\partial m_n}{x^\nu_{n'}}\delta_{nn'} , 
\quad \frac{\partial m_n}{\dot{x}^\nu_{n'}} =0
\nonumber \\
&\frac{\partial A^\nu}{\partial \dot{x}^\mu_{n}} = \frac{\partial A^\nu}{\partial \dot{x}^\mu_{n}} \delta^\mu_\nu, \quad
\frac{\partial A^0}{\partial \dot{x}^\nu_{n}} = 0
, \quad
\frac{\partial A^\nu}{\partial x^\mu_{n}} \equiv \frac{\partial A^\nu}{\partial x^\mu_{n}} \delta^\mu_\nu \quad \forall \;\nu = 0,1,2,3
\end{align}
or 
\begin{align}
&\frac{\partial m_n}{x^\nu_{n'}} \equiv \frac{\partial m_n}{x^\nu_{n'}}\delta_{nn'} , 
\quad \frac{\partial m_n}{\dot{x}^\nu_{n'}} =0
\nonumber \\
&\frac{\partial A^i}{\partial \dot{x}^j_{n}} = \frac{\partial A^i}{\partial \dot{x}^j_{n}} \delta^i_j , \quad
\frac{\partial A^0}{\partial \dot{x}^\nu_{n}} = 0
, \quad
\frac{\partial A^\nu}{\partial s_{nn'}} 
\neq \frac{\partial A^\mu}{\partial s_{nn'}} \quad \forall \;i=1,2,3 \quad \nu = 0,1,2,3.
\end{align}
Both of them treat the particle mass as a function depending on its particle position, and the fourth component of the vector field carry only the contributions of fixed particles. However, the first approach set each vector field's component depending only on the same component of the position and velocity of each particle while in the second case, the vector field components $A^i$ is a function of the fourth component $x_n^0$ of every particle and the distance between particles $s_{nn'}$.

Since the beginning, we try to adhere to the primary assumption of this work of considering the particle masses and field as entirely unknown variables, being the conservation of linear and angular momentums and the principle of least action their only restrictions. However, we can not avoid assuming some forms for the derivatives of those functions with particle position and velocities, so the system remains solvable. However, even the initial assumption has lost generality, the number of degrees of freedom related to the mass of the field is far higher than choosing a fixed form.

\subsection{Canonical transformations}
The advantage of using Hamilton theory lies not because of its contribution as a calculation tool, but on the more in-depth insight that it gives to the formal structure of the classical mechanic theory. The Hamilton theory basics ideas have an essential role in the construction of the modern theories as quantum mechanics. One of this concepts is the canonical transformation, which is the base to determine one of the main components in the modern quantum formalism: the operator. After obtaining the extended Hamilton equations, we are able then to define the canonical transformations for $n$-VMVF systems depending on the second-order derivative of generalized coordinates $\ddot{q}_i$. 

Canonical transformations are said to be the standard transformations of the system going from one set of coordinates to another while the extended Hamilton equations \ref{ExtHamiltonEq} are preserved. Under the Hamiltonian formulation, the transformation of the system involves the simultaneous transformations of the variables $q_i$, $p_i$ and $s_i$ into a new set $Q_i$, $P_i$ and $S_i$ with the following (invertible) transformations equations:
\begin{align}
Q_i&=Q_i(q_i,p_i,s_i) \nonumber \\
P_i&=P_i(q_i,p_i,s_i) \nonumber \\
S_i&=S_i(q_i,p_i,s_i) \label{newCordOldCord}
\end{align}
where $Q_i$, $P_i$ and $S_i$ satisfy:
\begin{align}
&\frac{\partial K}{\partial Q_i} =  -(\dot{P}_i-\ddot{S}_i)
\nonumber \\
&\frac{\partial K}{\partial P_i} = \dot{Q}_i
\nonumber \\
&\frac{\partial K}{\partial S_i} = \ddot{Q}_i \label{ExtHamiltonEqNew}
\end{align}
being $K$ the new transformed Hamiltonian. The transformation may include a factor $\lambda$ which describe a more global transformation known as ``scale transformation''. Here we assume $\lambda = 1$.

The function $K$ must also satisfies the least action principle:
\begin{equation}
\delta \int_{t_0}^{t_1} L(Q_i, \dot{Q}_i, \ddot{Q}_i) dt = \delta \int_{t_0}^{t_1} \sum_i P_i \dot{Q}_i + S_i \ddot{Q}_i - K(\bar{Q}_i, \bar{P}_i, \bar{S}_i,t)dt = 0
\end{equation}
where the bars symbols stand for the group of variables. Also,  $H$, $q_i$, $p_i$ and $s_i$ satisfy:
\begin{equation}
\delta \int_{t_0}^{t_1} L(q_i, \dot{q}_i, \ddot{q}_i) dt = \delta \int_{t_0}^{t_1} \sum_i p_i \dot{q}_i + s_i \ddot{q}_i - H(\bar{q}_i, \bar{p}_i, \bar{s}_i,t)dt = 0.
\end{equation}
Both integrand are not equals. Instead they are connected by the relation:
\begin{equation}
 \sum_i p_i \dot{q}_i + s_i \ddot{q}_i - H(\bar{q}_i, \bar{p}_i, \bar{s}_i,t) = \sum_i P_i \dot{Q}_i + S_i \ddot{Q}_i - K(\bar{Q}_i, \bar{P}_i, \bar{S}_i,t) + \frac{dF}{dt} \label{oldNewHamilrelations}
\end{equation}
where $F$ is any function depending on the coordinates of the phase space with continuous second derivatives. The contribution of function $F$ to the variation of the action integral occurs only at the endpoints. The time derivative
\begin{equation}
\int_{t_1}^{t_2} \frac{dF}{dt} \; dt = F(2)- F(1)
\end{equation}
shows that if function $F$ depends on the old and the new canonical variables, its variation is zero since canonical variables have zero variations at the endpoints. 

The relations \ref{newCordOldCord} connect the old and the new coordinates then, function $F$ shall depend on a combination of such type of coordinates up to the total value of $3n$. Let's suppose that the transformation function has the $F_1(q,Q,S)$ dependency. We can introduce, with no loss of generality, $2n$ more variables - $\dot{q}$ and $\dot{Q}$ -  to  $F_1$ function. Its dependency now is $F_1(q,\dot{q},Q,\dot{Q},S)$. $\dot{q}$ and $\dot{Q}$ variables are not independent on function $F$, in fact, we need $2n$ more relations for function $F_1(q,Q,S)$ keep its original $3n$ variables $(q,Q,S)$. We have $n$ relations from the straight time derivative of relations \ref{newCordOldCord}:
\begin{align}
&\dot{Q}_i=\frac{\partial Q_i}{\partial q_i}\dot{q}_i + \frac{\partial Q_i}{\partial p_i}\dot{p}_i+ \frac{\partial Q_i}{\partial s_i}\dot{s}_i
\nonumber \\
&\dot{P}_i=\frac{\partial P_i}{\partial q_i}\dot{q}_i + \frac{\partial P_i}{\partial p_i}\dot{p}_i+ \frac{\partial P_i}{\partial s_i}\dot{s}_i
\nonumber \\
&\dot{S}_i=\frac{\partial S_i}{\partial q_i}\dot{q}_i + \frac{\partial S_i}{\partial p_i}\dot{p}_i+ \frac{\partial P_i}{\partial S_i}\dot{s}_i.
\label{newCordOldCordDiff}
\end{align}
Others $n$ relations will be obtained later in the study of the identity transformation. 

Substituting $F_1(q,\dot{q},Q,\dot{Q},S)$ in equation \ref{oldNewHamilrelations} we obtain:
\begin{align}
\sum_i p_i \dot{q}_i + s_i \ddot{q}_i - H &= \sum_i P_i \dot{Q}_i + S_i \ddot{Q}_i -K
\nonumber \\
& + \frac{\partial F_1}{\partial q_i} \dot{q}_i + \frac{\partial F_1}{\partial \dot{q}_i} \ddot{q}_i + \frac{\partial F_1}{\partial Q_i} \dot{Q}_i + \frac{\partial F_1}{\partial \dot{Q}_i} \ddot{Q}_i + \frac{\partial F_1}{\partial S_i} \dot{S}_i +  \frac{\partial F_1}{\partial t} \label{F1HKrelation}
\end{align}

Since the old and new coordinates are separately independent, the equation holds if each coefficient of $\dot{q}_i$, $\ddot{q}_i$, $\dot{Q}_i$ and $\ddot{Q}_i$ vanish, from where we obtain:
\begin{align}
&p_i = \frac{\partial F_1}{\partial q_i}
,\;\;\;\; 
s_i = \frac{\partial F_1}{\partial \dot{q}_i}
,\;\;\;\;
P_i = -\frac{\partial F_1}{\partial Q_i}
,\;\;\;\;
S_i = - \frac{\partial F_1}{\partial \dot{Q}_i}
,\;\;\;\;
0 = \frac{\partial F_1}{\partial S_i}
\nonumber \\
&K=H+ \frac{\partial F_1}{\partial t}
\end{align}

Another transformation can be a different function depending on the new momentum $P$ as $F_2(q,P,S)$. We can obtain the new function $F_2$ from function $F_1$ using the D'Alembert transformation as
\begin{equation}
F_1 = F_2 - Q_i P_i
\end{equation} 
We can also expand the $F_2(q,P,S)$ function with the variables $\dot{q}$ and $\dot{Q}$ to the function $F_2(q,\dot{q},P,\dot{Q},S)$, being $\dot{q}_i$ and $\dot{Q}_i$ not independent variables. Again we will need 2-$n$ more relations for the added variables so the former function $F_2$ depend only on the $3n$ initials variables as $F_2(q,\dot{q},P,\dot{Q},S)$. Same as previous case, we have the $n$ relations given by the straight time derivative of the transformation relations \ref{newCordOldCord} shown in equation \ref{newCordOldCordDiff}. The others $n$ relations will be obtained once we study the Identity transformation for this type of functions.

The relation between the two Hamiltonians for this type of functions, equation \ref{F1HKrelation}, can be written as:
\begin{align}
\sum_i p_i \dot{q}_i + s_i \ddot{q}_i - H &= \sum_i -Q_i \dot{P}_i + S_i \ddot{Q}_i -K
\nonumber \\
& + \frac{\partial F_2}{\partial q_i} \dot{q}_i + \frac{\partial F_2}{\partial \dot{q}_i} \ddot{q}_i + \frac{\partial F_2}{\partial P_i} \dot{P}_i + \frac{\partial F_2}{\partial \dot{Q}_i} \ddot{Q}_i + \frac{\partial F_2}{\partial S_i} \dot{S}_i +  \frac{\partial F_2}{\partial t}.
\end{align}
The coefficient of the terms $\dot{q}_i$, $\ddot{q}_i$, $\dot{Q}_i$ and $\ddot{Q}_i$ must vanish, leading to equations:
\begin{align}
&p_i = \frac{\partial F_2}{\partial q_i}
,\;\;\;\; 
s_i = \frac{\partial F_2}{\partial \dot{q}_i}
,\;\;\;\;
Q_i = \frac{\partial F_2}{\partial P_i}
,\;\;\;\;
S_i = - \frac{\partial F_2}{\partial \dot{Q}_i}
,\;\;\;\;
0 = \frac{\partial F_2}{\partial S_i}
\nonumber \\
&K=H+ \frac{\partial F_2}{\partial t} \label{TransfEqF2}
\end{align}

We proceed now to define the identity transformation. Let us consider the this canonical transformations as a $F_2$ function type. In that case, the most straightforward Identity transformation has the form 
\begin{equation}
F_2 = \sum_i q_i P_i + \dot{q}_i S_i -  s_i \dot{Q}_i \label{Ident1}.
\end{equation}
From were the vanishing coefficients of equations \ref{TransfEqF2} result in
\begin{align}
&p_i = \frac{\partial F_2}{\partial q_i} = P_i
,\quad
s_i = \frac{\partial F_2}{\partial \dot{q}_i} = S_i
,\quad
Q_i = \frac{\partial F_2}{\partial P_i} = q_i
,\quad
S_i = - \frac{\partial F_2}{\partial \dot{Q}_i} = s_i \label{Ident2}
,\quad 
K=H.
\\
&0 = \frac{\partial F_2}{\partial S_i} = \dot{q}_i \label{Ident3}
\end{align}
The equations \ref{Ident2} shows that the old and the new coordinates are the same, probing function \ref{Ident1} being a suitable candidate for identity transformation. However, the last equation, \ref{Ident3}, set constraint for the generalized velocities. 
As we remember on the definition of functions $F_1$ and $F_2$, the variables $\dot{q}_i$ and $\dot{Q}_i$ were added as a dependent set of variables needed in the transformations and that it was needed $n$ more relations between these coordinates to describe the transformation successfully. Well, the equations \ref{Ident3} are the referred relations. Nevertheless, the obtained set of equations, $ \dot{q}_i=0$ are not acceptable solutions to our problem.

We instead, propose the identity transformation as:
\begin{equation}
F_2 = \sum_i q_i P_i + \mathcal{F}_{2_i}(\bar{\dot{q}}) S_i -  s_i \dot{Q}_i \label{Identity}.
\end{equation}
where $\mathcal{F}_{2_i}(\bar{\dot{q}})$ is the $i$-component of a function depending of all $\{\dot{q}_i\}$ that satisfied:
\begin{equation}
\frac{\partial \mathcal{F}_{2_i}(\bar{\dot{q}})}{\partial \dot{q}_j} = \delta_{ij},\;\;\;\;\;  \mathcal{F}_{2_i}(\bar{\dot{q}})\neq \dot{q}_i  + C_i\label{correlFunctDef}
\end{equation}
where $C_i$ are constant. For this transformation, we obtain the relations:
\begin{align}
&p_i = \frac{\partial F_2}{\partial q_i} = P_i
,\quad 
s_i = \frac{\partial F_2}{\partial \dot{q}_i} = \frac{\partial  \mathcal{F}_{2_i}(\bar{\dot{q}})}{\partial \dot{q}_i} S_i= S_i,
\nonumber \\
&Q_i = \frac{\partial F_2}{\partial P_i} = q_i
,\quad  
S_i = - \frac{\partial F_2}{\partial \dot{Q}_i} = s_i \label{Identity2}
,\quad 
K=H
\\
&0 = \frac{\partial F_2}{\partial S_i} = \mathcal{F}_{2_i}(\bar{\dot{q}}) \label{correlFunctCond}
\end{align}
The equations  \ref{correlFunctCond}
\begin{equation*}
 \mathcal{F}_{2_i}(\bar{\dot{q}}) = 0
\end{equation*}
are the $n$ remaining relations needed for variables $q, P, S$  being the only independent degrees of freedom in the function $F_2(q,\dot{q}, P, \dot{Q}, S)$ in agreeing with the number of equations and the number of old and new variables of the transformation. Once the forms of the correlation functions $F_2(q,\dot{q},P, \dot{Q},S)$ are defined, they set $n$ relations between the generalized velocities. These new constraints reduce the number of canonical variables equals the number of the variables on the Lagrange approach, removing the Ostrogradsky’s instability.

If there is only one particle in the system, the unique solution for $\mathcal{F}_{2_i}$ in equation \ref{correlFunctDef} is: $\mathcal{F}_{2} = \dot{q} + C$, being $C$ a constant. The transformation equation \ref{correlFunctCond} become $ \dot{q} = - C$, which is not an acceptable solutions to our problem. That means that we cannot define an identity transformation for a one particle system depending on $\ddot{q}$. This result is consistent with previous discussions since the $\ddot{q}$'s dependency appeared in our problem when we include mass variation as a new degree of freedom in the beginning. We showed that $n$-VMVF system must include at least two particles or the particle will violate the relativity principle under a Galilean transformation.

Similar correlation functions should be obtained for the others type of transformation function like $F_1(q,Q,S)$.

\subsection{Correlation functions $\mathcal{F}_{2_i}$}
The correlation functions $\mathcal{F}_{2_i}(\bar{\dot{q}})$ depend only on the $\dot{q}_i$ $n$-variables and they have no restrictions besides its definition. In fact, $F_2(q,\dot{q},P, \dot{Q},S)$ can be a sum or an integral. Also, they can be real or complex functions. 

For example, let's define a function $\mathcal{G}_{ij}$ from where each $k$-generalized velocity $\dot{q}_k$ can be obtained from the others as:
\begin{equation}
\dot{q}_i = \sum_j \dot{q}_j \mathcal{G}_{ij}. \label{corrFunction1}
\end{equation}
We obtain some properties for the function $ \mathcal{G}_{ik}$ if we apply the transformation twice:
\begin{equation}
\dot{q}_j = \sum_k \dot{q}_k \mathcal{G}_{jk}. \label{corrFunction2}
\end{equation}
By the inclusion of eq. \ref{corrFunction2} in eq. \ref{corrFunction1}, we obtain the relation:
\begin{equation}
\sum_j \mathcal{G}_{jk} \mathcal{G}_{ij} = \delta_{ik}. \label{coorNormal}
\end{equation}

The function $\mathcal{F}_{2_i}(\bar{\dot{q}})$ can be proposed then as:
\begin{equation}
\mathcal{F}_{2_i}(\bar{\dot{q}}) = \dot{q}_i + f_i = \dot{q}_i + \sum_j  f_j\mathcal{G}_{ij}.
\end{equation}
being $f_i$ any function depending on the particle velocities except $\dot{q}_i$ as $f_i(\dot{q}_1... \dot{q}_{i-1}, \dot{q}_{i+1}...\dot{q}_1)$

Applying correlation function definition, equation \ref{correlFunctDef}, we obtain:
\begin{equation}
\frac{\partial \mathcal{F}_{2_i}(\bar{\dot{q}})}{\partial \dot{q}_j} = \delta_{ij} + \sum_k  f_k \frac{\partial \mathcal{G}_{ik}}{\dot{q}_j} + \mathcal{G}_{ik} \frac{\partial  f_k }{\dot{q}_j} =  \delta_{ij}
\end{equation}
from where:
\begin{equation}
\sum_k  f_k \frac{\partial \mathcal{G}_{ik}}{\dot{q}_j} + \mathcal{G}_{ik} \frac{\partial  f_k }{\dot{q}_j}=0.
\end{equation}
Multiplying by $ \mathcal{G}_{ki}$ and summing by the $i$-index
\begin{equation}
\sum_{ik}  \mathcal{G}_{ki} f_k \frac{\partial \mathcal{G}_{ik}}{\dot{q}_j} + \mathcal{G}_{ki} \mathcal{G}_{ik} \frac{\partial  f_k }{\dot{q}_j}=0
\end{equation}
using normalization condition eq. \ref{coorNormal}, functions $f_i$ and $\mathcal{G}_{ij}$ should satisfied
\begin{equation}
\sum_{k} \Big [ \frac{\partial  f_k }{\dot{q}_j} + \sum_i \mathcal{G}_{ki} f_k \frac{\partial \mathcal{G}_{ik}}{\dot{q}_j} \Big ]= 0
\end{equation}

The previous development was about trying to obtain relations assuming the existence of a function from the velocities can be obtained from.
Others forms for the correlation functions should exist and each of them should provide a different interpretation of the phenomenon. We will explore over this topic in future works.

\subsection{Infinitesimal canonical transformations}
We study now the infinitesimal canonical transformations were new variables differ from the old ones just by infinitesimals. In that case, the transformation equations \ref{newCordOldCord} have the form:
\begin{align}
Q_i=q_i + \delta q_i 
\nonumber \\
P_i=p_i + \delta p_i 
\nonumber \\
S_i=q_i + \delta s_i, 
\end{align}
where $\delta q_i$, $\delta p_i$ and $\delta s_i$ are the real displacements of each variable, respectively. The canonical infinitesimal transformation can be written as the sum of the identity transformation plus an infinitesimal function. In the case of transformations describe with $F_2$ type functions, they have the form
\begin{equation}
F_2 = \sum_i q_i P_i + \mathcal{F}_{2_i}(\bar{\dot{q}}) S_i -  s_i \dot{Q}_i + \epsilon \mathcal{G}(q_i, p_i, s_i, t)
\end{equation}
were $\epsilon$ is an infinitesimal parameter for describing the magnitude of the transformation and $\mathcal{G}(q_i,\dot{q}_i, p_i,\dot{Q}_i, s_i, t)$ is a differentiable function with $3n+1$ arguments known as the generator of such transformation. After applying equations \ref{TransfEqF2} we obtain the transformation relations:
\begin{align}
&p_i = \frac{\partial F_2}{\partial q_i} = P_i + \epsilon \frac{\partial \mathcal{G}}{\partial q_i}
\;\;\;\; \text{or} \;\;\;\; \delta p_i = - \epsilon \frac{\partial \mathcal{G}}{\partial q_i}
\nonumber \\
&s_i = \frac{\partial F_2}{\partial \dot{q}_i} = S_i + \epsilon \frac{\partial \mathcal{G}}{\partial \dot{q}_i}
\;\;\;\; \text{or} \;\;\;\; \delta s_i = - \epsilon \frac{\partial \mathcal{G}}{\partial \dot{q}_i}
\nonumber \\
&Q_i = \frac{\partial F_2}{\partial P_i} = q_i + \epsilon \frac{\partial \mathcal{G}}{\partial P_i}
\;\;\;\; \text{or} \;\;\;\; \delta q_i = \epsilon \frac{\partial \mathcal{G}}{\partial P_i}
\nonumber \\
&S_i = - \frac{\partial F_2}{\partial \dot{Q}_i} = s_i + \epsilon \frac{\partial \mathcal{G}}{\partial \dot{Q}_i} 
\;\;\;\; \text{or} \;\;\;\; \delta s_i = \epsilon \frac{\partial \mathcal{G}}{\partial \dot{Q}_i}
\nonumber \\
&0 = \frac{\partial F_2}{\partial S_i} = \mathcal{F}_{2_i}(\bar{\dot{q}})  + \epsilon \frac{\partial \mathcal{G}}{\partial S_i}.
\end{align}

The infinitesimal canonical transformation generated by the generalized new momentum $P_i$
\begin{equation}
\mathcal{G} = P_i,
\end{equation}
result in the coordinates variations
\begin{align}
\delta q_j = \epsilon \delta_{ij}, 
\nonumber \\
\delta p_j = 0, 
\nonumber \\
\delta s_j = 0, 
\nonumber \\
\mathcal{F}_{2_j}(\bar{\dot{q}})=0.
\end{align}
The set of equations shows that the generator $\mathcal{G} = P_i = p_i +\delta p_i = p_i$ transforms the system displacing only of coordinate$q_i$ if parameter $\epsilon$ is the displacement value. This fact settles the generalized momentum as the generator of the displacement of its own coordinate, same as the former Hamilton theory. This result is expected since the second equation of the extended Hamilton equations \ref{ExtHamiltonEq} remains unaltered compared to the original Hamilton theory.

The transformation of the system with the new momentum $S_i$ as the generator of the transformation
\begin{equation}
\mathcal{G} = S_i,
\end{equation}
is described by the coordinates changes:
\begin{align}
\delta q_j = 0, 
\nonumber \\
\delta p_j = 0, 
\nonumber \\
\delta s_j = 0, 
\nonumber \\
\mathcal{F}_{2_j}(\bar{\dot{q}})=-\epsilon \delta_{ij}. \label{sGenEquations}
\end{align}
According to this results, the generator $s_i$ transforms the system keeping unaltered the variables $q_i,p_1$ and $s_i$, however, it changes the value of the right member of equation $\mathcal{F}_{2_i}(\bar{\dot{q}})=0$ to the infinitesimal displacement value. 

We define the new variable $f_j(\bar{\dot{q}})$ as the value of the correlation $j$-function at any time. In that case, the form of the correlation functions don't change across the evolution of the system, but their value will vary as $\mathcal{F}_{2_i}(\bar{\dot{q}}) = f_j(\bar{\dot{q}})$. 

The equations obtained \ref{correlFunctCond}, show that all resulting values $f_j(\bar{\dot{q}})$ are zero in the identity transformation, $f_{0_j}(\bar{\dot{q}})=0$. Then, last equation of \ref{sGenEquations} can be written as
\begin{equation}
\mathcal{F}_{2_j}(\bar{\dot{q}})= f_j(\bar{\dot{q}})-0 = f_j(\bar{\dot{q}}) - f_{0_j}(\bar{\dot{q}})= \delta f_j(\bar{\dot{q}})= -\epsilon \delta_{ij}.\label{corrFunctionValue}
\end{equation}

The second order momentum $s$ can be interpreted, then, as the generator of a negative displacement of the value of correlation functions, $\mathcal{F}_{2_i}(\bar{\dot{q}})$. Being the correlation function a constraint involving all particle of the system, we can conclude that the new momentum $s$ is the generator of a collective motion of the system. This collective evolution is in agreement with our initial supposition where the violation of the Newton second law, introduced by the terms proportional to $\ddot{q}$, will be suppressed by the coordinate action of all the particles of the system.

Another important canonical transformation is
\begin{equation}
\mathcal{G} = H + \sum_i \dot{s}_i\dot{Q}_i - \dot{s}_i\dot{q}_i-\ddot{s}_i q_i. \label{timeGenerator}
\end{equation}
The infinitesimal changes of the variables of the system are
\begin{align}
&\delta p_i = - \epsilon \frac{\partial \mathcal{G}}{\partial q_i}= -\epsilon \Big(\frac{\partial H}{\partial q_i} -\ddot{s}_i \Big ) = \epsilon \dot{p}_i
\nonumber \\
&\delta s_i = - \epsilon \frac{\partial \mathcal{G}}{\partial \dot{q}_i} = \epsilon \dot{s}_i
\nonumber \\
&\delta q_i = \epsilon \frac{\partial \mathcal{G}}{\partial P_i} \sim \epsilon \frac{\partial \mathcal{G}}{\partial p_i} = \epsilon \Big(\frac{\partial H}{\partial p_i} \Big) = \epsilon \dot{q}_i
\nonumber \\
&\delta s_i = \epsilon \frac{\partial \mathcal{G}}{\partial \dot{Q}_i} = \epsilon \dot{s}_i
\nonumber \\
&0 = \frac{\partial F_2}{\partial S_i} = \mathcal{F}_{2_i}(\bar{\dot{q}})  + \epsilon \frac{\partial \mathcal{G}}{\partial S_i}  \sim \mathcal{F}_{2_i}(\bar{\dot{q}})  + \epsilon \frac{\partial \mathcal{G}}{\partial s_i}  = \mathcal{F}_{2_i}(\bar{\dot{q}})  + \epsilon \frac{\partial H}{\partial s_i} 
\nonumber \\
& \mathcal{F}_{2_i}(\bar{\dot{q}}) \equiv \delta f_j(\bar{\dot{q}}) =- \epsilon \ddot{q}_i. \label{extCanonEqTime}
\end{align}
On the other side, the time derivative of function $\mathcal{F}_{2_i}(\bar{\dot{q}}) $ is
\begin{equation}
\frac{d f_i(\bar{\dot{q}})}{dt} 
= \frac{d \mathcal{F}_{2_i}(\bar{\dot{q}})}{dt} 
=  \sum_j \frac{\partial \mathcal{F}_{2_i}(\bar{\dot{q}})}{\partial q_j}\ddot{q}_j 
= \sum_j \delta_{ij}\ddot{q}_j
= \ddot{q}_i \label{fTimeDerivative}
\end{equation}
where we use the definition of function $\mathcal{F}_{2_i}(\bar{\dot{q}})$ \ref{correlFunctDef}.
The last relation of equations \ref{extCanonEqTime} can be rewritten then as:
\begin{equation}
\delta f_j(\bar{\dot{q}}) = -  \dot{f}_j(\bar{\dot{q}}) \epsilon
\end{equation}

If parameter $\epsilon$ is the infinitesimal time interval $dt$, then the generator function of equation \ref{timeGenerator}, evolves all variables of the system $q_i,p_i,s_i$ with time and also evolve the value of correlation functions, $ f_j(\bar{\dot{q}})$, in the negative direction.

The negative time evolution for quantities $ f_i(\bar{\dot{q}})$ is consistent with previous results where momentum $s_i$ generate a negative displacement for the value of the correlation $i$-function. According to that, in a time interval $dt$, $s$, as part of previous time generator,  evolve with time from value $s_{i_0}$ to $s_{i_0} + ds_i$. These $s_i$ values generate the negative displacement of the value of the correlation functions $- \delta f_{2_i}(\bar{\dot{q}})$ and $- (\delta f_{i}(\bar{\dot{q}}) + \frac{\partial (ds_i)}{\partial S_i} )$respectively, with an effective displacement of $ d(f_{i}(\bar{\dot{q}})) \equiv -\frac{\partial (ds_i)}{\partial S_i} $ or $ d(f_{2_i}(\bar{\dot{q}})) \equiv -\frac{\partial (\dot{s}_i dt)}{\partial S_i} $. Then, being $dt$ positive, the positive evolution of momentum $s_i$ evolve the quantity $f_{i}(\bar{\dot{q}})$ negatively.

The generator \ref{timeGenerator}, expressed in the old set of coordinates have the form:
\begin{align}
\mathcal{G} &= H + \sum_i  \dot{s}_i\dot{Q}_i - \dot{s}_i\dot{q}_i - \ddot{s}_i q_i = H + \sum_i  \dot{s}_i(\dot{Q}_i - \dot{q}_i) - \ddot{s}_i q_i 
\nonumber \\
&= H+\sum_i \dot{s}_i\delta \dot{q}_i - \ddot{s}_i q_i. \label{timeGenerator1}
\end{align}
We can approach $\delta \dot{q}_i \sim \ddot{q}_i t$. In this case, the system time generator \ref{timeGenerator1} is written as
\begin{equation}
\mathcal{G} \sim H + \sum_i  \dot{s}_i \ddot{q}_i t - \ddot{s}_i q_i. 
\end{equation}
or using equation \ref{fTimeDerivative}
\begin{equation}
\mathcal{G} \sim H + \sum_i  \dot{s}_i \dot{f}_i(\bar{\dot{q}}) t - \ddot{s}_i q_i. \label{timeGeneratorApx1}
\end{equation}

\subsection{Infinitesimal canonical transformations for n-VMVF systems}
The last section defined different canonical transformations for the extended Hamilton theory in general. We exploit these results to the problem that brought us here. On $n$-VMVF systems, the canonical transformation have a singular form because it's associated with two independent set of extended Hamilton equations, needed to solve the problem. We have then, two independent set of equations for two different extended Hamiltonians $H_T$ and $H_R$. The set of equations in the the Lorentzian set of coordinates include the generalized canonical variables of every particle and the values for the correlation functions:
\begin{equation}
q_i \equiv x_n^\nu, \qquad p_i \equiv p_{T_n}^\nu, \qquad s_i \equiv s_{T_n}^\nu \qquad f_{i}(\bar{\dot{q}}) \equiv f_{T_n}^\mu(\bar{\dot{x}}_l^\nu).
\end{equation}
The canonical variables of the set of equations in the angular coordinates are
\begin{equation}
q_i \equiv \xi_{n;i}, \qquad p_i \equiv l_{R_{n;i}}, \qquad s_i \equiv b_{R_{n;i}} \qquad f_{i}(\bar{\dot{q}}) \equiv f_{R_{n;j}}(\bar{\dot{\xi}}_{l;i}).
\end{equation}

The canonical transformation must be considered as one unified transformation. For example, the generator of translation in the  $\nu$- component of the position of the particle $n$ is the momentum $p_{T_n}^\nu$, while the generator of rotation on the $i$-direction of particle $n$ is angular momentum $l_{R_{n;i}}$. This generator shouldn't be seen as two distinct generators for separated transformations but as a single bi-dimensional generator of a more global transformation composed of one translation and one rotation. In that case, we are in the presence of a motion generator. All generator must act in this way because of, only the combined action of both will provide the solution of the system. This point of view is the cornerstone of the expansion of the complex numbers and the extension of quantum mechanics for $n$-VMVF systems. We rewrite the classical generators obtained before, applied to the studied systems.

The generator of a motion of the particle $n$ of the system
$n$ of the system
\begin{equation}
\mathcal{G}_{\text{motion of particle } n } \equiv
\classoperator {\mathcal{G}_T}{\mathcal{G}_R} =
\classoperator {p_{T_{n'}}^\nu}{l_{R_{n';i}}}
\end{equation}
transform the variables of the system as:
\begin{align}
&\classoperator {\delta x^\mu_n} {\delta \xi_{n;j} }
=\classoperator {\epsilon_1 \delta_\nu^\mu \delta_{nn'}}{\epsilon_2 \delta_{ij} \delta_{nn'}}
,\quad
\classoperator {\delta p_{T_n}^\mu}{\delta l_{R_{n;j}}}
= \classoperator {0}{0}
\nonumber \\
&\classoperator {\delta s_{T_n}^\mu}{\delta b_{R_{n;j}}}
= \classoperator {0}{0}
\;\;\;, \quad
\classoperator {\delta f_{T_n}^\mu(\bar{\dot{x}}_l^\nu)}{\delta f_{R_{n;j}}(\bar{\dot{\xi}}_{l;i})}
= \classoperator {0}{0}
\end{align}
The generator of a displacement of the value of correlation functions, $f_{i}(\bar{\dot{q}})$
\begin{equation}
\mathcal{G}_{ f_{i}(\bar{\dot{q}}) \text{'s displacement}} \equiv
\classoperator {\mathcal{G}_T}{\mathcal{G}_R} 
= \classoperator {s_{T_{n'}}^\nu}{b_{R_{n';i}}}
\end{equation}
transform the variables of the system as:
\begin{align}
&\classoperator {\delta x^\mu_n }{\delta \xi_{n;j}}
= \classoperator {0}{0}
\;\;,\quad
\classoperator {\delta p_{T_n}^\mu}{ \delta l_{R_{n;j}}}
= \classoperator {0}{0}
\nonumber \\
&\classoperator {\delta s_{T_n}^\mu }{ \delta b_{R_{n;j}}}
= \classoperator {0}{0}
, \quad
\classoperator {\delta f_{T_n}^\mu(\bar{\dot{x}}_l^\nu)}{\delta f_{R_{n;j}}(\bar{\dot{\xi}}_{l;i})}
= \classoperator {-\epsilon_1 \delta_\nu^\mu \delta_{nn'} }{-\epsilon_2 \delta_{ij} \delta_{nn'}}
\end{align}
Finally, the generator of time evolution of the system according eq. \ref{timeGenerator} and \ref{timeGeneratorApx1} is
\begin{equation}
\mathcal{G}_{\text{System Evolution}} \equiv
\classoperator {\mathcal{G}_T }{ \mathcal{G}_R } 
=
\classoperator {H_T + \sum_{n,\nu} \dot{s}_{T_{n_\nu}}  \dot{f}_{T_n}^\nu(\bar{\dot{x}}_l^\mu) \tau-\ddot{s}_{T_{n_\nu}} x^\nu_n}
{H_R + \sum_{n,i} \dot{b}_{R_{n_i}} \dot{f}_{R_{n;i}}(\bar{\dot{\xi}}_{l;j}) \tau - \ddot{b}_{R_{n_i}} \xi_{n;i} },
\end{equation}
which transforms the variables of the system as:
\begin{align}
&\classoperator {\delta x^\mu_n }{\delta \xi_{n;j}}
=
\classoperator {\epsilon_1 \dot{x}^\mu_n }{ \epsilon_2 \dot{\xi}_{n;j}}
\;\;,\quad
\classoperator {\delta p_{T_n}^\mu} {\delta l_{R_{n;j}} }
=
\classoperator {\epsilon_1 \dot{p}_{T_n}^\mu}{\epsilon_2 \dot{l}_{R_{n;j}}}
\nonumber \\
&\classoperator {\delta s_{T_n}^\nu}{\delta b_{R_{n;j}}}
=
\classoperator {\epsilon_1 \dot{s}_{T_n}^\nu }{ \epsilon_2 \dot{b}_{R_{n;j}}}
, \quad
\classoperator {\delta f_{T_n}^\mu(\bar{\dot{x}}_l^\nu)}{\delta f_{R_{n;j}}(\bar{\dot{\xi}}_{l;i})}
=-
\classoperator {\epsilon_1 \dot{f}_{T_n}^\mu(\bar{\dot{x}}_l^\nu) }{ \epsilon_2 \dot{f}_{R_{n;j}}(\bar{\dot{\xi}}_{l;i}) }
\end{align}

\subsection{Final extended Hamilton equations}
From the starter analysis in this chapter, we found that the Hamilton extended equations \ref{ExtHamiltonEq} have the form 
\begin{align*}
&\frac{\partial H}{\partial q_i} =  -(\dot{p}_i-\ddot{s}_i) &
&\frac{\partial H}{\partial p_i} = \dot{q}_i
\nonumber \\
&\frac{\partial H}{\partial s_i} = \ddot{q}_i &
&\frac{\partial H}{\partial t} = -  \frac{\partial L}{\partial t}.
\end{align*}
However, they aren't enough for describing the evolution of the system, as the canonical transformations shows. In that case, we introduce the correlation functions constrained the time derivative of the canonical variable $q_i$. We replace previous equations by
\begin{align}
&\frac{\partial H}{\partial q_i} =  -(\dot{p}_i-\ddot{s}_i) &
&\frac{\partial H}{\partial p_i} = \dot{q}_i
\nonumber \\
&\frac{\partial H}{\partial s_i} = \dot{f}_i  &
&f_i = \mathcal{F}_{2_i}(\bar{\dot{q}})
\nonumber \\
&\frac{\partial H}{\partial t} = -  \frac{\partial L}{\partial t}
\end{align}
where the values of the correlation functions ${f}_i$ are included as a canonical variable of the system.

\subsection{Constrained second derivative classical systems}
The classic theory we present to solve the problem of  $n$-VMVF systems, reveals a set of constrained functions. They appear after demand the isolated system satisfies the linear and angular momentum conservation laws. Also, there were other constraints added, in this case, because of the inclusion of Lorentz transformations which relate the components of the particle vector positions, the masses, and field first derivatives. Along the development of the classical approach for the referred systems, we include the first group of constraints in the Lagrange function using the Lagrange multipliers method while the relativistic constraints were suggested to treat them as ``weak conditions'', according to the methodology at Goldstein textbook textbook \cite{goldstein321}. The ``weak conditions'' terminology was introduced by Dirac and it means this constraint should be imposed after all derivation processes have been executed, to the detriment of the Lagrangian equations become inconsistent. As the main objective of this work is not to solve the problem but to propose a new quantum approach for the $n$-VMVF systems, we won't require, at that time, to go further on this direction. However, we are now forced to express the constraints as functions of the canonical coordinates.

The constraints obtained by the two momentum conservations laws are added to the single particle former Lagrangian: 
\begin{equation*}
L = \frac{1}{2}m_n \dot{x}^\nu_n \dot{x}_{n;\nu} -A^\nu \dot{x}_{n;\nu}
\end{equation*}
using the Lagrange multipliers method, which leads to the construction of two independent extended Lagrange functions depending on the second derivative of particle position $\ddot{\vec{r}}$.  We set this two Lagrangian functions as our starting point for the subsequent development in the Hamilton approach.

The corresponding Hamiltonians were obtained from each Lagrange function with no further complications. We were able to obtain the relation between velocities and momentum, and even there are no relations between the generalized accelerations $\ddot{q}_i$ and the second order momentum $s$, the terms  $\sum_i s_i \ddot{q}_i$  from our Lagrangian cancels because of the extended Hamilton definitions. This issue makes the obtained extended Hamiltonians independent of $s_i$ variables. However, as such relation does not exist, we can not express the constraints as functions of the canonical variables, or more specifically, we can not replace the $\ddot{q}_i$ variables on our constraints. Because of that, we are then in the presence of a second order singular system.

It is well known that the singularities lead to constraints that reduce the phase space of the physical system from which are obtained the so-called ``Gauge identities''. Singular Lagrangian is the base of the dynamic of the so-called Gauge theories.

As an introduction to this topic, let us take a look at the first order singular problem. A singular Lagrangian has been defined as whether the determinant of the Hessian is zero or not. On the ordinary Lagrangian problem, the Euler-Lagrange equations read as
\begin{equation*}
\frac{d}{dt} \frac{\partial L}{\partial \dot{q}_i}  - \frac{\partial L}{\partial q_i} = 0,
\end{equation*}
from where it follows that
\begin{equation}
\sum_j \frac{\partial^2 L}{\partial \dot{q}_i \partial \dot{q}_j} \ddot{q}_j + \frac{\partial^2 L}{\partial \dot{q}_i q_j} \dot{q}_j - \frac{\partial L}{\partial q_i} = 0
\end{equation}
or 
\begin{equation}
\sum_j W_{ij}^{\rom{1}} \ddot{q}_j = \frac{\partial L}{\partial q_i} -\sum_j \frac{\partial^2 L}{\partial \dot{q}_i q_j} \dot{q}_j 
\end{equation}
where the matrix 
\begin{equation}
W_{ij}^{\rom{1}} \equiv \frac{\partial^2 L}{\partial \dot{q}_i \partial \dot{q}_j}
\end{equation}
is the Hessian of the system. The Hessian $W_{ij}^{\rom{1}} $ also determine the relations between the generalized momentum and the generalized velocities, since, using the definition of the first order momentum, we have
\begin{equation}
W_{ij}^{\rom{1}} = \frac{\partial^2 L}{\partial \dot{q}_i \partial \dot{q}_j} 
= \frac{\partial}{\partial \dot{q}_i} \Big( \frac{\partial L}{ \partial \dot{q}_j} \Big)
= \frac{\partial p_j}{\partial \dot{q}_i}. \label{firstDerivHessian}
\end{equation}
If the Hessian determinant $\det|W_{ij}^{\rom{1}}| \neq 0$ then we will find $n$ well-defined relations of the acceleration as function of the velocities and particle position: $\ddot{q}_i = \ddot{q}_i(\dot{q}, q)$ and also $n$ well-defined relations of the momentum as a function of the velocities and particle position $p_i = p_i(\dot{q}, q)$. However, is the determinant of the Hessian vanish, new constraints appear and with them, new Gauge symmetries.

At the Lagrangian level, there are well-known methods to find the constraints which led to the so-called Gauge transformations. Those transformations are parametrized by a set of arbitrary functions of time, whose number equals the number of the Gauge identities. Lagrangian based algorithms can be found in literature \cite{sudarshan1974classical, rothe2010classical}, and they generate the transformation laws in the configuration space involving the independent set of arbitrary functions of time which characterize the symmetry transformations. The method to find the constraints at the Hamiltonian level was first introduced by Dirac \cite{dirac2001lectures} and developed for others authors \cite{rothe2010classical, hanson1976constrained, sundermeyer1982constrained}.

\subsubsection{Algorithm for detecting second-order local symmetries at the Lagrangian level}
Among the classic equations obtained to solve the $n$-VMVF systems, we have constraints between the variables $s_i$ and the generalized accelerations $\ddot{q}_j$ from the extended Lagrange functions. Because of that we are in present of a constrained second order system. In this section we attempt the initials ideas to treat such systems.

The extended Lagrange equations \ref{extLagrangeEq2Order} have the form
\begin{equation*}
\frac{d^2}{d t^2}\Big( \frac{\partial L}{\partial \ddot{q}_i}\Big) - \frac{d}{dt}\Big( \frac{\partial L}{\partial \dot{q}_i}\Big) + \frac{\partial L}{\partial q_i} =0.
\end{equation*}
Expanding the equation we have
\begin{align}
& \frac{d}{d t}\Big[ 
  \frac{\partial^2 L}{\partial \ddot{q}_i \partial q_j} \dot{q}_j
+ \frac{\partial^2 L}{\partial \ddot{q}_i \partial \dot{q}_j} \ddot{q}_j
+ \frac{\partial^2 L}{\partial \ddot{q}_i \partial \ddot{q}_j} \dddot{q}_j 
\Big]
- \frac{\partial^2 L}{\partial \dot{q}_i \partial q_j} \dot{q}_j
- \frac{\partial^2 L}{\partial \dot{q}_i \partial \dot{q}_j} \ddot{q}_j
- \frac{\partial^2 L}{\partial \dot{q}_i \partial \ddot{q}_j} \dddot{q}_j = 0
\nonumber \\
&=\frac{\partial^2 L}{\partial \ddot{q}_i \partial q_j} \ddot{q}_j
+ \frac{\partial^2 L}{\partial \ddot{q}_i \partial \dot{q}_j} \dddot{q}_j
+ \frac{\partial^2 L}{\partial \ddot{q}_i \partial \ddot{q}_j} \ddddot{q}_j 
+ \frac{\partial^3 L}{\partial \ddot{q}_i \partial q_j \partial q_k} \dot{q}_j \dot{q}_k
+ \frac{\partial^3 L}{\partial \ddot{q}_i \partial q_j \partial \dot{q}_k} \dot{q}_j \ddot{q}_k
\nonumber \\
&+\frac{\partial^3 L}{\partial \ddot{q}_i \partial q_j \partial \ddot{q}_k} \dot{q}_j \dddot{q}_k
+ \frac{\partial^3 L}{\partial \ddot{q}_i \partial \dot{q}_j \partial q_k} \ddot{q}_j \dot{q}_k
+ \frac{\partial^3 L}{\partial \ddot{q}_i \partial \dot{q}_j \partial \dot{q}_k} \ddot{q}_j \ddot{q}_k
+ \frac{\partial^3 L}{\partial \ddot{q}_i \partial \dot{q}_j \partial \ddot{q}_k} \ddot{q}_j \dddot{q}_k
+ \frac{\partial^3 L}{\partial \ddot{q}_i \partial \ddot{q}_j \partial {q}_k} \dddot{q}_j \dot{q}_k 
\nonumber \\
&+\frac{\partial^3 L}{\partial \ddot{q}_i \partial \ddot{q}_j \partial \dot{q}_k} \dddot{q}_j \ddot{q}_k 
+ \frac{\partial^3 L}{\partial \ddot{q}_i \partial \ddot{q}_j \partial \ddot{q}_k} \dddot{q}_j \dddot{q}_k 
- \frac{\partial^2 L}{\partial \dot{q}_i \partial q_j} \dot{q}_j
- \frac{\partial^2 L}{\partial \dot{q}_i \partial \dot{q}_j} \ddot{q}_j
- \frac{\partial^2 L}{\partial \dot{q}_i \partial \ddot{q}_j} \dddot{q}_j 
+ \frac{\partial L}{\partial q_i}= 0
\end{align}
or
\begin{align}
&\;W^{(\rom{2})}_{ij} \ddddot{q}_j  = \Big[
- \frac{\partial^2 L}{\partial \ddot{q}_i \partial q_j} \ddot{q}_j
- \frac{\partial^2 L}{\partial \ddot{q}_i \partial \dot{q}_j} \dddot{q}_j
- \frac{\partial^3 L}{\partial \ddot{q}_i \partial q_j \partial q_k} \dot{q}_j \dot{q}_k
- \frac{\partial^3 L}{\partial \ddot{q}_i \partial q_j \partial \dot{q}_k} \dot{q}_j \ddot{q}_k
-\frac{\partial^3 L}{\partial \ddot{q}_i \partial q_j \partial \ddot{q}_k} \dot{q}_j \dddot{q}_k
\nonumber \\
&- \frac{\partial^3 L}{\partial \ddot{q}_i \partial \dot{q}_j \partial q_k} \ddot{q}_j \dot{q}_k
- \frac{\partial^3 L}{\partial \ddot{q}_i \partial \dot{q}_j \partial \dot{q}_k} \ddot{q}_j \ddot{q}_k
- \frac{\partial^3 L}{\partial \ddot{q}_i \partial \dot{q}_j \partial \ddot{q}_k} \ddot{q}_j \dddot{q}_k
- \frac{\partial^3 L}{\partial \ddot{q}_i \partial \ddot{q}_j \partial {q}_k} \dddot{q}_j \dot{q}_k 
-\frac{\partial^3 L}{\partial \ddot{q}_i \partial \ddot{q}_j \partial \dot{q}_k} \dddot{q}_j \ddot{q}_k 
\nonumber \\
&- \frac{\partial^3 L}{\partial \ddot{q}_i \partial \ddot{q}_j \partial \ddot{q}_k} \dddot{q}_j \dddot{q}_k 
+ \frac{\partial^2 L}{\partial \dot{q}_i \partial q_j} \dot{q}_j
+ \frac{\partial^2 L}{\partial \dot{q}_i \partial \dot{q}_j} \ddot{q}_j
+ \frac{\partial^2 L}{\partial \dot{q}_i \partial \ddot{q}_j} \dddot{q}_j 
- \frac{\partial L}{\partial q_i} \Big] = 0. 
\label{ExtLagranExpansion}
\end{align}
where 
\begin{equation}
W^{\rom{2}}_{ij} \equiv \frac{\partial^2 L}{\partial \ddot{q}_i \partial \ddot{q}_j} 
\end{equation}
is the Hessian for the second derivative.

We explore the constrained systems restricted only to $n$-VMVF systems, which means we are going to study the system such that 
\begin{equation}
\det |W^{\rom{1}}_{ij}| \neq 0
\qquad \text{and} \qquad 
\det |W^{\rom{2}}_{ij}|  =0,\; \forall i,j = 1... n.
\end{equation}

At the Lagrangian level, we follow a similar algorithm used on the well-known algorithm for detecting all gauge symmetries of an ordinary Lagrangian \cite{sudarshan1974classical, rothe2010classical} so that we can find all gauge symmetries of the obtained extended Lagrangian. 

The four derivative term on equation \ref{ExtLagranExpansion} can be isolated as
\begin{equation}
\ddddot{q}_j = W^{\rom{2}-}_{ij} \Psi(q,\dot{q}, \ddot{q}, \dddot{q}) \label{ExtLagranExpansion0}
\end{equation}
if the determinant of the matrix  $W^{\rom{2}}_{ij}$ not vanish and we have the four derivatives of the generalized coordinate as a function of the others derivatives terms, like equation \ref{ExtLagranExpansion0}. 
However, if the determinant of $\det W^{\rom{2}}_{ij} =0$ is zero i.e  $W^{\rom{2}}_{ij}$ is not invertible, then the four derivative term $\ddddot{q}_j$ is not uniquely determined by the positions $q$, velocities $\dot{q}$, accelerations $\ddot{q}$ and $\dddot{q}$. The obtained set of Lagrange equations for the $n$-VMVF is an example of singular equations where the Hessian $W^{\rom{2}}_{ij}$ is not invertible. Fortunately, in our case, the determinant of matrix  $W^{\rom{1}}_{ij}$ not vanish, which means we can assume the existence of the expression of first-time derivative $\dot{q}_i$ as functions of the others time derivatives.

The infinitesimal variations of the coordinates, velocities, and accelerations have the form:
\begin{align*}
q_i(t) &\to q_i(t) + \delta q_i(t)
\nonumber \\
\dot{q}_i(t) &\to \dot{q}_i(t) + \frac{d}{dt}\delta q_i(t)
\nonumber \\
\ddot{q}_i(t) &\to \ddot{q}_i(t) + \frac{d^2}{dt^2}\delta q_i(t)
\end{align*}
where $ i = 1 \dots N$, and  we has made use of the property
\begin{equation}
\delta \dot{q}_i(t) =  \frac{d}{dt}\delta q_i(t) \qquad \text{and } 
\qquad \delta \ddot{q}_i(t) =  \frac{d}{dt}\delta \dot{q}_i(t).
\end{equation}
They are part of a transformation which is a local symmetry for the extended Lagrange equations if action
\begin{equation*}
S[q] = \int_{t_1}^{t_2} L(q,\dot{q}, \ddot{q}) dt
\end{equation*}
remain invariant under such variations. If the variation of the Lagrangian is a total derivative of an arbitrary function $\delta L = \frac{d F}{dt}$ and also the variation of the function $\delta F$ vanish at the upper and lower limit of integration, then we are in the presence of a local symmetry for the physical system.

Under the above transformation, the variation of the action of an extended Lagrangian $L(q,\dot{q}, \ddot{q})$ have the form:
\begin{equation}
\delta S[q] = \int_{t_1}^{t_2}  dt E_i^{(0)}(q,\dot{q}, \ddot{q}) \delta q,
\end{equation}
where $E_i^{(0)}$ is the Euler derivative
\begin{equation}
E_i^{(0)} = 
\frac{d^2}{d t^2}\Big( \frac{\partial L}{\partial \ddot{q}_i}\Big) 
- \frac{d}{dt}\Big( \frac{\partial L}{\partial \dot{q}_i}\Big) 
+ \frac{\partial L}{\partial q_i}
\end{equation}
and $\delta q_i (t_1) = \delta q_i (t_2)  = 0$. On the shell composed by the physical path, we have
\begin{equation}
E_i^{(0)} = 0, \qquad i = 1, ... , N \;(on\; shell).
\end{equation}
Given the above considerations, we have to find the variations $\delta q_i$ for which $\delta S = 0$.

We can express $E_i^{(0)}$ like 
\begin{equation}
E_i^{(0)}(q,\dot{q}, \ddot{q}, \dddot{q}, \ddddot{q}) = W^{\rom{2}(0)}_{ij}(q,\dot{q}, \ddot{q}, \dddot{q}) \ddddot{q} +  K^{\rom{2}(0)}(q,\dot{q}, \ddot{q}, \dddot{q}),
\end{equation}
where the superscript ``$n$'' stands for ``$n$-level'' or ``$n$-generation''. In the future, we suppress, for simplicity, the explicit dependence of $W^{\rom{2}(0)}_{ij}$ and $ K^{\rom{2}(0)}$ with variables $q,\dot{q}, \ddot{q}, \dddot{q}$ and $ E^{\rom{2}(0)}_{i}$ with variables $q,\dot{q}, \ddot{q}, \dddot{q}, \ddddot{q}$.

As mentioned before, the Lagrange equations of motion for a singular Lagrangian of a second order, cannot be solved for all of the $\ddddot{q}$s terms. However, if the rank of the second derivative Hessian matrix is  $R_W$ then, there are $N-R_W$ constraints in the theory which connect the quantities $q,\dot{q}, \ddot{q}, \dddot{q}$ and $\ddddot{q}$. Indeed, for the singular Lagrangian with a second order Hessian, there are $N-R_W$ independent left (or right), zero-mode eigenvectors $\vec{w}^{\rom{2}(0;k)}(q,\dot{q}, \ddot{q}, \dddot{q})$ satisfying
\begin{equation}
\sum_{i=1}^N w^{\rom{2}(0;k)}_i  W^{\rom{2}(0)}_{ij} = 0 ; \qquad k = 1,...,N-R_W.
\end{equation}
The ``zero level'' or ``zero generation'' constraints $\Phi^{\rom{2}(0;k)} $ then have the form
\begin{equation}
\Phi^{\rom{2}(0;k)} \equiv  \sum_{i=1}^N w^{\rom{2}(0;k)}_i E^{\rom{2}(0)}_{i} =0 \; \qquad (on \; shell).
\end{equation}

All constraints $\Phi^{\rom{2}(0;k)} $ may not be linearly independent. In that case, we can find a linear combination of the zero-mode eigenvectors such as
\begin{equation}
v^{\rom{2}(0;n_0)}_i  = \sum_{k} c_k^{\rom{2}(n_0)} w^{\rom{2}(0;k)}_i,
\end{equation}
such that we have identically
\begin{equation}
G^{\rom{2}(0;n_0)} \equiv  \sum_{i=1}^N v^{\rom{2}(0;n_0)}_i  E^{\rom{2}(0)}_{i} =0
; \qquad n_0  = 1,..., N_0.
\end{equation}
These constraints are called Gauge Identities, and they also hold off shell. 

One consequence is that the variation of the coordinates in the form
\begin{equation*}
\delta q_i = \sum_{n_0} \epsilon_{n_0}(t) v^{\rom{2}(0;n_0)}_i
\end{equation*}
will leave the action invariant which is the desired result for of the method. However, others ``zero-generation'' zero modes, which are denoted by $\vec{u}^{\rom{2}(0;\bar{n}_0)} $, can appear. They are genuine constraints that vanish on shell
\begin{equation}
\phi^{\rom{2}(0;\bar{n}_0)} = \vec{u}^{\rom{2}(0;\bar{n}_0)} \cdot \vec{E}^{\rom{2}(0)} ; \qquad \bar{n}_0  = 1,..., \bar{N}_0.
\end{equation}

It is also required the persistence of the constraints with time, which means 
\begin{equation}
\frac{d}{d t}\phi^{\rom{2}(0;\bar{n}_0)} = 0.
\end{equation}

The kernel of the algorithm is to gather all Gauss Identities until there are no more non-trivial constraints as pointed out on references \cite{sudarshan1974classical, rothe2010classical}. The genuine constraints are obtained by searching for further functions of $q,\dot{q}, \ddot{q}, \dddot{q},\ddddot{q}$ which vanish on the subspace of the physical paths. To find them is convenient to construct an $N + \bar{N}_0$ composed with the component vector $\vec{E}{\rom{2}(0)}$ and the time derivative of the non-trivial constraints:
\begin{equation}
\Big( \vec{E}^{\rom{2}(1)}\Big) =
\left(
\begin{array}{c}
 \vec{E}^{\rom{2}(0)}\\
 \frac{d}{dt}(\phi^{\rom{2}(0;\bar{n}_0)})
 \end{array}
\right) 
=
\left(
\begin{array}{c}
 \vec{E}^{\rom{2}(0)}\\
 \frac{d}{dt}(\vec{u}^{\rom{2}(0;1)} \cdot  \vec{E}^{\rom{2}(0)})\\
 \vdots \\
 \frac{d}{dt}(\vec{u}^{\rom{2}(0;\bar{N}_0)} \cdot  \vec{E}^{\rom{2}(0)})\\
\end{array}
\right).
\end{equation}

Because the constraints are only functions of $q,\dot{q}, \ddot{q}, \dddot{q}$, the component vector from $\vec{E}^{\rom{1}(0)}$ can be written as
\begin{equation}
E_i^{(1)} = W^{\rom{2}(1)}_{ij} \ddddot{q} +  K^{\rom{2}(1)},
\end{equation}
where $ W^{\rom{2}(1)}$ is a ``level 1'' $(N+\bar{N}_0) \times N$ matrix.

The process repeats and the (left) zero-modes of $ W^{\rom{2}(1)}$ are found. The already founded zero modes are also reproduced generating the same constraints. However, the new zero modes, if there exist, lead to others constraints when contracted with vector $\vec{E}^{\rom{2}(1)}$. The new constraints can be either written as a linear combination of the others constraints, which led to new gauge identities at level 1:
\begin{equation}
G^{\rom{2}(1;n_1)} = \vec{v}^{\rom{2}(1;n_1)} \cdot  \vec{E}^{\rom{2}(1)} 
- \sum_{\bar{n}_0 = 1}^{\bar{N}_0} M_{n_1, \bar{n}_0}^{\rom{2}(1,0)}
\Big( \vec{u}^{\rom{2}(0;\bar{n}_0)} \cdot \vec{E}^{\rom{2}(0)} \Big)
=0
; \qquad n_1  = 1,..., N_1.
\end{equation}
alternatively, they may represent new ``level 1'' genuine constraints:
\begin{equation}
\phi^{\rom{2}(1;\bar{n}_1)} = \vec{u}^{\rom{2}(1;\bar{n}_1)} \cdot \vec{E}^{\rom{2}(1)} ; \qquad \bar{n}_1  = 1,..., \bar{N}_1.
\end{equation}

The method continues constructing a new vector $\vec{E}^{\rom{2}(2)}$ and adding the time derivative of the new constraints $\phi^{\rom{2}(1;\bar{n}_1)} $ to the previous  $\vec{E}^{\rom{2}(1)}$ and with it the functions $W^{\rom{2}(2)}$ and $\vec{K}^{\rom{2}(2)}$. The iterative process will finish at one $M$ level if either:
\begin{itemize}
\item there are no further zero modes for matrix $W^{\rom{2}(M)}_{ij}$
\item there is no further genuine new constraints $\phi^{\rom{2}(M;\bar{n}_M)}$.
\end{itemize}

\subsubsection{Algorithm for detecting second-order local symmetries at the Hamiltonian level}
There is an alternative method to treat such constrained systems within the Hamilton theory, first described by Dirac. We expose what we think are the first ideas for the constrained systems with a second order. The starting point for obtaining the extended Hamiltonian is the set of definitions
\begin{equation*}
p_i = \frac{\partial L}{\partial \dot{q}_i}, \qquad 
s_i = \frac{\partial L}{\partial \ddot{q}_i}
\end{equation*}
Also, new constraints were founded on the extension of the Hamilton theory related to the displacement of the poles of the correlation functions $\mathcal{F}_{2_i}(\bar{\dot{q}})$ which also must be taken into account.

The first order Hessian can also be written from equation \ref{firstDerivHessian}as
\begin{equation*}
W_{ij}^{\rom{1}} = \frac{\partial^2 L}{\partial \dot{q}_i \partial \dot{q}_j} 
= \frac{\partial}{\partial \dot{q}_i} \Big( \frac{\partial L}{ \partial \dot{q}_j} \Big)
= \frac{\partial p_j}{\partial \dot{q}_i}.
\end{equation*}
If $ W^{\rom{1}}_{ij}$ is invertible, it will exist $n$ expressions for $p_j$ depending on the others derivative terms. In the present analysis, and according to the obtained extended Hamiltonians for the $n$-VMVF systems, we assume that assumption e.i. $\det W^{\rom{1}}_{ij} \neq 0$.

On the other side, the second order Hessian can be written as
\begin{equation}
 W^{\rom{2}}_{ij} = \frac{\partial^2 L}{\partial \ddot{q}_i \partial \ddot{q}_j} = 
 \frac{\partial}{\partial \ddot{q}_i } \Big( 
 \frac{\partial L}{\partial \ddot{q}_j}  \Big) 
 = \frac{\partial s_i}{\partial \ddot{q}_j}. 
\end{equation}
Also, if $ W^{\rom{2}}_{ij}$ is invertible, it will exist $n$ expressions for $s_j$ depending on the four derivative terms $\ddot{q}_i$. However, if $\det W^{\rom{2}}_{ij} =0$, then such functions will not exist. Instead, we have some specific relations connecting the $s$ momentum, the generalized momentum $p$, the position and also the pole displacement $f$ of the type
\begin{equation}
\psi_l(s, p,q,f) = 0,
\end{equation}
which reduce the phase space of the physical system.

The canonical extended Hamiltonian has the form
\begin{equation}
H = \sum_i p_i \dot{q}_i + s_i\ddot{q}_i  - L.
\end{equation}
One method to obtain the constraints uses the zero-mode eigenvector of the Hessian matrix $ W^{\rom{2}}_{ij}$. However, we need to know the form of how the extended Lagrange function depends on $\ddot{q}$, Ex.
\begin{equation}
L = \sum_{ij} W^{\rom{2}}_{ij} \ddot{q}_i \ddot{q}_j + 
\eta_0  \ddot{q}_i \dot{q}_j  + \eta_1  \ddot{q}_i q_j 
+ \eta_2 ( q, \dot{q}).
\end{equation}
We can instead, assume that from the momentum $s$ definitions, we can have a relation in the form:
\begin{equation}
\phi_\alpha = s_\alpha - g_\alpha(q,p,f,\{s_a\}) \label{ConstrSMom}
\end{equation}
If the rank of the second order Hessian $ W^{\rom{2}}_{ij}$ is $R_W$, then we will have the same number of accelerations $\ddot{q}_a$ expressed as:
\begin{equation}
\ddot{q}_a = h_a(q,p,f,\{s_b\}, \{\ddot{q}_\beta\}), \qquad a, b = 1, ...,R_W; \qquad \alpha,\beta = R_W + 1, ..., N, \label{ConstrAccel}
\end{equation}
where $\ddot{q}_\beta$ is the remaining accelerations. Inserting this expression into $s_i$ variable definition we have
\begin{equation}
s_j = d_j(q,p,f,\{s_a\}, \{\ddot{q}_\alpha\}),
\end{equation}
which is reduced to an identity for $j = a\;(a=1,...,R_W)$. The rest of the equations reads
\begin{equation}
s_\alpha = d_\alpha(q,p,f,\{s_a\}, \{\ddot{q}_\beta\}).
\end{equation}
However, the r.h.s. cannot depend on the accelerations $\ddot{q}_\beta$, because it will imply the existence of more relations between $\ddot{q}_\alpha$ and the rest of the canonical variables.

Following the ideas of reference \cite{rothe2010classical}, the construction of the extended Hamilton equation follows from the proof of the following three propositions:
\begin{description}

\item {Proposition 1}

On the subspace $\gamma_P$ given by the phase space restricted by the constraints of the system, the canonical Hamiltonian does not depend on the accelerations $\ddot{q}_\alpha$.

\textsf{Proof}
	
Let us consider the canonical extended Hamiltonian on the subspace $\Gamma_p$ 
\begin{equation}
H_0 \equiv H_{c}\eval_{\Gamma_P} = \sum_a s_a h_a + \sum_\alpha q_\alpha \ddot{q}_\alpha + \sum_i p_i \dot{q}_i - L(q,p,f,\{h_b\}, \{\ddot{q}_\beta\})
\end{equation}
where it was used expressions \ref{ConstrSMom} and \ref{ConstrAccel}. On the the subspace $\gamma_P$ defined by the constraints, the Hamiltonian won't depend on accelerations $ \ddot{q}_\alpha$.

The partial derivative of the constrained Hamiltonian $H_0$ have the form
\begin{align}
\frac{\partial H_0}{\partial \ddot{q}_\beta} &= 
\sum_a s_a \frac{\partial H_a}{\partial \ddot{q}_\beta} + g_\beta 
- \Big( \sum_a \frac{\partial L}{\partial \ddot{q}_a}\eval_{\ddot{q}_a = h_a}
\frac{\partial h_a}{\partial \ddot{q}_\beta} \Big)
-\Big( \frac{\partial L}{\partial \ddot{q}_\beta} \Big)\eval_{\ddot{q}_a = h_a} \Big)
\nonumber \\
&= \sum_a \Big(  s_a - \frac{\partial L}{\partial \ddot{q}_a}\eval_{\ddot{q}_a = h_a} \Big) \frac{\partial h_a}{\partial \ddot{q}_\beta}
+ \Big( g_\beta - \frac{\partial L}{\partial \ddot{q}_\beta} \eval_{\ddot{q}_a = h_a} \Big).
\end{align}
The expressions inside parentheses vanish on the subspace $\Gamma_p$, therefore $\frac{\partial H_0}{\partial \ddot{q}_\beta} = 0$, hence
\begin{equation}
H_0 = H_0 (q,p,f,\{s_a\}).
\end{equation}

\item {Proposition 2}

In the presence of constraints, the equation of motions have the form
\begin{align*}
\frac{\partial H_0}{\partial q_i} &= -\dot{p}_i + \ddot{s}_i - \sum_\beta \ddot{q}_\beta \frac{\partial \phi_\beta}{\partial q_i}
\nonumber \\
\frac{\partial H_0}{\partial p_i} &=  \dot{q}_i  - \sum_\beta \ddot{q}_\beta \frac{\partial \phi_\beta}{\partial p_i}
\nonumber \\
\frac{\partial H_0}{\partial s_i} &=  \ddot{q}_i  - \sum_\beta \ddot{q}_\beta \frac{\partial \phi_\beta}{\partial s_i},
\end{align*}
being $ \ddot{q}_\beta$ the undetermined accelerations.

\textsf{Proof}
	
As Hamiltonian $H_0$ is independent of the undetermined accelerations $ \ddot{q}_\beta$, its partial derivatives over the canonical variables are
\begin{align}
\frac{\partial H_0}{\partial q_i} &= \sum_b s_b \frac{\partial h_b}{\partial q_i}
+ \sum_\beta\frac{\partial g_\beta}{\partial q_i} \ddot{q}_\beta 
-  \frac{\partial L}{\partial q_i}\eval_{\ddot{q}_a = h_a} 
-\sum_b  \frac{\partial L}{\partial \ddot{q}_b}\eval_{\ddot{q}_b = h_b} \frac{\partial h_b}{\partial q_i}
\nonumber \\
&= -  \frac{\partial L}{\partial q_i}\eval_{\ddot{q}_a = h_a}
+ \sum_\beta\frac{\partial g_\beta}{\partial q_i} \ddot{q}_\beta 
= -  (\dot{p}_i - \ddot{s}_i)
+ \sum_\beta\frac{\partial g_\beta}{\partial q_i} \ddot{q}_\beta,
\end{align}
\begin{align}
\frac{\partial H_0}{\partial p_i} &= \sum_b s_b \frac{\partial h_b}{\partial p_i}
+ \sum_\beta\frac{\partial g_\beta}{\partial p_i} \ddot{q}_\beta 
+ \dot{q}_i 
-\sum_b  \frac{\partial L}{\partial \ddot{q}_b}\eval_{\ddot{q}_b = h_b} \frac{\partial h_b}{\partial p_i}
\nonumber \\
&=  \dot{q}_i + \sum_\beta\frac{\partial g_\beta}{\partial p_i} \ddot{q}_\beta,
\end{align}
and 
\begin{align}
\frac{\partial H_0}{\partial s_a} &= h_a + \sum_b s_b \frac{\partial h_b}{\partial s_a}
+ \sum_\beta\frac{\partial g_\beta}{\partial s_a} \ddot{q}_\beta 
-\sum_b  \frac{\partial L}{\partial \ddot{q}_b}\eval_{\ddot{q}_b = h_b} \frac{\partial h_b}{\partial s_a}
\nonumber \\
&=  h_a + \sum_\beta\frac{\partial g_\beta}{\partial s_a} \ddot{q}_\beta
=  \ddot{q}_a + \sum_\beta\frac{\partial g_\beta}{\partial s_a} \ddot{q}_\beta.
\label{constrPrep2Eq3}
\end{align}
As 
\begin{equation}
\frac{\partial H_0}{\partial s_\alpha} =0 , \qquad \text{and } \qquad \frac{\partial g_\beta }{\partial s_\alpha}  = \delta_{\alpha \beta},
\end{equation}
we can add these terms to the equation \ref{constrPrep2Eq3} and rewritten it as
\begin{equation}
\frac{\partial H_0}{\partial s_i} =  \ddot{q}_a + \sum_\beta\frac{\partial g_\beta}{\partial s_a} \ddot{q}_\beta.
\end{equation}

From $\phi_\alpha$'s definition we have then the wanted equations:
\begin{align}
\frac{\partial H_0}{\partial q_i} &= -(\dot{p}_i - \ddot{s}_i) - \sum_\beta \ddot{q}_\beta \frac{\partial \phi_\beta}{\partial q_i}
\nonumber \\
\frac{\partial H_0}{\partial p_i} &=  \dot{q}_i  - \sum_\beta \ddot{q}_\beta \frac{\partial \phi_\beta}{\partial p_i}
\nonumber \\
\frac{\partial H_0}{\partial s_i} &=  \ddot{q}_i  - \sum_\beta \ddot{q}_\beta \frac{\partial \phi_\beta}{\partial s_i}. \label{ConstrPrep2Eq}
\end{align}

It possible, however, that these equations lead to inconsistencies with the primary constraints equations \ref{ConstrSMom}. Because of that, we must demand that
\begin{equation}
\ddot{s}_\alpha = \frac{d^2}{dt^2} g_\alpha(q,p,f\{s_a\}),
\end{equation}
where $s_\alpha$ is given by the $s_i$ term in the r.h.s. of the first equation of \ref{ConstrPrep2Eq}. These two set of equations may imply the appearance of new constraints, which according to Dirac's terminology are called secondary constraints. They are not independent constraints, and they should not be added to the extended Hamiltonian since they are hidden within the primary constraints.
\end{description}

For the next proposition, it is useful to formally introduce the weak concept defined by Dirac and that we have mentioned a few times in the previous sections. The notions of \textit{weak equality} and also the \textit{strong equality} come with the appearance of constraints or what it is the same, with the reduction of the canonical physical space. It is said that two functions on phase space, $f$ and $g$, are weakly equal, $f \approx g$, if they are equal when in the subspace defined by the primary constraints while they are strongly equal, $f = g$, if they are equal all over the configuration space. Under this definition, the extended Hamilton equations can be written as
\begin{align}
\frac{\partial H_T}{\partial q_i} &\approx -(\dot{p}_i - \ddot{s}_i), \qquad &
\frac{\partial H_T}{\partial p_i} &\approx  \dot{q}_i 
\nonumber \\
\frac{\partial H_T}{\partial s_i} &\approx  \ddot{q}_i \qquad \qquad \text{where }&
H_T &= H_0 + \sum_\beta a_\beta \phi_\beta.
\end{align}
We must note that derivatives do not act over $a_\beta$ since the expression is valid \textit{on shell}. In fact, the derivative over the ``accelerations'' $a_\beta$ is not defined, since they are a priori not given functions of the canonical variables.

Now, using the \textit{weak equality} concept, we can expand the definition of $H_T$. That is shown by proving the proposition:

\begin{description}

\item {Proposition 3}

Let $d(q,p,f,s)$ and $h(q,p,f,s)$ being two functions defined over all the entire phase space $\Gamma$. If $d(q,p,f,s)|_{\Gamma_P} = h(q,p,f,s)|_{\Gamma_P}$ then 
\begin{align}
\frac{\partial }{\partial q_i} \bigg[ 
d(q,p,f,s) - \sum_\beta \phi_\beta \frac{\partial d(q,p,f,s)}{\partial s_\beta}
\bigg] &= 
\frac{\partial }{\partial q_i} \bigg[ 
h(q,p,f,s) - \sum_\beta \phi_\beta \frac{\partial h(q,p,f,s)}{\partial s_\beta}
\bigg]
\nonumber \\
\frac{\partial }{\partial p_i} \bigg[ 
d(q,p,f,s) - \sum_\beta \phi_\beta \frac{\partial d(q,p,f,s)}{\partial s_\beta}
\bigg] &= 
\frac{\partial }{\partial p_i} \bigg[ 
h(q,p,f,s) - \sum_\beta \phi_\beta \frac{\partial h(q,p,f,s)}{\partial s_\beta}
\bigg]
\nonumber \\
\frac{\partial }{\partial s_i} \bigg[ 
d(q,p,f,s) - \sum_\beta \phi_\beta \frac{\partial d(q,p,f,s)}{\partial s_\beta}
\bigg] &= 
\frac{\partial }{\partial s_i} \bigg[ 
h(q,p,f,s) - \sum_\beta \phi_\beta \frac{\partial h(q,p,f,s)}{\partial s_\beta}
\bigg]. \label{prepos3Equations}
\end{align}

\textsf{Proof}

To prove this proposition, we suppose the functions $h,d$ depend on $d(q,p,f,s_a, s_\alpha)$ and $h(q,p,f,s_a, s_\alpha)$. By assumption and from equation \ref{ConstrSMom} $d(q,p,f,s_a, g_\alpha) = h(q,p,f,s_a, g_\alpha)$ from where it follows:
\begin{align}
\bigg[ 
\frac{\partial d}{\partial q_i}  
- \sum_\beta \frac{\partial d}{\partial s_\beta} \frac{\partial g_\beta}{\partial q_i}
\bigg]_{\Gamma_P} &= 
\bigg[ \frac{\partial h}{\partial q_i}
 - \sum_\beta \frac{\partial h}{\partial s_\beta} \frac{\partial g_\beta}{\partial q_i}
\bigg]_{\Gamma_P}
\nonumber \\
\bigg[ 
\frac{\partial d}{\partial p_i}  
- \sum_\beta \frac{\partial d}{\partial s_\beta} \frac{\partial g_\beta}{\partial p_i}
\bigg]_{\Gamma_P} &= 
\bigg[ \frac{\partial h}{\partial p_i}
 - \sum_\beta \frac{\partial h}{\partial s_\beta} \frac{\partial g_\beta}{\partial p_i}
\bigg]_{\Gamma_P}
\nonumber \\
\bigg[ 
\frac{\partial d}{\partial s_a}  
- \sum_\beta \frac{\partial d}{\partial s_\beta} \frac{\partial g_\beta}{\partial s_a}
\bigg]_{\Gamma_P} &= 
\bigg[ \frac{\partial h}{\partial s_a}
 - \sum_\beta \frac{\partial h}{\partial s_\beta} \frac{\partial g_\beta}{\partial s_a}
\bigg]_{\Gamma_P},
\end{align}
or, using the \textit{weak equality} definition
\begin{align}
\frac{\partial }{\partial q_i}  
\bigg[ d
- \sum_\beta \frac{\partial d}{\partial s_\beta} \phi_\beta
\bigg] &\approx 
\frac{\partial }{\partial q_i}
\bigg[ h 
 - \sum_\beta \frac{\partial h}{\partial s_\beta} \phi_\beta
\bigg]
\nonumber \\
\frac{\partial }{\partial p_i}  
\bigg[ d
- \sum_\beta \frac{\partial d}{\partial s_\beta} \phi_\beta
\bigg] &\approx 
\frac{\partial }{\partial p_i}
\bigg[ h 
 - \sum_\beta \frac{\partial h}{\partial s_\beta} \phi_\beta
\bigg]
\nonumber \\
\frac{\partial }{\partial s_a}  
\bigg[ d
- \sum_\beta \frac{\partial d}{\partial s_\beta} \phi_\beta
\bigg] &\approx 
\frac{\partial }{\partial s_a}
\bigg[ h 
 - \sum_\beta \frac{\partial h}{\partial s_\beta} \phi_\beta
\bigg],
\end{align}
As $\frac{\phi_\beta}{s_\alpha} = \delta_{\alpha \beta} $, the last equations can be replace by
\begin{equation*}
\frac{\partial }{\partial s_i}  
\bigg[ d
- \sum_\beta \frac{\partial d}{\partial s_\beta} \phi_\beta
\bigg] \approx 
\frac{\partial }{\partial s_i}
\bigg[ h 
 - \sum_\beta \frac{\partial h}{\partial s_\beta} \phi_\beta
\bigg]
\end{equation*}
obtaining equations \ref{prepos3Equations}.

As a corollary, we have that if $d \equiv H_0$ and $h \equiv H$, where $H(\{q_i\}, \{p_i\},\{s_i\} )$ and $H_0(\{q_i\}, \{p_i\},\{s_a\})$ such that $H(\{q_i\}, \{p_i\},\{s_i\} )\approx H_0(\{q_i\}, \{p_i\},\{s_a\})$, then 
\begin{align}
\frac{\partial H_0}{\partial q_i}  
 &\approx 
\frac{\partial }{\partial q_i}
\bigg[ H 
 - \sum_\beta \frac{\partial H}{\partial s_\beta} \phi_\beta
\bigg]
\nonumber \\
\frac{\partial H_0}{\partial p_i}  
 &\approx 
\frac{\partial }{\partial p_i}
\bigg[ H 
 - \sum_\beta \frac{\partial H}{\partial s_\beta} \phi_\beta
\bigg]
\nonumber \\
\frac{\partial H_0}{\partial s_i}  
 &\approx 
\frac{\partial }{\partial s_i}
\bigg[ H 
 - \sum_\beta \frac{\partial H}{\partial s_\beta} \phi_\beta
\bigg]
\end{align}
Replacing equations \ref{ConstrPrep2Eq}, we can rewrite previous equations as,
\begin{align}
-(\dot{p}_i - \ddot{s}_i) \approx  \frac{\partial }{\partial q_i}
\bigg[ H 
+ \sum_\beta \Big( \ddot{q}_\beta - \frac{\partial H}{\partial s_\beta}
  \Big) \phi_\beta
\bigg] \equiv 
\frac{\partial }{\partial q_i}
\bigg[ H 
+ \sum_\beta a_\beta \phi_\beta
\bigg]
\nonumber \\
\dot{q}_i \approx  \frac{\partial }{\partial p_i}
\bigg[ H 
+ \sum_\beta \Big( \ddot{q}_\beta - \frac{\partial H}{\partial s_\beta}
  \Big) \phi_\beta
\bigg] \equiv 
\frac{\partial }{\partial p_i}
\bigg[ H 
+ \sum_\beta a_\beta \phi_\beta
\bigg]
\nonumber \\
\ddot{q}_i \approx  \frac{\partial }{\partial s_i}
\bigg[ H 
+ \sum_\beta \Big( \ddot{q}_\beta - \frac{\partial H}{\partial s_\beta}
  \Big) \phi_\beta
\bigg] \equiv 
\frac{\partial }{\partial s_i}
\bigg[ H 
+ \sum_\beta a_\beta \phi_\beta
\bigg] \label{ConstrHamEquations}
\end{align}
where 
\begin{equation}
a_\beta \equiv \ddot{q}_\beta - \frac{\partial H}{\partial s_\beta}.
\end{equation}
These equations can also be written as
\begin{equation}
-(\dot{p}_i - \ddot{s}_i) \approx  \frac{\partial H_T}{\partial q_i}, \qquad
\dot{q}_i \approx  \frac{\partial H_T}{\partial p_i}, \qquad
\ddot{q}_i \approx  \frac{\partial H_T}{\partial s_i},
\end{equation}
where 
\begin{equation}
H_T = H_0 + \sum_\beta a_\beta \phi_\beta, \qquad H_T \approx H_0. \label{ConstrGeneralHamiltonian}
\end{equation}
Note that for $H_T = H_0$, $a_\beta = \ddot{q}_\beta$.
\end{description}

The iterative procedure for generating all the constraint is based on the above equations, the primary constraint and also the conditions for the consistency of the equations 
\begin{equation}
\dot{\phi}_\beta = 0 \qquad \text{and} \qquad \ddot{\phi}_\beta = 0.
\label{ConstHamiltConsistencyEq}
\end{equation}
The method was developed by Dirac, and it reveals, if exist, the hidden or secondary constraints that restrict the system. 

In general, both relations \ref{ConstHamiltConsistencyEq} must be satisfied because of each one secure that the newly obtained constraint is consistent with the Lagrange equations by correctly relate the configuration variable that cannot be obtained from the canonical variables definition. For example, the first relation, $\dot{\phi}_\beta=0$, connects the variables $\dot{q}_\beta$, which cannot be isolated from the linear momentum definition's $p_\beta = \frac{\partial L}{\partial \dot{q}_\beta}$, with the all the canonical variables, while the consistency equation $\ddot{\phi}_\beta$ relates the variables $\ddot{q}_\beta$, from the definition of the second order momentum  $s_\beta = \frac{\partial L}{\partial \ddot{q}_\beta}$, to the canonical variables. In our case, and following the main objective of this work, we are only considering in this section that the first order Hessian is invertible, $e.i.$ $\det W^{\rom{1}}_{ij} \neq 0$; so the first order consistency equation, should return an identity; otherwise, we will have more constraint equation than variables.

The explicit time derivative of the constraint is
\begin{equation}
\dot{\phi}_\alpha = 
  \frac{\partial \phi_\alpha}{\partial q_i}\dot{q}_i
+ \frac{\partial \phi_\alpha}{\partial f_i}\dot{f}_i
+ \frac{\partial \phi_\alpha}{\partial p_i}\dot{p}_i
+ \frac{\partial \phi_\alpha}{\partial s_i}\dot{s}_i,
\end{equation}
while its second derivative with time is
\begin{align}
&\ddot{\phi}_\beta 
= \frac{\partial \phi_\alpha}{\partial q_i}\ddot{q}_i
+ \frac{\partial \phi_\alpha}{\partial f_i}\ddot{f}_i
+ \frac{\partial \phi_\alpha}{\partial p_i}\ddot{p}_i
+ \frac{\partial \phi_\alpha}{\partial s_i}\ddot{s}_i
+ \frac{\partial^2 \phi_\alpha}{\partial q_i \partial q_j}\dot{q}_i \dot{q}_j
+ \frac{\partial^2 \phi_\alpha}{\partial f_i \partial f_j}\dot{f}_i \dot{f}_j
+ \frac{\partial^2 \phi_\alpha}{\partial p_i \partial p_j}\dot{p}_i \dot{p}_j
+ \frac{\partial^2 \phi_\alpha}{\partial s_i \partial s_j}\dot{s}_i \dot{s}_j
\nonumber \\
&2\Big[
  \frac{\partial^2 \phi_\alpha}{\partial q_i \partial f_j}\dot{q}_i \dot{f}_j
+ \frac{\partial^2 \phi_\alpha}{\partial q_i \partial p_j}\dot{q}_i \dot{p}_j
+ \frac{\partial^2 \phi_\alpha}{\partial q_i \partial s_j}\dot{q}_i \dot{s}_j
+ \frac{\partial^2 \phi_\alpha}{\partial p_i \partial f_j}\dot{p}_i \dot{f}_j
+ \frac{\partial^2 \phi_\alpha}{\partial p_i \partial s_j}\dot{p}_i \dot{s}_j
+ \frac{\partial^2 \phi_\alpha}{\partial f_i \partial s_j}\dot{f}_i \dot{s}_j
\Big].
\end{align}
Including the Hamilton equations \ref{ConstrHamEquations}, we obtain
\begin{equation}
\dot{\phi}_\alpha = 
  \frac{\partial \phi_\alpha}{\partial q_i}\frac{\partial H_T}{\partial p_i} 
- \frac{\partial \phi_\alpha}{\partial p_i}\frac{\partial H_T}{\partial q_i} 
+ \frac{\partial \phi_\alpha}{\partial p_i}\ddot{s}_i 
+ \frac{\partial \phi_\alpha}{\partial f_i}\dot{f}_i
+ \frac{\partial \phi_\alpha}{\partial s_i}\dot{s}_i,
\end{equation}
or 
\begin{equation}
\dot{\phi}_\alpha =  [\phi_\alpha,H_T] + \mathcal{G}(\phi_\alpha), \label{ConstrFirstDeriv}
\end{equation}
where
\begin{equation}
[A,B] \equiv 
  \frac{\partial A}{\partial q_i}\frac{B}{\partial p_i} 
- \frac{\partial A}{\partial p_i}\frac{B}{\partial q_i} 
\qquad \text{and} \qquad
 \mathcal{G}(A) \equiv 
  \frac{\partial A}{\partial p_i}\ddot{s}_i 
+ \frac{\partial A}{\partial f_i}\dot{f}_i
+ \frac{\partial A}{\partial s_i}\dot{s}_i.
\end{equation}
The second derivative using the Hamilton equations  \ref{ConstrHamEquations} have the form
\begin{align}
\ddot{\phi}_\alpha& =
\Big[
\frac{\partial^2 \phi_\alpha}{\partial q_i \partial q_j} 
\frac{\partial H_T}{\partial p_i} 
\frac{\partial H_T}{\partial p_j} 
+
\frac{\partial^2 \phi_\alpha}{\partial p_i \partial p_j} 
\frac{\partial H_T}{\partial q_i} 
\frac{\partial H_T}{\partial q_j} 
-
2\frac{\partial^2 \phi_\alpha}{\partial q_i \partial p_j} 
\frac{\partial H_T}{\partial p_i} 
\frac{\partial H_T}{\partial q_j}
\Big] 
\nonumber\\
&+
2\Big[
\frac{1}{2}
\frac{\partial \phi_\alpha}{\partial q_i} 
\frac{\partial H_T}{\partial s_i}
-
\frac{\partial^2 \phi_\alpha}{\partial p_i \partial p_j} 
\frac{\partial H_T}{\partial q_i} 
\ddot{s}_j
+
\frac{\partial^2 \phi_\alpha}{\partial q_i \partial p_j} 
\frac{\partial H_T}{\partial p_i} 
\ddot{s}_j
+ 
\frac{\partial^2 \phi_\alpha}{\partial q_i \partial f_j} 
\frac{\partial H_T}{\partial p_i} 
\dot{f}_j
+ 
\frac{\partial^2 \phi_\alpha}{\partial q_i \partial s_j} 
\frac{\partial H_T}{\partial p_i} 
\dot{s}_j
\nonumber\\
&- 
\frac{\partial^2 \phi_\alpha}{\partial p_i \partial f_j} 
\frac{\partial H_T}{\partial q_i} 
\dot{f}_j
- 
\frac{\partial^2 \phi_\alpha}{\partial p_i \partial s_j} 
\frac{\partial H_T}{\partial q_i} 
\dot{s}_j
\Big]
+
\frac{\partial \phi_\alpha}{\partial p_j} 
\ddot{p}_j
+
\frac{\partial \phi_\alpha}{\partial f_j} 
\ddot{f}_j
+
\frac{\partial \phi_\alpha}{\partial s_j} 
\ddot{s}_j
+
\frac{\partial^2 \phi_\alpha}{\partial p_i \partial p_j} 
\ddot{s}_i \ddot{s}_j
\nonumber \\
&+
\frac{\partial^2 \phi_\alpha}{\partial f_i \partial f_j} 
\dot{f}_i
\dot{f}_j
+
\frac{\partial^2 \phi_\alpha}{\partial s_i \partial s_j} 
\dot{s}_i
\dot{s}_j
+
\frac{\partial^2 \phi_\alpha}{\partial p_i \partial s_j} 
\ddot{s}_i
\dot{s}_j
+
\frac{\partial^2 \phi_\alpha}{\partial f_i \partial s_j} 
\dot{f}_i
\dot{s}_j.
\end{align}
or
\begin{equation}
\ddot{\phi}_\alpha = \Psi(\phi_\alpha, H_T, H_T) + 2\Upsilon(\phi_\alpha, H_T) + \Omega(\phi_\alpha), \label{ConstrSecondDeriv}
\end{equation}
being
\begin{equation}
\Psi(A, B, C) \equiv 
\frac{\partial^2 A}{\partial q_i \partial q_j} 
\frac{\partial B}{\partial p_i} 
\frac{\partial C}{\partial p_j} 
+
\frac{\partial^2 A}{\partial p_i \partial p_j} 
\frac{\partial B}{\partial q_i} 
\frac{\partial C}{\partial q_j} 
-
2\frac{\partial^2 A}{\partial q_i \partial p_j} 
\frac{\partial B}{\partial p_i} 
\frac{\partial C}{\partial q_j},
\end{equation}
\begin{align}
\Upsilon(A,B) &\equiv
\frac{1}{2}
\frac{\partial A}{\partial q_i} 
\frac{\partial B}{\partial s_i}
-
\frac{\partial^2 A}{\partial p_i \partial p_j} 
\frac{\partial B}{\partial q_i} 
\ddot{s}_j
+
\frac{\partial^2 A}{\partial q_i \partial p_j} 
\frac{\partial B}{\partial p_i} 
\ddot{s}_j
+ 
\frac{\partial^2 A}{\partial q_i \partial f_j} 
\frac{\partial B}{\partial p_i} 
\dot{f}_j
\nonumber\\
&+ 
\frac{\partial^2 A}{\partial q_i \partial s_j} 
\frac{\partial B}{\partial p_i} 
\dot{s}_j
- 
\frac{\partial^2 A}{\partial p_i \partial f_j} 
\frac{\partial B}{\partial q_i} 
\dot{f}_j
- 
\frac{\partial^2 A}{\partial p_i \partial s_j} 
\frac{\partial B}{\partial q_i} 
\dot{s}_j
\end{align}
and
\begin{align}
\Omega(A) &\equiv
\frac{\partial A}{\partial p_j} 
\ddot{p}_j
+
\frac{\partial A}{\partial f_j} 
\ddot{f}_j
+
\frac{\partial A}{\partial s_j} 
\ddot{s}_j
+
\frac{\partial^2 A}{\partial p_i \partial p_j} 
\ddot{s}_i \ddot{s}_j
+
\frac{\partial^2 A}{\partial f_i \partial f_j} 
\dot{f}_i
\dot{f}_j
\nonumber \\
&+
\frac{\partial^2 A}{\partial s_i \partial s_j} 
\dot{s}_i
\dot{s}_j
+
\frac{\partial^2 A}{\partial p_i \partial s_j} 
\ddot{s}_i
\dot{s}_j
+
\frac{\partial^2 A}{\partial f_i \partial s_j} 
\dot{f}_i
\dot{s}_j.
\end{align}

While it is straightforward show the linearity for the B factor on functions
$[A,B]$ and $\Upsilon(A, B)$, function $\Psi(A, B, C)$ satisfy:
\begin{equation}
\Psi(A, B_1 + B_2, C_1 + C_2) = \Psi(A, B_1, C_1) + \Psi(A, B_2, C_2) + 2\Psi'(A, B_1,B_2 C_1,C_2),
\end{equation}
where
\begin{equation}
\Psi'(A, B_1,B_2 C_1,C_2) = 
\frac{\partial^2 A}{\partial q_i \partial q_j} 
\frac{\partial B_1}{\partial p_i} 
\frac{\partial C_2}{\partial p_j} 
+
\frac{\partial^2 A}{\partial p_i \partial p_j} 
\frac{\partial B_2}{\partial q_i} 
\frac{\partial C_1}{\partial q_j}
-
\frac{\partial^2 A}{\partial q_i \partial p_j} 
\frac{\partial B_1}{\partial p_i} 
\frac{\partial C_2}{\partial q_j}
-
\frac{\partial^2 A}{\partial q_i \partial p_j} 
\frac{\partial B_2}{\partial p_i} 
\frac{\partial C_1}{\partial q_j}.
\end{equation}

Applying last properties when substituting equation \ref{ConstrGeneralHamiltonian}, $H_T = H + \sum_\beta a_\beta \phi_\beta$, first and the second derivative in equations \ref{ConstrFirstDeriv} and \ref{ConstrSecondDeriv} have the form, respectively:
\begin{equation}
\dot{\phi}_\alpha = [\phi_\alpha,\phi_\beta]a_\beta 
+ [\phi_\alpha,H] 
+ \mathcal{G}(\phi_\alpha),
\end{equation}
and
\begin{align}
\ddot{\phi}_\alpha &= 
\Psi(\phi_\alpha, \phi_\beta, \phi_\gamma) a_\beta a_\gamma
+ 2 [\Psi'(\phi_\alpha, H, \phi_\beta, H, \phi_\beta)
+ \Upsilon(\phi_\alpha, \phi_\beta)]a_\beta
\nonumber \\
&+\Psi(\phi_\alpha, H, H) 
+ 2\Upsilon(\phi_\alpha, H)
+ \Omega(\phi_\alpha),
\end{align}
where we use the Einstein notation.
The matrix form of the previous equation is
\begin{equation}
\ddot{\phi} = \mathbf{B}^{(1)} \mathbf{a} + \mathbf{C}^{(1)} 
\qquad \text{and} \qquad
\ddot{\phi} = \mathbf{a} \mathbf{A}^{(2)} \mathbf{a} + \mathbf{B}^{(2)} \mathbf{a} + \mathbf{C}^{(2)}.
\end{equation}

The equations of consistency for the constraints required that the constraints remain constant and equal to zero at all time like
\begin{equation}
\dot{\phi} = \mathbf{B}^{(1)} \mathbf{a} + \mathbf{C}^{(1)}\approx 0
\qquad \text{and} \qquad
\ddot{\phi} = \mathbf{a} \mathbf{A}^{(2)} \mathbf{a} + \mathbf{B}^{(2)} \mathbf{a} + \mathbf{C}^{(2)} \approx 0
\end{equation}

According to Dirac \cite{DiracInterpQM}, three possible scenarios may come from the application of the consistency condition. One is that we arrive at the equation $0=0$, which means that the primary constraint is consistent with the Lagrange equations and the extended Hamilton equations are identically satisfied when the primary constraints are included. Another possibility is when the equation reduces to an equation independent of coefficients $a_\alpha$ involving only the canonical variables in the form $\chi(q,p,s,f)=0$. Such equations are also independent of the primary constraints; otherwise, it will be of the first kind. Finally, the constraint equations and its consistency may, instead, imposes a relation between the coefficients $a_\alpha$.

While in the first case, all equations are explicit and there is nothing else to add, the analysis related to the other cases involves specific difficulties. If new equation $\chi(q,p,s,f)=0$ is found as the result of the consistency of the primary constraints, they will be considered a new type of constraints called secondary constraints. While the primary constraints are direct consequences of the canonical variables definitions, only taking into account the secondary constraints, the full set of Lagrange equation completely manifest itself. The new constraint $\chi(q,p,s,f)=0$ implies a new set of consistency relations:
\begin{align}
\dot{\chi} &= B^{(1)}_\beta (\chi, \phi_\beta, H_T) a_\beta + C^{(1)}(\chi, H_T)\approx 0,
\nonumber \\
\ddot{\chi} &= A^{(2)}_{\beta,\gamma}(\chi, \phi_\beta, \phi_\gamma)  a_\beta a_\gamma + B^{(2)}_\beta (\chi, \phi_\beta, H_T) a_\beta + C^{(2)}(\chi, H_T)\approx 0. \label{ConstrHamiltEqConsistEqCoeff}
\end{align}

We must treat these equations, the secondary constraints, and their consistency equations, in the same form as the primary constraints which is developing the equations and find one of the above scenario the result will generate. If another independent equation of the second kind is obtained, we should do the same procedure again, until we have at the end no independent consistency conditions so at the end we have a set of secondary constraints together with a set of the initials consistent conditions for coefficients $a_\alpha$.

On the third scenario where we need to find the conditions for the coefficients $a_\beta$ that are imposed by the constraints and its persistence with time. We are in presence of a set of quadratic equations, eq. \ref{ConstrHamiltEqConsistEqCoeff}, and its solution or solvents existence should be studied on future works. 
However, the solution should give us the $a$'s as a function of the variable of the system:
\begin{equation}
a_\beta  = U_{\beta}(q,p,s,f).
\end{equation}
There must be solutions of this kind, so the extended Lagrange equations of motion are consistent. Also, the solution is not unique, which means that the solution must include all the independent solutions of equation \ref{ConstrHamiltEqConsistEqCoeff}. Dirac proposes, for the ordinary Hamilton constrained theory, to classify the coefficients whether they are solutions of the homogeneous, $U_\beta$, or the inhomogeneous equation, $V_{a \beta}$, respectively. The general solution of the coefficient is then
\begin{equation}
a_\beta = U_\beta + v_a V_{a \beta}
\end{equation}
where $v_a$ is arbitrary. The general Hamiltonian of equation \ref{ConstrGeneralHamiltonian} can be written as 
\begin{equation}
H_T = H + a_\beta \phi_\beta =  H + \big[ U_\beta \phi_\beta + v_a V_{a \beta} \phi_\beta \big]
\end{equation}
or
\begin{equation}
H_T = H' +  v_a \phi_a, 
\quad \text{where} \quad 
H' = H +  U_\beta \phi_\beta, 
\;\text{and}\; 
\phi_a =  V_{a \beta} \phi_\beta.
\end{equation}

This new definition is useful for the variation of a dynamical variable $g$.
The value of a general dynamical variable $g$ with time interval $\delta t$ with initial value $g_0$ is:
\begin{equation}
g(\delta t) = g_0 + \dot{g} \delta t + \frac{1}{2} \ddot{g} (\delta t)^2.
\end{equation}
The inclusion of Hamiltonian $H_T = H' + v_a \phi_a$ modify equation like
\begin{align}
g(\delta t) &= g_0 + 
\{
[g,\phi_\beta]v_a
+ [g,H] 
+ \mathcal{G}(g)
\}\delta t
+ 
\frac{1}{2}
\{
\Psi(g, \phi_a, \phi_b) v_a v_b
\nonumber \\
&+ 2[\Psi'(g, H', \phi_a, H', \phi_a)
+ \Upsilon(g, \phi_a)]v_a
+ \Psi(g, H', H') 
+ 2\Upsilon(g, H')
+ \Omega(g)
\}(\delta t)^2
\end{align}

As coefficients $v_a$ are entirely arbitrary, the variation of the general dynamical variable g with time interval $\delta t$ for two different sets of coefficients $v_a$ and $v'_a$ is
\begin{align}
\Delta g(\delta t) &= [g,\phi_\beta](v_a - v'_a)\delta t
+ \frac{1}{2}
\{
\Psi(g, \phi_a, \phi_b)( v_a v_b - v_a' v_b')
\nonumber \\
&+ 2[\Psi'(g, H', \phi_a, H', \phi_a)
+ \Upsilon(g, \phi_a)](v_a - v_a').
\}(\delta t)^2
\end{align}

We can write the variation as
\begin{align}
\Delta g(\delta t) &= [g,\phi_\beta]\epsilon_a
+ \frac{1}{2}
\{
\Psi(g, \phi_a, \phi_b)( v_a \epsilon_b - \epsilon_a v_b')
\nonumber \\
&+ 2[\Psi'(g, H', \phi_a, H', \phi_a)
+ \Upsilon(g, \phi_a)]\epsilon_a.
\}\delta t
\end{align}
where $\epsilon_a\equiv (v_a - v_a')\delta t$ is a small arbitrary number: small because of the time interval $\delta t$ and arbitrary because of the arbitrary property of $v_a, v_a'$. To solve the equation we can then change all Hamiltonian variables according to $\Delta g(\delta t)$ by an applying infinitesimal with the generating function $\epsilon_a \phi_a$ so the new Hamiltonian represent the same physical state.

According to Dirac's preposition, for the quantization, we must carry a second contact transformation with the generating function $\epsilon_{a'} \gamma_{a'}$, where $\gamma_{a'}$ so we can obtain the condition the constraint must satisfy for the dynamical variable remains unchanged. We let the calculations of this topic to future works.

\subsubsection{Constrained $n$-VMVF systems} \label{extContrainedSection}
The constraints play a fundamental role in the development of the present proposal. The first constraints were obtained from the conservation of linear and angular momentums which depend on the second derivative of the position. Later was added the relativistic relations for position, field and mass derivatives. We use the first group of constraints to construct the second order derivative Lagrangians using the Lagrange undetermined coefficient method and, with it, to obtain the second order derivative Hamiltonians for the linear and angular coordinates respectively. Recalling the previous development, the extended linear Hamiltonian, equation \ref{extLinearHamiltonian}, is 
\begin{align*}
H_T = \sum_{nn'}&\Big[ 
\Big ( {H_{T_1}}_n g_{\mu \nu}
+ {{H_{T_2}}_n}_\gamma X_n^\gamma g_{\mu \nu}
+ {{H_{T_3}}_n}_{\mu \nu}\Big )\delta_{nn'}
+ {{H_{T_4}}_{n n'}}_{\mu \nu}
\Big] \times
\nonumber \\ 
&\Big[ 
{{P_T}_{n}}^\alpha  
- \sum_l {{\mathcal{C}_T}_{nl}}^\alpha 
+  {{\mathcal{D}_T}_{nl}}^\alpha_{\delta } X^\delta_l
\big] 
\big[ \sum_l {{\mathcal{A}_T}_{nl}}^\alpha_\mu  
+  {{\mathcal{B}_T}_{nl}}^\alpha_{\mu \delta } X^\delta_l \Big]^- \times
\nonumber \\
&\Big[ 
{{P_T}_{n'}}^\beta 
- \sum_l {{\mathcal{C}_T}_{n'l}}^\beta 
+  {{\mathcal{D}_T}_{n'l}}^\beta_{\delta } X^\delta_l
\big] 
\big[ \sum_l {{\mathcal{A}_T}_{n'l}}^\beta_\nu  
+  {{\mathcal{B}_T}_{n'l}}^\beta_{\nu \delta } X^\delta_l \Big]^-
\end{align*}
while the extended angular Hamiltonian, equation \ref{extAngularHamiltonian}, is
\begin{align*}
H_R = \sum_{nn'}\Big[ 
&\Big ( \sum_{i}G_{n;i\mu}^{\;\;\;\;\;\gamma} D_{n;i\gamma}^{-} S_{R n;\nu}
+{H_{R_1}}_n g_{\mu \nu}
+ {{H_{R_2}}_n}_\gamma X_n^\gamma g_{\mu \nu}
+ {{H_{R_3}}_n}_{\mu \nu}\Big )\delta_{nn'}
\nonumber \\
&+ {{H_{R_4}}_{n n'}}_{\mu  \nu}
\Big] \Big[ 
{{P_R}_{n}}^\alpha  
- \sum_l {{\mathcal{C}_R}_{nl}}^\alpha 
+  {{\mathcal{D}_R}_{nl}}^\alpha_{\delta } X^\delta_l
\big] 
\big[ \sum_l {{\mathcal{A}_R}_{nl}}^\alpha_\mu  
+  {{\mathcal{B}_R}_{nl}}^\alpha_{\mu \delta } X^\delta_l \Big]^- \times
\nonumber \\
&\Big[ 
{{P_R}_{n'}}^\beta 
- \sum_l {{\mathcal{C}_R}_{n'l}}^\beta 
+  {{\mathcal{D}_R}_{n'l}}^\beta_{\delta } X^\delta_l
\big] 
\big[ \sum_l {{\mathcal{A}_R}_{n'l}}^\beta_\nu  
+  {{\mathcal{B}_R}_{n'l}}^\beta_{\nu \delta } X^\delta_l \Big]^- .
\end{align*} 

The constraints depending on the second derivative of the position and that are already included in the both extended Hamiltonians are shown in equations \ref{TransConstraintEqApprx} and \ref{RotConstraintEqApprx} and the have the form:
\begin{align*}
\Phi_{\mu_n}^{(L)} = \sum_{{n'} \neq n}  &\Big[
\Big(\square_{n'}^\alpha m_n' \dot{x}_{n';\alpha}\Big)g^\nu_\mu
- \frac{1}{2}\frac{\partial m_{n'}}{\partial x_{n'}^\mu}\dot{x}^\nu_{n'}
+ \frac{d}{d\tau} \Big(\frac{\partial A^\nu}{\partial \dot{x}_{n}^\mu}
- \frac{\partial A^\nu}{\partial \dot{x}_{n'}^\mu}
\Big) 
+ \frac{\partial A^\nu}{\partial x_{n'}^\mu}
- \frac{\partial A^\nu}{\partial x_{n}^\mu}
\Big]\dot{x}_{n';\nu}
\nonumber \\
&+ \Big[ 
\Big(m_{n'}(0) + \frac{\partial m_{n'}}{\partial x^\mu_{n'}}\Big) g^\nu_\mu + \frac{\partial A^\nu}{\partial x_{n}^\mu}
- \frac{\partial A^\nu}{\partial x_{n'}^\mu}
+ \frac{\partial A_\mu}{\partial x_{n'}^\nu}
\Big] \ddot{x}_{n';\nu}
\end{align*}
and
\begin{align*}
\Psi_{i_n}^{(L)}  = \sum_{{n'} \neq n} &\Big\{
 D_{\;\xi_{n',i}}^\mu \Big[ 
\Big(\square_{n'}^\alpha m_n' \dot{x}_{n';\alpha}\Big)g^\nu_\mu
- \frac{1}{2}\frac{\partial m_{n'}}{\partial x_{n'}^\mu}\dot{x}^\nu_{n'}
-\frac{d}{d\tau} \Big(
\frac{\partial A^\nu}{\partial \dot{x}_{n'}^\mu}
\Big) 
- \frac{\partial A_\mu}{\partial x_{n';\nu}}
+ \frac{\partial A^\nu}{\partial x_{n'}^\mu}
 \Big]
 \nonumber \\
&+D_{\;\xi_{n,i}}^\mu \Big[
\frac{d}{d\tau} \Big(
\frac{\partial A^\nu}{\partial \dot{x}_{n}^\mu}
\Big) 
+ \frac{\partial A_\mu}{\partial x_{n';\nu}}
- \frac{\partial A^\nu}{\partial x_{n}^\mu} 
  \Big]
\Big\} \dot{x}_{n';\nu}
\nonumber \\
+& \Big\{
 D_{\;\xi_{n',i}}^\mu \Big[ 
\Big(m_{n'}(0) + \frac{\partial m_{n'}}{\partial x^\mu_{n'}} \Big) g^\nu_\mu - \frac{\partial A^\nu}{\partial x_{n'}^\mu}
  \Big]
 +D_{\;\xi_{n,i}}^\mu \Big[ 
\frac{\partial A^\nu}{\partial x_{n}^\mu}
+\frac{\partial A_\mu}{\partial x_{n';\nu}}
\Big]
\Big\} \ddot{x}_{n';\nu}.
\end{align*}
Note that, from the second order momentum $s$ definition, there was no possible way to establish a relation between $\ddot{x}$ and the canonical variables, pointing out the existence of new constraints, according to the above exposed.

The remaining constraints are the relativistic constraints from equations \ref{relConstraintFinal}:
\begin{align*}
&x^\nu_n x_{\nu;n} =R^2_n , \qquad \partial_\nu A^\nu=0\;\; \forall\; n 
\\
&\frac{1}{2}\frac{\partial m_n}{\partial x^0_n}\dot{x}^0_n \dot{x}_\mu \dot{x}^\mu + m_n\ddot{x}_{\mu;n}\dot{x}^\mu_n=0. 
\end{align*}
which were left to treat them as weak conditions in the Hamiltonian level.

The extended Hamiltonian equations add new constraints for the generalized velocities related to the correlation functions and the new canonical variables $f_{x_n}$ and $f_{\xi_n}$ like:
\begin{equation}
\mathcal{F}_{x_n}(\{\dot{x}_m\}) = f_{x_n} \qquad \text{and} \qquad {\mathcal{F}_{\xi_n}(\{\dot{\xi}_m\})} = f_{\xi_n},  \label{relConstraintCorr}
\end{equation}
where correlation functions $\mathcal{F}_{x_n}(\{\dot{x}_m\}$ and ${\mathcal{F}_{\xi_n}(\{\dot{\xi}_m\})}$ are well-defined functions.

We are in the presence of a second-order derivative constrained system to describe the two constrained Hamiltonians whose second-order Hessians are not invertible. The phase space has 4-$n$ canonical variables, which are correlated with the previous constraint functions. Because both second order Hessian determinant is zero, others constraints should be found for each Hamiltonian, following the above procedure and added the equations \ref{relConstraintFinal} and \ref{relConstraintCorr}.

As mentioned before, the solution to the problem is not the objective of this work, but to show another approach to solve the problem. In that case, we suppose all constraints, including equations \ref{relConstraintFinal} and \ref{relConstraintCorr}, as found and we represented them as:
\begin{equation}
\Omega_{T_a} (q,p,s,f) = 0 \qquad \text{and } \qquad \Omega_{R_b} (q,p,s,f)=0.
\end{equation}
Using Dirac's \textit{weak} equality definition, the Hamiltonians have the form
\begin{align}
H_{T_T} &\approx H_T + \sum_a \lambda_a \Omega_{T_a} (q,p,s,f) \label{TransConstraintHamilt}
\\
H_{T_R} &\approx H_R + \sum_a \lambda_b \Omega_{R_b} (q,p,s,f), \label{RotConstraintHamilt}
\end{align}
which not include constraints of equations \ref{TransConstraintEqApprx} and \ref{RotConstraintEqApprx} since they are already included.

\newpage

\section{Extended complex domain}\label{ExtendedNumSection}
We proposed an extension of the classical theory for $n$-VMVF systems $e.i$ systems where the particle masses and fields are considered as variables of the system as functions depend on the particle's position and velocities with no predefined form. Specifically, the derivatives of the particle masses and field are included in the problem as new degrees of freedom. Because of the increased number of variables, we defined two sets of second-order Lagrange equations coming from expressing particle position in the Lorentzian and angular coordinates.

Different from previous classic theories, we have two variational problems that take place at the same time. Under the Lagrange approach, the solution for the $n$-VMVF systems problem at any time is given by solving two sets of extended Lagrange equations for two different Lagrangians each one, while the extension of the Hamilton theory provides one set of extended Hamilton equations also for two different Hamiltonian. From each function of Hamilton, we obtain a set of the equations describing the canonical transformations for $n$-VMVF systems. It is the concept of two simultaneous canonical transformations acting over the physical system that led us to think that the action of a quantum operator over a quantum state describing the $n$-VMVF system, must also have two components. We presume that the representation of that action of the quantum operator acting over the two-component physical system has the form:
\begin{equation}
\classoperator
{\mathcal{G}_T}
{\mathcal{G}_R}
\Bigg| 
\alpha
\Bigg\rangle
\equiv
\classoperator
{\mathcal{G}_T}
{\mathcal{G}_R}
\Bigg| 
\begin{array}{c}
\alpha_T \\ \\
\alpha_R
\end{array}
\Bigg\rangle
= 
\Bigg| 
\begin{array}{c}
\beta_T \\ \\
\beta_R
\end{array}
\Bigg\rangle. \label{extOperatorAction}
\end{equation}

The proposed bi-dimensional model for operators and physical states indicates that the quantum theory of $n$-VMVF systems might be developed on a vector space different from the quantum space which is a Hilbert space taken over the complex numbers. Such space should also have an infinite number of dimensions and also need to be an abstract vector space possessing the structure of an inner product that allows defining the length of a vector and the angle between two vectors. The main differences between the new space and the well-known ordinary quantum space should be in the algebraic structure and in the numbers on which the Hilbert is based. As the present approach increases the complexity of the actual theories, and also because the possibility of include physical concepts like the negative probability, we propose the initial study of the extension of the complex numbers. We will discuss in the next chapter more about why we think there is a real need for a new space. 

\subsection{The extension of complex numbers}

Numbers are fundamentals not only in physics but all science. They are one of the cornerstones of study concepts like time, space, matter, fields, among others. Physics theories like classical mechanics, electromagnetism, quantum theory, among others are developed using analytic entities like metrics space, tensors, fields, whose operations between them are based on the properties of the numbers that the theory lays on. The number type chain can be written as 
\begin{enumerate}
\item natural
\item integer
\item rational
\item real
\item complex
\end{enumerate}
Complex numbers are defined because they were needed as the solution of unsolved equations in the real domain, such as $x^2 + 1 = 0$. Indeed, it is from equation $x^2=-1$ where the complex unit is defined by $\imagi = \sqrt{-1}$. 

The fundamental theorem of algebra states that every non-constant single-variable polynomial with complex coefficients has at least one complex root, or from the algebraic point of view, the field of complex numbers is algebraically closed for sum and multiplication operations. Because of that, the search for extending the given above classification to numbers of high dimensionality may seem fruitless. However, based on advanced concepts of modern algebra, number systems called quaternions, tessarines, coquaternions, biquaternions, and octonions was developed in the nineteenth century, as an extension of complex numbers. They are all covered by the concept of hypercomplex number. 

The study of hypercomplex numbers began in 1872 when Benjamin Peirce and, later, his son Charles Sanders Peirce published the Linear Associative Algebra \cite{10.2307/2369153}. They identified the nilpotent and the idempotent elements to classify the hypercomplex numbers. Later involutions where used in the Cayley–Dickson construction to generate complex numbers, quaternions, and octonions.  Notorious theorems were proved about the limit of hyper-complexity such as Hurwitz (normed division algebras) and Frobenius's theorems (associative division algebras). Also, in 1958 J. Frank Adams, prove that there exist only four finite-dimensional real division algebras: the reals $\mathbb{R}$, the complexes $\mathbb{C}$, the quaternions $\mathbb{Q}$, and the octonions $\mathbb{O}$ \cite{10.2307/1970147}. Later Clifford develops what is known in the literature as Clifford algebra that generalizes the real numbers, complex numbers, quaternions, and several other hypercomplex number systems and is intimately connected with the theory of quadratic forms and orthogonal transformations.

One of the most famous are the quaternions, which were first described by William Rowan Hamilton in 1843 as the quotient of two directed lines in a three-dimensional space or equivalently as the quotient of two vectors \cite{hamilton1866elements}. A quaternion is usually represented as
\begin{equation}
a + b\mathbf{i} + c\mathbf{j} + d\mathbf{k}
\end{equation}
where $a, b, c$, and $d$ are real numbers, and $\mathbf{i}, \mathbf{j}$, and $\mathbf{k}$ are the fundamental quaternion units. A notable property of a quaternion is that the multiplication of two quaternions is noncommutative.

All the extension for the complex numbers have one feature in common and it that they are generated over an underlying vector space equipped with a quadratic form, but what if space has a different metric?

\subsection{The complex extended unit $\extk$}
Different from the others attempts for extending the complex numbers, we define the new number domain in the same way complex number was defined: as the solution of an unsolvable equation in the domain about to be extended. In the complex numbers domain, an unsolvable equation on the real domain $\mathbb{R}$ which led to the definition of the complex unit, $\imagi$, is:
\begin{equation}
x^2 = -1. \label{complexUnitDef}
\end{equation}
From this equation, the complex unit $\imagi$ was defined as the number from a new domain that satisfies 
\begin{equation}
\imagi^2=-1,  \;\;\;\;\;\;\;\;\;\;\;\; \imagi \in \mathbb{C}. 
\end{equation}

We can define the unit for the new set of numbers, named Extended Complex and represented by $\mathbb{E}$, from the unsolvable equation on complex numbers domain $\mathbb{C}$:
\begin{equation}
|z|^2 = \imagi.
\end{equation}
The extended complex unit $\extk$ can be defined then as
\begin{equation}
|\extk|^2\equiv \extk^* \extk = \imagi \;\;\;\;\;\;\;\;\;\;\;\; \extk \in \mathbb{E}
\end{equation}
being $\mathbb{E}$ the set of the extended complex numbers.

That is our starting point, which we try to apply it to the already known algebraic concepts. The present proposal is just a preliminary study that it should be enhanced. We are aware that an improved analysis of this topic may significantly change what we expose in here. 

We start by proposing a standard algebra for the extended numbers with the sum and multiplication that we represent as:
\begin{equation}
(\mathbb{E},+,\cdot).
\end{equation}
During the first development, we will find some issues that led us to reconsider some aspects of this algebra, particularly the former algebraic structure, that will upgrade all that is coming up next.

We can express a general extended complex number as
\begin{equation}
 \alpha = x\extk + y   \;\;\;\;\;\;\;\;\;\;\;\;\;\;\;\;\;\;\; \forall x,y \in \mathbb{C}.
\end{equation}
or, replacing the complex numbers with their real components:
\begin{equation}
\alpha=a  \imagi  \extk + b \extk + c \imagi + d, \;\;\;\;\;\;\;\; \forall a,b,c,d \in \mathbb{R}.
\end{equation}
We use the first expression, and we refer $x$ as the extended and $y$ as the imaginary part of the extended number $\alpha$, respectively. We need now to define the inner product $\langle w, v \rangle$ according to axioms \ref{vectorSpaceAxioms} for vector spaces. 

The absolute value for the extended unit must be equal to $1$. Because of that, the $\extk$'s definition ($\extk^* \extk = \imagi$) implies the need to define a new conjugated map, so the absolute value of the extended number is real. The new map, $ \mathcal{O}_{\text{new}}(\extk)$ , should behave in the same way when the conjugated  map over real numbers; this means that the referred map when acting on a pure complex number must keep the number invariant:
\begin{equation}
\mathcal{O}_{\text{new}}(x) = x \qquad \forall \quad x \in \mathbb{C}.
\end{equation}
We note then that an absolute value is a product which should have at least four factors. Indeed, if we proposed the absolute value for the extended unit as
\begin{equation}
1 = \mathcal{O}_{\text{new}}(k) \extk^* \extk = \mathcal{O}_{\text{new}}(k) \imagi \qquad \text{then} \qquad \mathcal{O}_{\text{new}}(k)=\imagi^*,
\end{equation}
which according to the invariant property of the new map over complex numbers led to $k=\imagi^*$. We can propose then four terms for the absolute value of the extended unit like
\begin{align}
\mathcal{O}_{\text{new}}(\extk^*) \mathcal{O}_{\text{new}}(\extk)=\imagi^* \quad \text{so}\quad \mathcal{O}_{\text{new}}(\extk^*) \mathcal{O}_{\text{new}}(\extk)\;\;\extk^*\extk= \imagi^*\imagi=1
\end{align}
or
\begin{align}
\mathcal{O}_{\text{new}}^*(\extk) \mathcal{O}_{\text{new}}(\extk)=\imagi^* \quad \text{so}\quad \mathcal{O}_{\text{new}}^*(\extk) \mathcal{O}_{\text{new}}(\extk)\;\;\extk^*\extk= \imagi^*\imagi=1,
\end{align}
where we have subtly use the associativity property of the multiplication.

We represent this operation $\mathcal{O}_{\text{new}}(\extk)$ as $\extk^\bullet$, so we have the first definition:
\begin{align}
&\extk^* \extk =\imagi
\nonumber \\
&(\extk^*)^\bullet \extk^\bullet = \imagi^* \label{extUnitDef}
\end{align}
or
\begin{align}
&\extk^* \extk =\imagi
\nonumber \\
&(\extk^\bullet)^* \extk^\bullet = \imagi^* \label{extUnitDef1}
\end{align}
Let us, for now, assume the first option $(\extk^*)^\bullet \extk^\bullet = \imagi^*$.

The inner product of four extended numbers $\alpha,\beta,\gamma,\delta$ and represented $\langle \alpha \cdot\beta \cdot \gamma \cdot\delta \rangle$ is a general case of the form of the absolute value for the extended unit. The most straightforward way to define the inner product is
\begin{equation}
\langle \alpha \cdot\beta \cdot \gamma \cdot\delta  \rangle \equiv  (\alpha^*)^\bullet \beta^\bullet \gamma^* \delta \qquad \forall \quad \alpha,\beta,\gamma,\delta \in \mathbb{E}.
\end{equation}

The inner product should satisfy some axioms that we intuitively adapt from the complex inner product:
\begin{itemize}
\item 
Positive-definiteness.

For every $\alpha \in \mathbb{E}$, it must be satisfied
\begin{align}
\langle \alpha \cdot \alpha \cdot \alpha \cdot \alpha \rangle \geq 0
\nonumber \\
\langle \alpha \cdot \alpha \cdot \alpha \cdot \alpha \rangle =0\Rightarrow \alpha =\mathbf {0} 
\end{align}
\item Conjugate symmetry

For every $\alpha,\beta,\gamma,\delta \in \mathbb{E}$, it must be satisfied
\begin{equation}
\langle \alpha \cdot \beta \cdot \gamma \cdot \delta\rangle ={\overline {\langle \gamma \cdot \delta \cdot \alpha \cdot \beta \rangle }}
\end{equation}

\item Linearity

For every $\alpha,\beta,\gamma,\delta, \epsilon,\theta,\lambda \in \mathbb{E}$, it must be satisfied
\begin{align}
\langle \epsilon \alpha \cdot \beta \cdot \gamma \cdot \delta\rangle &=(\epsilon^*)^\bullet \langle \alpha \cdot \beta \cdot \gamma \cdot \delta\rangle 
\nonumber \\
\langle \alpha \cdot \epsilon \beta \cdot \gamma \cdot \delta\rangle &=(\epsilon)^\bullet \langle \alpha \cdot \beta \cdot \gamma \cdot \delta \rangle 
\nonumber \\
\langle \alpha \cdot \beta \cdot \epsilon \gamma \cdot \delta\rangle &=(\epsilon)^* \langle \alpha \cdot \beta \cdot \gamma \cdot \delta\rangle 
\nonumber \\
\langle \alpha \cdot \beta \cdot \gamma \cdot \epsilon \delta\rangle &=(\epsilon) \langle \alpha \cdot \beta \cdot \gamma \cdot \delta \rangle 
\quad \epsilon \in \mathbb{E}
\nonumber \\
\langle (\alpha + \theta ) \cdot (\beta + \lambda )\cdot \gamma \cdot \delta\rangle &=\langle \alpha \cdot \beta \cdot \gamma \cdot \delta \rangle + \langle \theta \cdot \lambda \cdot v \cdot w\rangle 
\nonumber \\
\langle \alpha \cdot \beta  \cdot (\gamma + \theta) \cdot (\delta + \lambda) \rangle &=\langle \alpha \cdot \beta \cdot \gamma \cdot \delta \rangle + \langle \alpha \cdot \beta \cdot \theta \cdot \lambda \rangle \label{ExtLinearitAxiom}
\end{align}
\end{itemize}

The extended factor of the inner product can be expressed as the sum of their extended and imaginary parts like 
\begin{align}
\langle \alpha \cdot \beta \cdot \gamma \cdot \delta  \rangle &=  (\alpha^*)^\bullet \beta^\bullet \gamma^* \delta
\nonumber \\
&= (\alpha_Ek + \alpha_I)^{* \bullet} (\beta_Ek + \beta_I)^\bullet (\gamma_Ek + \gamma_I)^* (\delta_Ek + \delta_I),
\end{align}
where $\alpha^{* \bullet}\equiv (\alpha^*)^\bullet$. Then, the absolute value of extended number $\alpha=x \extk +y$ is defined as 
\begin{equation}
|\alpha| = \sqrt[4]{\langle \alpha  \cdot \alpha \cdot \alpha \cdot \alpha \rangle} \label{extAbsValueDef}
\end{equation}
and must satisfy  the positive-definiteness axiom, which means:
\begin{equation}
\langle \alpha  \cdot \alpha \cdot \alpha \cdot \alpha \rangle   = (x \extk + y)^{* \bullet} (x \extk + y)^\bullet (x \extk + y)^* (x \extk + y)= |\alpha|^4 > 0, \;\;\; \forall x,y \in \mathbb{C}.
\end{equation}
From this axiom, we can extract the relations between $\extk^{* \bullet}$, $\extk^{\bullet}$, $\extk^{ *}$ and $\extk$. 

By straightforward multiplication, we obtain
\begin{align}
|\alpha|^4  =& |x|^4 + \extk|x|^2x^*y + \extk^*|x|^2xy^* + \extk^*\extk|x|^2|y|^2 
\nonumber \\
& + \extk^\bullet|x|^2x^*y + \extk^\bullet \extk(x^* y)^2 + \extk^\bullet \extk^* |x|^2|y|^2 + \extk^\bullet \extk^* \extk|y|^2 x^*y
\nonumber \\
& + \extk^{* \bullet} |x|^2 xy^* + \extk^{* \bullet}\extk |x|^2|y|^2  + \extk^{* \bullet}\extk^* (x y^*)^2 + \extk^{* \bullet}\extk^* \extk |y|^2 xy^*
\nonumber \\
& + \extk^{* \bullet} \extk^\bullet |x|^2 |y|^2 + \extk^{* \bullet} \extk^\bullet \extk|y|^2 x^* y + \extk^{* \bullet} \extk^\bullet \extk^* |y|^2 x y^* + \extk^{* \bullet} \extk^\bullet \extk^* \extk |y|^4. \label{extInnerProd}
\end{align}
A convenient way to write the inner product in Eq. \ref{extInnerProd}, using definitions \ref{extUnitDef}, is:
\begin{align}
&|\alpha|^4   = |x|^4 + |y|^4 + |x|^2|y|^2 \Big \{ \imagi  \phi \Big [e^{-\imagi\theta}(\extk^{*\bullet} - \extk^*) + e^{\imagi\theta}(\extk^\bullet - \extk) \Big ] 
\nonumber \\
&+  \frac{1}{\phi} \Big [e^{-\imagi\theta}(\extk^{*\bullet} + \extk^*) + e^{\imagi\theta}(\extk^\bullet + \extk) \Big ] + e^{-2i\theta}\extk^{*\bullet}  \extk^* + e^{2i\theta}\extk^{\bullet} \extk + \extk^\bullet \extk^* + \extk^{* \bullet}\extk 
 \Big \} \label{extInnerProd1}
\end{align}
where $\phi = \frac{|x|}{|y|}$ and $\theta = \theta_x - \theta_y$, being $|x|, |y|, \theta_x $ and $ \theta_y$ the absolute values and angles of complex number $x$ and $y$ respectively. 

Positive-definiteness request that equation \ref{extInnerProd}
\begin{align}
 & x^*y(|x|^2 -\imagi |y|^2)\extk+ xy^*(|x|^2 -\imagi |y|^2) \extk^* + x^*y(|x|^2 + \imagi|y|^2)\extk^\bullet + xy^*(|x|^2 +  \imagi |y|^2)\extk^{* \bullet}+
 \nonumber \\
 & \extk^\bullet \extk(x^* y)^2 + \extk^\bullet \extk^* |x|^2|y|^2 + \extk^{* \bullet}\extk |x|^2|y|^2  + \extk^{* \bullet}\extk^* (x y^*)^2 +  |x|^2|y|^2( \extk^\bullet \extk^* + \extk^{* \bullet}\extk ) = {R}^4, \qquad R \in \mathbb{R}\label{realAxiomRequest},
\end{align}
or that equation \ref{extInnerProd1}
\begin{align}
&\imagi  \phi \Big [e^{-\imagi\theta}(\extk^{*\bullet} - \extk^*) + e^{\imagi\theta}(\extk^\bullet - \extk) \Big ] 
+  \frac{1}{\phi} \Big [e^{-\imagi\theta}(\extk^{*\bullet} + \extk^*) + e^{\imagi\theta}(\extk^\bullet + \extk) \Big ] 
\nonumber \\
&+ e^{-2i\theta}\extk^{*\bullet}  \extk^* + e^{2i\theta}\extk^{\bullet} \extk+ \extk^\bullet \extk^* + \extk^{* \bullet}\extk  = R,\qquad \qquad R \in \mathbb{R}.  \label{realAxiomRequest1}
\end{align}

We consider that the operations $\extk^*$ and $\extk^\bullet$ are isomorphisms between the elements in the Extended domain, meaning
\begin{equation}
f. X \to Y \quad X,Y \in \mathbb{E}.
\end{equation}
The extended maps of $\extk$ are expressed in general form as:
\begin{align}
\extk^*= z_1 \extk+w_1
\nonumber \\
\extk^\bullet = z_2 \extk+w_2 \label{extMappProp}
\end{align}
where $z_i$ and $w_i$ are complex numbers to be determined. The number $\extk^{*\bullet}$ is then
\begin{equation}
\extk^{*\bullet}\;\;=(\extk^*)^\bullet =  (z_1 \extk+w_1)^\bullet =  z_1 \extk^\bullet+w_1=  z_1 (z_2 \extk+w_2) +w_1
\end{equation}

Equations \ref{realAxiomRequest} or \ref{realAxiomRequest1}, together with the $\extk$'s definitions and the relations of Eq.\ref{extUnitDef} set the equations needed to find the values of $\extk^\bullet $ and $ \extk^*$, or what is the same, to determine the values of numbers $z_1, z_2, w_1, w_2$ in equation \ref{extMappProp}. Each extended equation results in two complex equations; one for the extended and other for imaginary part respectively. The equations are
\begin{align}
&\extk^* \extk =\imagi
\nonumber \\
&\extk^{*\bullet} \extk^\bullet = \imagi^* 
\nonumber \\
&\imagi  \phi \Big [e^{-\imagi\theta}(\extk^{*\bullet} - \extk^*) + e^{\imagi\theta}(\extk^\bullet - \extk) \Big ] 
+  \frac{1}{\phi} \Big [e^{-\imagi\theta}(\extk^{*\bullet} + \extk^*) + e^{\imagi\theta}(\extk^\bullet + \extk) \Big ] 
\nonumber \\
&+ e^{-2i\theta}\extk^{*\bullet}  \extk^* + e^{2i\theta}\extk^{\bullet} \extk + \extk^\bullet \extk^* + \extk^{* \bullet}\extk= R,\qquad \qquad R \in \mathbb{R}. \label{kMappEqSyst}
\end{align}
$\extk^2$ is another conjugated map on the extended domain that also needs to be determined. It can be calculated by substituting $\extk^*$ form on the first equation:
\begin{equation}
\extk^2 = -\frac{w_1}{z_1} \extk + \frac{\imagi}{z_1}. \label{extksquare}
\end{equation}

The complex numbers $z_1,z_2, w_1,w_2$ are included into the relations \ref{kMappEqSyst} by substituting equation \ref{extMappProp}. The finals equations depend only on parameters $\phi$, $\theta$ and $R$ like:
\begin{align}
&(2z_1 z_2 w_2-z_2^2 w_1 + z_2w_1)\extk + iz_2^2 + z_1w_2^2 + w_1w_2+\imagi=0
 \\
&\Big[ \imagi\phi e^{-\imagi\theta}z_1(z_2-1) + \imagi\phi e^{\imagi\theta} (z_2-1) + \frac{e^{-\imagi\theta}}{\phi}z_1(z_2+1) + \frac{e^{\imagi\theta}}{\phi}(z_2+1) + e^{-2i\theta}(z_1^2w_2 + z_1 w_1) 
\nonumber \\
&+  e^{2i\theta}(w_2 - \frac{w_1z_2}{z_1} )  + 2z_1w_2 -z_2w_1 + w_1 \Big]\extk +  \imagi\phi e^{-\imagi\theta}z_1w_2 +  \imagi\phi e^{\imagi\theta}w_2 +  \frac{e^{-\imagi\theta}}{\phi}(z_1w_2 + 2w_1) 
\nonumber \\
&+  \frac{e^{\imagi\theta}}{\phi}w_2 + e^{-2i\theta}( iz_1z_2 + z_1w_1w_2 + w_1^2) + ie^{2i\theta}\frac{z_2}{z_1} + 2iz_2 + w_1w_2 -R=0
\end{align}
We consider that an extended number is zero if its extended and imaginary part are both nulls. We obtain then, four equations
\begin{align}
&2z_1 z_2 w_2-z_2^2 w_1 + z_2w_1=0
\nonumber \\
&iz_2^2 + z_1w_2^2 + w_1w_2+\imagi=0
\nonumber \\
& \imagi\phi e^{-\imagi\theta}z_1(z_2-1) + \imagi\phi e^{\imagi\theta} (z_2-1) + \frac{e^{-\imagi\theta}}{\phi}z_1(z_2+1) + \frac{e^{\imagi\theta}}{\phi}(z_2+1) + e^{-2i\theta}(z_1^2w_2 + z_1 w_1) 
\nonumber \\
&+  e^{2i\theta}(w_2 - \frac{w_1z_2}{z_1} )  + 2z_1w_2 -z_2w_1 + w_1 =0
\nonumber \\
&\imagi\phi e^{-\imagi\theta}z_1w_2 +  \imagi\phi e^{\imagi\theta}w_2 +  \frac{e^{-\imagi\theta}}{\phi}(z_1w_2 + 2w_1) 
+  \frac{e^{\imagi\theta}}{\phi}w_2 + e^{-2i\theta}( iz_1z_2 + z_1w_1w_2 + w_1^2) 
\nonumber \\
&+ ie^{2i\theta}\frac{z_2}{z_1} + 2iz_2 + w_1w_2 -R=0  \label{kMappEqSyst1}
\end{align}
that are sufficient to compute the unknown quantities  $z_1,z_2,w_1$ and $w_2$. Unfortunately, those values or, what is the same $\extk^*$ and $\extk^\bullet$, are not so simple to solve. Nevertheless, the fact that we have two extended or four complex equations and that we have four complex variables, then the systems is solvable. We are in the presence of a system of equations with a degree 3 that have three solutions for any combination of $\phi$, $\theta$ and $R$ parameters. If we had established different conditions for the terms of the absolute value, we might obtain more equations than variables that will turn the solution for the extended maps non-viable. 

From our proposition for absolute value of an extended number, we can extract two conclusions:
\begin{enumerate}
\item The metric of the proposed space is quartic and  have the form 
\begin{equation}
|\alpha|^4 = |x\extk + y|^4 = |x|^4 + |y|^4 + R|x|^2|y|^2 \qquad \forall x,y \in \mathbb{C}
\end{equation}
where $R$ is a real number.

\item For an extended number $\alpha = x \extk + y$, the maps $k^* = z_1 \extk + w_1$ and $\extk^\bullet= z_2 \extk + w_2$ from $\alpha*$ and $\alpha^\bullet$ respectively, depend on the extended number which the maps modify. Specifically, the complex values $= z_1,\;z_2,\; w_1,\; w_2$, depend on the ratio between the absolute value of the extended and imaginary parts of the extended number $\phi = \frac{|x|}{|y|}$ and also on the difference between the angles of both parts $\theta = \theta_{x} - \theta_{y}$, where ${x}$ and ${y}$ are the extended and imaginary parts of the number respectively.
\end{enumerate}

The second conclusion led to some inconsistencies, once we analyse the product between extended numbers for example:
\begin{itemize}
\item According to the definition of the extended unit, $\extk^2$ will depend on $z_1$ and $w_1$ as shown in equation \ref{extksquare} which depends on the complex conjugate of the number. If that is the case then, in the product of two extended numbers $(x\extk + y)(u\extk + v)$, what it would be the number which should consider as the complex conjugated for extract the values $z_1$ and $w_1$?
\item The expression $\extk^2$  is extracted using $\extk^*\extk = \imagi$ definition, however, from $\extk^{* \bullet}\extk^\bullet = \imagi^*$, we can obtain a different relation. What will be then the correct expression to use? Note that if we force both expressions of $\extk^2$ to be equal, we will be imposing then constraint to both maps, $\extk^*$ and $ \extk^\bullet$, $e.i.$, one map will depend on the other.
\item The last example and maybe the most important is that, because of previous issues, we cannot find a proper way to define the inner product and the absolute value for extended complex numbers.
\end{itemize}

In the next section, we start the study of the algebraic structure of the extended numbers that we hope help to clarify the above issues.

\subsection{Introduction to the extended numbers}
In the previous section, we proposed the extension of complex numbers because of the necessity, from our point of view, of including it as the vector space for developing the quantum mechanics for $n$-VMVF systems. We started defining the extended unit as $\extk^*\extk = \imagi$, which came from an unsolvable equation in the complex field. It is contradictory since the fundamental theorem of algebra states that the set of complex numbers is algebraically closed for the sum and multiplication operations, so it should not exist an unsolvable equation for an $n$-degree polynomial. Well, there aren't.  The equations obtained from the positive-definiteness axiom result in algebraic contradictions if we consider only the standard sum and multiplication operations. However, $|x|^2 = x^* x = i$ only appears if the complex conjugate numbers are included in the polynomial, which, in turn, can be done if others mathematical operations are added to the standard sum and multiplication: the sum and multiplication with the complex conjugate. Indeed, from a polynomial like
\begin{equation}
a_n x^n + a_{n-1} x^{n-1} + ... + a_2 x^2 + a_1 x + a_0 = \sum_n a_n x^n=0 \qquad \forall a_n \in \mathbb{C}
\end{equation}
could never be extracted the unsolvable equation used for defining the extended unit $\extk$. This equation without solution can be extracted from a polynomial which includes the operations addition and multiplication by a complex conjugate such as
\begin{equation}
a_2 x^2 x^{*} + a_1 x^* + a'_1 x + a_0 =0 \qquad \forall a_n \in \mathbb{C}.
\end{equation}

We represent the conjugated sum of two complex numbers $a,b$ as 
\begin{equation}
a\oplus b \equiv a^* + b = b + a^*
\end{equation}
and the conjugated  product as
\begin{equation}
a\odot b \equiv a^*  b = b  a^*.
\end{equation}

A polynomial function that includes the conjugated sum and multiplication will have unsolvable equations like $x\odot x =x x^*= -1$. Therefore it cannot be considered closed, signalizing the possibility of expansion without violating the fundamental theorem in algebra. If the complex field is algebraically represented like 
\begin{equation}
(\mathbb{C},+,\cdot),
\end{equation}
then we are in the presence of a new algebraic structure which can be represented as
\begin{equation}
(\mathbb{E},+,\cdot,\oplus, \odot) \label{extNumberStruc}
\end{equation}
which should be studied with precision. Perhaps, another way to treat this problem is to keep the standard sum and multiplication as the only operations and add the set of complex numbers to the algebraic structure like 
\begin{equation}
(\mathbb{E},\mathbb{E}^*,+,\cdot).
\end{equation}
We present an initial analysis of the extended domain using an structure like \ref{extNumberStruc}, where new operations are included.

We must assume the possibility of the the isomorphism depend on the number which is applied on. However, the definition of the extended unit should keep invariant for all the extended numbers, and it can be written as
\begin{equation}
\extk \odot \extk \equiv  \extk^*  \extk= \imagi \qquad \text{ or } \qquad \extk = \sqrt[*]{\imagi}
\end{equation}
where we define the conjugated square root, $\sqrt[*]{()}$,  of a number $x$ as the operation that results in a number $y$ such that its product over its complex conjugate is  $x$: $y^*y=x$. The extended unit can be defined as the conjugated square root of the imaginary unit $\imagi$:
\begin{equation}
\extk = \sqrt[*]{\imagi} \qquad \text{ if } \qquad \extk \odot \extk =  \extk^*  \extk= \imagi. \label{extConjProdDef}
\end{equation}
If $\extk^* = z_1 \extk + w_1$, we obtain 
\begin{equation}
\extk^*  \extk= \imagi \qquad \to \qquad \extk^2 = -\frac{w_1}{z_1} \extk + \frac{\imagi}{z_1} \label{extConjProdDef1}
\end{equation}
where $z_1, w_1$ corresponds to the extended and the imaginary parts of the complex conjugated map of the extended number.

On the other side, the standard product of two extended units $\extk^2 = \extk \extk$ must also remain invariant for all the set of extended numbers. Different from the definition we gave to the extended unit, the square operation won't arise from an unsolvable equation as we see before. Nevertheless, it must be defined for all extended numbers. We define $\extk^2$ map as:
\begin{equation}
\extk^2 = z_0 \extk + w_0 \qquad \forall \;\mathbb{E}. \label{extStandProdDef}
\end{equation}
where $z_0, w_0$ are two complex numbers which need to be determined in future studies. In this case, the expression 
\begin{equation}
\extk^*  \extk = (z_1 \extk + w_1)\extk = (z_0z_1 + w_1)\extk + z_1 w_0 \neq \imagi \label{extStandProdDef1}
\end{equation}

Resuming, within the extended numbers we define not one product but two: the standard and the conjugated product. Their definitions are
\begin{equation}
\extk \odot  \extk = \imagi \qquad \text{ and } \qquad \extk\cdot \extk = z_0 \extk + w_0. 
\end{equation}

Being $\extk^* = z_1 \extk + w_1$, the expressions $\extk^2$ and $\extk^* \extk$ can be related on to the other as showed before. However, it must be clear that for the standard product the $\extk^2$ and $\extk^* \extk$ operations must compute as equations \ref{extStandProdDef} and \ref{extStandProdDef1} while for the conjugated product the expression for these equations are \ref{extConjProdDef1}  and \ref{extConjProdDef}. The table \ref{extDefTable} resumes the different expressions for every case.
\begin{table}[h]
\caption{Expressions for $\extk \cdot\extk$ and $\extk^*\; \extk$ for the two defined extended product} \label{extDefTable}
\begin{center}
\begin{tabular}{@{} c  c  c @{}}
\hline \hline
 & $\cdot$ & $\odot$	\\ 
\hline
$\extk \cdot\extk$ & $z_0 \extk + w_0 $ & $ -\frac{w_1}{z_1} \extk + \frac{\imagi}{z_1} $  \\

$ \extk^*\; \extk$ &$  (z_0z_1 + w_1)\extk + z_1w_0$& $  \imagi$ \\
\hline
\end{tabular} 
\end{center}
\end{table}

As $\oplus$ and $\odot$ are a type of sum and multiplication operations respectively; the order of priority of the operations then is similar. That means that any multiplication is granted a higher precedence than any type of addition. However, the order of priority between standard and conjugated operations, being multiplication or addition, should be emphasized with parentheses $(\;)$ or brackets $[\;]$. We also consider the absence of operator as the standard multiplication.

Lets explicitly show both types of extended products in two extended numbers:
\begin{itemize}
\item The standard product of two extended can be computed using the equation \ref{extStandProdDef}, $\extk^2 = z_0 \extk + w_0$. For the extended numbers $\alpha =x \extk + y$ and $\beta = u \extk + v$:
\begin{align}
\alpha \beta =(x \extk + y)(u \extk + v) &= xu \extk^2 + (xv + yu)\extk + yv
\nonumber \\
&=  xu (z_0 \extk + w_0) + (xv + yu)\extk + yv
\nonumber \\
&=  (xu z_0 + xv + yu)\extk + xuw_0+ yv
\end{align}
One of the factors, let's say $\alpha$, can be expressed in its complex conjugated form using its inverse transformation of the map $\extk^* = z_1^{(\alpha)} \extk + w_1^{(\alpha)}$:
\begin{equation}
\extk = \frac{1}{z_1^{(\alpha)}}(\extk^* - w_1^{(\alpha)}), \label{extComplexInv}
\end{equation}
where $z_1^{(\alpha)}, w_1^{(\alpha)}$ are the extended and imaginary part of the extender number \ref{extStandProdDef1}. In this case, the standard product can be founded using equation \ref{extStandProdDef1}. Substituting equation \ref{extComplexInv} in $\alpha=x \extk + y$, we get
\begin{equation}
x\extk + y = x\big[  \frac{1}{z_1^{(\alpha)}}(\extk^* - w_1^{(\alpha)})\big]+ y = \frac{x}{z_1^{(\alpha)}}\extk^* +y - \frac{x w_1^{(\alpha)}}{z_1^{(\alpha)}}.
\end{equation}
In this case, the product has the form
\begin{align}
(x \extk + y)(u \extk + v) = \big(  \frac{x}{z_1^{(\alpha)}}\extk^* +y - \frac{x w_1^{(\alpha)}}{z_1^{(\alpha)}}\big) (u \extk + v)
\nonumber \\
\frac{xu}{z_1^{(\alpha)}}\extk^* \extk + \frac{xv}{z_1^{(\alpha)}}\extk^* - \frac{xu w_1^{(\alpha)}}{z_1^{(\alpha)}}\extk + yu \extk - \frac{xv w_1^{(\alpha)}}{z_1^{(u)}} + yv.
\end{align}
Substituting equation \ref{extStandProdDef1}  
\begin{equation*}
\extk^*  \extk = (z_0z_1^{(\alpha)} + w_1^{(\alpha)})\extk + z_1^{(\alpha)} w_0,
\end{equation*}
and $\extk^* = z_1^{(\alpha)} \extk + w_1^{(\alpha)}$, we obtain the same result. In effect,
\begin{align}
&(x \extk + y)(u \extk + v) =\frac{xu}{z_1^{(\alpha)}}\big[(z_0z_1^{(\alpha)} 
+ w_1^{(\alpha)})\extk + z_1^{(\alpha)} w_0\big] 
+ xv\extk + \frac{xv}{z_1^{(\alpha)}}w_1^{(\alpha)} 
- \frac{xu w_1^{(\alpha)}}{z_1^{(\alpha)}}\extk 
\nonumber \\
= & + yu \extk - \frac{xv w_1^{(\alpha)}}{z_1^{(\alpha)}} + yv
\nonumber \\
= &(xu z_0 + \frac{xuw_1^{(\alpha)}}{z_1^{(\alpha)}} + xv - \frac{xuw_1^{(\alpha)}}{z_1^{(\alpha)}} + yu )\extk + xuw_0+ \frac{xv w_1^{(\alpha)}}{z_1^{(\alpha)}} - \frac{xv w_1^{(\alpha)}}{z_1^{(\alpha)}} + yv
\nonumber \\
&=  (xu z_0 + xv + yu)\extk + xuw_0+ yv
\end{align}

\item The conjugate product can be computed using the definition $\extk \odot \extk =\imagi$. For the extended numbers $\alpha=x \extk + y$ and $\beta = u \extk + v$ we have
\begin{align}
\alpha \odot \beta = (x \extk + y)\odot(u \extk + v) &= x^*u \extk^* \extk + x^*v\extk^* + y^*u\extk + y^*v
\nonumber \\
&=   x^*u \imagi + x^*v(z_1^{(\alpha)} \extk + w_1^{(\alpha)}) + y^*u\extk + y^*v
\nonumber \\
&=   (x^*vz_1^{(\alpha)}  +  y^*u )\extk + x^*u \imagi +x^*vw_1^{(\alpha)}+ y^*v.
\end{align}

The expression can be treated as a standard product, only that $\extk^* \extk$ and $\extk^2$ must have the definitions according to table \ref{extDefTable}, then
\begin{equation}
\extk^2 = -\frac{w_1^{(\alpha)}}{z_1^{(\alpha)}} \extk + \frac{\imagi}{z_1^{(\alpha)} }
\end{equation}

\begin{align}
(x \extk + y)\odot(u \extk + v) &=(x^* \extk^* + y^*)(u \extk + v) 
\nonumber \\
&=(x^*z_1^{(\alpha)} \extk + x^*w_1^{(\alpha)} +  y^*)(u \extk + v)  
\nonumber \\
&= x^*u z_1^{(\alpha)}\extk^2 + (x^*v z_1^{(\alpha)} + x^*uw_1^{(\alpha)} +  y^*u)\extk + x^*vw_1^{(\alpha)} +  y^*v
\nonumber \\
&= (x^*v z_1^{(\alpha)} + x^*uw_1^{(\alpha)} +  y^*u - x^*uw_1^{(\alpha)})\extk + x^*vw_1^{(\alpha)} +  y^*v +  x^*u \imagi
\nonumber \\
&=   (x^*vz_1^{(\alpha)}  +  y^*u )\extk + x^*u \imagi +x^*vw_1^{(\alpha)}+ y^*v ,
\end{align}
\end{itemize}
obtaining the same result.

The standard and the conjugated sum of two extended number are
\begin{align}
\alpha + \beta = (x \extk + y) + (x \extk + y) &= (x + u)\extk +( y+ v)
\end{align}
and
\begin{align}
\alpha \oplus \beta &=  (x \extk + y) \oplus (u \extk + v) 
\nonumber \\
&= (x \extk + y)^* + (u \extk + v) = (x^* \extk^* + y^*) + (u \extk + v)
\nonumber \\
&= \big[x^* (z_1^{(\alpha)} \extk + w_1^{(\alpha)}) + y^*\big] + (u \extk + v) = (x^*z_1^{(\alpha)} \extk +  x^*w_1^{(\alpha)} + y^*) + (u \extk + v) 
\nonumber \\
&= (x^*z_1^{(\alpha)} + u)\extk +(  x^*w_1^{(\alpha)} + y^*+ v)
\end{align}
respectively.

\subsection{Algebraic properties of extended numbers } \label{extAlgebraSection}

In Abstract Algebra theory, the classification of any set of numbers on which are defined the binary operations sum and multiplication are described by the compliance or not of the following properties:
\begin{enumerate}
\item Associativity of addition and multiplication
\item Commutativity of addition and multiplication
\item Existence of additive and multiplicative identity elements
\item Existence of additive inverses and multiplicative inverses
\item Distributivity of multiplication over addition
\end{enumerate}
We start the study of the properties of the standard and conjugated sum and product of the extended numbers, using the definitions of the standard and the conjugated operations for the extended numbers and taking into account that values like $z_0,w_0$ for the standard multiplication operation are parameters still to be determined.

\subsubsection*{The standard summation ``$+$''}
\begin{enumerate}
\item Associativity:
For all $\alpha_1 = x_1\extk + y_1, \alpha_2= x_2\extk + y_2, \alpha_3= x_3\extk + y_3 \; \in \mathbb{E}$ then
\begin{align}
&(\alpha_1 + \alpha_2) + \alpha_3  = \big[( x_1\extk + y_1) +  (x_2\extk + y_2)\big] +  (x_2\extk + y_3) 
\nonumber \\
&= \big[(x_1 + x_2)\extk + (y_1 + y_2)\big] +  (x_2\extk + y_3)
\nonumber \\
&= (x_1 + x_2 + x_3)\extk + (y_1 + y_2 + y_3) = x_1\extk + y_1 + \big[(x_2 + x_3)\extk + (y_2 + y_3)\big]
\nonumber \\
&= \alpha_1 + (\alpha_2 + \alpha_3)
\end{align}
\item Commutativity
For all $\alpha_1 = x_1\extk + y_1, \alpha_2= x_2\extk + y_2\; \in \mathbb{E}$ then
\begin{equation}
\alpha_1 + \alpha_2 = (x_1 + x_2)\extk + (y_1 + y_2) = \alpha_2 + \alpha_1 \qquad \forall \alpha_1,\alpha_2 \in \mathbb{E}
\end{equation}
\item Existence of additive identity element $0_E$ in $\mathbb{E}$ such that 
\begin{equation}
\alpha + 0_E = (x + 0)\extk + (y +0) = 0_E + \alpha = \alpha
\end{equation}
\item Existence of additive inverse such that for every $\alpha \in \; \mathbb{E}$, there exists an element $-\alpha\in\; \mathbb{E}$, such that $\alpha + (-\alpha) = 0$. If $\alpha = x \extk + y$ then $-\alpha = - x\extk -y$. It can be verified that
\begin{equation}
y \extk + y + (- x \extk -y) = (x - x)\extk + (y - y) = 0.
\end{equation}
\end{enumerate}

\subsubsection*{ The standard product ``$\cdot$''} 

\begin{enumerate}
\item Associativity:

Standard product is associative if the extended numbers $\alpha_1,\alpha_2,\alpha_3 \; \in \mathbb{E}$ satisfy
\begin{equation}
\alpha_1 (\alpha_2\alpha_3) = (\alpha_1 \alpha_2)\alpha_3.
\end{equation}
Being $\alpha_1 = x_1\extk + y_1, \alpha_2= x_2\extk + y_2$ and $\alpha_3= x_3\extk + y_3 $. The left member of the above equation is
\begin{align}
& (x_1 \extk + y_1)\big[(x_2 \extk + y_2)(x_3 \extk + y_3)\big] = (x_1 \extk + y_1)\big[(x_2 x_3 z_0 + x_2 y_3 + y_2 x_3 )\extk + x_2 x_3 w_0 + y_2 y_3  \big]
\nonumber \\
&= (x_1 x_2 x_3 z_0 + x_1 x_2 y_3 + x_1 y_2 x_3 )\extk^2 + ( x_1 x_2 x_3 w_0 + x_1 y_2 y_3 + y_1 x_2 x_3 z_0 + y_1 x_2 y_3 + y_1 y_2 x_3 )\extk
\nonumber \\
&\quad + y_1 x_2 x_3 w_0 + y_1 y_2 y_3  
\nonumber \\
&= \big[x_1 x_2 x_3 z_0^2 + (x_1 x_2 y_3  + x_1 y_2 x_3 + y_1 x_2 x_3)z_0 + x_1 x_2 x_3 w_0 + x_1 y_2 y_3  + y_1 x_2 y_3 + y_1 y_2 x_3 \big]\extk 
\nonumber \\
& \quad + x_1 x_2 x_3 w_0 z_0  + (x_1 x_2 y_3  + x_1 y_2 x_3  + y_1 x_2 x_3) w_0 + y_1 y_2 y_3.
\end{align}
The right member of the axiomatic equation is
\begin{align}
& \big[(x_1 \extk + y_1)(x_2 \extk + y_2)\big](x_3 \extk + y_3) = \big[ (x_1 x_2 z_0 + x_1 y_2 + y_1 x_2 )\extk + x_1 x_2 w_0 + y_1 y_2 \big](x_3 \extk + y_3)
\nonumber \\
&= ( x_1 x_2 x_3 z_0 + x_1 y_2 x_3 + y_1 x_2 x_3 )\extk^2 + ( x_1 x_2 y_3 z_0 + x_1 y_2 y_3 + y_1 x_2 y_3  + x_1 x_2 x_3 w_0 + y_1 y_2 x_3)\extk
\nonumber \\
& \quad + x_1 x_2 y_3 w_0 + y_1 y_2 y_3 
\nonumber \\
&= \big[x_1 x_2 x_3 z_0^2 + (x_1 x_2 y_3  + x_1 y_2 x_3 + y_1 x_2 x_3)z_0 + x_1 x_2 x_3 w_0 + x_1 y_2 y_3  + y_1 x_2 y_3 + y_1 y_2 x_3 \big]\extk 
\nonumber \\
& \quad + x_1 x_2 x_3 w_0 z_0  + (x_1 x_2 y_3  + x_1 y_2 x_3  + y_1 x_2 x_3) w_0 + y_1 y_2 y_3.
\end{align}
which is the same result of above.
\item Distributivity of the standard multiplication over the standard addition

The standard extended inner product is distributive over the standard addition if for all $ \alpha_1,\alpha_2,\alpha_3 \in \mathbb{E}$ it is satisfied the relation 
\begin{equation}
\alpha_1( \alpha_2 + \alpha_3) = \alpha_1 \alpha_2 + \alpha_1 \alpha_3.
\end{equation}
If the numbers $\alpha_i = x_i \extk + y_i$, for all $i=1,2$, the right member of the preposition is
\begin{align}
&\alpha_1( \alpha_2 + \alpha_3)  = ( x_1 \extk + y_1 ) \big[ ( x_2 \extk + y_2 ) + ( x_3 \extk + y_3 )\big] = ( x_1 + \extk y_1 ) \big[ ( x_2 + x_3 )\extk  + y_2 + y_3 )\big]
\nonumber \\
&= \big[ x_1 ( x_2 + x_3 )z_0 + x_1 ( y_2 + y_3 ) + y_1 ( x_2 + x_3 ) \big]\extk + x_1 ( x_2 + x_3 )w_0 
\nonumber \\
& \quad + y_1 ( y_2 + y_3 )
\nonumber \\
&= \big[ ( x_1 x_2 z_0 + x_1 y_2 + y_1 x_2 )\extk + x_1 x_2 w_0 + y_1 y_2 \big] + \big[ ( x_1 x_3 z_0 + x_1 y_3 + y_1 x_3 )\extk 
\nonumber \\
& \quad + x_1 x_3 w_0 + y_1 y_3 \big]
\nonumber \\
& = ( x_1 \extk + y_1 )( x_2 \extk + y_2 ) + ( x_1 \extk + y_1 )( x_3 \extk + y_3 ) = \alpha_1 \alpha_2 + \alpha_1 \alpha_3,
\end{align}
as stated before.
\item Commutativity:

The standard extended product is commutative if relation $\alpha_1 \alpha_2 = \alpha_2 \alpha_1$ is satisfied for all $ \alpha_1,\alpha_2 \in \mathbb{E}$. Substituting numbers $\alpha_i = x_i \extk + y_i$, for all $i=1,2$, standard product has the form:
\begin{align}
&\alpha_1 \alpha_2 = (x_1 \extk + y_1)(x_2 \extk + y_2) = (x_1 x_2 z_0 + x_1 y_2 + x_2 y_1)\extk + x_1 x_2 w_0 + y_1 y_2 
\nonumber \\
&=  (x_2 \extk + y_2)(x_1 \extk + y_1) = \alpha_2 \alpha_1
\end{align}
\item Zero-product property

In algebra, the zero-product property states that the product of two nonzero elements is nonzero. In other words, if:
\begin{equation}
\alpha \beta = 0, \qquad \text{only if} \qquad \alpha = 0 \quad \text{or} \quad \beta =0.
\end{equation}
The standard product of these two extended numbers, where $\alpha=x\extk + y$ and $\beta =u \extk + v$, show the existence of nontrivial zero divisors on the standard product. Setting zero the standard product of two extended numbers we have
\begin{equation}
\alpha \beta = (x\extk + y)(u \extk + v)= \big[x(u z_0 + v) + y u\big]\extk + x u w_0 + y v=0
\end{equation}
which led to two complex equations
\begin{equation}
x(u z_0 + v) + y u =0, \qquad x u w_0 + y v=0 \label{zeroProEq}
\end{equation}
Without lost generality, we can analyze this set of equations for different cases of $\beta$ number:
\begin{enumerate}
\item $u=0, \;v\neq 0$.
In this case, the equations are
\begin{equation}
x v =0, \qquad y v=0
\end{equation}
which its satisfied if $x=y=0$.
\item $u\neq 0, \;v= 0$.
In this case, the equation \ref{zeroProEq} take the form
\begin{equation}
(x z_0 + y) u =0, \qquad x u w_0 = 0
\end{equation}
which its satisfied if $x=y=0$, for $w_0 \neq 0$. If $w_0 = 0$ then a nontrivial root from the equation $x z_0 + y=0$ is included.
\item $u\neq 0, \;v\neq 0$. In this case, multiplying first equation \ref{zeroProEq} by $v$, the second one by $u$ and subtracting one from the other we obtain
\begin{equation}
x(v^2 + u v z_0 - u_2 w_0)=0,
\end{equation}
which introduce the nontrivial root coming from the equation:
\begin{equation}
v^2 + u v z_0 - u_2 w_0=0
\end{equation}
\end{enumerate}
\item Existence of multiplicative identity element:

For every $\alpha \in \mathbb{E}$, exist a number   $1_E\equiv 1$ in $\mathbb{E}$ such that 
\begin{equation}
\alpha\; 1_E = 1_E \; \alpha = \alpha
\end{equation}

\end{enumerate}

On the other side, the conjugate sum nor product is not associative, commutative and have no additive identity nor inverse element. The following discussion is referred to extended numbers whose maps are defined. That means we exclude purely extended and complex numbers. For these cases, we should proceed using the extended unit definition.
\subsubsection*{The conjugated sum ``$\oplus$''}
\begin{enumerate}
\item Associativity

The associativity property implies that
\begin{equation}
(\alpha_1 \oplus \alpha_2)\oplus \alpha_3 =  \alpha_1 \oplus (\alpha_2 \oplus \alpha_3).
\end{equation}
The left member of the last relation
\begin{align}
&(\alpha_1 \oplus \alpha_2)\oplus \alpha_3 = \big[(x_1 \extk + y_1) \oplus (x_2 \extk + y_2)\big] \oplus (x_3 \extk + y_3)
\nonumber \\
&= \big[(x_1^* z_1^{(\alpha_1)} + x_2)\extk + x_1^* w_1^{(\alpha_1)} + y_1^* + y_2 \big]\oplus (x_3 \extk + y_3)
\nonumber \\
&= \big[ (x_1 z_1^{*(\alpha_1)} + x_2^*) z_1^{(\alpha_1\oplus \alpha_2)} + x_3 \big] \extk + (x_1 z_1^{*(\alpha_1)} + x_2^*) w_1^{(\alpha_1 \oplus \alpha_2)} + x_1^* w_1^{(\alpha_1)} + y_1^* + y_2
\end{align}
is not equal to the right member since
\begin{align}
& \alpha_1 \oplus (\alpha_2 \oplus \alpha_3) = (x_1 \extk + y_1) \oplus \big[(x_2 \extk + y_2) \oplus (x_3 \extk + y_3)\big] 
\nonumber \\
& = (x_1 \extk + y_1) \oplus \big[(x_2^* z_1^{(\alpha_2)} + x_3)\extk + x_2^* w_1^{(\alpha_2)} + y_2^* + y_3 \big]
\nonumber \\
&= \big[ x_1^* z_1^{(\alpha_1)} + x_2^* z_1^{(\alpha_2)} + x_3 \big] \extk + x_1^* w_1^{(\alpha_1)} +  x_2^* w_1^{(\alpha_2)} + y_1^* + y_2^* + y_3
\end{align}

\item Commutativity

The expression 
\begin{align}
&\alpha_1 \oplus \alpha_2 = (x_1 \extk + y_1) \oplus (x_2 \extk + y_2) 
\nonumber \\
&= \big[x_1^*z_1^{(\alpha_1)} + x_2\big]\extk + x_1^*w_1^{(\alpha_1)} + y_1^* + y_2
\end{align}
while
\begin{align}
&\alpha_2 \oplus \alpha_1 = (x_2 \extk + y_2) \oplus (x_1 \extk + y_1) 
\nonumber \\
&= \big[x_1 + x_2^*z_1^{(\alpha_2)}\big]\extk + x_2^*w_1^{(\alpha_2)} + y_1 + y_2^*,
\end{align}
which means that the conjugated sum doesn't satisfy the commutative property.
\item There is no conjugated additive identity element  $0_E$ in $\mathbb{E}$ because 
\begin{equation}
0_E \oplus \alpha =  x \extk + y  = \alpha
\end{equation}
while
\begin{equation}
\alpha \oplus 0_E = x^* z_1^{(\alpha)}\extk + x^* w_1^{(\alpha)} + y \neq \alpha
\end{equation}

\item The conjugated additive inverse element also doesn't exist. That can be probed by straight substitution but also by noting that the property won't be satisfied if the conjugated sum is non-commutative.
\end{enumerate}

\subsubsection*{The conjugate product ``$\odot$''}
\begin{enumerate}
\item Associativity

The associativity property implies that
\begin{equation}
\alpha_1 \odot (\alpha_2 \odot \alpha_3) = (\alpha_1 \odot \alpha_2)\odot \alpha_3.
\end{equation}
Computing the left member we have
\begin{align}
&\alpha_1 \odot (\alpha_2\odot \alpha_3) = (x_1 \extk + y_1) \odot \big[(x_2 \extk + y_2) \odot (x_3 \extk + y_3)\big]
\nonumber \\
&= (x_1 \extk + y_1) \odot \big[(x_2^* y_3 z_1^{(\alpha_2)} + x_3 y_2^*)\extk + \imagi x_2^* x_3 + x_2^* y_3 w_1^{(\alpha_2)} + y_2^* y_3\big]
\nonumber \\
&=\big[x_1^*(\imagi x_2^* x_3 + x_2^* y_3 w_1^{(\alpha_2)} + y_2^* y_3) z_1^{(\alpha_1)} + y_1^*(x_2^* y_3 z_1^{(\alpha_2)} + x_3 y_2^*) \big] \extk 
\nonumber \\
& \quad + \imagi x_1^*(x_2^* y_3 z_1^{(\alpha_2)} + x_3 y_2^*) + x_1^*(\imagi x_2^* x_3 + x_2^* y_3 w_1^{(\alpha_2)} + y_2^* y_3) w_1^{(\alpha_1)} 
\nonumber \\ 
& \quad + y_1^*(\imagi x_2^* x_3 + x_2^* y_3 w_1^{(\alpha_2)} + y_2^* y_3)
\nonumber \\ 
&=\big[\imagi x_1^* x_2^* x_3 z_1^{(\alpha_1)}+ x_1^* x_2^* y_3 z_1^{(\alpha_1)} w_1^{(\alpha_2)} + x_1^* y_2^* y_3z_1^{(\alpha_1)}  + x_2^* y_1^* y_3 z_1^{(\alpha_2)} + x_3 y_1^*y_2^* \big] \extk 
\nonumber \\
& \quad + x_1^*x_2^* y_3 z_1^{(\alpha_2)} + x_1^* x_3 y_2^* + x_1^*\imagi x_2^* x_3w_1^{(\alpha_1)} + x_1^*x_2^* y_3 w_1^{(\alpha_1)} w_1^{(\alpha_2)} + x_1^*y_2^* y_3 w_1^{(\alpha_1)} 
\nonumber \\ 
& \quad + \imagi x_2^* x_3 y_1^*+ x_2^* y_1^* y_3 w_1^{(\alpha_2)} + y_1^*y_2^* y_3.
\end{align}
while the computation of the right member is 
\begin{align}
&(\alpha_1 \odot \alpha_2)\odot \alpha_3 = \big[(x_1 \extk + y_1) \odot (x_2 \extk + y_2)\big] \odot (x_3 \extk + y_3)
\nonumber \\
&=\big[(x_1^* y_2 z_1^{(\alpha_1)} + x_2 y_1^*)\extk  + \imagi x_1^* x_2 + x_1^* y_2 w_1^{(\alpha_1)} + y_1^* y_2\big]\odot (x_3 \extk + y_3)
\nonumber \\
&= \big[ (x_1^* y_2 z_1^{(\alpha_1)} + x_2 y_1^*)^*y_3z_1^{(\alpha_1 \odot \alpha_2)} + (\imagi x_1^* x_2 + x_1^* y_2 w_1^{(\alpha_1)} + y_1^* y_2)^*x_3\big]\extk
\nonumber \\
&\quad + \imagi (x_1^* y_2 z_1^{(\alpha_1)} + x_2 y_1^*)^* x_3  + (x_1^* y_2 z_1^{(\alpha_1)} + x_2 y_1^*)^*y_3w_1^{(\alpha_1 \odot \alpha_2)}  
\nonumber \\
&\quad +  (\imagi x_1^* x_2 + x_1^* y_2 w_1^{(\alpha_1)} + y_1^* y_2)^*y_3
\nonumber \\
&= \big[ x_1 y_2^*y_3 z_1^{*(\alpha_1)}z_1^{(\alpha_1 \odot \alpha_2)} 
+ x_2^* y_1 y_3z_1^{(\alpha_1 \odot \alpha_2)} 
- \imagi x_1 x_2^*x_3 + x_1 x_3 y_2^* w_1^{*(\alpha_1)} + x_3y_1 y_2^*\big]\extk
\nonumber \\
&\quad + \imagi x_1 x_3 y_2^* z_1^{*(\alpha_1)} + \imagi x_2^* x_3 y_1 + x_1 y_2^*y_3 z_1^{*(\alpha_1)}w_1^{(\alpha_1 \odot \alpha_2)} 
+ x_2^* y_1 y_3 w_1^{(\alpha_1 \odot \alpha_2)} 
\nonumber \\
&\quad   \imagi x_1 x_2^*y_3 + x_1 y_2^*y_3 w_1^{*(\alpha_1)} + y_1 y_2^*y_3,
\end{align}
showing that
\begin{equation}
\alpha_1 \odot (\alpha_2 \odot \alpha_3) \neq (\alpha_1 \odot \alpha_2)\odot \alpha_3,
\end{equation}
or what is the same, it not comply with the associative property.
\item Commutativity

The commutative property, $\alpha_1 \odot \alpha_2 = \alpha_2 \odot \alpha_1$, $ \forall \alpha_1,\alpha_2 \in \mathbb{E}$ is not satisfied. For numbers  $\alpha_i = x_i \extk + y_i$, where $i=1,2$, we have:
\begin{align}
\alpha_1 \odot \alpha_2 = (x_1^* y_2 z_1^{(\alpha_1)} + x_2 y_1^*)\extk + x_1^* y_2 w_1^{(\alpha_1)} + \imagi x_1^* x_2 + y_1^* y_2 
\end{align}
while 
\begin{align}
\alpha_2 \odot \alpha_1 = (x_2^* y_1 z_1^{(\alpha_2)} + x_1 y_2^*)\extk + x_2^* y_1 w_1^{(\alpha_2)} + \imagi x_2^* x_1 + y_1 y_2^*
\end{align}

Due to the noncommutative property of the conjugated product, it's convenient to specify the order of the conjugated multiplication. Then, we can assume that the left conjugate multiplication of an extended number  $\alpha$ by other extended $\beta$ stands for  $\alpha  \odot \beta$, while the right multiplication of  $\alpha$ by $\beta$ means $ \beta \odot \alpha$.

\item Distributivity of the conjugated multiplication over the standard addition

We study the distributive property for the right and left conjugated multiplication of the standard sum. In the first case, this property is satisfied if:
\begin{equation}
\alpha_1 \odot ( \alpha_2 + \alpha_3) = \alpha_1 \odot \alpha_2 + \alpha_1 \odot \alpha_3.
\end{equation}
Computing the left member we have
\begin{align}
&\alpha_1 \odot ( \alpha_2 + \alpha_3)  = (x_1 \extk + y_1) \odot \big[(x_2 \extk + y_2) + (x_3 \extk + y_3)\big]
\nonumber \\
&= \big[ x_1^*(y_2 + y_3) z_1^{(\alpha_1)} + (x_2 + x_3)y_1^* \big] \extk + x_1^*(y_2 + y_3) w_1^{(\alpha_1)} + \imagi  x_1^*(x_2 + x_3) +  y_1^*(y_2 + y_3)
\nonumber \\
&=  \big[ x_1^*y_2  z_1^{(\alpha_1)} + x_2 y_1^* \big] \extk + x_1^* y_2 w_1^{(\alpha_1)} + \imagi  x_1^*x_2 +  y_1^* y_2
\nonumber \\
&\quad + \big[ x_1^*y_3  z_1^{(\alpha_1)} + x_3 y_1^* \big] \extk + x_1^* y_3 w_1^{(\alpha_1)} + \imagi  x_1^*x_3 +  y_1^* y_3
\nonumber \\
&=  \alpha_1 \odot \alpha_2 + \alpha_1 \odot \alpha_3,
\end{align}
showing that the right conjugated product is distributive. Instead, we can't arrive at the same conclusion for the left conjugated multiplication of the sum. The distributivity property for the left conjugated multiplication of the sum is verified if:
\begin{equation}
(\alpha_1 + \alpha_2) \odot \alpha_3 = \alpha_1 \odot \alpha_3 + \alpha_2 \odot \alpha_3.
\end{equation}
The left member of previous expression
\begin{align}
&(\alpha_1 + \alpha_2) \odot \alpha_3 = \big[(x_1 \extk + y_1) + (x_2 \extk + y_2)\big] \odot (x_3 \extk + y_3)
\nonumber \\
&=\big[(x_1  + x_2 )\extk + (y_1 + y_2)\big] \odot (x_3 \extk + y_3)\big]
\nonumber \\
&= \big[(x_1^*  + x_2^* )y_3 z_1^{(\alpha_1 + \alpha_2)}  + x_3(y_1^* + y_2^*)\big]\extk + (x_1^*  + x_2^* )y_3 w_1^{(\alpha_1 + \alpha_2)} 
\nonumber \\
&\quad + \imagi (x_1^*  + x_2^* )x_3 +  (y_1^* + y_2^*)y_3
\end{align}
is different from the right's
\begin{align}
&\alpha_1\odot \alpha_3 + \alpha_2 \odot \alpha_3 = (x_1 \extk + y_1) \odot (x_3 \extk + y_3) + (x_2 \extk + y_2) \odot (x_3 \extk + y_3)
\nonumber \\
&=\big[(x_1^*z_1^{(\alpha_1)}  + x_2^*z_1^{(\alpha_2)} )y_3 + x_3(y_1^* + y_2^*) \big]\extk +  (x_1^*w_1^{(\alpha_1)}  + x_2^*w_1^{(\alpha_2)} )y_3 
\nonumber \\
& \quad + \imagi (x_1^* + x_2^*)x_3 + (y_1^* + y_2^*)y_3.
\end{align}
The distributive property is then satisfied if :
\begin{align}
\mathcal{D}^{(\alpha_1 + \alpha_2)}(\alpha_1,\alpha_2,\alpha_3) &\equiv \big[ (x_1^*  + x_2^* )y_3 z_1^{(\alpha_1 + \alpha_2)} - (x_1^*z_1^{(\alpha_1)}  + x_2^*z_1^{(\alpha_2)} )y_3 \big ] \extk + 
\nonumber \\
& \qquad (x_1^*  + x_2^* )y_3 w_1^{(\alpha_1 + \alpha_2)} - (x_1^*w_1^{(\alpha_1)}  + x_2^*w_1^{(\alpha_2)} )y_3 = 0  \label{extDfuntion}
\end{align}
is zero. The function $\mathcal{D}^{(\alpha_1 + \alpha_2)}(\alpha_1,\alpha_2,\alpha_3)$ measure the magnitude of the differences of between the factors of the maps. Subtracting the explicit product of both members, the distribution law can be expressed as:
\begin{equation}
(\alpha_1 + \alpha_2) \odot \alpha_3 = \alpha_1 \odot \alpha_3 + \alpha_2 \odot \alpha_3 + 
\mathcal{D}^{(\alpha_1 + \alpha_2)}(\alpha_1,\alpha_2,\alpha_3). \label{distPropRight}
\end{equation}

The function $\mathcal{D}^{(\alpha_1 + \alpha_2)}(\alpha_1,\alpha_2,\alpha_3)$ is null if
\begin{itemize}
\item the maps  for $\alpha_1$ and $\alpha_2$ numbers satisfy:
\begin{equation}
z_1^{(\alpha_1)} = z_1^{(\alpha_2)} =  z_1^{(\alpha_1 + \alpha_2)}, \qquad
w_1^{(\alpha_1)} = w_1^{(\alpha_2)} =  w_1^{(\alpha_1 + \alpha_2)} 
\end{equation}
\item $\alpha_1$ and $\alpha_2$  are both pure complex numbers, $e.i.$ $x_1 = x_2 = 0$
\item  $\alpha_3$ is a pure complex numbers,  $e.i.$ $y_3 =0$
\end{itemize}
It can also be verified that
\begin{equation}
\mathcal{D}^{(\alpha_1 + \alpha_2)}(\alpha_1,\alpha_2,\alpha_3) 
+ \mathcal{D}^{(\alpha_1 + \alpha_2)}(\alpha_1,\alpha_2,\alpha_4) 
= \mathcal{D}^{(\alpha_1 + \alpha_2)}(\alpha_1,\alpha_2,\alpha_3 + \alpha_4).\label{distPropRight1}
\end{equation}

\item Existence of multiplicative identity element.

If the conjugate product has an identity element $1_E^{(\oplus)}$, it must satisfy
\begin{equation}
\alpha \odot 1_E^{(\oplus)} = 1_E^{(\oplus)} \odot \alpha = \alpha.
\end{equation}
Just by noting the noncommutative property we can see that such identity element doesn't exist. 
\end{enumerate}

We don't attempt in here to classify the set of the extended numbers, however, according to Abstract Algebra, they behave like a commutative ring with the existence of nontrivial zero divisors. We note that, same as extended number, whose satisfy the Associativity, Commutativity and Distributivity axioms for the standard sum and product, the pair $\alpha \odot \beta$ will also satisfy the same properties. That means, for example, that pairs satisfy the distribution property:
\begin{equation}
(\alpha_1 \odot \beta_1)\big[(\alpha_2 \odot \beta_2) + (\alpha_3 \odot \beta_3)\big] = (\alpha_1 \odot \beta_1)(\alpha_2 \odot \beta_2) + (\alpha_1 \odot \beta_1)(\alpha_3 \odot \beta_3).\label{ExtVectorDistrProp}
\end{equation}

Also, a product like 
\begin{equation}
(\alpha_1^\bullet \odot \beta_1^\bullet) (\alpha_2 \odot \beta_2)
\end{equation}
satisfies the the distributive properties
\begin{equation}
(\alpha_1^\bullet \odot \beta_1^\bullet)\big[(\alpha_2 \odot \beta_2) + (\alpha_3 \odot \beta_3)\big] = (\alpha_1^\bullet \odot \beta_1^\bullet)(\alpha_2 \odot \beta_2) + (\alpha_1^\bullet \odot \beta_1^\bullet)(\alpha_3 \odot \beta_3) \label{ExtProdDistrProp1}
\end{equation}
and
\begin{equation}
\big[(\alpha_1^\bullet \odot \beta_1^\bullet) + (\alpha_2^\bullet \odot \beta_2^\bullet) \big] (\alpha_3 \odot \beta_3) = (\alpha_1^\bullet \odot \beta_1^\bullet)(\alpha_3 \odot \beta_3) + (\alpha_2^\bullet \odot \beta_2^\bullet)(\alpha_3 \odot \beta_3).\label{ExtProdDistrProp2}
\end{equation}
The properties for quantities $(\alpha \odot \beta)$, together with the intuition of the how of the quantum operator should act over a two-component quantum state for $n$-VMVF systems, give us a hint for finding the form of the inner product in the extended domain.

\subsection{The extended inner product.}
We are now in the position to propose the absolute value for an extended number using the new operations.

The inner product should be a set of operations applied on four extended numbers, according the first section of this chapter, and also connected with the two types of products. The most straightforward possible definitions for the inner product for the extended numbers $\alpha , \beta , \gamma , \delta$ are, regardless the order of priority:
\begin{align}
&1.\quad \alpha^\bullet \cdot \beta^\bullet \cdot \gamma \cdot \delta   &&5.\quad \alpha^\bullet \odot \beta^\bullet \cdot \gamma \cdot \delta 
\nonumber \\
&2.\quad \alpha^\bullet \cdot \beta^\bullet \cdot \gamma \odot \delta   &&6.\quad \alpha^\bullet \odot \beta^\bullet \cdot \gamma \odot \delta 
\nonumber \\
&3.\quad \alpha^\bullet \cdot \beta^\bullet \odot \gamma \cdot \delta   &&7.\quad \alpha^\bullet \odot \beta^\bullet \odot \gamma \cdot \delta 
\nonumber \\
&4.\quad \alpha^\bullet \cdot \beta^\bullet \odot \gamma \odot \delta   &&8.\quad \alpha^\bullet \odot \beta^\bullet \odot \gamma \odot \delta, 
\end{align}
where we include the extended conjugate map $()^\bullet$.

Our proposal for the inner product will provide the algebraic structure for defining the new Hilbert space in which the extension of the quantum mechanic should be based on. The extended kets should have a bi-dimensional form, indicate that there should exist some pair-pair symmetry. The cases 1,3,6 and 8 are the only ones who have that kind of symmetry. Also, the superposition principle points out that the states kets must satisfy the Associativity, Commutativity and Distributivity properties. In that case, the product operation that relates the kets should be the standard multiplication. In that case, the only cases remaining is 1 and 6. Our first proposition for the inner product, which results on arithmetical inconsistencies, was a first case's type proposition. Also, from the extended unit definition, we see that for the absolute value of $\extk$, we propose the computation in first place of the quantity $\extk^* \extk$. 

Based on this weak explanation embedded with intuition and the need to adjust the inner product to the expected approach for the quantum theory, we propose the definition of the inner product of the four extended numbers $\alpha , \beta , \gamma , \delta$ as:
\begin{equation}
\langle \alpha , \beta , \gamma , \delta \rangle \equiv (\alpha^\bullet \odot \beta^\bullet) \cdot (\gamma \odot \delta). \label{ExtInnerProdDef}
\end{equation}
The four power of the absolute value for an extended number $\alpha = x \extk + y$ can be then written as:
\begin{align}
|\alpha|^4 &= ( \alpha^\bullet \odot \alpha^\bullet )\cdot (\alpha \odot \alpha )=
\\
&=\big[(x \extk + y)^{\bullet *} (x \extk + y)^\bullet\big] \cdot \big[(x \extk + y)^* (x \extk + y)\big], \;\;\; \forall x,y \in \mathbb{C}.
\end{align}
for  $x=1$ and $y=0$ we rewrite the first equation for the absolute value of the extended unit, which we supposed is equal to $1$:
\begin{equation}
\big(\extk^\bullet \odot \extk^\bullet\big)\big(\extk \odot \extk\big)=1
\end{equation}

It's also useful to check the expression $\extk^\bullet \odot \extk^\bullet$ since it set constraints to the map of an extended number. Indeed, the extended unit definition set this expression equal to $\imagi^*$. In this case, we have:
\begin{align}
\extk^\bullet \odot \extk^\bullet &= (z_2 \extk + w_2) \odot (z_2 \extk + w_2)
\nonumber \\
&= (z_1 z_2^* w_2 + z_2 w_2^*)\extk + \imagi |z_2|^2 + z_2^* w_1 w_2 + |w_2|^2 = \imagi^* \label{extConjDangerous}
\end{align}
which lead to the relations
\begin{align}
z_1 z_2^* w_2 + z_2 w_2^* &= 0
\nonumber \\
\imagi |z_2|^2 + z_2^* w_1 w_2 + |w_2|^2 &= \imagi^* \label{extConjRelations}
\end{align}

Following the rules of the standard and conjugated product we have the expression of the absolute value:
\begin{align}
|\alpha|^4 =& \Big[ |x|^2|y|^2(z_2^*w_1 + w_2^* + z_1 w_2 + z_0 z_1z_2^* + z_0 z_1 z_2 + z_2 w_1) 
+ |x|^2 x^*y(\imagi z_1 z_2^* + \imagi^*z_1)
\nonumber \\
&+ |x|^2 xy^*(\imagi z_2 + \imagi^*) + |y|^2 x^*y(z_1 z_2^* + z_1) + |y|^2 xy^*(z_2 + 1) + (xy^*)^2(z_0 z_2 + w_2)
\nonumber \\
& + (x^*y)(z_0 z_1^2 z_2^* + 2 z_1 z_2^* w_1 + z_1 w_2^*)
\Big] \extk 
+ |x|^4 + |y|^4 + |x|^2|y|^2(z_1 z_2^* w_0  + z_1 z_2 w_0 
\nonumber \\
&+ w_1 w_2 ) + |x|^2 x^*y(\imagi z_2^2 w_1 + \imagi w_2^* + \imagi^* w_1) +  |x|^2 xy^*(\imagi w_2) + |y|^2 xy^*(w_2)
\nonumber \\
&+ |y|^2 x^*y(z_2^* w_1 + w_2^* + w_1)  + (x^*y^2)(z_2^* w_1^2 + w_1 w_2^* + z_1^2 z_2^* w_0 ) 
+ (xy^*)^2(z_2 w_0)\label{extFinalAbsRelations}
\end{align}
where relations \ref{extConjRelations} replace some expressions to easy the final form of the absolute value. Similar to the first proposal, we group addends using the parameters  $\phi = \frac{|x|}{|y|}$ and $\theta = \theta_x - \theta_y$. Using these parameters, the following terms have the form:
\begin{align}
(x^*y)^2   &= |x|^2 |y|^2 e^{- 2\imagi \theta}
\nonumber \\
(xy^*)^2   &= |x|^2 |y|^2 e^{2\imagi \theta}
\nonumber \\
|x|^2 x^*y &= |x|^2 |y|^2 \phi e^{-\imagi \theta}
\nonumber \\
|x|^2 xy^* &= |x|^2 |y|^2 \phi e^{ \imagi \theta}
\nonumber \\
|y|^2 x^*y &= |x|^2 |y|^2 \phi^- e^{-\imagi \theta}
\nonumber \\
|y|^2 xy^* &= |x|^2 |y|^2 \phi^- e^{\imagi \theta} \label{extFinalParam}
\end{align}

The positive-definiteness condition of the absolute value state that the extended part of the extended equation \ref{extFinalAbsRelations} must be zero and the imaginary part must be real and greater than zero. Substituting relations \ref{extFinalParam} into equations \ref{extFinalAbsRelations}, extracting the common factor $|x|^2 |y|^2$ and putting together with relations \ref{extConjRelations} we have the final set of four complex equations for obtaining the four $z_1,z_2,w_1, w_2$ values, which characterize the maps of every extended number:
\begin{align}
&\mathbf{1}. \; z_1 z_2^* w_2 + z_2 w_2^* = 0
\nonumber \\
&\mathbf{2}. \; \imagi |z_2|^2 + z_2^* w_1 w_2 + |w_2|^2 = \imagi^*
\nonumber \\
&\mathbf{3}. \; z_2^*w_1 + w_2^* + z_1 w_2 + z_0 z_1z_2^* + z_0 z_1 z_2 + z_2 w_1 
+ \phi e^{-\imagi \theta}(\imagi z_1 z_2^* + \imagi^*z_1)
\nonumber \\
&\quad + \phi e^{ \imagi \theta}(\imagi z_2 + \imagi^*) + \phi^- e^{-\imagi \theta}(z_1 z_2^* + z_1) + \phi^- e^{\imagi \theta}(z_2 + 1) + e^{ 2\imagi \theta}(z_0 z_2 + w_2)
\nonumber \\
&\quad + e^{- 2\imagi \theta}(z_0 z_1^2 z_2^* + 2 z_1 z_2^* w_1 + z_1 w_2^*) = 0
\nonumber \\
&\mathbf{4}. \; z_1 z_2^* w_0  + z_1 z_2 w_0 + w_1 w_2 + \phi e^{-\imagi \theta}(\imagi z_2^2 w_1 + \imagi w_2^* + \imagi^* w_1) + \phi e^{ \imagi \theta}(\imagi w_2) + \phi^- e^{\imagi \theta}(w_2)
\nonumber \\
&\quad + \phi^- e^{-\imagi \theta}(z_2^* w_1 + w_2^* + w_1)  +  e^{- 2\imagi \theta}(z_2^* w_1^2 + w_1 w_2^* + z_1^2 z_2^* w_0 ) + e^{2\imagi \theta}(z_2 w_0)=R \label{extFinalAbsRelations1}
\end{align}
where $R$ is a real non-negative number. The expression of the four power of the absolute value of extended numbers is then:
\begin{equation}
|\alpha|^4 = ( \alpha^\bullet \odot \alpha^\bullet )\cdot (\alpha \odot \alpha ) = |x|^4 + |y|^4 + R|x|^2|y|^2 \label{extAbsValueDef1}
\end{equation}
whose maps are determined by equations \ref{extFinalAbsRelations1}, which depend on the parameters $\phi$, $\theta$ and $R$. 

We deliberately set the complex map of the number $\alpha^\bullet$ equal to the complex map of the extended number $\alpha$ $e.i$, $z_1^{(\alpha)} = z_1^{(\alpha^\bullet)}$ and $w_1^{(\alpha)} = w_1^{(\alpha^\bullet)}$,  as used in equation \ref{extConjDangerous}. This is not correct since there is no way to know now the complex map of the number $\alpha^\bullet$ without involve another set of equations like \ref{extFinalAbsRelations1} for the number $\alpha^\bullet$, which includes another set of equations for the other complex map for the number $(\alpha^\bullet)^\bullet$. This continuous loop will finally over when we find a complex map which corresponds to the original $\alpha^\bullet$. A priori, we can say no further about this, until a more in-depth analysis is performed, which possibly change the set of equations to determine the map for every extended number correctly. However, we can conduct an initial study of extended number and its maps $()^\bullet, ()^*$ depending on quantities $\phi$, $\theta$, parameter $R$ and on values $z_0, w_0$ which define the form for the standard product. The search of parameters $z_0, w_0$ can help to minimize the numbers of equations \ref{extFinalAbsRelations1}. The parameter $R$ will be determined in next sections using a different approach. We even should consider the possibility of replacing our definition of the extended unit to a more general definition $\extk = \sqrt[*]{z},\;z\in \mathbb{C}$, $z\neq \pm 1$ in order to also reduce the complexity of the equations needed to obtain the maps $\extk^*,\extk^\bullet$.

\subsection{The R-parameter}

The real parameter $R$ can be founded using the isotropic property for linear spaces. The set of extended complex numbers $\mathbb{E}$, i.e., numbers that can be written as $a + ib + \extk c +i \extk d $ where  $a,b,c,d \in \mathbb{R}$ form a four-dimensional vector space over the real's whose length is the absolute value of the number. The isotropic property states that the length of a vector remains invariant under an axis rotation. If we look at the complex vector space, all numbers laying on the same centered circle as, for example, the complex number $z=a + \imagi b$, have the same absolute value, equal to $|z| = \sqrt{a^2 + b^2}$. Some of this numbers laying on the sphere are $z_1 = a + \imagi b$, $z_2 = -a + \imagi b$, $z_3 = a - \imagi b$, $z_4 = -a - \imagi b$, $z_5 = b + \imagi a$, $z_6 = -b + \imagi a$, $z_7 = b - \imagi a$ and $z_8 = -b - \imagi a$. 

The absolute value of extended number $\alpha = a + \imagi b + \extk c + \imagi \extk d$, where $a,b,c,d \in \mathbb{R}$ must, then, remain constant when the axis is rotated, which means that numbers:
\begin{align*}
\alpha_1 = \pm a \pm \imagi b \pm  \extk c \pm  \imagi \extk d \\
\alpha_2 = \pm b \pm \imagi c \pm  \extk d \pm  \imagi \extk a \\
\alpha_3 = \pm c \pm \imagi d \pm  \extk a \pm  \imagi \extk b \\
\alpha_4 = \pm d \pm \imagi a \pm  \extk b \pm  \imagi \extk c 
\end{align*}
should have same absolute value as $\alpha$.

The proposed absolute value for an extended number raised to the fourth power, as shown in equations  \ref{extAbsValueDef1}, is
\begin{equation}
|\alpha|^4  = |x|^4 + |y|^4 + R|x|^2|y|^2 \quad x,y \in  \mathbb{C}\quad \text{and} \quad R \in \mathbb{R}, R\geq 0
\end{equation}
being $x$ and $y$ the extended and imaginary part of the extended number. If we substitute $ x = a \imagi + b $ and $y =c \imagi +d$, being $a,b,c,d \in \mathbb{R}$, we have
\begin{align}
|\alpha|^4 =& (a^2+b^2)^2+ (c^2+d^2)^2 + R(a^2+b^2)(c^2+d^2)
\nonumber \\
=& a^4 + b^4 + c^4 + d^4 + 2a^2b^2  + 2c^2d^2 + R(a^2c^2 + a^2d^2 + b^2c^2 + b^2d^2).
\end{align}
$R=2$ is the only possible value for the absolute value 
\begin{equation}
|\alpha|^4 = a^4 + b^4 + c^4 + d^4 + 2( a^2b^2  + a^2c^2 + a^2d^2 + b^2c^2 + b^2d^2  + c^2d^2),
\end{equation}
remain constant under any permutation of $a,b,c,d$ with any combination of $\pm$ sign.

The final form for the absolute value of an extended number $\alpha = x \extk + y$ is then
\begin{equation}
|\alpha|^4  = |x|^4 + |y|^4 + 2|x|^2|y|^2.
\end{equation}

\subsection{Division between extended numbers}

In the complex numbers domain, the division of two complex numbers can be accomplished by multiplying the numerator and denominator by the complex conjugate of the denominator. To do a similar procedure for the division between extended numbers, we must review first the existences of extraneous and missing solutions when the same factor multiplies both members of an extended equation. The extraneous solution (or spurious solution) emerges from the process of solving the problem while a missing solution is a valid solution that of the original problem, but disappeared along with the solution. In this case, we study the necessity and sufficiency of the equality of an extended equation, before and after the multiplication, in the standard and conjugated way, of both members of the equation by the same factor. 

Let us consider the extended equation
\begin{equation}
\alpha=\beta \quad \forall \;\alpha,\beta \; \in \mathbb{E}. \label{extSameFactorMultEq}
\end{equation}
If we multiply both members of the equation \ref{extSameFactorMultEq} by a third extended factor $\gamma$, will the equality of the new equation still hold? And if it does, would it introduce or eliminate solutions to the original equations? 

Let us analyze different cases:
\begin{itemize}
\item The standard multiplication of both members by an extended number
\begin{equation}
\alpha \gamma \stackrel{?}{=} \beta \gamma \quad \forall \; \alpha,\beta, \gamma \; \in \mathbb{E}
\label{extSameFactorMultEq1}
\end{equation}
The equation $\alpha=\beta$ means that their extended and imaginary parts are equals respectively like $\alpha_E=\beta_E $ and $\alpha_I=\beta_I $. The members of the equation \ref{extSameFactorMultEq1} have the form:
\begin{align}
\alpha \gamma = \big( \alpha_E \gamma_E z_0 + \alpha_E \gamma_I + \alpha_I \gamma_E \big)\extk + \alpha_E \gamma_E w_0 + \alpha_I \gamma_I
\nonumber \\
\beta  \gamma = \big( \beta_E \gamma_E z_0 + \beta_E \gamma_I + \beta_I \gamma_E \big)\extk + \beta_E \gamma_E w_0 + \beta_I \gamma_I,
\end{align}
which led to the equations
\begin{align}
( \alpha_E - \beta_E )( \gamma_E z_0 + \gamma_I ) + ( \alpha_I - \beta_I) \gamma_E = 0
\nonumber \\
( \alpha_E - \beta_E ) \gamma_E w_0 + ( \alpha_I - \beta_I) \gamma_I = 0. \label{extSameFactorMultEq2}
\end{align}
If $\alpha_E=\beta_E $ and $\alpha_I=\beta_I $, the sufficiency of the statement is probed. However, the necessity or the inverted proposition is not true. Indeed, from equations \ref{extSameFactorMultEq2} we cannot extract the initial equality $\alpha=\beta$, and that is because of the appearances of extraneous solutions. Multiplying first complex equation of \ref{extSameFactorMultEq2} by $\gamma_I$, the second by $\gamma_E$ and subtract one equation from another we obtain:
\begin{equation}
( \alpha_E - \beta_E )( \gamma_I^2 + \gamma_E \gamma_I z_0 - \gamma_E^2 w_0)=0,
\end{equation}
which indicate the presence of a new solution related to the case 
\begin{equation}
\gamma_I^2 + \gamma_E \gamma_I z_0 - \gamma_E^2 w_0=0.
\end{equation}
The necessity can be easily probed for some specific cases like $(\forall \; \gamma_E=0, \gamma_I\neq 0)$ or $(\forall \; \gamma_E\neq 0, \gamma_I= 0, w_0 \neq 0 )$.

\item  The left conjugated multiplication of an extended number, 
\begin{equation}
\gamma \odot \alpha \stackrel{?}{=} \gamma \odot \beta, 
\end{equation}
the equation for every member are
\begin{align}
\gamma \odot \alpha = \big( \gamma_E^* \alpha_I z_1^{(\gamma)} + \gamma_I^* \alpha_E  \big)\extk + \gamma_E^* \alpha_I w_1^{(\gamma)} + \imagi \gamma_E^* \alpha_E + \gamma_I^* \alpha_I
\nonumber \\
\gamma \odot  \beta = \big( \gamma_E^* \beta_I z_1^{(\gamma)} + \gamma_I^* \beta_E  \big)\extk + \gamma_E^* \beta_I w_1^{(\gamma)} + \imagi \gamma_E^* \beta_E + \gamma_I^* \beta_I.
\end{align}
Grouping the extended and imaginary part and setting equal to zero, we have
\begin{align}
\gamma_E^*  z_1^{(\gamma)} (\alpha_I - \beta_I)  + \gamma_I^* (\alpha_E - \beta_E) = 0
\nonumber \\
(\gamma_E^* w_1^{(\gamma)} + \gamma_I^*) ( \alpha_I - \beta_I ) + \imagi \gamma_E^* (\alpha_E - \beta_E ) = 0  \label{extSameFactorMultEq3}
\end{align}
which probe the sufficiency of the statement. After multiplying first complex equation by Equation \ref{extSameFactorMultEq3}, the second by $\gamma_I^*$ and subtract one equation from another,led to the equation
\begin{equation}
\big[ (\gamma_I^*)^2 + \gamma_E^* \gamma_I^* w_1^{(\gamma)} - \imagi (\gamma_E^*)^2  z_1^{(\gamma)} \big ] ( \alpha_I - \beta_I ) = 0
\end{equation}
which shows the inclusion of a new solution from the extended number $\gamma$ that satisfy the equation
\begin{equation}
(\gamma_I^*)^2 + \gamma_E^* \gamma_I^* w_1^{(\gamma)} - \imagi (\gamma_E^*)^2  z_1^{(\gamma)} =0
\end{equation}

\item   The right conjugated multiplication of an extended number, 
\begin{equation}
\alpha \odot \gamma \stackrel{?}{=} \beta \odot \gamma , 
\end{equation}
led to equations:
\begin{align}
(\alpha_E^* z_1^{(\alpha)} - \beta_E^*  z_1^{(\beta)})\gamma_I  + ( \alpha_I^* - \beta_I^* )\gamma_E = 0
\nonumber \\
(\alpha_E^* w_1^{(\alpha)} - \beta_E^*  w_1^{(\beta)} +  \alpha_I^* - \beta_I^*)  \gamma_I + \imagi (\alpha_E^* - \beta_E^* )\gamma_E  = 0.  \label{extSameFactorMultEq4}
\end{align}

In this case, the sufficiency of the statement can be verified because if numbers $\alpha,\beta$ are equals then also their maps. However, while the necessity in the previous case of left conjugated multiplication was verified taking in account the presence of extraneous solutions, in this case, we can not verify the necessity due to the impossibility of factorizing the terms with $z_1$ and $w_1$.
\end{itemize}

We can define now the division between extended numbers. Our first proposal is similar to the division between complex numbers which is done by multiplying the numerator and denominator by the extended complex conjugates of the denominator. Follow this reasoning; our initial proposition is first, conjugately left multiply both numerator and denominator by the number like $ (\alpha \odot \alpha)$ and then standard multiply by the expression $\alpha^{\bullet}  \odot \alpha^{\bullet}$ like
\begin{equation}
\frac{\lambda}{\alpha} = \frac{(\alpha \odot \lambda) }{ (\alpha \odot \alpha)} 
= \frac{(\alpha \odot \lambda) \cdot (\alpha^{\bullet}  \odot \alpha^{\bullet} )}{ (\alpha \odot \alpha) \cdot (\alpha^{\bullet}  \odot \alpha^{\bullet} )}
 = \frac{(\alpha \odot \lambda)\cdot (\alpha^{\bullet}  \odot \alpha^{\bullet} )}{|\alpha|^4}.
\end{equation}
This proposal is incorrect because we cannot obtain $\alpha$ with an inverse process from the number resulting from the division. That means that we cannot get the former number by any multiplication:
\begin{equation}
\frac{\lambda}{\alpha} \odot \alpha = \alpha \odot \frac{(\alpha \odot \lambda)\cdot (\alpha^{\bullet}  \odot \alpha^{\bullet} )}{|\alpha|^4} \neq \lambda 
\qquad \text{ or } \qquad 
\frac{\lambda}{\alpha} \cdot \alpha = \alpha \cdot \frac{(\alpha \odot \lambda)\cdot (\alpha^{\bullet}  \odot \alpha^{\bullet} )}{|\alpha|^4} \neq \lambda
\end{equation}
In that case, we propose the standard multiplication of both numerator and denominator by the expression $(\sqrt[*]{\alpha})^\bullet \odot (\sqrt[*]{\alpha})^\bullet$ where we remember that the conjugated root is the inverse operation of the conjugated product. Indeed, any extended number can be expressed as the conjugated product of its conjugated root like $\alpha = \sqrt[*]{\alpha} \odot \sqrt[*]{\alpha}$. The new proposal for the division between extended numbers is then:
\begin{equation}
\frac{\lambda}{\alpha} 
= \frac{ \lambda}{ (\sqrt[*]{\alpha} \odot \sqrt[*]{\alpha})} 
= \frac{\lambda \cdot  [(\sqrt[*]{\alpha})^\bullet \odot (\sqrt[*]{\alpha})^\bullet]}
	{ { [\sqrt[*]{\alpha} \odot \sqrt[*]{\alpha}]} \cdot { [(\sqrt[*]{\alpha})^\bullet \odot (\sqrt[*]		{\alpha})^\bullet]}} 
= \frac{\lambda \cdot  [(\sqrt[*]{\alpha})^\bullet \odot (\sqrt[*]{\alpha})^\bullet]}{|\sqrt[*]{\alpha}|^4}.
\end{equation}
Now we can verify that:
\begin{equation}
\frac{\lambda}{\alpha} \cdot \alpha =\alpha \cdot \frac{\lambda \cdot  [(\sqrt[*]{\alpha})^\bullet \odot (\sqrt[*]{\alpha})^\bullet]}{|\sqrt[*]{\alpha}|^4} 
=  \frac{\lambda (\sqrt[*]{\alpha} \odot \sqrt[*]{\alpha}) \cdot  [(\sqrt[*]{\alpha})^\bullet \odot (\sqrt[*]{\alpha})^\bullet]}{|\sqrt[*]{\alpha}|^4} 
= \frac{\lambda |\sqrt[*]{\alpha}|^4}{|\sqrt[*]{\alpha}|^4} = \lambda
\end{equation}

For $\lambda=1$ we can probe the existence of multiplicative inverse $\alpha^-$ as one of the properties that the extended numbers satisfy. The multiplicative inverse have the form
\begin{equation}
\alpha^- =\frac{1}{\alpha} = \frac{ 1}{ (\sqrt[*]{\alpha} \odot \sqrt[*]{\alpha})} 
= \frac{[(\sqrt[*]{\alpha})^\bullet \odot (\sqrt[*]{\alpha})^\bullet]}
	{ { [\sqrt[*]{\alpha} \odot \sqrt[*]{\alpha}]} \cdot { [(\sqrt[*]{\alpha})^\bullet \odot (\sqrt[*]		{\alpha})^\bullet]}} 
= \frac{[(\sqrt[*]{\alpha})^\bullet \odot (\sqrt[*]{\alpha})^\bullet]}{|\sqrt[*]{\alpha}|^4}
\end{equation}

There is another possibility for the division between extended numbers since we can also define the extended conjugated root $\sqrt[\bullet]{\alpha}$ as the inverse operation of the conjugated product of the $()^\bullet$-map like
\begin{equation}
\alpha = (\sqrt[\bullet]{\alpha})^\bullet \odot (\sqrt[\bullet]{\alpha})^\bullet.
\end{equation}
In this case, we can propose other division like
\begin{equation}
\frac{\lambda}{\alpha} 
= \frac{ \lambda}{ (\sqrt[\bullet]{\alpha})^\bullet \odot (\sqrt[\bullet]{\alpha})^\bullet} 
= \frac{\lambda \cdot  [\sqrt[\bullet]{\alpha} \odot \sqrt[\bullet]{\alpha}]}
	{ [\sqrt[\bullet]{\alpha} \odot \sqrt[\bullet]{\alpha}] \cdot { [(\sqrt[\bullet]{\alpha})^\bullet \odot (\sqrt[\bullet]		{\alpha})^\bullet]}} 
= \frac{\lambda \cdot  [\sqrt[\bullet]{\alpha} \odot \sqrt[\bullet]{\alpha}]}{|\sqrt[\bullet]{\alpha}|^4}
\end{equation}
and the inverse like
\begin{equation}
\alpha^- =\frac{[\sqrt[\bullet]{\alpha} \odot \sqrt[\bullet]{\alpha}]}{|\sqrt[\bullet]{\alpha}|^4}
\end{equation}
Rest now to interpret the double existence of two division operation and with it, the existence of two inverse for every extended number.

\subsection{Extended domain properties. Absolute value.} \label{extPropSection}

The dependency of the maps $\extk^2$, $\extk^*$ and $\extk^\bullet$ on parameters $\phi$, $\theta$ and $R$, provide the extended domain with some properties:
\begin{itemize}
\item 
The absolute value raised to the fourth power of an extended number is:
\begin{equation}
\langle \alpha  , \alpha, \alpha , \alpha \rangle  = |x|^4 + |y|^4 + 2|x|^2|y|^2 \qquad x,y \in \mathbb{C},
\end{equation}
and the absolute value is defined as:
\begin{equation}
|\alpha| = \sqrt[4]{|x|^4 + |y|^4 + 2|x|^2|y|^2 }.
\end{equation}

\item 
The absolute value of an extended number is zero only if the extended and the imaginary part of the number are also zero.
\item 
It is straightforward to probe the property of the extended numbers that:
\begin{align*}
(c\alpha)^* = c^*\alpha^* \qquad \text{ and } \qquad (c\alpha)^\bullet = c\alpha^\bullet 
\qquad \forall \; c \in \mathbb{C}, \alpha \in \mathbb{E}.
\end{align*}

\item 
The relations for obtaining the extended and conjugated maps for pure extended and pure complex numbers result in undetermined values for numbers $z_1,z_2,w_1$ and $w_2$. This is because of the dependency of equations \ref{kMappEqSyst1} on the quantities $\phi$ and $\phi^-$, whose values turns zero or infinity for $|x|$ and/or $|y|$ being zero. However, both cases are trivial, since taking $\alpha = x \extk + y$, for $y=0$ we have
\begin{equation}
x\extk \odot x\extk=|x|^2i \qquad \text{and} \qquad [(x\extk)^\bullet \odot (x\extk)^\bullet][(x\extk) \odot (x\extk)] = |x|^4,
\end{equation}
while for $x=0$, we have the typical operations between complex numbers .

\item 
The extended numbers $\alpha \equiv x \extk + y$ and $c\alpha = c(x \extk + y)\;\; c \in \mathbb{C}$ have equals values of $\theta$ and $\phi$. From their definitions we have:  
\begin{equation}
\phi_{c\alpha} = \frac{|cx|}{|cy|} \frac{|c||x|}{|c||y|} = \frac{|x|}{|y|}\equiv \phi_{\alpha} \qquad \theta_{c\alpha} = (\theta_x+\theta_c) - (\theta_y + \theta_c) =  \theta_x - \theta_y \equiv \theta_{\alpha}. \label{extThetaPhiInvariant}
\end{equation}

\item 
From the expression of the absolute value in Eq. \ref{extAbsValueDef1} and the invariant character of $\theta$ and $\phi$ for extended numbers number $\alpha$ and $c\alpha$ where $c \in \mathbb{C}$, as shown in Eq. \ref{extThetaPhiInvariant}, the absolute value of $c\alpha$ have the expression:
\begin{align}
|c\alpha| =& \sqrt[4]{[(c\alpha^{\bullet})\odot( c\alpha^{\bullet} )] \cdot [(c\alpha \odot c\alpha)]} 
\nonumber \\
 =&\sqrt[4]{c^*c\;c^*c[(\alpha^{\bullet}  \odot \alpha^{\bullet} \cdot (\alpha \odot \alpha)]}
\nonumber \\
=& \sqrt[4]{|c|^4[(\alpha^{\bullet}  \odot \alpha^{\bullet} \cdot (\alpha \odot \alpha)]} 
\nonumber \\
=& |c||\alpha|
\end{align}

\item 
In general, we have 
\begin{align}
(\alpha+\beta)^\bullet \neq \alpha^\bullet + \beta^\bullet
\nonumber \\
(\alpha+\beta)^* \neq \alpha^* + \beta^* \label{extDistMappProp}
\end{align}
and
\begin{align}
(\alpha \beta)^\bullet &\neq \alpha^\bullet \beta^\bullet
\nonumber \\
(\alpha \beta)^* &\neq \alpha^* \beta^*.
\end{align}
However, we can find the conditions for the members of these equations are equal. Expanding the numbers as $\alpha = \alpha_E \extk + \alpha_I$  and $\beta = \beta_E \extk + \beta_I$ and substituting in every member of the equations, we obtain for each case, the equations for the condition being satisfied.
\begin{description}
\item $(\alpha \beta)^\bullet \stackrel{?}{=} \alpha^\bullet \beta^\bullet$\\
Both members of the equation have the form
\begin{align}
&[(\alpha_E \extk + \alpha_I)(\beta_E \extk + \beta_I)]^\bullet = \big[ 
(\alpha_E \beta_E z_0 + \alpha_E \beta_I + \alpha_I  \beta_E)z_2^{(\alpha \beta)} \big] \extk 
\nonumber \\
& \qquad + (\alpha_E \beta_E z_0 + \alpha_E \beta_I + \alpha_I  \beta_E)w_2^{(\alpha \beta)} + \alpha_E \beta_ E w_0 +  \alpha_I \beta_I
\nonumber \\
&(\alpha_E \extk + \alpha_I)^\bullet(\beta_E \extk + \beta_I)^\bullet = \big[ 
\alpha_E z_2^{(\alpha)} \beta_E z_2^{(\beta)}z_0  + \alpha_E z_2^{(\alpha)} \beta_E w_2^{(\beta)} +  \alpha_E z_2^{(\alpha)} \beta_I 
\nonumber \\
& \qquad + \alpha_E w_2^{(\alpha)} \beta_E z_2^{(\beta)} +  \alpha_I \beta_E z_2^{(\beta)}
\big] \extk
+ \alpha_E z_2^{(\alpha)} \beta_E z_2^{(\beta)}w_0 + \alpha_E w_2^{(\alpha)} \beta_E w_2^{(\beta)} 
\nonumber \\
& \qquad + \alpha_E w_2^{(\alpha)} \beta_I +  \alpha_I \beta_E w_2^{(\beta)} + \alpha_I \beta_I. \label{extPropLinearity0}
\end{align}
Setting the extended and imaginary parts equals, we obtain that the relation $(\alpha \beta)^\bullet = \alpha^\bullet \beta^\bullet$ holds if:
\begin{align}
&(\alpha_E \beta_E z_0 + \alpha_E \beta_I + \alpha_I  \beta_E)z_2^{(\alpha \beta)} = 
\alpha_E z_2^{(\alpha)} \beta_E z_2^{(\beta)}z_0  + \alpha_E z_2^{(\alpha)} \beta_E w_2^{(\beta)} +  \alpha_E z_2^{(\alpha)} \beta_I 
\nonumber \\
& \qquad + \alpha_E w_2^{(\alpha)} \beta_E z_2^{(\beta)} +  \alpha_I \beta_E z_2^{(\beta)}
\nonumber \\
&(\alpha_E \beta_E z_0 + \alpha_E \beta_I + \alpha_I  \beta_E)w_2^{(\alpha \beta)} + \alpha_E \beta_ E w_0 = 
+ \alpha_E z_2^{(\alpha)} \beta_E z_2^{(\beta)}w_0 + \alpha_E w_2^{(\alpha)} \beta_E w_2^{(\beta)} 
\nonumber \\
& \qquad + \alpha_E w_2^{(\alpha)} \beta_I +  \alpha_I \beta_E w_2^{(\beta)}
\label{extPropLinearity1}
\end{align}
\item $(\alpha \beta)^* \stackrel{?}{=} \alpha^* \beta^*$\\
The relations needed for the equation holds, are obtained in the same way as before with some few changes:
\begin{align}
&(\alpha_E \beta_E z_0 + \alpha_E \beta_I + \alpha_I  \beta_E)^*z_1^{(\alpha \beta)} = 
\alpha_E^* z_1^{(\alpha)} \beta_E^* z_1^{(\beta)}z_0  + \alpha_E^* z_1^{(\alpha)} \beta_E^* w_1^{(\beta)} +  \alpha_E^* z_1^{(\alpha)} \beta_I^* 
\nonumber \\
& \qquad + \alpha_E w_1^{(\alpha)} \beta_E z_1^{(\beta)} +  \alpha_I \beta_E z_1^{(\beta)}
\nonumber \\
&(\alpha_E \beta_E z_0 + \alpha_E \beta_I + \alpha_I  \beta_E)^*w_1^{(\alpha \beta)} + \alpha_E^* \beta_E^* w_0^* = 
+ \alpha_E^* z_1^{(\alpha)} \beta_E^* z_1^{(\beta)}w_0^* + \alpha_E^* w_1^{(\alpha)} \beta_E^* w_1^{(\beta)} 
\nonumber \\
& \qquad + \alpha_E^* w_1^{(\alpha)} \beta_I^* + \alpha_I^* \beta_E^* w_1^{(\beta)}
\label{extPropLinearity2}
\end{align}

\end{description}
The above relations shows the difference between the conjugated map of the product of two extended numbers and the product of the conjugated maps. It depends on the same magnitude than the coefficients $z_i$ and $w_i$ of the product of the factors is different, for example $z_1^{(\alpha \beta)}$,  to the same quantities of the product of the factors, $z_1^{(\alpha)}$ and $z_1^{(\beta)}$ and also on the difference between these last factors.
\end{itemize}

\subsection{Linearity}

Linearity is another property that the inner product must satisfied since it is the base property of the superposition principle. From the algebraic properties of the extended numbers, specifically those related to the standard product, we found that the complex products $\alpha \odot \beta$ satisfy the distribution and associative property like shown on equation \ref{ExtVectorDistrProp}:
\begin{equation*}
(\alpha_1 \odot \beta_1)\big[(\alpha_2 \odot \beta_2) + (\alpha_3 \odot \beta_3)\big] = (\alpha_1 \odot \beta_1)(\alpha_2 \odot \beta_2) + (\alpha_1 \odot \beta_1)(\alpha_3 \odot \beta_3).
\end{equation*}
and products like Eq.$(\alpha_1^\bullet \odot \beta_1^\bullet) (\alpha_2 \odot \beta_2)$
satisfies the distributive properties of equation \ref{ExtProdDistrProp1}
\begin{equation*}
(\alpha_1^\bullet \odot \beta_1^\bullet)\big[(\alpha_2 \odot \beta_2) + (\alpha_3 \odot \beta_3)\big] = (\alpha_1^\bullet \odot \beta_1^\bullet)(\alpha_2 \odot \beta_2) + (\alpha_1^\bullet \odot \beta_1^\bullet)(\alpha_3 \odot \beta_3)
\end{equation*}

Even when the two-dimensional vector satisfies the property of Linearity, we also study what conditions the addends of the factors of the inner product should satisfy for the product remain linear. 

For simplicity, we express the inner product of the equation \ref{ExtInnerProdDef} in the form of a two-row matrix
\begin{equation}
\extproduct{\alpha}{\beta}{\gamma}{\delta}
\equiv (\alpha^{\bullet}  \odot \beta^{\bullet}) \cdot (\gamma \odot \delta).
\end{equation}

The Linearity property of the inner factors imply that:
\begin{align}
\extproduct{\alpha_1 + \alpha_2}{\beta_1 + \beta_2}{\gamma}{\delta}  &=
\extproduct{\alpha_1}{\beta_1}{\gamma}{\delta} +  \extproduct{\alpha_2}{\beta_2}{\gamma}{\delta} 
\nonumber \\
\extproduct{\alpha}{\beta}{\gamma_1 + \gamma_2}{\delta_1 + \delta_2}  &=
\extproduct{\alpha}{\beta}{\gamma_1}{\delta_1} + \extproduct{\alpha}{\beta}{\gamma_2}{\delta_2}
 \label{extLinearitySum}
\end{align}
and also
\begin{align}
\extproduct{\alpha_1 \cdot \alpha_2}{\beta_1 \cdot \beta_2}{\gamma}{\delta}  &=
(\alpha_1^\bullet \odot \beta_1^\bullet)  \extproduct{\alpha_2}{\beta_2}{\gamma}{\delta} 
= (\alpha_2^\bullet \odot \beta_2^\bullet)  \extproduct{\alpha_1}{\beta_1}{\gamma}{\delta}. 
\nonumber \\
\extproduct{\alpha}{\beta}{\gamma_1 \cdot \gamma_2}{\delta_1 \cdot \delta_2}  &=
(\gamma_1 \odot \delta_1 ) \; \extproduct{\alpha}{\beta}{\gamma_2}{\delta_2}
 = (\gamma_2 \odot \delta_2 ) \; \extproduct{\alpha}{\beta}{\gamma_1}{\delta_1}
\label{extLinearityEscalar}
\end{align}

The relations needed for the linearity axiom of equations \ref{extLinearitySum} being satisfied are extracted using the algebraic properties of the pairs in the inner product for the standard sum and multiplication operations. Both equations are satisfied then if 
\begin{align}
(\alpha_1 + \alpha_2)^\bullet \odot (\beta_1 + \beta_2)^\bullet = \alpha_1^\bullet \odot \beta_1^\bullet  + \alpha_2^\bullet \odot \beta_2^\bullet. \label{extLinearitySumCond1}
\end{align}
and
\begin{equation}
(\gamma_1 + \gamma_2) \odot (\delta_1 + \delta_2) = \gamma_1 \odot \delta_1 + \gamma_2 \odot \delta_2
\label{extLinearitySumCond2}
\end{equation}
respectively.

Both extended conditions can be expanded into two pure complex conditions by setting equals the extended and the imaginary part of the explicit multiplication of both members of each equation. We can also use the properties of the distribution law described before on equation \ref{distPropRight} and \ref{distPropRight1}. Let's use those equations in the last case \ref{extLinearitySumCond2}. For the previous linearity condition, we have
\begin{align}
(\gamma_1 + \gamma_2) \odot (\delta_1 + \delta_2) &= 
\gamma_1 \odot \delta_1 + \gamma_2 \odot \delta_2 
+ \gamma_2 \odot \delta_1  + \gamma_1 \odot \delta_2
\nonumber \\
& \quad + \mathcal{D}^{(\gamma_1 + \gamma_2)}(\gamma_1,\gamma_2,\delta_1 + \delta_2).
\label{extLinearitySumExpress}
\end{align}
The linearity axiom is then satisfied if
\begin{equation}
\gamma_2 \odot \delta_1  + \gamma_1 \odot \delta_2
+ \mathcal{D}^{(\gamma_1 + \gamma_2)}(\gamma_1,\gamma_2,\delta_1 + \delta_2) =0 
\label{extLinearitySumCond}
\end{equation}

The equations \ref{extLinearityEscalar} hold if
\begin{equation}
(\alpha_1 \cdot \alpha_2)^\bullet \odot (\beta_1 \cdot \beta_2 )^\bullet \stackrel{?}{=}
(\alpha_1^\bullet \odot \beta_1^\bullet ) (\alpha_2^\bullet \odot \beta_2^\bullet)
\end{equation}
and 
\begin{equation}
({\gamma_1 \cdot \gamma_2})\odot ({\delta_1 \cdot \delta_2}) \stackrel{?}{=} (\gamma_1 \odot \delta_1 ) (\gamma_2 \odot \delta_2)
\end{equation}
respectively. By straightforward multiplication and setting the extended and imaginary parts of the result equals, we get to the conditions for first relations being true
\begin{align}
\Phi_E^* \Psi_I z_1^{(\Phi)} + \Phi_I^* \Psi_E = \Theta_E \Gamma_E z_0 + \Theta_E \Gamma_I + \Theta_I \Gamma_E
\nonumber \\
\imagi \Phi_E^* \Psi_E + \Phi_E^* \Psi_I w_1^{(\Phi)} + \Phi_I^* \Psi_I =  \Theta_E \Gamma_E w_0 + \Theta_I \Gamma_I \label{extLinearityEscalarCond}
\end{align}
where 
\begin{equation}
\Phi = (\alpha_1 \cdot \alpha_2)^\bullet, \qquad \Psi = (\beta_1 \cdot \beta_2 )^\bullet, \qquad \Theta =\alpha_1^\bullet \odot \beta_1^\bullet, \qquad \Gamma= \alpha_2^\bullet \odot \beta_2^\bullet.
\label{extLinearityEscalarCond1}
\end{equation}

The equations of \ref{extLinearityEscalarCond} are also the conditions for the second relation of \ref{extLinearityEscalar} being correct with the few changes in the functions $\Phi,Psi,\Theta$ and $\Gamma$ like:
\begin{equation}
\Phi = \gamma_1 \cdot \gamma_2, \qquad \Psi = \delta_1 \cdot \delta_2, \qquad \Theta = \gamma_1 \odot \delta_1, \qquad \Gamma= \gamma_2 \odot \delta_2.\label{extLinearityEscalarCond2}
\end{equation}

We can, same as before, express the equation 
\begin{equation}
(\alpha_1 \cdot \alpha_2)^\bullet \odot (\beta_1 \cdot \beta_2 )^\bullet =
(\alpha_1^\bullet \odot \beta_1^\bullet ) (\alpha_2^\bullet \odot \beta_2^\bullet)
+\mathcal{F}(\alpha_1,\alpha_2,\beta_1, \beta_2)
\end{equation}
where 
\begin{align}
\mathcal{F}_E(\alpha_1,\alpha_2,\beta_1, \beta_2) = \Phi_E^* \Psi_I z_1^{(\Phi)} + \Phi_I^* \Psi_E - \Theta_E \Gamma_E z_0 - \Theta_E \Gamma_I - \Theta_I \Gamma_E
\nonumber \\
\mathcal{F}_I(\alpha_1,\alpha_2,\beta_1, \beta_2) = \imagi \Phi_E^* \Psi_E + \Phi_E^* \Psi_I w_1^{(\Phi)} + \Phi_I^* \Psi_I -  \Theta_E \Gamma_E w_0 - \Theta_I \Gamma_I 
\label{extFfuntion}
\end{align}
The form of the function $\mathcal{F}$ depending on numbers $\alpha_1,\alpha_2,\beta_1$ and $\beta_2$ is quite extensive and is obtained by the explicit substitution of functions $\Phi,Psi,\Theta$ and $\Gamma$.

Also, because of the properties of extended complex numbers from section \ref{extPropSection}: 
\begin{align*}
(c\alpha)^* = c^*\alpha^* \qquad \text{ and } \qquad (c\alpha)^\bullet = c\alpha^\bullet 
\qquad \forall \; c \in \mathbb{C}, \alpha \in \mathbb{E}.
\end{align*}
we have that
\begin{equation}
(c \cdot \alpha)^\bullet \odot (d \cdot \beta )^\bullet =
(c\odot d ) (\alpha^\bullet \odot \beta^\bullet) \qquad \forall  \; c,d \in \mathbb{C}, \alpha,\beta \in \mathbb{E}.
\end{equation}

We see that on the extended domain we can define two different cases for the Linearity property for inner products. One is applied to the pairs of the complex multiplication of the extended product, which can be related to a physical entity that satisfies the superposition principle. On the other side, the inner factors of an inner product, as complex numbers which satisfy the linearity properties for the standard sum and multiplication, are also related to a new type of linearity which is constrained to the conditions shown in this section.

\subsection{Conjugate symmetry} \label{ConjugateSymmetry}

The inner product should satisfy the positive definiteness axiom and the linearity property for the bidimensional vectors over the standard summation and multiplication. Also, a vector whose upper or downer components is a sum or product of two extended numbers can be divided into two vectors or factorized as a vector and a scalar under certain conditions. The inner product space definition is completed with a third axiom known as the conjugate symmetry property. 

In the space of complex vectors, the property of the conjugate symmetry has the form:
\begin{equation}
\langle u,v \rangle = \overline{\langle v,u \rangle} \qquad \forall \; u,v \in \mathbb{C}
\end{equation}
where $\overline{()}$ is referred to the map $J: V \to V^*$. For real inner product space, the conjugate symmetry is like
\begin{equation}
\langle u,v \rangle = \langle u,v \rangle \qquad \forall \; u,v \in \mathbb{R}
\end{equation}
where the map is the identity $J: V \to V$.

Let us see, from our point of view, how can be obtained the form of the map on the complex space. Applying  the linearity axioms we have
\begin{equation}
\langle u,v \rangle = u^*v \langle 1,1 \rangle \qquad \text{ and } 
\qquad \langle v,u \rangle = v^*u \langle 1,1 \rangle.
\end{equation}
Let us also presume the existence of an, until now, not defined map $J: V \to V$, whose action over the number $u^*v$ return the former quantity like
\begin{equation}
v^*u  = J(u^*v). \label{comInnPrdMap}
\end{equation}
Taking into account that $\langle 1,1 \rangle = 1$ and replacing the first equation into the second, we have the equation, 
\begin{align}
\langle v,u \rangle &=  v^*u \langle 1,1 \rangle = J(u^*v) 
\nonumber \\
&= J(\langle u,v \rangle).
\end{align}
From the properties of the product between complex numbers, we know that the map $J: V \to V$ on equation \ref{comInnPrdMap} is referred to the complex conjugation operator $()^*$. 

The extended inner product must be symmetric to its extended conjugated, and its conjugated map should be also defined. We apply the later reasoning to the extended complex numbers. From the linearity axioms we studied before:
\begin{align}
\extproduct{\alpha}{\beta}{\gamma}{\delta} &= 
(\alpha^{\bullet}  \odot \beta^{\bullet}) \cdot (\gamma \odot \delta)
\extproduct{1}{1}{1}{1} 
\nonumber \\ 
\extproduct{\gamma}{\delta}{\alpha}{\beta} &= 
(\gamma^{\bullet} \odot \delta^{\bullet}) \cdot (\alpha  \odot \beta) 
\extproduct{1}{1}{1}{1}.
\end{align} 
We need to find a map that
\begin{equation}
\extproduct{\alpha}{\beta}{\gamma}{\delta} = J \left( \extproduct{\gamma}{\delta}{\alpha}{\beta} \right)
\end{equation}
In this case we will have the property
\begin{equation}
(\gamma^{\bullet} \odot \delta^{\bullet}) \cdot (\alpha  \odot \beta)  = 
J\big ( 
(\alpha^{\bullet}  \odot \beta^{\bullet}) \cdot (\gamma \odot \delta) 
\big ). 
\label{extInnPrdMap}
\end{equation}

We can represent the map $J \equiv ()^\cbullet$ so: 
\begin{equation}
\big [ (\alpha  \odot \beta) \cdot (\gamma^{\bullet} \odot \delta^{\bullet}) \big ]^\cbullet = 
(\alpha^{\bullet}  \odot \beta^{\bullet}) \cdot (\gamma \odot \delta).
\end{equation}

We suppose that any extended number can be expressed as a standard product of two complex products like
\begin{equation}
\lambda = (\alpha^{\bullet}  \odot \beta^{\bullet}) \cdot (\gamma \odot \delta) 
\end{equation}
where $\alpha, \beta, \gamma, \delta$ are extended numbers. The extended conjugate operator of an extended number is obtained then, by just setting the $()^\bullet$ map on the factor which do not have it and suppress it to the ones who do have it, like
\begin{equation}
\lambda^\cbullet = (\alpha  \odot \beta) \cdot (\gamma^\bullet \odot \delta^\bullet). 
\end{equation}
Note also that 
\begin{equation}
(\lambda^\cbullet)^\cbullet = \lambda
\end{equation}

We call then, the conjugate-symmetry of the extended inner product space by demanding the relation:
\begin{equation}
\extproduct{\alpha}{\beta}{\gamma}{\delta} = \extproduct{\gamma}{\delta}{\alpha}{\beta}^\cbullet. \label{conjSymmEq}
\end{equation}
We can redefine the inner product like:
\begin{equation}
\extproduct{\alpha}{\beta}{\gamma}{\delta} \equiv 
(\alpha^{\bullet}  \odot \beta^{\bullet}) \cdot (\gamma \odot \delta) \delta_D \left( 
\extproduct{\alpha}{\beta}{\gamma}{\delta} - \extproduct{\gamma}{\delta}{\alpha}{\beta}^\cbullet\right)
\end{equation}
where $\delta_D$ is the Delta Dirac function.

One particular expression for any extended number can be written using the roots operator as the inverse operation for the defined products between extended numbers. We can have three roots defined:
\begin{itemize}
\item 
the standard square root $\sqrt{\alpha}$ too as the inverse operation for the standard product like
\begin{equation}
\alpha = \sqrt{\alpha} \cdot \sqrt{\alpha} = (\sqrt{\alpha})^2.
\end{equation}
\item 
the complex conjugated root $\sqrt[*]{\alpha}$ as the inverse operation of the conjugated product
\begin{equation}
\alpha = \sqrt[*]{\alpha} \odot \sqrt[*]{\alpha}.
\end{equation}
\item we can define the extended conjugated root $\sqrt[\bullet]{\alpha}$ as the inverse operation of the conjugated product of the $()^\bullet$-map like
\begin{equation}
\alpha = (\sqrt[\bullet]{\alpha})^\bullet \odot (\sqrt[\bullet]{\alpha})^\bullet.
\end{equation}
\end{itemize} 
Note that $\sqrt[\bullet]{\alpha}$ is not an new type of inverse operation for the complex product since it can be written as function of the complex conjugated root. We can express any number as
\begin{equation}
\alpha = \sqrt{\alpha} \cdot \sqrt{\alpha} \label{extNumbRootExpr0}
\end{equation}
and express both factors as a complex product of the complex and extended conjugated and roots respectively
as
\begin{equation}
\sqrt{\alpha} = \left(\sqrt[\bullet]{\sqrt{\alpha}}\right)^\bullet \odot \left(\sqrt[\bullet]{\sqrt{\alpha}} \right)^\bullet \label{extNumbRootExpr1}
\end{equation}
and also
\begin{equation}
\sqrt{\alpha} = \left( \sqrt[*]{\sqrt{\alpha}} \right) \odot \left( \sqrt[*]{\sqrt{\alpha}} \right).
\label{extNumbRootExpr2}
\end{equation}

Putting all together, an extended number can be expressed then as
\begin{equation}
\alpha = \left [ \left(\sqrt[\bullet]{\sqrt{\alpha}}\right)^\bullet \odot \left(\sqrt[\bullet]{\sqrt{\alpha}} \right)^\bullet \right]
\cdot
\left[ \left( \sqrt[*]{\sqrt{\alpha}} \right) \odot \left( \sqrt[*]{\sqrt{\alpha}} \right) \right]
\label{extNumbConjSymm0}
\end{equation}
Until a complete study of the extended numbers is performed, we can't ensure that all numbers can be expressed like that. Assuming that, the extended conjugate operator of an extended number expressed on its roots is obtained just putting the $()^\bullet$ map on the factor who don't have it and suppress it to the ones who do have it like
\begin{equation}
\alpha^\cbullet = \left[ \left( \sqrt[\bullet]{\sqrt{\alpha}} \right) \odot \left( \sqrt[\bullet]{\sqrt{\alpha}} \right) \right]
\cdot
\left [ \left(\sqrt[*]{\sqrt{\alpha}}\right)^\bullet \odot \left(\sqrt[*]{\sqrt{\alpha}} \right)^\bullet \right]
\label{extNumbConjSymm}
\end{equation}

We can use the previous expression to study the conjugate property for symmetric extended inner products
\begin{equation}
\alpha = \extproduct{\mu}{\nu}{\mu}{\nu} \label{conjSymmProd}.
\end{equation}
If we express this inner product using its extended roots, it can be expressed in the form of equation \ref{extNumbConjSymm0} like
\begin{equation}
\alpha = (\gamma^{\bullet}  \odot \gamma^{\bullet}) \cdot (\delta \odot \delta) 
\end{equation}
where 
\begin{equation}
\gamma = \sqrt[\bullet]{\sqrt{\alpha}} \qquad \text{ and } \qquad \delta = \sqrt[*]{\sqrt{\alpha}}.
\label{extRootsDef}
\end{equation}
The conjugate symmetry imply then
\begin{equation}
\alpha  = \alpha^\cbullet \qquad \text{or} \qquad (\gamma^{\bullet}  \odot \gamma^{\bullet}) \cdot (\delta \odot \delta) = (\gamma  \odot \gamma) \cdot (\delta^{\bullet} \odot \delta^{\bullet}).
\label{conjSymmProd1}
\end{equation}
According to the standard product properties, the equations \ref{extNumbRootExpr0},\ref{extNumbRootExpr1} and \ref{extNumbRootExpr2}, also its satisfied the relation
\begin{equation}
(\gamma^{\bullet}  \odot \gamma^{\bullet}) =\pm (\delta \odot \delta). \label{conjSymmProd2}
\end{equation}
For initials calculations, we use the $+$ sign.

We have then two independent extended equations involving two extended variables $\gamma, \delta$ or four complex equations depending on four complex variables replacing the equation \ref{conjSymmProd1}. The expression of the extended and imaginary parts of both members of equation \ref{conjSymmProd1} are extensive, and we expose them on the appendix \ref{ApxConjSymm}. We need to find the conditions for equation \ref{conjSymmProd1} always hold. One of this case is the trivial solution $\gamma_E = \delta_E = 0$, which has no use for us. Other is $\gamma_E = \delta_E = 0$ which reduces equations \ref{conjSymmProd1} to
\begin{equation}
\gamma^*_{I} \gamma_{I} \delta_{I}^* \delta_{I} = \gamma^*_{I} \gamma_{I} \delta_{I}^* \delta_{I}
\end{equation}
and 
\begin{equation}
\gamma^*_{I} \gamma_{I}  =\delta_{I}^* \delta_{I}  
\end{equation}
respectively. The first equation is an absolute truth and the other set a new condition for all the extended numbers satisfy the complex conjugated symmetry. 

From the expansion of the extended equations, it can be seen that the result are real and nonnegative like
\begin{equation}
\extproduct{\alpha}{\beta}{\alpha}{\beta} = \gamma^*_{I} \gamma_{I} \delta_{I}^* \delta_{I}.
\end{equation}

The previous analysis shows that the absolute values of those roots are equals $|\gamma_{I}|^2  = |\delta_{I}|^2$.

These results are not entirely accurate. That is because, similar to the square root, where the number has a positive and a negative root like $a = \pm \sqrt{(a)^2}$, the complex root also has different roots whose complex product result on the same number. Indeed, from the complex product definition, it can be seen that all extended numbers $\alpha, - \alpha, \imagi \alpha, - \imagi \alpha$ are complex root of the same number since
\begin{equation}
\alpha \odot \alpha = (-\alpha) \odot (-\alpha) = (\imagi \alpha) \odot (\imagi \alpha) 
= ( - \imagi \alpha) \odot (- \imagi\alpha)
\end{equation}
in this case, the correct for represent the complex root of a number is
\begin{equation}
\beta  = \substack{\pm 1\\\pm \imagi} \sqrt[*]{\alpha}
\end{equation}
where the symbols $\substack{\pm 1\\\pm \imagi}$ are referred to one of options $+1,-1,+\imagi,-\imagi$.
The correct form of expressing an extended number of its extended and complex roots in substitution of equation \ref{extNumbConjSymm0} then is
\begin{equation}
\alpha = \substack{\pm 1\\\pm \imagi} \left [ \left(\sqrt[\bullet]{\sqrt{\alpha}}\right)^\bullet \odot \left(\sqrt[\bullet]{\sqrt{\alpha}} \right)^\bullet \right]
\cdot
\left[ \left( \sqrt[*]{\sqrt{\alpha}} \right) \odot \left( \sqrt[*]{\sqrt{\alpha}} \right) \right] \label{extNumbRootRep}
\end{equation}
That shows that the symmetric inner product
\begin{equation}
\extproduct{\alpha}{\beta}{\alpha}{\beta} 
= (\alpha^{\bullet}  \odot \beta^{\bullet}) \cdot (\alpha \odot \beta)
=  \substack{\pm 1\\\pm \imagi} (\gamma^{\bullet}  \odot \gamma^{\bullet}) \cdot (\delta \odot \delta) 
= \substack{\pm 1\\\pm \imagi} \gamma^*_{I} \gamma_{I} \delta_{I}^* \delta_{I} 
\end{equation}
is, in general, a pure complex number. 

Also, based on the standard product properties, we see that the symmetric inner product $\extproduct{\alpha}{\beta}{\alpha}{\beta}$ is different from zero if one root of \ref{extRootsDef}
\begin{equation}
\gamma = \sqrt[\bullet]{\sqrt{\extproduct{\alpha}{\beta}{\alpha}{\beta}}} \qquad \text{ or } \qquad \delta = \sqrt[*]{\sqrt{\extproduct{\alpha}{\beta}{\alpha}{\beta}}} \label{extRootsDef1}
\end{equation}
is different from zero.

A special case of a symmetric inner product is
\begin{equation}
\extproduct{\alpha}{\alpha}{\alpha}{\alpha},
\end{equation}
which is equal to the quartic absolute value of number $\alpha$, and is a real number, which is consistent with our starting definitions.


\subsection{Associative property of inner products}\label{associativePropertyChapter}
We study now the expansion of the extended inner product when its internal components or factors are also expanded. The associative property states that the result of an expression containing two or more occurrences in a row connected by the same associative binary operator, in this case, the standard sum, don't change its value by the rearranging of its elements. We study how the associative property for the inner factors are reflected in the addends decomposition of inner products.

Resuming, if both of the upper and the bottom components of the inner product are expanded
\begin{equation}
\extproduct{\alpha}{\beta}{\gamma}{\delta}  = \extproduct{\alpha}{\beta}{\sum_i \gamma_i}{\sum_i \delta_i}, \label{associationAxiom0}
\end{equation}
then what relations should the addends, $\gamma_i$ and $\delta_i$,  must satisfy so the inner product are also expanded like:
\begin{equation}
\extproduct{\alpha}{\beta}{\sum_i \gamma_i}{\sum_i \delta_i} = \sum_i \extproduct{\alpha}{\beta}{\gamma_i}{\delta_i}?
\end{equation}

Also, the expansion of an inner product like:  
\begin{equation}
\extproduct{\alpha}{\beta}{\gamma_1 + \gamma_2 + \gamma_3}{\delta_1 + \delta_2 + \delta_3} = \extproduct{\alpha}{\beta}{\gamma_1}{\delta_1} 
+ \extproduct{\alpha}{\beta}{\gamma_2}{\delta_2} 
+ \extproduct{\alpha}{\beta}{\gamma_3}{\delta_3}, \label{associationAxiom}
\end{equation}
would depend on the way how the inner product is expanded? For example, would the final result of the expansion on equation \ref{associationAxiom} depend on the chosen path like
s\begin{align}
\extproduct{\alpha}{\beta}{\gamma_1 + \gamma_2 + \gamma_3}{\delta_1 + \delta_2 + \delta_3} &=  \extproduct{\alpha}{\beta}{\gamma_1 + \gamma_2}{\delta_1 + \delta_2 } 
+ \extproduct{\alpha}{\beta}{\gamma_3}{\delta_3}
\nonumber \\
&= \left(\extproduct{\alpha}{\beta}{\gamma_1}{\delta_1} + \extproduct{\alpha}{\beta}{\gamma_2}{\delta_2}\right) 
+ \extproduct{\alpha}{\beta}{\gamma_3}{\delta_3}
\nonumber \\
&= \extproduct{\alpha}{\beta}{\gamma_1}{\delta_1} 
+ \extproduct{\alpha}{\beta}{\gamma_2}{\delta_2} 
+ \extproduct{\alpha}{\beta}{\gamma_3}{\delta_3} \label{associationAxiom1}
\end{align}
,
\begin{align}
\extproduct{\alpha}{\beta}{\gamma_1 + \gamma_2 + \gamma_3}{\delta_1 + \delta_2 + \delta_3} &=  \extproduct{\alpha}{\beta}{\gamma_1 + \gamma_3}{\delta_1 + \delta_3 } 
+ \extproduct{\alpha}{\beta}{\gamma_2}{\delta_2}
\nonumber \\
&= \left(\extproduct{\alpha}{\beta}{\gamma_1}{\delta_1} + \extproduct{\alpha}{\beta}{\gamma_3}{\delta_3}\right) 
+ \extproduct{\alpha}{\beta}{\gamma_2}{\delta_2}
\nonumber \\
&= \extproduct{\alpha}{\beta}{\gamma_1}{\delta_1} 
+ \extproduct{\alpha}{\beta}{\gamma_2}{\delta_2} 
+ \extproduct{\alpha}{\beta}{\gamma_3}{\delta_3}\label{associationAxiom2}
\end{align}
or
\begin{align}
\extproduct{\alpha}{\beta}{\gamma_1 + \gamma_2 + \gamma_3}{\delta_1 + \delta_2 + \delta_3} &=  \extproduct{\alpha}{\beta}{\gamma_2 + \gamma_3}{\delta_2 + \delta_3 } 
+ \extproduct{\alpha}{\beta}{\gamma_1}{\delta_1}
\nonumber \\
&= \left(\extproduct{\alpha}{\beta}{\gamma_2}{\delta_2} + \extproduct{\alpha}{\beta}{\gamma_3}{\delta_3}\right) 
+ \extproduct{\alpha}{\beta}{\gamma_1}{\delta_1}
\nonumber \\
&= \extproduct{\alpha}{\beta}{\gamma_1}{\delta_1} 
+ \extproduct{\alpha}{\beta}{\gamma_2}{\delta_2} 
+ \extproduct{\alpha}{\beta}{\gamma_3}{\delta_3}?\label{associationAxiom3}
\end{align}

According to condition \ref{extLinearitySumCond}, the requirements for the first sum expansion case are given by the equations
\begin{align}
&(\gamma_1 + \gamma_2) \odot \delta_3  + \gamma_3 \odot (\delta_1 + \delta_2)
+ \mathcal{D}^{(\gamma)}(\gamma_1 + \gamma_2,\gamma_3,\delta) 
\nonumber \\
& = \gamma_1 \odot \delta_3 + \gamma_2 \odot \delta_3 + \gamma_3 \odot \delta_1 + \gamma_3 \odot \delta_2 + \mathcal{D}^{(\gamma_1 + \gamma_2)}(\gamma_1,\gamma_2,\delta_3) 
+ \mathcal{D}^{(\gamma)}(\gamma_1 + \gamma_2,\gamma_3,\delta) 
\nonumber \\
&=0 \label{associationAxiom4}
\end{align}
where $\gamma = \gamma_1 + \gamma_2 +\gamma_3$ and $\delta = \delta_1 + \delta_2+ \delta_3$ while the second step on the first expansion is possible if
\begin{align}
\gamma_1 \odot \delta_2 + \gamma_2 \odot \delta_1 + \mathcal{D}^{(\gamma_1 + \gamma_2)}(\gamma_1,\gamma_2,\delta_1 + \delta_2)=0. \label{associationAxiom4_1}
\end{align}
Following the same procedure, the second sum expansion case is characterized by the equations
\begin{align}
&\gamma_1 \odot \delta_2 + \gamma_2 \odot \delta_3 + \gamma_2 \odot \delta_1 + \gamma_3 \odot \delta_2 + \mathcal{D}^{(\gamma_1 + \gamma_3)}(\gamma_1,\gamma_3,\delta_2) 
+ \mathcal{D}^{(\gamma)}(\gamma_1 + \gamma_3,\gamma_2,\delta) = 0
\nonumber \\ 
&\gamma_1 \odot \delta_3 + \gamma_3 \odot \delta_1 + \mathcal{D}^{(\gamma_1 + \gamma_3)}(\gamma_1,\gamma_3,\delta_1 + \delta_3)=0 \label{associationAxiom5}
\end{align}
and the third case are constrained by the equations
\begin{align}
&\gamma_1 \odot \delta_2 + \gamma_1 \odot \delta_3 + \gamma_2 \odot \delta_1 + \gamma_3 \odot \delta_1 + \mathcal{D}^{(\gamma_2 + \gamma_3)}(\gamma_2,\gamma_3,\delta_1) 
+ \mathcal{D}^{(\gamma)}(\gamma_2 + \gamma_3,\gamma_1,\delta) = 0
\nonumber \\ 
&\gamma_2 \odot \delta_3 + \gamma_2 \odot \delta_1 + \mathcal{D}^{(\gamma_2 + \gamma_3)}(\gamma_2,\gamma_3,\delta_2 + \delta_3)=0. \label{associationAxiom6}
\end{align}

We conclude then that the expansion of inner products on inner products where the inner factors are the addends of the sum of the factors of the inner product like equation \ref{associationAxiom0}, depend on the path or the way of the expansion is carry out. The conditions for each one depending on the $\mathcal{D}^{(\epsilon)}(\gamma,\beta,\delta)$'s functions. For example, if all the terms of the series satisfy
\begin{equation}
\extproduct{\alpha}{\beta}{\gamma_i + \gamma_j}{\delta_i + \delta_j} =  \extproduct{\alpha}{\beta}{\gamma_i}{\delta_i} + \extproduct{\alpha}{\beta}{\gamma_j}{\delta_j} 
\end{equation}
and all $\mathcal{D}^{(\epsilon)}(\gamma,\beta,\delta)$ functions are zero for all intermediate expansion; we have then the series expansion as
\begin{equation}
\extproduct{\alpha}{\beta}{\sum_i \gamma_i}{\sum_i \delta_i} = \sum_i \extproduct{\alpha}{\beta}{\gamma_i}{\delta_i}.
\end{equation}

\subsection{Inner commutation}
The phenomenology in Physics is strongly related to the algebra of vector space where the theory is developed on. That is the case of the uncertainty principle, which has no classical analogue and it is proved to be a consequence of quantum mechanic being developed on a complex vector space \cite{sakurai}.

There is a property for the extended inner products that may, or may not, turn into a new physical event and its based on the switching between terms of the multiplication of two inner products. The referred property may look simple, but it can be related to a new type of physical phenomena for any theory developed on an extended vector space, which cannot be observed on a theory which vector space is complex, such as the standard quantum mechanic. We refer this property as the ``inner commutative property'' of the extended vector space. 

The standard multiplication of two extended inner products have the form
\begin{align}
\extproduct{\alpha_1}{\beta_1}{\gamma_1}{\delta_1}  \extproduct{\alpha_2}{\beta_2}{\gamma_2}{\delta_2}  =& \big[(\alpha_1^{\bullet}  \odot \beta_1^{\bullet}) \cdot (\gamma_1 \odot \delta_1) \big]
\cdot \big[(\alpha_1^{\bullet}  \odot \beta_2^{\bullet}) \cdot (\gamma_2 \odot \delta_2) \big]
\nonumber \\
=& (\alpha_1^{\bullet}  \odot \beta_1^{\bullet}) \cdot (\gamma_1 \odot \delta_1) 
\cdot (\alpha_1^{\bullet}  \odot \beta_2^{\bullet}) \cdot (\gamma_2 \odot \delta_2).
\end{align}

From the associativity property of the standard product of extended numbers, it is possible to rearranged the four terms of the multiplication and get the relation
\begin{equation}
\extproduct{\alpha_1}{\beta_1}{\gamma_1}{\delta_1} \extproduct{\alpha_2}{\beta_2}{\gamma_2}{\delta_2} 
=\extproduct{\alpha_1}{\beta_1}{\gamma_2}{\delta_2}\extproduct{\alpha_2}{\beta_2}{\gamma_1}{\delta_1}.
\label{innerComm00}
\end{equation}
if the inner product on the right member of the equation is valid; which means they satisfy all required axioms for extended inner products.

It's also possible to switch any factor for its equivalent of the other product if the result is also two valid inner products and the switched terms satisfy some conditions. Let's assume, for instance, that the inner products
\begin{equation}
\extproduct{\alpha_1}{\beta_1}{\gamma_2}{\delta_1},\extproduct{\alpha_2}{\beta_2}{\gamma_1}{\delta_2},
\extproduct{\alpha_1}{\beta_1}{\gamma_1}{\delta_2},\extproduct{\alpha_2}{\beta_2}{\gamma_2}{\delta_1}
\end{equation}
are valid. In this case, we can study under what conditions its possible to switch the inner factors as 
\begin{equation}
\extproduct{\alpha_1}{\beta_1}{\gamma_1}{\delta_1} \extproduct{\alpha_2}{\beta_2}{\gamma_2}{\delta_2} 
=\extproduct{\alpha_1}{\beta_1}{\mathbf{\gamma_2}}{\delta_1}\extproduct{\alpha_2}{\beta_2}{\mathbf{\gamma_1}}{\delta_2}
\label{innerCommProp}
\end{equation} 
or 
\begin{equation}
\extproduct{\alpha_1}{\beta_1}{\gamma_1}{\delta_1} \extproduct{\alpha_2}{\beta_2}{\gamma_2}{\delta_2} 
=\extproduct{\alpha_1}{\beta_1}{\gamma_1}{\mathbf{\delta_2}}\extproduct{\alpha_2}{\beta_2}{\gamma_2}{\mathbf{\delta_1}}.
\end{equation} 
where terms $\gamma_1 \leftrightarrow \gamma_2$ and  $\delta_1 \leftrightarrow \delta_2$ have been switched off. 

If there is some quantity with a physical meaning, which we can represent as $\Psi_{\alpha,\beta}(x',x'')$, and is related to the inner product
\begin{equation}
\extproduct{x'}{x''}{\alpha}{\beta} \equiv \Psi_{\alpha,\beta}(x',x''),
\end{equation}
then the inner commutation property between two of these quantities can be written as:
\begin{equation}
\Psi_{\alpha_1,\beta_1}(x',x'')\Psi_{\alpha_2,\beta_2}(x',x'') 
= \Psi_{\alpha_1,\beta_2}(x',x'')\Psi_{\alpha_2,\beta_1}(x',x'').
\end{equation}
For example, if  $\Psi_{\alpha,\beta}(x',x'')$ is a wave function depending on variables $x',x''$ and the standard product of two wave functions is studied, being $\alpha_1,\beta_1,\alpha_2,\beta_2$ the quantum numbers which describe the quantum states, then the property of the inner commutation will algebraically allow the switching of quantum numbers between states.

The condition for the first statement being true is
\begin{align}
 (\gamma_1 \odot \delta_1) (\gamma_2 \odot \delta_2) =   (\gamma_2 \odot \delta_1)(\gamma_1 \odot \delta_2). \label{innerComm0}
\end{align}

Multiplying left member we have:
\begin{align}
&(\gamma_1 \odot \delta_1) (\gamma_2 \odot \delta_2) = \big [ \big ( \gamma^*_{1_E} \delta_{1_I} z_1^{(\gamma_1)} + \gamma^*_{1_I} \delta_{1_E}  \big ) \big ( \gamma^*_{2_E} \delta_{2_I} z_1^{(\gamma_2)} + \gamma^*_{2_I} \delta_{2_E}  \big )  z_0
\nonumber \\
& \qquad + \big ( \gamma^*_{1_E} \delta_{1_I} z_1^{(\gamma_1)} + \gamma^*_{1_I} \delta_{1_E}  \big ) 
\big ( \gamma^*_{2_E} \delta_{2_I} w_1^{(\gamma_2)} + \imagi \gamma^*_{2_E} \delta^*_{2_E} + \gamma^*_{2_I} \delta_{2_I}  \big )
\nonumber \\
& \qquad + \big ( \gamma^*_{1_E} \delta_{1_I} w_1^{(\gamma_1)} + \imagi \gamma^*_{1_E} \delta^*_{1_E}  + \gamma^*_{1_I} \delta_{1_I}  \big )
\big ( \gamma^*_{2_E} \delta_{2_I} z_1^{(\gamma_2)} + \gamma^*_{2_I} \delta_{2_E}  \big ) 
\big ] \extk
\nonumber \\
& \qquad + \big ( \gamma^*_{1_E} \delta_{1_I} z_1^{(\gamma_1)} + \gamma^*_{1_I} \delta_{1_E}  \big ) \big ( \gamma^*_{2_E} \delta_{2_I} z_1^{(\gamma_2)} + \gamma^*_{2_I} \delta_{2_E}  \big )  w_0
\nonumber \\
& \qquad \big ( \gamma^*_{1_E} \delta_{1_I} w_1^{(\gamma_1)} + \imagi \gamma^*_{1_E} \delta^*_{1_E}  + \gamma^*_{1_I} \delta_{1_I}  \big ) \big ( \gamma^*_{2_E} \delta_{2_I} w_1^{(\gamma_2)} + \imagi \gamma^*_{2_E} \delta^*_{2_E} + \gamma^*_{2_I} \delta_{2_I}  \big ) \label{innerComm1}
\end{align}
Explicitly, the extended part of the above equation has the form
\begin{align}
&\big [ (\gamma_1 \odot \delta_1) (\gamma_2 \odot \delta_2)\big ]_E = 
\gamma^*_{1_E} \delta_{1_I} \gamma^*_{2_E} \delta_{2_I} z_1^{(\gamma_1)} z_1^{(\gamma_2)} z_0 
+ 		\gamma^*_{1_E} \delta_{1_I} \gamma^*_{2_I} \delta_{2_E} z_1^{(\gamma_1)} z_0
+ 		\gamma^*_{1_I} \delta_{1_E} \gamma^*_{2_E} \delta_{2_I} z_1^{(\gamma_2)} z_0
\nonumber \\
& \qquad 
+ 		\gamma^*_{1_I} \delta_{1_E} \gamma^*_{2_I} \delta_{2_E} z_0
+ 		\gamma^*_{1_E} \delta_{1_I} \gamma^*_{2_E} \delta_{2_I} z_1^{(\gamma_1)} w_1^{(\gamma_2)}
+\imagi \gamma^*_{1_E} \delta_{1_I} \gamma^*_{2_E} \delta_{2_E} z_1^{(\gamma_1)}
+ 		\gamma^*_{1_E} \delta_{1_I} \gamma^*_{2_I} \delta_{2_I} z_1^{(\gamma_1)}
\nonumber \\
& \qquad 
+ 		\gamma^*_{1_I} \delta_{1_E} \gamma^*_{2_E} \delta_{2_I} w_1^{(\gamma_2)}
+\imagi \gamma^*_{1_I} \delta_{1_E} \gamma^*_{2_E} \delta_{2_E}
+		\gamma^*_{1_I} \delta_{1_E} \gamma^*_{2_I} \delta_{2_I}
+ 		\gamma^*_{1_E} \delta_{1_I} \gamma^*_{2_E} \delta_{2_I} w_1^{(\gamma_2)} z_1^{(\gamma_2)}
\nonumber \\
& \qquad 
+		\gamma^*_{1_E} \delta_{1_I} \gamma^*_{2_I} \delta_{2_E} w_1^{(\gamma_1)}
+\imagi \gamma^*_{1_E} \delta_{1_E} \gamma^*_{2_E} \delta_{2_I} z_1^{(\gamma_2)}
+\imagi \gamma^*_{1_E} \delta_{1_E} \gamma^*_{2_I} \delta_{2_E}
+		\gamma^*_{1_I} \delta_{1_I} \gamma^*_{2_E} \delta_{2_I} z_1^{(\gamma_2)}
\nonumber \\
& \qquad 
+		\gamma^*_{1_I} \delta_{1_I} \gamma^*_{2_I} \delta_{2_E}. \label{innerComm2}
\end{align}

The extended part of the right member of equation \ref{innerComm0} $(\gamma_1 \odot \delta_2) (\gamma_2 \odot \delta_1)$ is easily obtained by swapping $\gamma_1 \leftrightarrow \gamma_2$ on equations \ref{innerComm2}. Subtracting the extended parts of both members, we obtain the relation between numbers $\gamma_i, \delta_j$:
\begin{equation}
(\delta_{2_E} \delta_{2_I} - \delta_{2_I} \delta_{2_E}) \big [ 
  \gamma^*_{1_E} \gamma^*_{2_I} ( w_1^{(\gamma_1)} + z_0 z_1^{(\gamma_1)} )
- \gamma^*_{1_I} \gamma^*_{2_E} ( w_1^{(\gamma_2)} + z_0 z_1^{(\gamma_2)} )
+ \gamma^*_{1_E} \gamma^*_{2_E} ( z_1^{(\gamma_1)} - z_1^{(\gamma_2)} )
\big ] = 0 \label{innerComm3}
\end{equation}
On the other hand, the imaginary part of equation \ref{innerComm1} have the form
\begin{align}
&\big [ (\gamma_1 \odot \delta_1) (\gamma_2 \odot \delta_2)\big ]_I = 
\gamma^*_{1_E} \delta_{1_I} \gamma^*_{2_E} \delta_{2_I} z_1^{(\gamma_1)} z_1^{(\gamma_2)} w_0 
+ 		\gamma^*_{1_E} \delta_{1_I} \gamma^*_{2_I} \delta_{2_E} z_1^{(\gamma_1)} w_0
+ 		\gamma^*_{1_I} \delta_{1_E} \gamma^*_{2_E} \delta_{2_I} z_1^{(\gamma_2)} w_0
\nonumber \\
& \qquad 
+ 		\gamma^*_{1_I} \delta_{1_E} \gamma^*_{2_I} \delta_{2_E} w_0
+ 		\gamma^*_{1_E} \delta_{1_I} \gamma^*_{2_E} \delta_{2_I} w_1^{(\gamma_1)} w_1^{(\gamma_2)}
+\imagi \gamma^*_{1_E} \delta_{1_I} \gamma^*_{2_E} \delta_{2_E} w_1^{(\gamma_1)}
+		\gamma^*_{1_E} \delta_{1_I} \gamma^*_{2_I} \delta_{2_I} w_1^{(\gamma_1)}
\nonumber \\
& \qquad 
+\imagi \gamma^*_{1_E} \delta_{1_E} \gamma^*_{2_E} \delta_{2_I} w_1^{(\gamma_2)}
- 		\gamma^*_{1_E} \delta_{1_E} \gamma^*_{2_E} \delta_{2_E}
+\imagi \gamma^*_{1_E} \delta_{1_E} \gamma^*_{2_I} \delta_{2_I}
+ 		\gamma^*_{1_E} \delta_{1_I} \gamma^*_{2_E} \delta_{2_I} w_1^{(\gamma_1)}
\nonumber \\
& \qquad 
+\imagi \gamma^*_{1_I} \delta_{1_I} \gamma^*_{2_E} \delta_{2_E}
+		\gamma^*_{1_I} \delta_{1_I} \gamma^*_{2_I} \delta_{2_I} \label{innerComm31}
\end{align}

If same switching procedure is followed for the imaginary to obtain the imaginary part of right member and subtracting, we get the equation 
\begin{equation}
(\delta_{2_E} \delta_{2_I} - \delta_{2_I} \delta_{2_E}) \big [ 
  \gamma^*_{1_E} \gamma^*_{2_I} ( w_0 z_1^{(\gamma_1)} - \imagi )
- \gamma^*_{1_I} \gamma^*_{2_E} ( w_0 z_1^{(\gamma_2)} - \imagi )
+ \gamma^*_{1_E} \gamma^*_{2_E} ( w_1^{(\gamma_1)} - w_1^{(\gamma_2)} )
\big ] = 0 \label{innerComm4}
\end{equation}

Putting all together, the switching of the inner factors on the multiplication of two inner products like \ref{innerCommProp}
\begin{equation*}
\extproduct{\alpha_1}{\beta_1}{\gamma_1}{\delta_1} \extproduct{\alpha_2}{\beta_2}{\gamma_2}{\delta_2} 
=\extproduct{\alpha_1}{\beta_1}{\mathbf{\gamma_2}}{\delta_1}\extproduct{\alpha_2}{\beta_2}{\mathbf{\gamma_1}}{\delta_2}
\end{equation*}
are given by the conditions \ref{innerComm3} and \ref{innerComm4}:
\begin{align}
&(\delta_{2_E} \delta_{2_I} - \delta_{2_I} \delta_{2_E}) \big [ 
  		\gamma^*_{1_E} \gamma^*_{2_I} ( w_1^{(\gamma_1)} + z_0 z_1^{(\gamma_1)} )
- 		\gamma^*_{1_I} \gamma^*_{2_E} ( w_1^{(\gamma_2)} + z_0 z_1^{(\gamma_2)} )
+\imagi \gamma^*_{1_E} \gamma^*_{2_E} ( z_1^{(\gamma_1)} - z_1^{(\gamma_2)} )
\big ] = 0 
\nonumber \\
&(\delta_{2_E} \delta_{2_I} - \delta_{2_I} \delta_{2_E}) \big [ 
  		\gamma^*_{1_E} \gamma^*_{2_I} ( w_0 z_1^{(\gamma_1)} - \imagi )
- 		\gamma^*_{1_I} \gamma^*_{2_E} ( w_0 z_1^{(\gamma_2)} - \imagi )
+\imagi \gamma^*_{1_E} \gamma^*_{2_E} ( w_1^{(\gamma_1)} - w_1^{(\gamma_2)} ) \label{innerComm5}
\big ] = 0 
\end{align}
Note that if ${\delta_2}$ is purely a complex or an extended number, the condition is automatically satisfied.

The commutation of the downer inner factors
\begin{equation*}
\extproduct{\alpha_1}{\beta_1}{\gamma_1}{\delta_1} \extproduct{\alpha_2}{\beta_2}{\gamma_2}{\delta_2} 
=\extproduct{\alpha_1}{\beta_1}{\gamma_1}{\mathbf{\delta_2}}\extproduct{\alpha_2}{\beta_2}{\gamma_2}{\mathbf{\delta_1}}
\end{equation*}
are right if it's satisfied the equation
\begin{align}
 (\gamma_1 \odot \delta_1) (\gamma_2 \odot \delta_2) =  (\gamma_1 \odot \delta_2) (\gamma_2 \odot \delta_1),
\end{align}
which is the same equation than \ref{innerComm0} and is also constrained by the same conditions. That is an expected result, because if conditions \ref{innerComm5} are satisfied, and it is possible to commute the top inner factors, then the downer inner factor pair should also be commutative, this time under no extra conditions. This is consistent with the fact that pair $\extscalar{\gamma_1}{\delta_1} \to \extscalar{\gamma_2}{\delta_2}$ can be switched with no condition because of the associative property as showed in equation \ref{innerComm00}. 

The inner commutation of elements in the left column factors
\begin{equation*}
\extproduct{\alpha_1}{\beta_1}{\gamma_1}{\delta_1} \extproduct{\alpha_2}{\beta_2}{\gamma_2}{\delta_2} 
=\extproduct{\mathbf{\alpha_2}}{\beta_1}{\gamma_1}{\delta_1} \extproduct{\mathbf{\alpha_1}}{\beta_2}{\gamma_2}{\delta_2} 
\end{equation*}
is given by the condition
\begin{align}
({\alpha_1^\bullet}\odot{\beta_1}^\bullet) ({\alpha_2^\bullet}\odot{\beta_2}^\bullet)
= ({\alpha_2^\bullet}\odot{\beta_1}^\bullet) ({\alpha_1^\bullet}\odot{\beta_2}^\bullet)
\label{innerComm000}
\end{align}
which have a similar expansion than equation \ref{innerComm5} if $\gamma_i$ are replaced by $\alpha_i^\bullet$ and $\delta_i$ by $\beta_i^\bullet$. 

The product of two non-null symmetric extended inner products can be rearranged using the last property as
\begin{align}
\extproduct{\alpha}{\beta}{\alpha}{\beta}  \extproduct{\beta}{\alpha}{\beta}{\alpha}  =  \extproduct{\alpha}{\alpha}{\alpha}{\alpha} \extproduct{\beta}{\beta}{\beta}{\beta}
\end{align}
if they are valid and satisfy the conditions \ref{innerComm5}. The last expression is the standard product of two non-negative real numbers whose result is null only if one or both of its earlier factors are null. As shown, in the positive-definiteness section, each of these factors is null only if its inner factors are null, in this case, if $\alpha$ or $\beta$ are null. The product of two non-negative real number is also a non-negative real number, then 
\begin{align}
 \extproduct{\alpha}{\beta}{\alpha}{\beta}  \extproduct{\beta}{\alpha}{\beta}{\alpha}  =  \extproduct{\alpha}{\alpha}{\alpha}{\alpha} \extproduct{\beta}{\beta}{\beta}{\beta} 
 = R  \geq 0 \qquad R \in \mathbb{R}. \label{innerProdPropInvRow1}
\end{align}

At the same time, both inner products of the left member of equation \ref{innerProdPropInvRow1} are complex numbers, then they also satisfy the relation
\begin{equation}
 \extproduct{\alpha}{\beta}{\alpha}{\beta}^*  \extproduct{\alpha}{\beta}{\alpha}{\beta} \geq 0 \qquad \extproduct{\beta}{\alpha}{\beta}{\alpha}^* \extproduct{\beta}{\alpha}{\beta}{\alpha} \geq 0. \label{innerProdPropInvRow2}
\end{equation}
If they are written with their exponential form
\begin{equation}
\extproduct{\alpha}{\beta}{\alpha}{\beta} \equiv \Delta = R_\Delta \exp^{\imagi \theta_\Delta},\qquad 
\extproduct{\beta}{\alpha}{\beta}{\alpha} \equiv \nabla = R_\nabla \exp^{\imagi \theta_\nabla},
\end{equation}
then equations \ref{innerProdPropInvRow1} and \ref{innerProdPropInvRow2} have the form
\begin{equation}
 \extproduct{\alpha}{\beta}{\alpha}{\beta}  \extproduct{\beta}{\alpha}{\beta}{\alpha} =
 R_\Delta R_\nabla \exp^{\imagi (\theta_\Delta + \theta_\nabla)} = R  \geq 0 \qquad R \in \mathbb{R}
 \label{innerProdPropInvRow4}
\end{equation}
and 
\begin{equation}
 \extproduct{\alpha}{\beta}{\alpha}{\beta}^*  \extproduct{\alpha}{\beta}{\alpha}{\beta}  = |R_\Delta|^2,\qquad 
\extproduct{\beta}{\alpha}{\beta}{\alpha}^* \extproduct{\beta}{\alpha}{\beta}{\alpha} = |R_\nabla|^2 , \label{innerProdPropInvRow5}
\end{equation}
respectively. The fact that the standard product on equation \ref{innerProdPropInvRow4} is a nonnegative real number implies that $\exp^{\imagi (\theta_\Delta + \theta_\nabla)} = 1$. The division of equation \ref{innerProdPropInvRow4} by equation \ref{innerProdPropInvRow5}, we obtain the relations
\begin{equation}
R_\Delta \extproduct{\beta}{\alpha}{\beta}{\alpha} =  R_\nabla \extproduct{\alpha}{\beta}{\alpha}{\beta}^*  
\qquad \text{ and } \qquad 
\theta_\Delta + \theta_\nabla = 2\pi n \qquad \qquad n = 0,1,2... \label{innerProdPropInvRow}.
\end{equation}

For $\alpha=\beta$, the last relation become
\begin{equation}
\extproduct{\alpha}{\alpha}{\alpha}{\alpha} =  \extproduct{\alpha}{\alpha}{\alpha}{\alpha}^*, \label{innerProdPropInvRow3}
\end{equation}
which means that the symmetric inner product $\extproduct{\alpha}{\alpha}{\alpha}{\alpha}$ is a real number as expected.

\newpage
\section{Quantum mechanics for  $n$-VMVF systems}\label{ExtendedQMSection}

The approach we develop in this work can be considered as an extension of the modern quantum mechanic theory. The modern construction of the theory of quantum mechanic embraces the definition a complex vector space where the algebraic objects like the operators and the state vectors are defined and also the inclusion of the corresponding classical theory to find how operators act over the state vectors. The complex vectors on the Hilbert space represent physical states while the physics magnitudes, known as operators, like position, or linear momentum are defined as the operators that modify the states. Here is where classical theory plays its primary role: to obtain the form of the classic action of the contact transformations over the canonical variables for use it as the base of the analytical form of the quantum actions of the operators over the state vector. Others concepts included in the quantum theory were obtained from experimentation, logic or merely from common sense, such as measurement, the normalization relations or the concepts of eigenvalues and eigenstates.

We studied the classical theory for $n$-VMVF systems starting proposing a relativistic Lagrangian with the universal form of the potential energy like $U = A_\nu x^\nu$. The assumption that the second law of Newton cannot be applied over each particle but over the whole particle system, defines new constraint functions for the variables of the system. Those constraints were added to the original Lagrangian depending on particle positions, particle's mass and field derivatives using the  method of the undetermined multipliers of Lagrange. The constraints show an $\ddot{x}^\nu$ dependency that force to extend the Lagrange and Hamilton equations with a new canonical variable that we defined as $s$. The inclusion of this new variable result in a new Lagrange's, and therefore, a new Hamilton's set of equations, which permits the definitions of new generators in the canonical transformations of the system. This new set of generators are the cornerstone of the quantization of all quantities in the standard Quantum Mechanics. 

Another result obtained in the classical theory of the $n$-VMVF systems is related to the number of equations need to find the solutions of the degree of freedom of the system. The masses of the particles and field derivatives were added in theory as unknown quantities to be determine or variables. The initial Lagrange function is split into two independent Lagrangians by the addition of two set of constraints from the conservation law of the linear and the angular momentum. Because the number of variables surpasses the number of possible equations, the solution of the problem becomes the resolution of 2 independent Lagrange's equations on two independent representation for particle position: rectangular and angular.

Every second order function of Lagrange led to one extended function of Hamilton according to the second-order Hamilton theory that we propose in here. The classical theory reveals the existence of a new type of canonical transformation and with it, a new type of generator related to the new second-order momentum $s$. Also, the classical development exposes the two bi-dimensional structure for the canonical transformations and for their generators to modify the system in the phase space. These results accomplish the primary objective of the classical development of obtaining the classical generators of the canonical transformation for using them in the quantum approach of the $n$-VMVF systems to find the form of the action of quantum operators over the state vectors. 

On the other side, the classical phenomenon of the $n$-VMVF systems where masses continuously vary as a function of the position of the particles do not exist, at least to the best of our knowledge. However, besides of providing the classical generators, the classical approach offers the classical physics that the quantum mechanic should reproduce in the limit of large quantum numbers in agreement with the correspondence principle for $n$-VMVF systems.

The remaining piece of the puzzle for developing the quantum theory for $n$-VMVF systems is the mathematical frame on where the theory is developed. 

A quantum state should be represented with a unique vector whose dimensionality depends on the nature of the problem. In our case, the state vector should depend on the particle positions $\{x_n^\mu\}$, the poles of the correlation functions $\{f_{x_n^\mu}\}$ and the particle masses and field derivatives. In particular, the set of variables $\{x_n^\mu\}$ and  $\{f_{x_n^\mu}\}$ are the canonical variables modified by the canonical transformations obtained from classical Lagrangians depending on $\{ \ddot{x}_n^\mu\}$. An extended quantum state can be represented in its first option using the Dirac's bracket representation as
\begin{equation}
\ket{
\{x_n^\mu\}, 
\{f_{x_n^\mu}\}, 
\{\frac{\partial m_n}{\partial x_n^\mu }\}, 
\{\frac{\partial A^\nu}{\partial x_n^\mu }\},
\{\frac{\partial A^\nu}{\partial \dot{x}_n^\mu }\} \label{extStateVector1}
}
\end{equation}
where we assume some approximations for particle mass and field derivatives as shown in section \ref{massFieldApprox}. For example, we choose that for any system $\frac{\partial m_n}{\partial \dot{x}_n^\mu }=0$.

In ordinary quantum mechanics, the action of one operator over a state vector is sufficient for describing the system. However, we verified that, in the treatment of $n$-VMVF systems, we need a two-set of equations for describing the system. In that way, the quantification of the action of two classical canonical transformations over a state in the phase space in a quantum mechanic approach, following the ket representation like \ref{extStateVector1}, can be written as two independent quantum equations:
\begin{align}
\mathbf{A} \ket{\alpha_{a_0}} = \ket{\alpha_a}
\qquad \qquad
\mathbf{B} \ket{\alpha_{b_0}} = \ket{\alpha_b}.
\end{align}
The equations describe the transformation of a quantum system which is at the initial state describe by $\ket{\alpha}_{a_0}$ and $\ket{\alpha}_{b_0}$ by the operators $\mathbf{A}$ and $\mathbf{B}$ into two different states $\ket{\alpha}_{a}$ and $\ket{\alpha}_{b}$. However, even the state given by $\ket{\alpha_a}$ and $\ket{\alpha_b}$ are the result of the action of different operators over the same state, they represent the same physical state. In that case, what are the relations between both vectors? For example, we cannot construct, following the rules of the ordinary quantum mechanics, a general state resembling equation \ref{extStateVector1} like
\begin{equation}
\ket{\alpha_R} = c_a \ket{\alpha_a} + c_b \ket{\alpha_b},
\end{equation}
since this expression attends for the spanning of the state ket state in the two modified states. According to the quantum theory, each state vector has a real probability of being part of the general state. However, this not the case because $\ket{\alpha_a}$ and $\ket{\alpha_b}$ together \textbf{are} the state of the system. We believe that in our approach, we cannot go from two classical to two quantum equations using the state vector as expressed in equation \ref{extStateVector1}. We understand that the two classical equations correspond to two quantum quantities should form part as components of a single and more complex algebraic structure.

Along with the development of the present proposal, we base our assumptions and propositions mostly on mathematical concepts or analytical needs such as the fact that the number of variables must be equal to the number of equations. We avoid defining axioms based on experimental notions, either because of the lack of knowledge of a similar approach and because most of the well-established definitions from previous theories have a weak or none connection with the new concepts defined in here; for example, the concept of probability. Because of that, our reasoning is guided more by analytically needs instead of any expected physical behavior. 

We are aware that a significant problem of quantum field theory (QFT), as the main mathematical framework on which are developed most of the quantum mechanical models of subatomic particles in particle physics, is the problem of infinities. The emergence of infinities appears when calculations of higher orders of the perturbation series led to infinite results. The various forms of infinities suggested the divergences are related to a  more fundamental issue instead of some failure for specific calculations. One way of treat these infinities are solved include formal tricks truncating the integrals, or even ignoring the infinite contributions terms. However, some physicists propose alternative approaches that included changes in the basic concepts. Among them, we found the assumption of the existence of negative or incomplete probabilities.

The concept of negative energies and negative probabilities were introduced by Paul Dirac, in 1942, when he quote on the article "The Physical Interpretation of Quantum Mechanics" \cite{DiracInterpQM} that: ``Negative energies and probabilities should not be considered as nonsense. They are well-defined concepts mathematically, like a negative of money.'' Also, Richard Feynman argued on its work ``Negative probabilities'' \cite{Feynman1987FEYNP} how not only negative probabilities but probabilities different from unity could be useful in probability calculations. Negative or incomplete probabilities have later been proposed to solve several problems and paradoxes.

Although there is no mathematical reason for not to define the concepts of negative and incomplete probability, their physical interpretation is as complicated as diverse. If the probability in the quantum approach of any classical theory, which the quantum theory is based on, is proportional to some scalar quantity $\psi$ multiplied by its complex conjugate $\psi^*$, then, mathematically speaking, the assumption of a negative probability can be written as the existence in the theory of an expression like $\psi^* \psi < 0$. This event, together with the need for a structure of 2-set of equations is the principal motivation for the definition of the new space. As seen before, a non real and positive probability is the definition of the new space unit.

We name the new space as the extended complex space and we represent it by $\mathbb{E}$. Its extended unit $\extk$ is defined from an unsolvable equation in the complex space recognized as $\extk^* \extk = \imagi$ from where $\extk = \sqrt[*]{\imagi}$, being $\sqrt[*]{()}$ the complex square root operator. The complex square root of a number $z$ is a number $y$ such that $y^*y = z$. The identification of an unsolvable equation in the complex space adds the complex numbers to the algebra of the complex space, what is traduced as the existence of two new binary operations for the set of complex numbers: the complex sum ($z_1^* + z_2$) and the complex multiplication ($z_1^* \cdot z_2$). These new operations permit the extension of the complex numbers domain without contradicting the Fundamental Theorem of Algebra. The new space defines the inner product as a standard product of two complex products between of two extended numbers and includes a new defined isomorphism $\extk^{\bullet}$. We impose conditions for the inner product satisfies positive-definiteness and conjugate symmetry axioms, using the properties of standard and complex sum and multiplication.

With these ideas, we can adventure to set the first ideas for the quantum theory for $n$-VMVF systems.

\subsection{Kets, bras, scalars, inner products, and operators}
The classic theory for $n$-VMVF systems take us to consider a bi-dimensional structure for the algebraic magnitudes in the quantum theory, also compatible with the proposition for the inner product of the extended vector space. In this section, we formulate the underlying mathematics of Quantum mechanics for $n$-VMVF systems using the of bra and ket Dirac's notation. 

We assume that the kernel of the extended quantum mechanic is based on the ``cause and effect'' philosophy $e.i.$: one analytic entity, represented as the ``physical state'' change to another physical state when other entity, as the cause of the modification, act over the state. This kernel should have a bidimensional representation like:
\begin{equation}
\extoperator{A}{B} \extket{\alpha}{\beta} \xrightarrow{transformation} \extket{\gamma}{\zeta}.
\end{equation}
where 
$  \extket{\alpha}{\beta}$ is the physical state and $\extoperator{A}{B}$ is the cause of the modification, named operator. 

Another important entity is named scalars and it must also have a bi dimensional form $ \extscalar{\alpha}{\beta}$.
A scalar is just an extended number expressed as a complex product. The entity $\extscalar{\alpha}{\beta}$ is a two-column vector representing the complex product of two numbers:
\begin{equation}
\extscalar{\alpha}{\beta} \equiv {\alpha}\odot{\beta} = \gamma \qquad \forall \; \alpha, \beta, \gamma \in \mathbb{E}.
\end{equation}
By convention, we set the ``bottom'' component of the scalar as the extended component of the complex product while the ``upper'' position stands as its complex component. Because of that, the vectors
\begin{equation}
\extket{\alpha}{\beta} \quad \text{and} \quad \extket{\beta}{\alpha}
\end{equation}
represent different states.

All the bi-dimensional algebraic structures of this quantum approach like vectors, operators, scalars and others are related to the complex products ${\alpha}\odot{\beta}$ or ${\alpha^\bullet}\odot{\beta^\bullet}$. This association is the base for assuming the axioms of Associativity, Commutativity and the Distributivity of the standard multiplication over the standard addition for all the bi-dimensional algebraic structures used in the theory.

Rest only to settle how the form bi-dimensional operators act over bi-dimensional vectors. The classical solution of the $n$-VUMF systems involves solving two independent set of equations. The general transformation of the canonical variables is also composed of two canonical transformations in both sets of coordinates. From our point of view, this feature set a different form in the action of an extended operator over the extended ket such as that one component of the operator should act over the same component of the vector. For example, the motion operator composed on the translation and the rotation operators should act over the rectangular and the angular components of the vector, respectively. We use the column representation for a better association. We can write the above kernel of the action of an operator over a state ket, using the infinitesimal motion operator as an example, like
\begin{equation}
\extoperator{\mathcal{T}(dx)}{\mathcal{R}(d\xi)} \extket{x'}{\xi'} \equiv
\extket{\mathcal{T}(dx)\ket{x'} \to \ket{x'+dx} }{\mathcal{R}(d\xi) \ket{\xi'} \to \ket{\xi' + d\xi} } \equiv
\extket{x' + dx}{\xi' + d\xi}
\end{equation}

We proceed now to the definition of the algebraic magnitudes for the quantum mechanic theory of $n$-VMVF systems.
\subsubsection{Ket vector space}
Similar to the ordinary quantum mechanic, we consider that a vector represents a physical state whose dimensionality is specified according to the nature of the physical system itself and is given by the number of degree of freedom of the system. A physical state is represented like:
\begin{equation}
\extket{\alpha}{\beta}
\end{equation}
and it contains all the information related to the state itself. The Linearity axiom of the state vectors states that two kets can be added by a sum operation resulting in another state:
\begin{equation}
\extket{\alpha_1}{\beta_1} + \extket{\alpha_2}{\beta_2} = \extket{\alpha_3}{\beta_3}.
\end{equation}
The standard sum operation between extended vectors satisfies the associativity and commutativity axioms. These axioms have their support on the properties of complex product $\alpha \odot \beta$ studied in the previous chapter. 

%

The product of an scalar and a vector ket the resulting product is a new ket and is commutative
\begin{equation}
\extscalar{a}{b} \extket{\alpha}{\beta} =  \extket{\alpha}{\beta} \extscalar{a}{b},
\end{equation}
being $a,b \in \mathbb{E}$ extended numbers.

\subsubsection{Bra vector space}
The vector space ``bra'' is defined as the dual space vector needed for the extended Hilbert space satisfies axioms like Positive-definiteness and Conjugated symmetry. The theory should postulate the one-to-one correspondence between one ket and a bra with same numbers
\begin{align}
\extket{\alpha}{\beta} \leftrightarrow \extbra{\alpha}{\beta} 
\end{align}
The scalars transform into bra space as
\begin{equation}
\extscalar{a}{b} \extket{\alpha}{\beta} 
\leftrightarrow \extscalarBra{a}{b} \extbra{\alpha}{\beta} 
\end{equation}
where $()^\bullet$ stands for the ${}^\bullet$ extended conjugated map for both, up and down $i.e$
\begin{equation}
\extscalarBra{a}{b} = (\alpha^\bullet \odot \beta^\bullet).
\end{equation}
In general, linearity is expressed for bra vector space as
\begin{equation}
\extscalar{a_1}{b_1} \extket{\alpha_1}{\beta_1} + \extscalar{a_2}{b_2} \extket{\alpha_2}{\beta_2}
\extscalarBra{a_1}{b_1} \leftrightarrow \extbra{\alpha_1}{\beta_1} + \extscalarBra{a_2}{b_2} \extbra{\alpha_2}{\beta_2} 
\end{equation}

Same as the ket space, two bra vectors are added by a sum operation resulting in another state in the bra space like
\begin{equation}
\extbra{\alpha_1}{\beta_1} + \extbra{\alpha_2}{\beta_2} = \extbra{\alpha_3}{\beta_3}.
\end{equation}
The sum operation between bra vectors satisfy the associativity and commutativity axioms. 

%

\subsubsection{Inner product}
We define the inner product as the standard multiplication of a bra by a ket. The inner product is written in the standard form by the bra standing on the left and a ket standing on the right like
\begin{equation}
\left(
\extbra{\alpha_1}{\beta_1} \right) \cdot
\left(
\extket{\alpha_2}{\beta_2} \right)
\equiv
\extinnerprod{\alpha_1}{\beta_1}{\alpha_2}{\beta_2}
\end{equation}
This product is, in general, an extended number. The inner product must satisfy the inner product axiom of extended complex space. In that case, all factor from inner product must have the same extended conjugated maps and also satisfy equations for conjugated symmetry holds \ref{conjSymmEq}. From now on, we refer only to valid inner products.

From study in the domain of the extended numbers, we can postulate the property
\begin{equation}
\extinnerprod{\alpha}{\alpha}{\alpha}{\alpha} =R \geq 0  \qquad \forall\; \alpha \in \mathbb{E}, \; R \in \mathbb{R},
\end{equation}
which resembles the Positive-definiteness property. Also, we can postulated another property, based on the relation that inner factor must satisfied:
\begin{equation}
\extinnerprod{\alpha_1}{\beta_1}{\alpha_2}{\beta_2} = \extinnerprod{\alpha_1}{\beta_1}{\alpha_2}{\beta_2}^\cbullet.
\end{equation}
The complex conjugated $()^\cbullet$ of an extended number is the operation who
puts the $\bullet$ map on the factor who do not have it and suppresses it to the ones who do have it, in the extended inner product which the extended number can be expressed. If $\alpha = (\gamma^{\bullet}  \odot \gamma^{\bullet})(\delta \odot \delta)$ then
$\alpha^\cbullet = (\gamma \odot \gamma)(\delta^{\bullet}  \odot \delta^{\bullet} )$ where $\gamma $ and $\delta$ are the complex and extended roots of the number
\begin{equation*}
\gamma = \sqrt[\bullet]{\sqrt{\alpha}} \qquad \text{ and } \qquad \delta = \sqrt[*]{\sqrt{\alpha}}.
\end{equation*}

It is important to note that at this point we cannot say that all quantum numbers satisfy the conjugate symmetry, in fact, the inner product in the domain of the extended numbers should satisfy the equation \ref{conjSymmEq}
\begin{equation*}
\extproduct{\alpha}{\beta}{\gamma}{\delta} = \extproduct{\gamma}{\delta}{\alpha}{\beta}^\cbullet.
\end{equation*}
Then, we can redefine the quantum inner product like:
\begin{equation}
\extinnerprod{\alpha}{\beta}{\gamma}{\delta} \equiv 
\extinnerprod{\alpha}{\beta}{\gamma}{\delta} \delta_D \left( 
\extproduct{\alpha}{\beta}{\gamma}{\delta} - \extproduct{\gamma}{\delta}{\alpha}{\beta}^\cbullet\right)
\end{equation} 

From the analysis of section \ref{ConjugateSymmetry}, we can also postulate that the symmetric inner products
\begin{equation}
\extinnerprod{\alpha}{\beta}{\alpha}{\beta}
\end{equation}
are complex numbers, in other words, theirs extended part is zero. We can also add to this postulate that the symmetric inner product are zero only if the complex products $(\alpha^{\bullet}  \odot \beta^{\bullet})$ or $(\alpha \odot \beta)$ are zero.

The conjugate symmetry conjugate imply then that expressions like
\begin{equation}
 \extinnerprod{\alpha}{\beta}{\alpha}{\beta} \odot
\extinnerprod{\alpha}{\beta}{\alpha}{\beta} \equiv  
\extinnerprod{\alpha}{\beta}{\alpha}{\beta}^* \cdot
\extinnerprod{\alpha}{\beta}{\alpha}{\beta} \label{extQMUnitaryCond}
\end{equation}
and 
\begin{equation}
\extinnerprod{\alpha}{\beta}{\alpha}{\beta} \oplus
\extinnerprod{\alpha}{\beta}{\alpha}{\beta} \equiv 
\extinnerprod{\alpha}{\beta}{\alpha}{\beta}^* +
\extinnerprod{\alpha}{\beta}{\alpha}{\beta} \label{extQMUnitaryCond1}
\end{equation}
are real numbers. 

About this postulate, same as in the study of extended numbers which is incomplete, we should take into account the existence of numbers $\gamma_0,\delta_0$ that satisfy equations \ref{conjSymmProd1}
and  \ref{conjSymmProd2}:
\begin{align}
(\gamma^{\bullet}_0  \odot \gamma^{\bullet}_0) &=\pm (\delta_0 \odot \delta_0)
\nonumber \\ 
(\gamma^{\bullet}_0  \odot \gamma^{\bullet}_0) \cdot (\delta_0 \odot \delta_0) &= (\gamma  \odot \gamma_0) \cdot (\delta^{\bullet}_0 \odot \delta^{\bullet}_0).
\end{align}
which are exceptions of this rule and they satisfy the conjugate symmetry being extended numbers and so its associated inner product $\extinnerprod{\gamma_0}{\gamma_0}{\delta_0}{\delta_0} $.

We also postulate, based on the named ``inner Commutative'' property described on \ref{innerCommProp}, the inner Commutative axiom for inner space. Let being four valid inner product:
\begin{equation}
\extinnerprod{\alpha_1}{\beta_1}{\gamma_2}{\delta_1},\extinnerprod{\alpha_2}{\beta_2}{\gamma_1}{\delta_2},
\extinnerprod{\alpha_1}{\beta_1}{\gamma_1}{\delta_2},\extinnerprod{\alpha_2}{\beta_2}{\gamma_2}{\delta_1}
\end{equation}
then, on the product of two inner product, equivalent factors, for example, ${\gamma_1}$ and ${\gamma_2}$, can be commuted as
\begin{equation}
\extinnerprod{\alpha_1}{\beta_1}{\gamma_1}{\delta_1} \extinnerprod{\alpha_2}{\beta_2}{\gamma_2}{\delta_2} 
=\extinnerprod{\alpha_1}{\beta_1}{\gamma_2}{\delta_1}\extinnerprod{\alpha_2}{\beta_2}{\gamma_1}{\delta_2} \label{QmInnerCommProp}
\end{equation}
if the quantum numbers that describe the state ${\gamma_1},{\gamma_2}, \delta_1,\delta_2$ satisfy the condition \ref{innerComm0}
\begin{align*}
 (\gamma_1 \odot \delta_1) (\gamma_2 \odot \delta_2) =  (\gamma_1 \odot \delta_2) (\gamma_2 \odot \delta_1)
\end{align*}
is satisfied. Also ${\alpha_1}$ and ${\alpha_2}$ numbers can be commuted on the inner product as
\begin{equation}
\extinnerprod{\alpha_1}{\beta_1}{\gamma_1}{\delta_1} 
\extinnerprod{\alpha_2}{\beta_2}{\gamma_2}{\delta_2} 
=
\extinnerprod{\alpha_2}{\beta_1}{\gamma_1}{\delta_1}
\extinnerprod{\alpha_1}{\beta_2}{\gamma_2}{\delta_2} \label{QmInnerCommProp0}
\end{equation}

if ${\alpha_1},{\alpha_2},{\beta_1},{\beta_2}$ satisfy the condition \ref{innerComm000}
\begin{align*}
 (\alpha_1^\bullet \odot \beta_1^\bullet) (\alpha_2^\bullet \odot \beta_2^\bullet) =  (\alpha_1^\bullet \odot \beta_2^\bullet) (\alpha_2^\bullet \odot \beta_1^\bullet)
\end{align*}
is satisfied. Also, because of the associative properties of kets and bra we have that
\begin{equation}
\extinnerprod{\alpha_1}{\beta_1}{\gamma_1}{\delta_1} 
\extinnerprod{\alpha_2}{\beta_2}{\gamma_2}{\delta_2} 
=
\extinnerprod{\alpha_2}{\beta_2}{\gamma_1}{\delta_1} 
\extinnerprod{\alpha_1}{\beta_1}{\gamma_2}{\delta_2}
=
\extinnerprod{\alpha_1}{\beta_1}{\gamma_2}{\delta_2} 
\extinnerprod{\alpha_2}{\beta_2}{\gamma_1}{\delta_1}.
\end{equation}

We set this axiom for inner products because the need we have to deal with fundamental relations involving the multiplication of inner products like the normalization relation that we propose in later sections.
 
There is a connection between the algebra of the space on which the theory is developed and the observed phenomena on Quantum Mechanics. One example of such connection is the uncertainty relations in quantum mechanics, which is obtained using only theorems and properties of the complex vector space: the Schwarz's lemma, and the pure real and pure complex character of the expectation value of Hermitian and anti-Hermitian operators respectively \citep{sakurai}. This relations have no analogue in classical mechanic. We also believe in the connection of the extended complex algebra and the phenomenology of a quantum theory for $n$-VMVF systems. Let us assume that the inner product is related with some quantity which may have a physical meaning. Looking at the ordinary quantum mechanics, we can represent this quantity as
\begin{equation*}
\Psi_{\alpha, \beta}(x',x'') \equiv \extinnerprod{x'}{x''}{\alpha}{\beta}.
\end{equation*}
In that case, the inner commutation property imply the exchange of quantum numbers $\alpha$ and $\gamma$ of the referred quantities like
\begin{equation*}
\Psi_{\alpha, \beta}(x',x'') \Phi_{\gamma,\delta}(x',x'') = 
\Psi_{\gamma, \beta}(x',x'') \Phi_{\alpha,\delta}(x',x'').
\end{equation*}
if the numbers satisfy
\begin{equation}
 (\alpha \odot \beta) (\gamma \odot \delta) =  
 (\alpha \odot \delta) (\gamma \odot \beta).
\end{equation}
If confirmed, this property of exchanging the quantum numbers might describe a new phenomena with no classical or standard quantum analog.

Supposing that the quantum numbers satisfy the inner commutation conditions are satisfied, we can apply the inner commutation property on the multiplication of symmetric inner products
\begin{equation}
 \extinnerprod{\alpha}{\beta}{\alpha}{\beta} \extinnerprod{\beta}{\alpha}{\beta}{\alpha}  = \extinnerprod{\alpha}{\alpha}{\alpha}{\alpha} \extinnerprod{\beta}{\beta}{\beta}{\beta}  = R  \geq 0 \qquad R \in \mathbb{R}. \label{QmInnerCommProp1}
\end{equation}
where we use the positive definiteness axiom. 

The symmetric inner products are pure complex numbers that satisfy:
\begin{equation}
\extinnerprod{\alpha}{\beta}{\alpha}{\beta}^*\extinnerprod{\alpha}{\beta}{\alpha}{\beta}\geq 0,
\qquad
\extinnerprod{\beta}{\alpha}{\beta}{\alpha}^*\extinnerprod{\beta}{\alpha}{\beta}{\alpha}\geq 0.
\label{QmInnerCommProp2}
\end{equation}

Also as complex numbers, they can be represented in its exponential form like
\begin{equation}
\extinnerprod{\alpha}{\beta}{\alpha}{\beta} \equiv \Delta = R_\Delta \exp^{\imagi \theta_\Delta},
\qquad 
\extinnerprod{\beta}{\alpha}{\beta}{\alpha} \equiv \nabla = R_\nabla \exp^{\imagi \theta_\nabla},
\end{equation}
relations \ref{QmInnerCommProp1} and \ref{QmInnerCommProp2} are satisfied for any numbers pair of extended numbers $\alpha,\beta$, if
\begin{equation}
R_\Delta \nabla =  R_\nabla \Delta^*  
\qquad \text{ and } \qquad 
\theta_\Delta + \theta_\nabla = 2\pi n \qquad \qquad n = 0,1,2... \label{QmInnerCommProp3}.
\end{equation}
These relations are the extended quantum version of the calculation obtained of extended vector space on equation \ref{innerProdPropInvRow1}.

\subsubsection{Operator}
Another important part of the theory is the observable. They are related to physical magnitudes such as position, linear momentum, the derivative of the particle mass function, etc.. and are represented in the quantum theory by arithmetical entities named operators. The operator present a bi-dimensional structure and act to over the ket from the left resulting in other ket. The upper(downer) component of the operator act over the upper(downer) component of the state vector as:
\begin{equation}
\extoperator{A}{B} \extket{\alpha}{\beta} \equiv \extket{(\mathbf{A}^\dagger \alpha)}{(\mathbf{B}\;\beta)} = \extket{\gamma}{\zeta}.
\end{equation}
The top and the bottom components of the operator are represented as $\mathbf{A}^\dagger$ and $\mathbf{B}$, and they act over the top and bottom component of the space vector respectively.

If the action of two operator over a ket result on the same ket, its said operators are equals
\begin{equation}
\extoperator{A}{B} \extket{\alpha}{\beta} = \extoperator{C}{D} \extket{\alpha}{\beta} = \extket{\gamma}{\zeta}
\qquad \therefore \qquad \extoperator{A}{B}  = \extoperator{C}{D}.
\end{equation}
Operator are said to be null if for any arbitrary ket
\begin{equation}
\extoperator{A}{B} \extket{\alpha}{\beta} = \extket{0}{0} \qquad \therefore \qquad \extoperator{A}{B}  = \extoperator{0}{0}.
\end{equation}
Operators can be added, being summation operations commutative and associative:
\begin{align}
\extoperator{A}{B}  +  \extoperator{C}{D} =  \extoperator{C}{D} + \extoperator{A}{B} 
\nonumber \\
\left[\extoperator{A}{B}  +  \extoperator{C}{D} \right]+ \extoperator{E}{F} = \extoperator{A}{B}  +  \left[ \extoperator{C}{D} + \extoperator{E}{F} \right]
\end{align}
Operator we treat in here are also linear:
\begin{equation}
\extoperator{A}{B} \left[ \extscalar{a_1}{b_1} \extket{\alpha_1}{\beta_1} + \extscalar{a_2}{b_2} \extket{\alpha_2}{\beta_2} \right]=  \extscalar{a_1}{b_1} \extoperator{A}{B}  \extket{\alpha_1}{\beta_1} + \extscalar{a_2}{b_2} \extoperator{A}{B}  \extket{\alpha_2}{\beta_2}
\end{equation}
Also, the sum of two linear operators operating on an arbitrary ket produce the sum of each operator acting over the ket like:
\begin{equation}
\left [ \extoperator{A}{B}  +  \extoperator{C}{D} \right] \extket{\alpha}{\beta}  =
\extoperator{A}{B}  \extket{\alpha}{\beta} +  \extoperator{C}{D}  \extket{\alpha}{\beta} 
\end{equation}

The last property implies that each component of the operator must satisfy some linearity related to the inner factors of the scalar pair. From quantum mechanic, we can see the relation between the linearity of scalars and operators more clear. Let us have a Hamiltonian which a sum of two terms $\mathcal{H}_T = \mathcal{H}_1 + \mathcal{H}_2$. If the total Hamiltonian act over a energy state ket, we have the same ket plus the total energy eigenvalue
\begin{equation}
\mathcal{H}_T \ket{E} = E_T \ket{E}
\end{equation}

The linearity imply
\begin{equation}
\mathcal{H}_T \ket{E} = (\mathcal{H}_1  + \mathcal{H}_2)\ket{E} = \mathcal{H}_1 \ket{E} + \mathcal{H}_2 \ket{E} = E_1 \ket{E} +  E_2 \ket{E},
\end{equation}
which led to $E_T = E_1 + E_2$ which is consistent with the linear character of the energy.

By the same analysis, we can restrict an extended operator which represent physical observables to satisfy the same relations that scalars, equation \ref{extLinearitySumExpress}. Then we have that:
\begin{align}
\extoperator{(A_1 + A_2)}{(B_1 + B_2)} &= \extoperator{A_1}{B_1} + \extoperator{A_1}{B_2} + \extoperator{A_2}{B_1} + \extoperator{A_2}{B_2}
\nonumber \\
 &+ \extoperator{(\sqrt[*]{\mathbf{D}^{(A_1 + A_2)}(A_1,A_2,B_1 + B_2) })}{(\sqrt[*]{\mathbf{D}^{(A_1 + A_2)}(A_1,A_2,B_1 + B_2)})} \label{extLinearSumOpe}
\end{align}
where operator $ \extoperator{(\sqrt[*]{\mathbf{D}^{(A_1 + A_2)}(A_1,A_2,B_1 + B_2) })}{(\sqrt[*]{\mathbf{D}^{(A_1 + A_2)}(A_1,A_2,B_1 + B_2)})} $ satisfy
\begin{equation}
 \extoperator{(\sqrt[*]{\mathbf{D}^{(A_1 + A_2)}(A_1,A_2,B_1 + B_2) })}{(\sqrt[*]{\mathbf{D}^{(A_1 + A_2)}(A_1,A_2,B_1 + B_2)})} \extket{a_1 + a_2}{b_1 + b_2}  = 
\extscalar{\sqrt[*]{\mathcal{D}^{(a_1 + a_2)}(a_1,a_2,b_1 + b_2)}}
{\sqrt[*]{\mathcal{D}^{(a_1 + a_2)}(a_1,a_2,b_1 + b_2)}}
\extket{a_1 + a_2}{b_1 + b_2} 
\end{equation}
where $\mathcal{D}^{(a_1 + a_2)}(a_1,a_2,a_3)$ function is the term appearing on the distributivity property of extended numbers, equation \ref{extDfuntion}, $\extscalar{a_1}{b_1}$ and $\extscalar{a_2}{b_2}$ are the extended eigenvalues for operators $\extoperator{A_1}{B_1} $ and $\extoperator{A_2}{B_2}$ respectively.

Keeping the same reasoning of operator should follow the scalars algebra we set the property:
\begin{align}
\extoperator{(A_1  A_2)}{(B_1  B_2)} &= \extoperator{A_1}{B_1} \extoperator{A_2}{B_2} + \extoperator{(\sqrt[*]{\mathbf{F}(A_1,A_2,B_1,B_2) }}{(\sqrt[*]{\mathbf{F}(A_1,A_2,B_1,B_2) }} \label{extLinearProdOpe}
\end{align}
where operator $ \extoperator{(\sqrt[*]{\mathbf{F}(A_1,A_2,B_1,B_2) }}{(\sqrt[*]{\mathbf{F}(A_1,A_2,B_1,B_2)}} $ satisfy
\begin{equation}
\extoperator{(\sqrt[*]{\mathbf{F}(A_1,A_2,B_1,B_2) }}{(\sqrt[*]{\mathbf{F}(A_1,A_2,B_1,B_2)}} \extket{a_1 a_2}{b_1 b_2}  = 
\extscalar{\sqrt[*]{\mathcal{F}(a_1,a_2,a_1,a_2) }}
{\sqrt[*]{\mathcal{F}(a_1,a_2,a_1,a_2) }}
\extket{a_1  a_2}{b_1  b_2} 
\end{equation}
being $\mathcal{F}(A_1,A_2,B_1,B_2)$ the function defined on \ref{extFfuntion}, which we show is zero if $a_i, b_i$ are pure complex numbers.

An operator act to bra vector from the right side returning a new bra:
\begin{equation}
 \extbra{\alpha}{\beta} \extoperator{A}{B} = \extbra{\gamma}{\zeta}.
\end{equation}
The dual correspondent of an operator is defined as the extended adjoint operator
\begin{equation}
\extoperator{A}{B} \extket{\alpha}{\beta} \leftrightarrow \extbra{\alpha}{\beta} \extoperator{A}{B}^\ddagger \equiv  \extbra{\alpha}{\beta} \extcolumn{\mathbf{A}^{\dagger \ddagger}\;}{\mathbf{B}^{ \ddagger}}.
\end{equation}
Operators have other properties as: 
\begin{itemize}
\item In general operators are noncommutative:
\begin{equation}
\extoperator{A}{B}  \extoperator{C}{D} \neq  \extoperator{C}{D} \extoperator{A}{B} 
\end{equation}
\item The standard multiplication between operators is associative
\begin{equation}
\left[\extoperator{A}{B}  \extoperator{C}{D} \right] \extoperator{E}{F} = \extoperator{A}{B}  \left[\extoperator{C}{D}  \extoperator{E}{F}\right]
\end{equation}
\item  The standard multiplication between operators and vector bra and ket is associative
\begin{equation}
\extoperator{A}{B}  \left[\extoperator{C}{D} \extket{\alpha}{\beta} \right] =
\left[ \extoperator{A}{B} \extoperator{C}{D} \right] \extket{\alpha}{\beta} =
\extoperator{A}{B} \extoperator{C}{D} \extket{\alpha}{\beta}
\end{equation}
and
\begin{equation}
\left[ \extbra{\alpha}{\beta} \extoperator{C}{D} \right]\extoperator{A}{B} =
\extbra{\alpha}{\beta} \left[ \extoperator{C}{D}\extoperator{A}{B} \right] =
\extbra{\alpha}{\beta} \extoperator{C}{D} \extoperator{A}{B}
\end{equation}
\item Using the correspondence between bras and kets and the associative axiom, the adjoint of a product of operators can be determined as
\begin{align}
\extoperator{A}{B}  \left[\extoperator{C}{D} \extket{\alpha}{\beta} \right]
\leftrightarrow 
\left[ \extbra{\alpha}{\beta} \extoperator{C}{D}^\ddagger \right]\extoperator{A}{B}^\ddagger,
\end{align}
from where can be extracted the relation
\begin{align}
\extoperator{A}{B} \extoperator{C}{D} 
\leftrightarrow
\extoperator{C}{D}^\ddagger \extoperator{A}{B}^\ddagger, \label{extOperProdDualRel}
\end{align}
\end{itemize}
\subsection{Expectation Value and normalization relations for physical states} \label{NormExpValueSection}
	
We postulate the normalization relations for inner products when vectors represent physical states. The point of view discussed on section \ref{SectNorm}, consider this postulate as a normalization relation, rather than a postulate or axiom. On that section, we propose that the well-known quantum mechanic postulate which imposes the normalization condition for physical states is related to the expectation value of operator Identity $\mathbf{I}$ stated as
\begin{quote}
``The expectation value of the operator identity defined on a vector space \textbf{1} must be 1 for any physical vector of such space''.
\end{quote}
We apply the same reasoning to the extended domain. 

What is exposed below is entirely intuitive and only the solution of problems and the verification with experiment can validate the arguments exposed in here.

For applying this type of reasoning, the concepts of normalization and expectation value must be consistently defined. The definition of the expectation value of the quantum operators and the normalization condition for any extended vector that represent physical states must take into consideration that:
\begin{enumerate}
\item The expectation value relations must be sufficient to fully estimate the physical magnitudes represented by a quantum operator. The bi-dimensional structure of the form of an operator acting over a ket
\begin{equation}
\extoperator{A}{B} \extket{\alpha}{\beta} = \extket{\gamma}{\zeta}.
\end{equation}
suggest that we need two normalization relations for the extended operator. For example, the position operator $\extoperator{x_T}{x_R}$ have two physical quantities: the vector position obtained by translation and the vector position obtained by rotation. Their values should be measured at any quantum state, and the expectation values relations are the relations needed for that  propose. 

\item The expectation value for an arbitrary operator that represents an observable should return a scalar, so the relations involve extended numbers as numerical values. The primary motivation to define the extended vector space is to describe processes for $n$-VMVF systems where those degrees of freedom remain on extended Lagrange(or Hamiltonian) equations as unknown quantities. Because of that, physical magnitudes as linear and angular momentum will depend on these quantities, and they can have different values from their typical values: for example, they can be complex numbers.

\item The expectation value of an extended operator $\mathbf{A}$ in a physical vector $V$ must be a function, named $\mathcal{F}_i$, depending on inner products. This dependency is because the inner products are the only defined algebraic operation between vectors and operators that return a scalar:
\begin{equation}
\langle \mathbf{A}\rangle_i = \mathcal{F}_i([V \times (\mathbf{A} V)])= \mathcal{F}_i([(V \mathbf{A} ) \times V]) \to F_i, \qquad F_i \in \mathbb{E}
\end{equation}
where we use the notation equation \ref{innerProdRepr}.
\item The expectation value of an operator, in whatever the form it is defined, are related to the complex product between the two components of the operator. For example, the inner product between a position state and the resulting ket of the position operator acting over a position state vector is equal to the standard product of the complex product of both the position values and an inner product like:
\begin{equation}
\extaverage{x'_T}{x'_R}{x_T}{x_R}{x'_T}{x'_R} = \extcolumn{x'_T}{x^{'}_R} \extinnerprod{x'_T}{x'_R}{x'_T}{x'_R} \equiv x'_T \odot x^{'}_R \extinnerprod{x'_T}{x'_R}{x'_T}{x'_R}.
\end{equation}
In that case, being the expectation value a function of extended inner products then they depend then on the expectation values of the components $x'_T \odot x^{'}_R$.
\item  The expectation values relations must converge to relations of symmetric inner products when the sample operator is the identity operator  $\mathbf{I}$, the result of the relations, whatever they are, should be real and also need to be consistent with the normalization relations for the state vector: 
\begin{equation}
\langle \mathbf{I}\rangle_i = \mathcal{F}_i([V \times (\mathbf{I} V)]) = \mathcal{F}_i([V \times V]) \to R_i,
\end{equation}
where $R_i$ is a real number.
\end{enumerate}

Using the properties of the extended inner product, we propose two independent relations for functions  $\mathcal{F}_i$ depending on the extended operator $\extoperator{A}{B}$ and the state vector $\extket{\alpha}{\beta}$:
\begin{equation}
\mathcal{F}_1([V \times (\mathbf{A} V)] )\equiv \extaverage{\alpha}{\beta}{A}{B}{\alpha}{\beta}
\odot
\extaverage{\alpha}{\beta}{A}{B}{\alpha}{\beta}  \label{expValueDef1}
\end{equation}
and 
\begin{equation}
\mathcal{F}_2([V \times (\mathbf{A} V)])\equiv \extaverage{\alpha}{\beta}{A}{B}{\alpha}{\beta} \oplus \extaverage{\alpha}{\beta}{A}{B}{\alpha}{\beta}. \label{expValueDef2}
\end{equation}
These two functions set the relations needed for obtaining the expectation values of the component of the operator in any state. 

The proposition of relations \ref{expValueDef1} and \ref{expValueDef2} is based on the fact that for identity operator, they converge to relations involving only symmetric inner products like
\begin{equation}
\mathcal{F}_1([V \times  (\mathbf{I} V)] )= \extinnerprod{\alpha}{\beta}{\alpha}{\beta} \odot \extinnerprod{\alpha}{\beta}{\alpha}{\beta}\label{expValueDef3}
\end{equation}
and 
\begin{equation}
\mathcal{F}_2([V \times  (\mathbf{I} V)])= \extinnerprod{\alpha}{\beta}{\alpha}{\beta} \oplus \extinnerprod{\alpha}{\beta}{\alpha}{\beta}. \label{expValueDef4}
\end{equation}
As product $\extinnerprod{\alpha}{\beta}{\alpha}{\beta}$ is a complex number the relation can be replaced by 
\begin{equation}
\mathcal{F}_1([V \times  (\mathbf{I} V)] )= \extinnerprod{\alpha}{\beta}{\alpha}{\beta}^* \extinnerprod{\alpha}{\beta}{\alpha}{\beta}
\end{equation}
and 
\begin{equation}
\mathcal{F}_2([V \times  (\mathbf{I} V)])= \extinnerprod{\alpha}{\beta}{\alpha}{\beta}^* + \extinnerprod{\alpha}{\beta}{\alpha}{\beta},
\end{equation}
so because of that, both proposed relations will result on a real number.

Expressions \ref{expValueDef3} and \ref{expValueDef4} set consistent relations for the identity operator. As the inner product is proportional to the complex product between the expected value of the components of the operator, the relations \ref{expValueDef1} and \ref{expValueDef2} for the Identity operator should have an approximated form like 
\begin{equation}
\extaverage{\alpha}{\beta}{I}{I}{\alpha}{\beta} \odot \extaverage{\alpha}{\beta}{I}{I}{\alpha}{\beta} 
= (\langle 1 \rangle \odot \langle 1 \rangle) \odot (\langle 1 \rangle \odot \langle 1 \rangle)
= \langle 1 \rangle \langle 1 \rangle \langle 1 \rangle \langle 1 \rangle \label{expValueDef5}
\end{equation}
and 
\begin{equation}
\extaverage{\alpha}{\beta}{I}{I}{\alpha}{\beta} \oplus \extaverage{\alpha}{\beta}{I}{I}{\alpha}{\beta}
=\langle 1 \rangle \odot \langle 1 \rangle +\langle 1 \rangle \odot \langle 1 \rangle= 
\langle 1 \rangle \langle 1 \rangle +\langle 1 \rangle \langle 1 \rangle,\label{expValueDef6}
\end{equation}
where we, intuitively, represent the expectation value of each components of the unitary operator like $\langle 1 \rangle$. Our understanding is that relations \ref{expValueDef5} and \ref{expValueDef6} should have the resulting values of 1 and 2, respectively.

According to previous discussion, we propose then two normalization conditions for an arbitrary state vector $\extket{\alpha}{\beta}$:
\begin{equation}
\extinnerprod{\alpha}{\beta}{\alpha}{\beta} \odot \extinnerprod{\alpha}{\beta}{\alpha}{\beta} =1 \label{extQMNormalCond}
\end{equation}
\begin{equation}
\frac{1}{2}\left( \extinnerprod{\alpha}{\beta}{\alpha}{\beta} \oplus \extinnerprod{\alpha}{\beta}{\alpha}{\beta}\right) =1 \label{extQMNormalCond1}
\end{equation}
These relations substitute the well-known postulate $\langle \alpha|\alpha \rangle=1$ in the standard quantum mechanic which is related to its probability character. In the studied case, however, we cannot based this postulated on the conservation of probability since, to the best of our knowledge, the physical interpretation of the phenomena, including variable particle masses and field where length scale is in the order to its de Broglie wavelength, is unclear.

The expectation value relations are defined then as
\begin{equation}
\biaverage{A}{B} \odot \biaverage{A}{B} \equiv \extaverage{\alpha}{\beta}{A}{B}{\alpha}{\beta}
\odot
\extaverage{\alpha}{\beta}{A}{B}{\alpha}{\beta} \label{expValueDef}
\end{equation}
and
\begin{equation}
\biaverage{A}{B}  \oplus \biaverage{A}{B} \equiv \extaverage{\alpha}{\beta}{A}{B}{\alpha}{\beta}
\oplus
\extaverage{\alpha}{\beta}{A}{B}{\alpha}{\beta} \label{expValueDef0},
\end{equation}
where value $\biaverage{A}{B}$ should be related to the measured quantities like 
\begin{equation}
\biaverage{A}{B} \equiv \expval{\mathbf{A}} \odot \expval{\mathbf{B}}.
\end{equation}
That is because the averaged values of the operator must tend to the eigenvalues relations if the state vector is one of its eigenvectors, so we extend the concept to all the quantities of the theory.

Being the symmetric extended inner product a complex number, it can be represented as
\begin{equation*}
 \extinnerprod{\alpha}{\beta}{\alpha}{\beta} = R_\Delta e^{i\theta_\Delta}.
\end{equation*}
From the first normalization relation \ref{extQMNormalCond}, we can obtain the absolute value of the vector $R_\Delta$
\begin{equation}
\extinnerprod{\alpha}{\beta}{\alpha}{\beta} \odot \extinnerprod{\alpha}{\beta}{\alpha}{\beta} =
\extinnerprod{\alpha}{\beta}{\alpha}{\beta}^* \extinnerprod{\alpha}{\beta}{\alpha}{\beta} = |R_\Delta|^2 =1, \; \therefore R_\Delta =1.
\end{equation}
Also, from the first normalization condition from the inner product
\begin{equation*}
 \extinnerprod{\beta}{\alpha}{\beta}{\alpha} = R_\nabla e^{i\theta_\nabla},
\end{equation*}
we obtain $ R_\nabla = 1$. If we substitute these results on equation \ref{QmInnerCommProp3}:
\begin{equation*}
R_\Delta \extinnerprod{\beta}{\alpha}{\beta}{\alpha} =  R_\nabla \extinnerprod{\alpha}{\beta}{\alpha}{\beta}^*, 
\end{equation*}
we obtain
\begin{equation}
\extinnerprod{\beta}{\alpha}{\beta}{\alpha} = \extinnerprod{\alpha}{\beta}{\alpha}{\beta}^* \label{QmInnerCommProp3_Norm}
\end{equation}

If state vectors $\extket{\alpha}{\beta}$ and $\extket{\beta}{\alpha}$ satisfy the inner commutation conditions, the normalization relations are the same for both vectors and they are
\begin{equation}
\extinnerprod{\alpha}{\beta}{\alpha}{\beta} \extinnerprod{\beta}{\alpha}{\beta}{\alpha} =1 \label{extQMNormalCondR_1}
\end{equation}
and
\begin{equation}
\frac{1}{2}\left( \extinnerprod{\alpha}{\beta}{\alpha}{\beta} + \extinnerprod{\beta}{\alpha}{\beta}{\alpha} \right) =1. \label{extQMNormalCond1R_1}
\end{equation}

\subsection{Eigenvalues and eigen-states}
The action of an operator over an extended ket results, in general, in a different ket. However, there are some particular cases where the action of a specific operator over a state vector results on the same ket times a scalar. Those ket are known as eigenkets, and the scalars pair are named eigenvalues. We can express the previous statement as:
\begin{equation}
\extoperator{A}{B} \extket{a'}{b'} = \extscalar{a'}{b'} \extket{a'}{b'}.
\end{equation}
On the bra space the relation have the form
\begin{equation}
 \extbra{a''}{b''}\extoperator{A}{B}^\ddagger  = \extscalarBra{a''}{b''} \extbra{a''}{b''}.
\end{equation}
In this section, we study the relations between the eigenvalues of operators with this characteristics. 

By left multiplying both sides of the first relation by $\extbra{a''}{b''}$, right multiplying by $\extket{a'}{b'}$ on both sides of the second relation, and subtract them, we obtain:
\begin{equation}
 \extbra{a''}{b''}\extoperator{A}{B}^\ddagger - \extoperator{A}{B} \extket{a'}{b'} =\left[  \extscalarBra{a''}{b''} - \extscalar{a'}{b'} \right] \extinnerprod{a''}{b''}{a'}{b'}. \label{eigenEq1}
\end{equation}

In our previous analysis about the axioms for inner products, we saw that the symmetric inner products $\extinnerprod{\alpha}{\beta}{\alpha}{\beta} $ and $\extinnerprod{\beta}{\alpha}{\beta}{\alpha}$ satisfy the relation \ref{QmInnerCommProp3_Norm}:
\begin{equation*}
\extinnerprod{\beta}{\alpha}{\beta}{\alpha} = \extinnerprod{\alpha}{\beta}{\alpha}{\beta}^* 
\end{equation*}
if the quantum numbers satisfy the conditions for the inner commutation property. In this case, the normalization relations for the state vector $\extket{\alpha}{\beta}$ is equal to the relations for the state vector $\extket{\beta}{\alpha}$. 
Let us designate the state vector $\extket{\beta}{\alpha}$ as the inverted state vector of $\extket{\alpha}{\beta}$ and vice versa.

That gives us the idea that a physical state can be related not only to the vector $\extket{\alpha}{\beta}$ but also to its inverted vector $\extket{\beta}{\alpha}$. Also, operator $\extoperator{A}{B}$ should be related to the inverted operator $\extoperator{B}{A}$. From our point of view, there is no reason for a physical state being represented with an extended vector or with its inverted's.
For example, there is no physical restriction for representing the position operator acting over the position space vector like
\begin{equation}
\extoperator{x_T}{x_R} \extket {x'_T}{x'_R}  \qquad \text{or} \qquad \extoperator{x_R}{x_T} \extket {x'_R}{x'_T},
\end{equation}
except that the operator component $\mathbf{x_T}$, $\mathbf{x_R}$ must act over the ${x_T}$, ${x_R}$ component of the ket, respectively. Because of that, the eigenvalue of the physical magnitude should have the same value for the states and the operators being represented in one way or the other.

We propose the study also for the inverted vector. In that case, we can verify the relation
\begin{equation}
\extoperator{B}{A} \extket{b'}{a'} = \extscalar{b'}{a'} \extket{b'}{a'}.
\end{equation}
and
\begin{equation}
 \extbra{b''}{a''}\extoperator{B}{A}^\ddagger  = \extscalarBra{b''}{a''} \extbra{b''}{a''}.
\end{equation}

The subtraction of the relation obtained from the left multiplication on both sides by $\extbra{b''}{a''}$ on the first relation, and the relation coming from right multiplying by $ \extket{b'}{a'}$ on both members of the second relation results in:
\begin{equation}
 \extbra{b''}{a''}\extoperator{B}{A}^\ddagger - \extoperator{B}{A} \extket{b'}{a'} 
 =\left[  \extscalarBra{b''}{a''} - \extscalar{b'}{a'} \right] \extinnerprod{b''}{a''}{b'}{a'} \label{eigenEq2}
\end{equation}

We, similar than standard quantum mechanics, study extended operators that satisfy:
\begin{equation}
\extoperator{A}{B}^\ddagger = \extoperator{A}{B} \qquad \text{and} \qquad \extoperator{B}{A}^\ddagger = \extoperator{B}{A}. \label{eigenOprProp}
\end{equation}
Both relations are independent. Also, as the action of every component of the operator over the corresponding component on the state vector is independent of the action of the other component, we think we can go further and assume that
\begin{equation}
\mathbf{A} = \mathbf{A}^\ddagger, \qquad
\mathbf{A}^\dagger = (\mathbf{A}^\dagger)^{ \ddagger}, \qquad
\mathbf{B} = \mathbf{B}^\ddagger \qquad 
\mathbf{B}^\dagger =  (\mathbf{B}^\dagger)^{ \ddagger}
\label{eigenOprProp0}
\end{equation}
which is a particular case of equations \ref{eigenOprProp}. We name the operators whose components satisfy relations \ref{eigenOprProp0} as the extended Hermitian operators. Because of the eigenvalues properties of this kind of operators, we can state that the physical magnitudes from the classical problem of the $n$-VMVF systems are represented in the quantum theory by extended hermitian operators.

For the extended hermitian operators, the left members of equations \ref{eigenEq1} and \ref{eigenEq2} are zero: 
\begin{align}
\left[  \extscalarBra{a''}{b''} - \extscalar{a'}{b'} \right] \extinnerprod{a''}{b''}{a'}{b'}&\equiv \Big[ (a''^\bullet \odot b''^{\bullet}) - (a' \odot b')  \Big ] \extinnerprod{a''}{b''}{a'}{b'} = 0
\nonumber \\
\left[  \extscalarBra{b''}{a''} - \extscalar{b'}{a'} \right] \extinnerprod{b''}{a''}{b'}{a'} &= \Big[ (b''^\bullet \odot a''^{\bullet}) - (b'\odot a')  \Big ] \extinnerprod{b''}{a''}{b'}{a'} = 0.\label{eigenMainEq}
\end{align}
There are four options for numbers $a'$, $b'$, $a''$ and $b''$:
\begin{enumerate}\label{eigenMainEqCases}
\item $a'=a''$ 		and $b'=b''$
\item $a'=a''$ 		and $b'\neq b''$
\item $a' \neq a''$ and $b'=b''$.
\item $a'\neq a''$ 	and $b'\neq b''$ 
\end{enumerate}

\begin{itemize}
\item 
Let suppose the first case where coefficients are equals. In ordinary quantum mechanics, using the positive-definiteness axiom for complex inner products, its shown that symmetric complex inner product of physical states $\langle a' | a' \rangle$ is different from zero. This fact allows proving the real property of for eigenvalue of a Hermitian operator. We presume that extended vectors that represent physical states should satisfy that its symmetric inner product condition is not null. For example, if we analyze the inner product like
\begin{equation}
\extbra {x'_T}{x'_R} \extoperator{x_T}{x_R} \extket {x'_T}{x'_R}  = \extscalar{x'_T}{x'_R} \extinnerprod{x'_T}{x'_R}{x'_T}{x'_R}
\end{equation}
which corresponds to the expectation value of the translational and rotational positions. We can expect that the expectation value should exist for all values of ${x'_T},{x'_R}$. If we apply this analysis to all physical objects, we can assume the non-nullity of symmetric inner products. This assumption should be considered appropriate, but also we can associate the non-nullity property for symmetric inner products to the non-nullity of its extended and complex roots as shown in equation \ref{extRootsDef1} assuming any inner product can be expressed using  its complex and extended roots like 
\begin{equation}
\extinnerprod{\alpha}{\beta}{\alpha}{\beta} = \extinnerprod{\gamma}{\gamma}{\delta}{\delta}.
\end{equation}

We assume then that symmetric inner products
\begin{equation}
\extinnerprod{a'}{b'}{a'}{b'} \neq 0 \qquad \extinnerprod{b'}{a'}{b'}{a'} \neq 0.
\end{equation}

If inner products on equations \ref{eigenEq1} and \ref{eigenEq2} are different to zero then
\begin{align}
(a'^\bullet \odot b'^{\bullet}) - (a' \odot b')&=0
\nonumber \\
(b'^\bullet \odot a'^{\bullet}) - (b'\odot a') &=0. \label{eigenEq4}
\end{align}
Substituting the extended numbers with $a' = a'_E \extk + a'_I$, $b' = b'_E \extk + b'_I$ and setting equals the imaginary  and extended part of equations \ref{eigenEq4} we have:
\begin{align}
&a^{'*}_E b^{'}_E \alpha_1 + a^{'*}_E b^{'}_I \beta_1 + a^{'*}_I b^{'}_E \gamma_1 = 0
\nonumber \\
&a^{'*}_E b^{'}_E \alpha_2 + a^{'*}_E b^{'}_I \beta_2 + a^{'*}_I b^{'}_E \gamma_2 = 0
\nonumber \\
&a^{'*}_E b^{'}_E \alpha_3^* + a^{'*}_E b^{'}_I \beta_3^* + a^{'*}_I b^{'}_E \gamma_3^* = 0
\nonumber \\
&a^{'*}_E b^{'}_E \alpha_4^* + a^{'*}_E b^{'}_I \beta_4^* + a^{'*}_I b^{'}_E \gamma_4^* = 0 \label{eigenEq5}
\end{align}
where
\begin{align}
&\alpha_1 = z_2^{(a')^*} w_2^{(b')} z_1^{(a'^\bullet)} + w_2^{(a')^*} z_2^{(b')} , 
\qquad 
\nonumber \\
&\alpha_2 =  z_2^{(a')^*} w_2^{(b')} w_1^{(a'^\bullet)} + \imagi z_2^{(a')^*} z_2^{(b')} + w_2^{(a')^*} w_2^{(b')} - \imagi, 
\qquad 
\nonumber \\
&\alpha_3 =  z_2^{(b')^*} w_2^{(a')} z_1^{(b'^\bullet)} + w_2^{(b')^*} z_2^{(a')} , 
\qquad 
\nonumber \\
&\alpha_4 =  z_2^{(b')^*} w_2^{(a')} w_1^{(b'^\bullet)} + \imagi z_2^{(b')^*} z_2^{(a')} + w_2^{(b')^*} w_2^{(a')} - \imagi,
\end{align} 
\begin{align}
&\beta_1 = z_2^{(a')^*} z_1^{(a'^\bullet)}  - z_1^{(a')}  , 
\qquad 
\nonumber \\
&\beta_2 = z_2^{(a')^*} w_1^{(a'^\bullet)} + w_2^{(a')^*}  - w_1^{(a')}, 
\qquad 
\nonumber \\
&\beta_3 = z_2^{(b')^*} z_1^{(b'^\bullet)}  - z_1^{(b')}  , 
\qquad 
\nonumber \\
&\beta_4 = z_2^{(b')^*} w_1^{(b'^\bullet)} + w_2^{(b')^*}  - w_1^{(b')}, 
\end{align} 
\begin{align}
&\gamma_1 = z_2^{(b')} -1
\qquad 
\nonumber \\
&\gamma_2 = w_2^{(b')}
\qquad 
\nonumber \\
&\gamma_3 = z_2^{(a')} -1
\qquad 
\nonumber \\
&\gamma_4 = w_2^{(a')}.
\end{align}

The set of four equations must hold for every possible extended conjugated maps for numbers $a^{'}$ and $b^{'}$ or what its the same, for any value of parameters $\alpha_i, \beta_i, \gamma_i$, where $i=1,2,3,4$. 

From the first equation we have
\begin{equation}
b^{'}_I = - \frac{(a_E^{'*}\alpha_1 + a_I^{'*}\gamma_1)b_E}{a_E^{'*} \beta_1},
\end{equation}
which, substituted on the others equations results in
\begin{align}
\Big [ a_E^{'*}(\alpha_2 - \frac{\alpha_1 \beta^*_1 \beta_2}{|\beta_1|^2}) + 
a_I^{'*}(\gamma_2 - \frac{\gamma_1 \beta^*_1 \beta_2}{|\beta_1|^2})
\Big ] b^{'}_E = 0
\nonumber \\
\Big [ a_E^{'*}(\alpha_3 - \frac{\alpha_1 \beta^*_1 \beta_3^*}{|\beta_1|^2}) + 
a_I^{'*}(\gamma_3 - \frac{\gamma_1 \beta^*_1 \beta_3^*}{|\beta_1|^2})
\Big ] b^{'}_E = 0
\nonumber \\
\Big [ a_E^{'*}(\alpha_4 - \frac{\alpha_1 \beta^*_1 \beta_4^*}{|\beta_1|^2}) + 
a_I^{'*}(\gamma_4 - \frac{\gamma_1 \beta^*_1 \beta_4^*}{|\beta_1|^2})
\Big ] b^{'}_E = 0.
\end{align}
We have two possible solutions to this complex system of equations. On one side we have $b'_E=0$ or the equation system
\begin{align*}
a_E^{'*}(\alpha_2 - \frac{\alpha_1 \beta^*_1 \beta_2}{|\beta_1|^2}) + 
a_I^{'*}(\gamma_2 - \frac{\gamma_1 \beta^*_1 \beta_2}{|\beta_1|^2})
 = 0
\nonumber \\
 a_E^{'*}(\alpha_3 - \frac{\alpha_1 \beta^*_1 \beta_3^*}{|\beta_1|^2}) + 
a_I^{'*}(\gamma_3 - \frac{\gamma_1 \beta^*_1 \beta_3^*}{|\beta_1|^2})
 = 0
\nonumber \\
 a_E^{'*}(\alpha_4 - \frac{\alpha_1 \beta^*_1 \beta_4^*}{|\beta_1|^2}) + 
a_I^{'*}(\gamma_4 - \frac{\gamma_1 \beta^*_1 \beta_4^*}{|\beta_1|^2}) = 0.
\end{align*}
The last equations system is unsolvable for every $\alpha_i, \beta_i, \gamma_i$ number since its compose of 3 independent equations with two variables. That left us with the solution $b'_E=0$. We substitute this value on equations of the initial system \ref{eigenEq5} and obtain the set of equations
\begin{equation}
 a^{'*}_E b^{'}_I \beta_i = 0 \qquad \forall \; i=1..4.
\end{equation}
The only possible solution for any value of $\beta_i$ is $a^{'*}_E=0$ since $b^{'}_I= b^{'}_E=0$ is referred to the case $b^{'}=0$ which contradict initial preposition $b^{'}\neq0$. Then, eigenvalues for physical observables are complex numbers or what is the same $a^{'*}_E=b^{'*}_E=0$.

Also, the fact that eigenvalues are pure complex numbers set the expressions
\begin{equation*}
(a''^\bullet \odot b''^{\bullet}) - (a' \odot b')  \equiv a''^* b'' - a'^* b'
\end{equation*}
\qquad \text{and} \qquad
\begin{equation*}
(b''^\bullet \odot a''^{\bullet}) - (b'\odot a')  \equiv b''^* a'' - b'^* a' 
\end{equation*}
linearly dependents, since one is the complex conjugated of the other. This an expected result since if eigenvalues are complex numbers, the inner commutation property for $\extinnerprod{a''}{b''}{a'}{b'}$ and $\extinnerprod{b''}{a''}{b'}{a'}$ is automatically satisfied. So we can work using one of the two case from above.

\item 
For the cases of \ref{eigenMainEqCases} where the eigenvalue pair have one component equal
\begin{enumerate}
\item $a'=a''$ 		and $b'\neq b''$
\item $a' \neq a''$ and $b'=b''$,
\end{enumerate}
we have that the correspondent inner product is zero
\begin{equation}
\extinnerprod{a'}{b''}{a'}{b'} =\extinnerprod{b''}{a'}{b'}{a'} =0, \qquad \extinnerprod{a''}{b'}{a'}{b'} = \extinnerprod{b'}{a''}{b'}{a'} = 0
\end{equation}
respectively. Indeed, the first equation from \ref{eigenMainEq}, considering eigenvalues as complex numbers and using the $a'=a''$ case as example, has the form 
\begin{align}
\Big[  a''^{*} b'' - a'^{*}b'   \Big ] \extinnerprod{a''}{b''}{a'}{b'} &=  \Big[ a'^{*}(b'' - b') \Big ]\extinnerprod{a''}{b''}{a'}{b'} = 0.
\label{eigenMainEqImg}
\end{align}
Expressions $a' (b''^{*} - b'^{*})$ must be different from zero since the expression $(b''^{*} - b'^{*})$ cannot vanish, by assumption. Then, the related inner products are zero:
\begin{equation}
 \extinnerprod{b''}{a''}{b'}{a'} = \extinnerprod{a''}{b''}{a'}{b'} = 0.  
\end{equation}
\item 
In the last case of options \ref{eigenMainEqCases}, where $a'$, $b'$, $a''$ and $b''$ are different, the relations for the eigenvalues may be equal or not to zero:
\begin{equation}
a''b''^{*} - a' b'^{*} = 0 \qquad\text{or} \qquad a''b''^{*} - a' b'^{*} \neq 0. \label{eigenEq7}
\end{equation}

For the case where the components of the eigenvalue satisfy
\begin{equation}
a''b''^{*} - a' b'^{*} \neq 0,
\end{equation}
the solution is straightforward and the correspondent inner products are then
\begin{align}
\extinnerprod{a''}{b''}{a'}{b'} = 0,
\qquad 
\extinnerprod{b''}{a''}{b'}{a'} = 0.
\end{align}
\end{itemize}

For study the case where $a''b''^{*} - a' b'^{*} = 0$, we need to look forward in the features of eigenvalues. There is no a priori reason, at least an algebraic one, to assume other properties for the eigenvalues of the studied operators. According to the algebra for extended numbers, they satisfy in general:
\begin{equation*}
\extscalar{a' + a''}{b'+ b''} = \extscalar{a'}{ b'} + \extscalar{a''}{ b''} + \extscalar{a''}{ b'} + \extscalar{a'}{ b''} + \mathcal{D}^{(a'' + a')}(a',a'',b'' + b').
\end{equation*}
where $\mathcal{D}^{(a'' + a')}(a',a'',b'' + b')=0$ because of the complex property of eigenvalues. All these results are obtained from algebraic properties and the fact they represent values of physical quantities and therefore are complex numbers. However, except the last statement, very few have been said about the physical meaning of the extended eigenvalues and specifically, about its bi-dimensional form.

To study the significance of the bidimensional form of the eigenvalue, we need to recall that the components of the extended scalar are related to two processes, in our case, processes described by rectangular and angular coordinates. One process is the dynamic that occurs through the translation of the system, and the other is through a rotation. Both processes are independent of the other even sharing the same parameters like particle masses and field derivatives. Also, the transformation of the system is independent through one way or the other. 

The physical meaning of the eigenvalue show up once we realize that ``both'' component of the scalar describe the system at the same time.

As exposed before, there is no reason to represent the translation nor rotation in the upper or the downer position of the extended representation of state vector and scalars, however, as an example, we put rectangular coordinates in the upper's position while the rotational stands in the down position. As an example, let us have the state vector $\extket{\vec{X}}{\vec{\xi}}$. It describes a state of the systems which is at two positions: $\vec{X}$ and $R\vec{\xi}$ where $R$ is the length of the rotation vector $\vec{\xi}$ so that we can relate the state with the sum vector position
\begin{equation*}
\vec{X}_{state} = \vec{X} + R\vec{\xi}.
\end{equation*}
The action of the motion operator $\extcolumn{\mathcal{T}^\dagger(dx')}{\mathcal{R}(d\xi')}$, which we will study later, translate and rotate both positions of the system as
\begin{equation}
\extcolumn{\mathcal{T}^\dagger(\Delta\vec{X}')}{\mathcal{R}(\Delta \vec{\xi}')} \extket{\vec{X}}{\vec{\xi}}
= \extket{\vec{X} + \Delta \vec{X}'}{\vec{\xi} + \Delta\vec{\xi}'}.
\end{equation}
Because of the action of the operator, the system move from 
\begin{equation*}
\vec{X}_{state}  = \vec{X} + R\vec{\xi} \quad \to \quad 
\vec{X}_{state}' = \vec{X} + \Delta\vec{X}' + R\vec{\xi} + R\Delta\vec{\xi}'
\end{equation*}

The concept is clearer when we analyze another fundamental quantity: the energy of the system whose shows similar behavior. Indeed, a state of the system described by the vector $\extket{E_T}{E_R}$ have a total energy $E_{state}' = E_T' + E_R'$. If the system evolves to a different state described by $E_{state}'' = E_T'' + E_R''$, we can say that the translation and rotation energy make a quantum leap of value $\Delta E_R = E_R'' - E_R'$ and $\Delta E_T = E_T'' - E_T'$, respectively. The total energy of the system changes like
\begin{equation*}
E_{state}'  = E_T' + E_R' \quad \to \quad 
\vec{X}_{state}' = E_T' + \Delta E_T + E_R' + \Delta E_R.
\end{equation*}

We can conclude that for $n$-VMVF systems if the physical observables of the components of the studied operator satisfy the superposition principle and also the values of the components can be added then the eigenvalues should satisfy the linearity principle pair to pair. That means that if each component of the eigenvalue of a state is a sum of the component of others eigenvalues, then the eigenvalue of the sum is the sum of the eigenvalue built with the addends of the component of the sum eigenvalue. Explicitly, the sum eigenvalue $\extscalar{a' + a''}{b'+ b''}$ can be separated only by two possibilities:
\begin{align}
\extscalar{a' + a''}{b'+ b''} =& \extscalar{a'}{ b'}
+ \extscalar{a''}{ b''}\label{eigenExpansion1}
\end{align} 
or
\begin{align}
\extscalar{a' + a''}{b'+ b''} =& \extscalar{a''}{ b'}
+ \extscalar{a'}{ b''}\label{eigenExpansion2}
\end{align}
which means that the pairs related to the eigenvalues must satisfied also the conditions \ref{extLinearitySumCond}:
\begin{equation}
a'' \odot b'  + a' \odot b''
+ \mathcal{D}^{(a'' + a')}(a'',a',b'' + b') =0. 
\end{equation}
or 
\begin{equation}
a' \odot b'  + a'' \odot b''
+ \mathcal{D}^{(a'' + a')}(a',a'',b'' + b') =0.
\end{equation}
respectively. As $a',a'', b', b''$ are complex numbers, the conditions for the vector expansion, for each case, have the form
\begin{equation}
a''b'^{*} + a' b''^{*} = 0 \label{eigenEq80}
\end{equation}
or 
\begin{equation}
a'b'^{*} + a'' b''^{*} = 0 \label{eigenEq81}.
\end{equation}
Note that we obtain the same conditions for the expansion
\begin{align}
\extscalar{b'+ b''}{a' + a''} =& \extscalar{ b'}{a'}
+ \extscalar{ b''}{a''}
\end{align} 
or
\begin{align}
\extscalar{b'+ b''}{a' + a''} =& \extscalar{ b'}{a''}
+ \extscalar{ b''}{a'}.
\end{align}

In the case of the component of the eigenvalues $a'$, $b'$, $a''$ being different, for the equality in equation \ref{eigenEq7}, $a''b''^{*} - a' b'^{*} = 0$, we have then two possible expansions as shown in equations \ref{eigenExpansion1} and \ref{eigenExpansion2}. Let's see both cases:
\begin{enumerate}
\item 
In the case of conditions of Eq. \ref{eigenEq80}, we have then 
\begin{equation}
a' b''^{*} + a''b'^{*} = 0 \qquad a''b''^{*} - a' b'^{*} = 0. \label{eigenEq8}
\end{equation}

Multiplying first equation by $a'$, the second by $a''$ and summing the equations, we obtain
\begin{equation}
b''^*(a'^2 + a''^2)=0
\end{equation}
As $b''^*\neq 0$, the equation holds if $a''=\pm \imagi a'$, which led to the solution $b'' = \pm \imagi b'$ by substitute the result on the first equation. 

Let us analyze the inner product
\begin{equation*}
 \extinnerprod{\imagi a'}{\imagi b'}{a'}{b'}.
\end{equation*}
During the development of the present proposal, we suggest the assumption of several axioms as the bases of the theory. All of them must be consistent with the results obtained because of the action of the operator over a state and also with the algebraic properties of the resulted scalar. The eigenvalues for any operator, representing physical observables or not and obtained from the eigenvalue equation
\begin{equation*}
\extoperator{A}{B} \extket{a'}{b'} = \extscalar{a'}{b'} \extket{a'}{b'},
\end{equation*}
remain invariant for the quantum numbers pair $(a',b')$, $(-a',-b')$, and $(\imagi a',\imagi b')$. Indeed, the eigenvalue equations referred to the ket with those quantum numbers are
\begin{equation*}
\extoperator{A}{B} \extket{\imagi a'}{\imagi b'} = \extscalar{\imagi a'}{\imagi b'} \extket{\imagi a'}{\imagi b'}
= \imagi^* \imagi \extscalar{a'}{b'} \extket{\imagi a'}{\imagi b'}
= \extscalar{a'}{b'} \extket{\imagi a'}{\imagi b'}
\end{equation*}
and 
\begin{equation*}
\extoperator{A}{B} \extket{-a'}{-b'} = \extscalar{-a'}{-b'} \extket{-a'}{-b'}
= \extscalar{a'}{b'} \extket{-a'}{-b'},
\end{equation*}
where we apply the properties of the complex extended products for the extended scalar. Being this behavior applicable to all kets, we an state that in our approach kets 
\begin{equation}
\extket{a'}{b'},\; \extket{-a'}{-b'},\;  \text{and}\; \extket{\imagi a'}{\imagi b'} 
\end{equation}
represent the same physical state. In that case, the inner products satisfy
\begin{equation}
\extinnerprod{\imagi a'}{\imagi b'}{a'}{b'} = \extinnerprod{a'}{b'}{\imagi a'}{\imagi b'} = \extinnerprod{ a'}{b'}{a'}{b'} \neq 0.
\end{equation}

\item For conditions of Eq. \ref{eigenEq81}, we have the equations
\begin{equation}
a'' b''^{*} + a'b'^{*} = 0 \qquad a''b''^{*} - a' b'^{*} = 0. \label{eigenEq9}
\end{equation}
which give the solution $a''b''^{*} = a' b'^{*} = 0$. As we suppose the eigenvalues different from zero, we conclude like for an expansion like \ref{eigenExpansion2}, the only possible solution is 
\begin{equation}
a''b''^{*} - a' b'^{*} \neq 0,
\end{equation}
Which means
\begin{align}
\extinnerprod{a''}{b''}{a'}{b'} = 0,
\qquad 
\extinnerprod{b''}{a''}{b'}{a'} = 0.
\end{align}
\end{enumerate}

In general, it is our understanding that all the cases for the values of the components of the eigenvalues, $a'$ and $b'$, should satisfy both relations showed in equation \ref{eigenEq8} referred to the fact that they are eigenvalues from operators like \ref{eigenOprProp}. For example, for the cases where $a''=a', b''=b'$ and $a''=a', b''\neq b'$ or $a''\neq a', b''= b'$, both conditions cannot be satisfied at the same time. 

We can conclude that the eigenvalues and eigenkets for operators that satisfy relations \ref{eigenOprProp} are:
\begin{itemize}
\item The eigenvalues are pure complex numbers
\item The inner product $\extinnerprod{a''}{b''}{a'}{b'}$ are different from zero only if
$a'' = a'$ and $b'' = b'$ or $a'' = \pm \imagi a'$ and $b'' = \pm \imagi b'$
\end{itemize}


The normalization relations \ref{extQMNormalCond} and \ref{extQMNormalCond1} for the extended eigenkets are
\begin{align}
\extinnerprod{a''}{b''}{a'}{b'}^* \extinnerprod{a''}{b''}{a'}{b'} = \delta_{a'',a'} \delta_{b'',b'}
+ \delta_{a'', \pm \imagi a'} \delta_{b'', \pm \imagi b'}
\\
\frac{1}{2}\left[ \extinnerprod{a''}{b''}{a'}{b'}^*  + \extinnerprod{a''}{b''}{a'}{b'} \right]= \delta_{a'',a'} \delta_{b'',b'}
+ \delta_{a'', \pm \imagi a'} \delta_{b'', \pm \imagi b'}.
\end{align}
The solution for the exponential expression of inner product $\extinnerprod{a''}{b''}{a'}{b'} = R_\Delta e^{\imagi \theta_\Delta}$ of the above equation system is
\begin{equation}
R_\Delta = \delta \qquad \text{and} \qquad \theta_\Delta = 2\pi n, \;\forall n=0,1,2...
\end{equation}
where $\delta \equiv \delta_{a'', \pm \imagi a'} \delta_{b'', \pm \imagi b'}$. From the inner commutation relation, we obtain then
\begin{equation}
\extinnerprod{b''}{a''}{b'}{a'} = \delta.
\end{equation}

We can find the same solution from the normalization relations for physical states written as
\begin{equation}
\extinnerprod{a''}{b''}{a'}{b'} \extinnerprod{b''}{a''}{b'}{a'} 
= \delta_{a'',a'} \delta_{b'',b'} \equiv \delta' 
\label{extEigenValueRel1}
\end{equation}
and
\begin{equation}
\frac{1}{2}\left( \extinnerprod{a''}{b''}{a'}{b'} + \extinnerprod{b''}{a''}{b'}{a'} \right)= \delta'
\label{extEigenValueRel2}.
\end{equation}
Indeed, isolating $\extinnerprod{b''}{a''}{b'}{a'} $ on equations \ref{extEigenValueRel1}
\begin{equation}
\extinnerprod{b''}{a''}{b'}{a'} =  \frac{\delta'}{\extinnerprod{a''}{b''}{a'}{b'}}
\end{equation}
and replacing on equation \ref{extEigenValueRel2}, we obtain
\begin{equation}
\extinnerprod{a''}{b''}{a'}{b'}^2 - 2 \delta' \extinnerprod{a''}{b''}{a'}{b'} + \delta' = 0.
\end{equation}
which using $\delta' = \delta'^2 \quad \forall \;a',b',a'',b''$, can be conveniently rewritten as
\begin{align}
\extinnerprod{a''}{b''}{a'}{b'}^2 - 2 \delta' \extinnerprod{a''}{b''}{a'}{b'} + \delta'^2=0
\nonumber \\
\left(\extinnerprod{a''}{b''}{a'}{b'} - \delta' \right)^2=0.
\end{align}
We finally obtain
\begin{equation}
\extinnerprod{a''}{b''}{a'}{b'} = \extinnerprod{b''}{a''}{b'}{a'} = \delta'
\end{equation}

\subsection{Series Expansion}
An arbitrary state vector, lets say $\extket{\alpha}{\beta}$, can be expanded in terms of the eigenkets of operators $\extoperator{A}{B}$ as:
\begin{equation}
\extket{\alpha}{\beta} = \sum_{a'b'} \extscalar{c_{a'}}{c_{b'}} \extket{a'}{b'}.
\end{equation}

Multiplying on the left both equations by $\extbra{a'''}{b'''}$ we have
\begin{align}
\extinnerprod{a''}{b''}{\alpha}{\beta}
= \sum_{a',b'} \extscalar{c_{a'}}{c_{b'}}  \extinnerprod{a''}{b''}{a'}{b'} 
&= \sum_{a',b'} (c_{a'} \odot c_{b'}) \extinnerprod{a''}{b''}{a'}{b'} 
\nonumber \\
&= (c_{a'}^{} \odot c_{b'}) + (c_{\imagi a'} \odot c_{\imagi b'})
\label{eqSystCoefExpan}
\end{align}
According the linearity axiom for vector state, $\extket{a'}{b'} = \extket{\imagi a'}{\imagi b'}$, which means that the expansion must also be the same or that the expansion coefficients satisfy
\begin{align}
c_{a'}^{} \odot c_{b'} = c_{\imagi a'} \odot c_{\imagi b'}
\end{align}
The solution for the inner product in equation \ref{eqSystCoefExpan} is
\begin{align}
\extinnerprod{a'}{b'}{\alpha}{\beta} = 2 c_{a'}^{} \odot c_{b'}
\label{eqSystCoefExpan1}
\end{align}

We can perform the same analysis for the bra $\extbra{\alpha}{\beta}$ and obtain
\begin{align}
\extinnerprod{\alpha}{\beta}{a'}{b'} = 2 c_{a'}^{\bullet} \odot c_{b'}^{\bullet}
\label{eqSystCoefExpan2}
\end{align} 

The complex equations of equations \ref{eqSystCoefExpan1} are
\begin{align}
c_{a'_E}^* c_{b'_I} z_1^{(a')} + c_{a'_I}^* c_{b'_E}  = \frac{1}{2} \extinnerprod{a'}{b'}{\alpha}{\beta}_E
\nonumber \\
\imagi c_{a'_E}^* c_{b'_E}  + c_{a'_E}^* c_{b'_I} w_1^{(a')} + c_{a'_I}^* c_{b'_I}  = \frac{1}{2} \extinnerprod{a'}{b'}{\alpha}{\beta}_I
\end{align}
while for  equations \ref{eqSystCoefExpan2}, we have
\begin{align}
c_{a'_E}^{\bullet*} c_{b'_I}^\bullet z_1^{(a')} + c_{a'_I}^{\bullet*}  c_{b'_E}^\bullet  = \frac{1}{2} \extinnerprod{\alpha}{\beta}{a'}{b'}_E
\nonumber \\
\imagi c_{a'_E}^{\bullet*}  c_{b'_E}^\bullet  + c_{a'_E}^{\bullet*}  c_{b'_I}^\bullet w_1^{(a')} + c_{a'_I}^{\bullet*}  c_{b'_I}^\bullet  = \frac{1}{2} \extinnerprod{\alpha}{\beta}{a'}{b'}_I
\end{align}
where the extended form of the coefficients are
\begin{equation}
c_{x_E}^{\bullet} = c_{x_E} z_2^{(c_x^\bullet)},\;
\qquad 
c_{x_I}^{\bullet} = c_{x_E} w_2^{(c_x^\bullet)} + c_{x_I},\;
\qquad \forall x = \{a',b' \}
\end{equation}

We can solve the equation system and obtain the expansion coefficient as functions of the inner products as
\begin{align}
c_{a'}=\mathcal{C}_{a'}\left( 
\extinnerprod{a'}{b'}{\alpha}{\beta}, 
\extinnerprod{\alpha}{\beta}{a'}{b'}
\right)
\nonumber \\
c_{b'}=\mathcal{C}_{b}\left( 
\extinnerprod{a'}{b'}{\alpha}{\beta}, 
\extinnerprod{\alpha}{\beta}{a'}{b'}
\right).
\end{align}

The state vectors $\extket{\alpha}{\beta}$ must satisfy the normalization conditions \ref{extQMNormalCond} and \ref{extQMNormalCond1} like:
\begin{align}
&\extinnerprod{\alpha}{\beta}{\alpha}{\beta} \odot \extinnerprod{\alpha}{\beta}{\alpha}{\beta} =
\nonumber \\
& \sum_{\substack{a'a''a'''a''''\\b'b''b'''b''''}} 
\left[ \extscalarBra{c_{a'}}{c_{b'}} \extscalar{c_{a''}}{c_{b''}}
\extinnerprod{a'}{b'}{a''}{b''}\right]^*
\extscalarBra{c_{a'''}}{c_{b'''}} \extscalar{c_{a''''}}{c_{b''''}}
\extinnerprod{a'''}{b'''}{a''''}{b''''} = 1 \label{coefNormRel}
\end{align}
and 
\begin{align}
&\frac{1}{2} \left(\extinnerprod{\alpha}{\beta}{\alpha}{\beta} \oplus \extinnerprod{\alpha}{\beta}{\alpha}{\beta}  \right) = 
\nonumber \\
&\frac{1}{2} \left[ \sum_{\substack{a'a''\\b'b''}} 
\left( \extscalarBra{c_{a'}}{c_{b'}} 
\extscalar{c_{a''}}{c_{b''}} 
\extinnerprod{a'}{b'}{a''}{b''} \right)^*+ 
\extscalarBra{c_{a'}}{c_{b'}} 
\extscalar{c_{a''}}{c_{b''}} 
\extinnerprod{a'}{b'}{a''}{b''}
 \right] = 1  \label{coefNormRel1}
\end{align}
Applying the normalization conditions for the inner product $\extinnerprod{\alpha}{\beta}{\alpha}{\beta} $ we have
\begin{align}
\extinnerprod{\alpha}{\beta}{\alpha}{\beta} \odot \extinnerprod{\alpha}{\beta}{\alpha}{\beta} =
\sum_{\substack{a'a''\\b'b''}} 
4\left[ \extscalarBra{c_{a'}}{c_{b'}} \extscalar{c_{a'}}{c_{b'}}
\right]^*
\extscalarBra{c_{a''}}{c_{b''}} \extscalar{c_{a''}}{c_{b''}}
 = 1 \label{coefNormRel_1}
\end{align}
and 
\begin{align}
\frac{1}{2}\left(\extinnerprod{\alpha}{\beta}{\alpha}{\beta} \oplus \extinnerprod{\alpha}{\beta}{\alpha}{\beta}  \right) = 
 \sum_{a',b'}
\left( \extscalarBra{c_{a'}}{c_{b'}} 
\extscalar{c_{a'}}{c_{b'}} \right)^*+ 
\extscalarBra{c_{a'}}{c_{b'}} 
\extscalar{c_{a'}}{c_{b'}} 
= 1.  \label{coefNormRel1_1}
\end{align}

Also, if the quantum numbers of the ket $\extket{\alpha}{\beta}$ and its inverted ket $\extket{\beta}{\alpha}$ satisfy the inner commutation conditions, we can relate both of their coefficients. Indeed, following the same procedure we can expand the inverted ket as
\begin{equation}
\extket{\beta}{\alpha} = \sum_{a''b''} \extscalar{d_{b''}}{d_{a''}} \extket{b''}{a''},
\end{equation}
where the coefficients can be written as function of the inner products:
\begin{align}
d_{a'}=\mathcal{D}_{a'}\left( 
\extinnerprod{b'}{a'}{\beta}{\alpha}, 
\extinnerprod{\beta}{\alpha}{b'}{a'} 
\right)
\nonumber \\
d_{b'}=\mathcal{D}_{b'}\left( 
\extinnerprod{b'}{a'}{\beta}{\alpha}, 
\extinnerprod{\beta}{\alpha}{b'}{a'} 
\right).
\end{align}

The normalization conditions for both kets are the same and they have the form
\begin{align}
&\extinnerprod{\alpha}{\beta}{\alpha}{\beta} \extinnerprod{\beta}{\alpha}{\beta}{\alpha} =
\nonumber \\
& \sum_{\substack{a'a''a'''a''''\\b'b''b'''b''''}} 
\extscalarBra{c_{a'}}{c_{b'}} \extscalar{c_{a''}}{c_{b''}}
\extscalarBra{d_{b'''}}{d_{a'''}} \extscalar{d_{b''''}}{d_{a''''}}
\extinnerprod{a'}{b'}{a''}{b''} \extinnerprod{b'''}{a'''}{b''''}{a''''} = 1 \label{coefNormRel2}
\end{align}
and 
\begin{align}
&\frac{1}{2} \left(\extinnerprod{\alpha}{\beta}{\alpha}{\beta} + \extinnerprod{\beta}{\alpha}{\beta}{\alpha}  \right) = 
\nonumber \\
&\frac{1}{2} \left( \sum_{\substack{a'a''\\b'b''}} 
\extscalarBra{c_{a'}}{c_{b'}} \extscalar{c_{a''}}{c_{b''}} \extinnerprod{a'}{b'}{a''}{b''} + 
\extscalarBra{d_{b'}}{d_{a'}} \extscalar{d_{b''}}{d_{a''}} \extinnerprod{b'}{a'}{b''}{a''} 
\right) = 1  \label{coefNormRel3}
\end{align}

Substituting equations \ref{eqSystCoefExpan}, the conditions \ref{coefNormRel} and \ref{coefNormRel1} reduce to
\begin{equation}
4 \sum_{\substack{a'a''\\b'b''}}  \extscalarBra{c_{a'}}{c_{b'}} \extscalar{c_{a'}}{c_{b'}}
\extscalarBra{d_{b''}}{d_{a''}} \extscalar{d_{b''}}{d_{a''}} = 1  \label{coefNormRel4}
\end{equation}
and 
\begin{equation}
\sum_{a',b'}  \extscalarBra{c_{a'}}{c_{b'}} \extscalar{c_{a'}}{c_{b'}}
+
\extscalarBra{d_{b'}}{d_{a'}} \extscalar{d_{b'}}{d_{a'}} = 1  \label{coefNormRel5}
\end{equation}
respectively. Also, from the relations between their inner products
\begin{equation*}
\extinnerprod{\alpha}{\beta}{\alpha}{\beta} = \extinnerprod{\beta}{\alpha}{\beta}{\alpha}^*,
\end{equation*}
we have
\begin{equation}
\sum_{a',b'}  \extscalarBra{c_{a'}}{c_{b'}} \extscalar{c_{a'}}{c_{b'}} =
\left[ \sum_{a',b'} \extscalarBra{d_{b'}}{d_{a'}} \extscalar{d_{b'}}{d_{a'}} 
\right]^*.
\end{equation}

The results of equations \ref{coefNormRel3} and \ref{coefNormRel4} indicate that we still can relate the expansion coefficients (or their product) to some probability. However, their physical meaning demands a more in-depth study of extended numbers and also, what it is more critical, the verification of the solution to a specific problem with experimental data.

\subsection{Measurements}
In this section, we introduce the concept of measurement in the quantum mechanic for $n$-VMVF systems. From the ordinary quantum mechanics this is an already complicated topic so, for the extended vector ket it will be even more difficult, mostly because there are several descriptions for the processes occurring on systems at length scale in order to its de Broglie wavelength. The measurement is by far, the most unconfirmed and unsustained theme treated in our work. We have no evidence, back up results or experimental data of how the measurement should occur in the quantum theory for $n$-VMVF. Nevertheless, Measurement is crucial in the quantum approach of any theory, and it should be discussed.

The experiment has been the starting and the end point of a theoretical proposition for describing phenomena in physics along the history of humankind. The observation and the recording of data that usually set the foundations of a theory and also the confirmation of all the results the theory predicts. One example is the set of well-known experiments which shows the particle-wave duality behavior of the matter. That was the base for the proposition of Erwin Schr\"odinger in 1926 of a partial differential equation for the wave functions of particles and also for the formulation of quantum mechanics created by Werner Heisenberg, Max Born, and Pascual Jordan in 1925.

Even when we are lack of experimental data for setting the bases of this theory, we assumed from the start of this work the quantum behavior of nature. Indeed, our proposal should be seen as an extension of the ordinary quantum mechanics for systems with variables masses, and we follow almost all the axioms of the theory. The reason behind this choice lies precisely that all the experiments with particles and all the theories in particle physics, even with its inconsistencies, predict this quantum behavior and others concepts like the probabilistic measurement, and nothing indicates otherwise. This assumption, together with the study of the extended domain allow us to introduce the first ideas that, from our point of view, can describe the measurement processes. 

The proposed relations of normalization for a state vector and the definition of the expectation value for an operator suggest that measurement in the present theory also have a probability association. However, according to the series expansion, the probabilistic meaning of the coefficients is different from the ordinary quantum mechanic. Following the quantum ``way of thinking'', we assume that the measurement in the present theory should be guided by the same principle existing in the conventional quantum mechanic. This principle was correctly explained by Dirac when he states ``A measurement always cause the system to jump to an eigenstate of the dynamical variable that is being measured'' and is  represented like
\begin{equation}
\extket{\alpha}{\beta} \xrightarrow{\text{measurement}} \extket{a'}{b'}.
\end{equation}

On our previous discussion about the relation between the normalization relations and the expected value on section \ref{NormExpValueSection}, we define the expectation values relations. Because of the exposed on that section, specifically on equations \ref{expValueDef} and \ref{expValueDef0}, we propose that the relations for the measured quantities have the form
\begin{equation}
\Big[\expval{\mathbf{A}} \odot \expval{\mathbf{B}}\Big]\odot \Big[\expval{\mathbf{A}} \odot \expval{\mathbf{B}}\Big] = 
\extaverage{\alpha}{\beta}{A}{B}{\alpha}{\beta}
\odot
\extaverage{\alpha}{\beta}{A}{B}{\alpha}{\beta} 
\end{equation}
and
\begin{equation}
\Big[\expval{\mathbf{A}} \odot \expval{\mathbf{B}} \Big]\oplus \Big[\expval{\mathbf{A}} \odot \expval{\mathbf{B}}\Big] =
 \extaverage{\alpha}{\beta}{A}{B}{\alpha}{\beta}
\oplus
\extaverage{\alpha}{\beta}{A}{B}{\alpha}{\beta}.
\end{equation}

%

\subsection{Classical mechanic on the Quantum mechanic for $n$-VMVF systems}
We proposed a Hilbert space over the extended numbers for developing the quantum theory for $n$-VMVF systems. Then, we defined the algebraic objects like vectors, scalars, and operators for the description of the magnitudes of the theory. Later, we settle some properties and relations according to logic, the properties of the extended vector space and others facts like the operators and state vectors represent physical objects. Among such properties, we have the fulfillment of the principle of superposition and the normalization relations for state vectors and also the pure complex nature of the eigenvalues for operator representing physical observables. 

As we have commented before, the modern construction of the quantum mechanics includes the laws physics using the classical theory. The inclusion is done by the borrowing of the concept of canonical transformations, and use them as the base to define the form of the action of quantum operators over the state vectors. From the theory of second-order Hamiltonian, we identify two quantities transformed because of the infinitesimal canonical transformations. One of them is the degree of freedom  $q_i$, and the other is the pole of the correlation function $\mathcal{F}_{2_i}(\bar{\dot{q}})$. The correlation function's definition relates them; however, they are independently transformed by the momentums $p_i$ and $s_i$ respectively. On the other side, the classical theory for $n$-VMVF systems involves the solution of two set of extended equations: one for the rectangular and other for the angular coordinates. As a consequence of the double set of equations, the transformations to modify the canonical variables, also have two components. Thereby the proposition for the extended space vector and the two-component structure for all the quantum algebraic objects.

Following the methodology for constructing the theory of modern quantum mechanic, we introduce the physics of $n$-VMVF systems to the already defined Hilbert space over the extended numbers, using all the axioms and relations exposed before.

\subsubsection{States vectors for n-VMVF systems}
The classical development of the theory for $n$-VMVF systems shows the particles position and the value of the pole of the correlation function, as the canonical variables that change along the evolution of the system. Therefore, they are the default degree of freedom for the quantum description of $n$-VMVF systems. The classical theory also shows the inclusion of the fourth coordinate of the position of particles constrained by the Lorentz relation and with it the fourth coordinate of all the rest of magnitudes of the theory. We treat this constraint as ``a weak condition'' according to the Dirac point of view, which means that all the coordinates $x^{\nu'}$ are considered as independent, and the Lorentz constraint will be imposed after all calculation has been carry through \cite{goldstein321}. For simplicity, we write the particle position as four component vector, and we denote them as $x$.

The classical description of $n$-VMVF system is given by the study of simultaneous evolution of the system by the translation and the rotation. They are naturally described using the rectangular and the angular coordinates respectively.
The system also depends on the value of the pole of functions $f_{x_n}(\bar{\dot{x}}^\nu_m)$ and $f_{\xi_n}(\bar{\dot{\xi}}_{m})$ defined on equation \ref{correlFunctCond}.
A general state vector for a $n$-VMVF particle systems can be written then as
\begin{equation}
\extket{x_1,...x_N, f_{x_1}...f_{x_N}}
{\xi_1,...\xi_N, f_{\xi_1}...f_{\xi_N}}.
\end{equation}

From the classical mechanics, the quantities $x_i, \xi_i$ and $f_{x_i}, f_{\xi_i}$ are related by the equations \ref{extEulerLagranEqV}:
\begin{equation}
\mathcal{F}_{x_n}(\bar{\dot{x}}^\nu_m) - f_{x_n}=0 \qquad \qquad \mathcal{F}_{\xi_n}(\bar{\dot{\xi}}_{m}) - f_{\xi_n} =0\label{extRelation1}
\end{equation}
were functions  $\mathcal{F}_{2_i}(\bar{\dot{q}})$ are already defined as in the extension of classical mechanics theory and they have a fixed form. Also, we have the constraints \ref{extClassHamConst}
\begin{equation}
\Phi_{\nu_n} (\bar{x}^\nu_m, \bar{\dot{x}}^\nu_m, \bar{\ddot{x}}^\nu_m) = 0 \qquad \qquad 
\Psi_{i_n}(\bar{\xi}_{m}, \bar{\dot{\xi}}_{m}, \bar{\ddot{\xi}}_{m})=0 \label{extRelation2}.
\end{equation}
Rest to see how this classical relations or constraints are included on the this proposal for the extension of the quantum mechanics for $n$-VMVF systems.

This approach considers observables with continuous spectra of eigenvalues like position and momentum. In this case, the sum is substitute by the integral taking into account the analysis must be on finite-dimensional vector space.


\subsubsection{Physical operators}
Because of the classical solution for the $n$-VMVF systems, we propose that the action of a quantum operator over a state ket is done component over component as
\begin{equation}
\extoperator{A}{B} \extket {\alpha}{\beta} \equiv \extcolumn{\mathbf{A}^\dagger \ket{\alpha}}{\mathbf{B} \; \ket{\beta}},
\end{equation}
so the action of one component operator will not infer over the other. When an operator acting over its eigenkets, the components of the eigenvalues are also the result of the action of each component of the operator over its respective component of the eigenket like
\begin{equation*}
\extoperator{A}{B} \extket{a'}{b'} = \extscalar{a'}{b'} \extket{a'}{b'}.
\end{equation*}
 
Also, following the same reasoning, if we have that each component of the operator is an arbitrary function, named $\mathcal{F}$ and $\mathcal{G}$ depending each one on operators $\mathbf{A}$ and $\mathbf{B}$ respectively, we state that
\begin{equation}
\extcolumn{\mathcal{F}^\dagger(\mathbf{A})}{\mathcal{G}(\mathbf{B})\;} \extket{a'}{b'} = \extscalar{\mathcal{F}(a')}{\mathcal{G}(b')} \extket{a'}{b'}.
\end{equation}
It seems logical to presume that there is a ``connection'' between the action of both components of the operator over the corresponding components of the vector, as they are part of a single algebraic structure. However, from our point of view, this ``connection'' have no effects in the moment of the action, and it takes place only once the scalar is obtained. The ``connection'' the is given by the algebra properties for the extended numbers and the normalization conditions. 

Following these ideas, we study the operators coming from the classical theory and their action over the states vectors. Among them, we have the translation and rotation of the position vectors and the displacements of the pole of the correlation functions.

\subsubsection{Infinitesimal operators}
Our proposal for the quantum solution for $n$-VMVF systems is in the continuous spectra. Because of that, before analysing a particular operator, we study the general form of the infinitesimal operators.

The action of an infinitesimal operator over a state  transform the former state into a new state with the values of its components adjusted by an infinitesimal quantity like
\begin{equation}
\extcolumn{\mathbf{A}^\dagger(da')}{\mathbf{B}\; (db')}\extket{a'}{b'} = \extket{a'+da'}{b'+db'}
\end{equation}

Infinitesimal operators must satisfy some properties:
\begin{enumerate}
\item \label{extOperProp1} The infinitesimal operator depends on the infinitesimal displacement value, and it will tend to identity operator is the infinitesimal displacement tend zero:
\begin{equation}
\lim_{\substack{da' \to 0\\db' \to 0}} \extcolumn{\mathbf{A}^\dagger(da')}{\mathbf{B}\; (db')} = \extcolumn{\mathbf{1}}{\mathbf{1}}.
\end{equation}

\item \label{extOperProp3}The action of the operator must preserve the normalization conditions for any state, which means
\begin{equation}
\extbra{a'}{b'}  
\extcolumn{\mathbf{A}^{\dagger\ddagger}(da')}{\mathbf{B}^\ddagger\; (db')} 
\extcolumn{\mathbf{A}^{\dagger\ddagger}(da')}{\mathbf{B}^\ddagger\; (db')}
\extket{a'}{b'}
\odot  
\extbra{a'}{b'}  
\extcolumn{\mathbf{A}^{\dagger\ddagger}(da')}{\mathbf{B}^\ddagger\; (db')} 
\extcolumn{\mathbf{A}^{\dagger\ddagger}(da')}{\mathbf{B}^\ddagger\; (db')}
\extket{a'}{b'}  = 1.
\end{equation}
and 
\begin{equation}
\frac{1}{2}\left[ \extbra{a'}{b'}  
\extcolumn{\mathbf{A}^{\dagger\ddagger}(da')}{\mathbf{B}^\ddagger\; (db')} 
\extcolumn{\mathbf{A}^{\dagger\ddagger}(da')}{\mathbf{B}^\ddagger\; (db')}
\extket{a'}{b'}
\oplus  
\extbra{a'}{b'}  
\extcolumn{\mathbf{A}^{\dagger\ddagger}(da')}{\mathbf{B}^\ddagger\; (db')} 
\extcolumn{\mathbf{A}^{\dagger\ddagger}(da')}{\mathbf{B}^\ddagger\; (db')}
\extket{a'}{b'}\right]  = 1.
\end{equation}

One infinitesimal operator which satisfies those relations for any state then
\begin{equation}
\extcolumn{\mathbf{A}^{\dagger\ddagger}(da')}{\mathbf{B}^\ddagger\; (db')} 
\extcolumn{\mathbf{A}^{\dagger}(da')}{\mathbf{B}\; (db')}
 = \extcolumn{1}{1}.
\end{equation}
\end{enumerate}

We review the linearity property of the quantum operators representing physical magnitudes under the frame of $n$-VMVF systems. For that, we analyze the dual transformation of the canonical variables on such systems one more time. The two transformations, $\mathbf{A}$ and $\mathbf{B}$, from the components of an extended operator transform the state vector of the system into a new state. That means that the set of quantum numbers are changed collectively transform into another. The bi-dimensional state vector appears as the substitute of one component vector depending on the particle positions and masses and field derivatives like
\begin{equation}
\ket{
\{x_n^\mu\}, 
\{f_{x_n^\mu}\}, 
\{\frac{\partial m_n}{\partial x_n^\mu }\}, 
\{\frac{\partial A^\nu}{\partial x_n^\mu }\},
\{\frac{\partial A^\nu}{\partial \dot{x}_n^\mu }\}}
\longrightarrow
\extket
{\{x_n^{\mu'}\},\{f_{x_n^\mu}'\}}
{\{x_n^{\mu''}\},\{f_{x_n^\mu}''\}}.
\end{equation}
As shown in the extension of the classical mechanics for $n$-VMVF systems, this replacement responds to the impossibility of solving the classical problem for those systems using only one set of Lagrange or Hamilton's equations. That is because there are no conservation laws for the hidden variables related to the mass of particles or the field.

For example, the operator of motion whose components are the translation and the rotation operators, represented by ${\mathcal{T}^\dagger(dx')}$ and ${\mathcal{R}(dx'')}$ respectively, corresponds to two independent canonical transformations, each one for a different set of Hamilton's extended equations. The action of such operator over the state vector has the form
\begin{equation}
\extcolumn{\mathcal{T}^\dagger(dx'_{n'})}{\mathcal{R}(dx''_{n'})}
\extket
{\{x_n^{\mu'}\},\{f_{x_n^\mu}'\}}
{\{x_n^{\mu''}\},\{f_{x_n^\mu}''\}}
=
\extket
{x_{n'}^{\mu'} + dx_{n'}^{\mu'},\{x_{n \neq n'}^{\mu'}\},\{f_{x_n^\mu}'\}}
{x_{n'}^{\mu''}+ dx_{n'}^{\mu''},\{x_{n \neq n'}^{\mu''}\},\{f_{x_n^\mu}''\}}
\end{equation}
and it replaces the action of a single operator that should include one of the components of the motion operator, let say $\mathcal{T}(dx')$, and an unknown operator which act over the hidden variables. If we represent $d\mathbf{a}$ like the differential displacement of the hidden variables:
\begin{equation}
d\mathbf{a} = 
d\Big( 
\{\frac{\partial m_n}{\partial x_n^\mu }\}, 
\{\frac{\partial A^\nu}{\partial x_n^\mu }\},
\{\frac{\partial A^\nu}{\partial \dot{x}_n^\mu }\}
\Big),
\end{equation}
the action of the motion operator replace a single component operator like
\begin{align}
&\extcolumn{\mathcal{T}^\dagger(dx'_{n'})}{\mathcal{R}(dx''_{n'})}
\extket
{\{x_n^{\mu'}\},\{f_{x_n^\mu}'\}}
{\{x_n^{\mu''}\},\{f_{x_n^\mu}''\}}
\equiv
\nonumber \\
& \mathrm{F} (\mathcal{T}(dx'_{n'}), \mathcal{A}(d\mathbf{a}))
\ket{
\{x_{n}^{\mu'}\}, 
\{f_{x_n^\mu}\}, 
\{\frac{\partial m_n}{\partial x_n^\mu }\}, 
\{\frac{\partial A^\nu}{\partial x_n^\mu }\},
\{\frac{\partial A^\nu}{\partial \dot{x}_n^\mu }\}}
\nonumber \\
&=\ket{
x_{n'}^{\mu'} + dx_{n'}^{\mu'},\{x_{n \neq n'}^{\mu'}\}, 
\{f_{x_n^\mu}\},
\Big\{ 
\{\frac{\partial m_n}{\partial x_n^\mu }\}, 
\{\frac{\partial A^\nu}{\partial x_n^\mu }\},
\{\frac{\partial A^\nu}{\partial \dot{x}_n^\mu }\}
\Big\} + d\mathbf{a}}.
\end{align}

We can see now that the motion operator also transforms the hidden variables. We cannot ensure that those variables have a linear behavior. Because of that, we can not expect the linear property for the dual transformation that an extended operator represents, which means that two consecutive transformations of a group of variables of the system will not be equal to a unique transformation as the sum of the previous transformations or what is the same
\begin{equation}
\extcolumn{\mathbf{A}^\dagger(da')}{\mathbf{B}\; (db')} 
\extcolumn{\mathbf{A}^\dagger(da'')}{\mathbf{B}\; (db'')} 
\neq \extcolumn{\mathbf{A}^\dagger (da'+da'' )}{\mathbf{B}(db' + db'')}.
\end{equation}

We can instead impose linearity for the infinitesimal transformation of each component of the operator in the form:
\begin{equation}
\extcolumn{\big[ \mathbf{A}(da')\mathbf{A}(da'')\big]^\dagger }{\big[ \mathbf{B} (db')\mathbf{B} (db'') \big]} 
= \extcolumn{\mathbf{A}^\dagger(da'+da'' )}{\mathbf{B} (db' + db'')},
\end{equation}
which is consistent with the linearity property of the classical canonical transformations.

A general canonical transformation on classic mechanic is the sum of the identity operator and the generator of the transformation times the infinitesimal quantity of the transformation. We take the form of classical transformations and apply it to each operator component. So, in the extended quantum mechanic we propose that the components of the operator is written like
\begin{equation}
\mathbf{O}^\dagger = (\mathbf{1} + \mathbf{J_O} do').
\end{equation}
Then, we can write any infinitesimal extended operators as:
\begin{equation}
\extcolumn{\mathbf{A}^\dagger(da')}{\mathbf{B}\;(db')}= \extcolumn{\mathbf{1} + a_0^* \mathbf{J_A}^\dagger da'}{\mathbf{1} + b_0\mathbf{J_B}\; db'}
\end{equation}
where the components of the operator ${\mathbf{J_A}^\dagger}$ and ${\mathbf{J_B}}$ satisfy equations \ref{eigenOprProp0}:
\begin{equation*}
\mathbf{J_A} = \mathbf{J_A}^\ddagger, \qquad
\mathbf{J_A}^\dagger = (\mathbf{J_A}^\dagger)^{ \ddagger}, \qquad
\mathbf{J_B} = \mathbf{J_B}^\ddagger \qquad 
\mathbf{J_B}^\dagger =  (\mathbf{J_B}^\dagger)^{ \ddagger}
\end{equation*}
We must note that the top component 
\begin{equation}
\mathbf{A}^\dagger = (\mathbf{1} + \mathbf{J_A} da')^\dagger 
\neq \mathbf{1} + a_0^* \mathbf{J_A}^\dagger da',
\end{equation}
since the complex map of an extended number was shown to be not associative. For our proposes, we can assume that for every operator $\mathbf{A}$ there is an operator $\mathbf{J_A}$ which permit to express the complex conjugated of $\mathbf{A}$ as $\mathbf{A}^\dagger = \mathbf{1} + a_0^* \mathbf{J_A}^\dagger da'$. This is the main reason for de inclusion of the constant $a_0$. However, as these operators are extracted from the classical mechanics where the transformation's functions are at most complex numbers, we expect that all operators, at least the exposed in here, can be written as
\begin{equation}
\mathbf{A}^\dagger = \mathbf{1} + a_0^* \mathbf{J_A}^\dagger da'.
\end{equation}

The proposed infinitesimal operators tend to zero if the displacement is zero. They are provided with the extended constants $a_0,b_0$ for guarantee the normalization relations. These constants should be determined after furthers studies of extended numbers and specifically with the relation of numbers $(1 + \alpha)^\bullet$ and $\alpha^\bullet$. For now on, we consider the constants as known values. Also, it is easy to show that each component satisfies the linearity property same like in standard quantum mechanics:
\begin{align}
\extcolumn{\big[\mathbf{A}(da')\mathbf{A}(da'')\big]^\dagger}
{\big[\mathbf{B}\;(db')\;\mathbf{B}\;(db'') \big]} 
&= \extcolumn{\big[(\mathbf{1} + a_0^* \mathbf{J_A}\; da')(\mathbf{1} + a_0^* \mathbf{J_A}\; da'')\big]^\dagger}{\big[(\mathbf{1} + b_0 \mathbf{J_B}\; db')(\mathbf{1} + b_0 \mathbf{J_B} \; db'')\big]}
\nonumber \\
& = \extcolumn{\big[\mathbf{1} + a_0^*\mathbf{J}_A (da' + da'') + \mathcal{O}(da^{'2})\big]^\dagger}{\big[\mathbf{1} + b_0\mathbf{J_B}\; (db' + db'') + \mathcal{O}(db^{'2})\big]}
\nonumber \\
& \simeq  \extcolumn{\mathbf{A}^\dagger (da'+da'' )}{\mathbf{B}(db' + db'')}.
\end{align}

In our case and according to the extended classical mechanics for $n$-VMVF systems, we build this theoretical proposition by studying the translation and rotation operators over the position vectors and also over the translation and rotation of the poles of the velocities correlation function. 


\subsubsection{Momentums \textbf{p} and \textbf{l} as generators of translation and rotation.}
One key aspect of the methodology for constructing the modern quantum theory is to find the expression for the quantum operators, specifically the expression for the action of the operator over a general ket. The methodology uses the action of canonical transformations over the canonical variables on the classical theory. The most typical examples are the momentum as the generator of infinitesimal translation, the angular momentum as the generator of infinitesimal rotations and the Hamiltonian, with its peculiarities, as the generator of the time evolution of the system.

According to the classical mechanics, the solution for $n$-VMVF systems is given by the simultaneous evolution of the system in two coordinate systems: the rectangular, $x^\nu$, and the rotational's, $\xi_i$. We defined on previous sections that the extended vectors that describe the states of $n$-VMVF systems have two independent components containing the position of the particle in each representation respectively. The action of the operator position for the $n$-particle over the position ket is
\begin{equation}
\extoperator{x}{\xi} \extket{x_n'}{\xi'_i} = \extscalar{x_n'}{\xi'_i}  \extket{x_n'}{\xi'_i}.
\end{equation}

Other fundamental physical observables are the linear and angular momentum $p_n$ and $l_n$ which, according to the extended classical mechanic, remain as the generators for the translation and rotation of particle $n$, respectively. We need to find how the operators $\mathbf{p}$ and $\mathbf{l}$ act over the position ket, or what is the same, we need to find the function $p_u(x_n')$ or $l_u(\xi'_i)$ that appears as result of the action of an operator which have the linear or the angular momentum at the top component like:
\begin{equation}
\extoperator{ \mathbf{p} }{ A } \extket{x_n'}{\gamma} = \extscalar{p_u(x_n')}{1} \extket{x_n'}{\gamma} 
\qquad 
\extoperator{ \mathbf{l} }{ A } \extket{\xi'_i}{\gamma} = \extscalar{l_u(\xi'_i)}{1} \extket{\xi'_i}{\gamma}.
\end{equation}
Also, we need to find the form of the functions $p_d(x_n')$ or $l_d(\xi'_i)$ if the operators linear or the angular momentum, are at the bottom component of the extended operator like
\begin{equation}
\extoperator{ A }{ \mathbf{p} } \extket{\gamma}{x_n'} = \extscalar{1}{p_d(x_n')} \extket{\gamma} {x_n'}
\qquad 
\extoperator{ A }{ \mathbf{l} } \extket{\gamma}{\xi'_i} = \extscalar{1}{l_d(\xi'_i)} \extket{\gamma}{\xi'_i}
\end{equation}

The method to obtain the $p_u(x_n')$ , $p_d(x_n')$, $l_u(\xi'_i)$ and $l_d(\xi'_i)$ functions is the same as the ordinary quantum mechanic. It is based on the form the state vector is modified because of the action of the transformation operator on each component whose generator is the wanted operator. We have for example that the translation and the rotation operator as part of the motion operator has the linear and the angular momentum as the generator of each component respectively like
\begin{equation}
\extcolumn{\mathcal{T}^\dagger(dx')}{\mathcal{R}(d\xi'')} = \extcolumn{(\mathbf{1} + a_0^* \hbar_1^- \mathbf{p} d\mathbf{x}_n')^\dagger}{\mathbf{1} + b_0 \;\hbar_2^-\mathbf{l} \;d\mathbf{\xi}_n''}.
\end{equation}


The motion operator act over a position ket transforming the ket by displacing the corresponding coordinate of each component with an infinitesimal value without change the values of the others quantum degree of freedom like:
\begin{align}
\extcolumn{\mathcal{T}^\dagger(dx_n')}{\mathcal{R}(d\xi_n'')} 
\extket{x_1,.. x_n',..x_N}{\xi_1,..\xi_{n}'',.. \xi_N} 
&= \extcolumn{(\mathbf{1} + a_0^* \hbar_1^- \mathbf{p} \;d\mathbf{x}_n')^\dagger}{\mathbf{1} + b_0 \hbar_2^-\mathbf{l} \;d\mathbf{\xi}_n''\;} 
\extket{x_1,.. x_n', ..x_N}{\xi_1,..\xi_{n}'',.. \xi_N}
\nonumber \\
&= \extket{x_1,..x_n' +d\mathbf{x}_n',..x_N }{\xi_1,..\xi_{n}'' + d\mathbf{\xi}_n'',.. \xi_N}
\end{align}
where $\hbar_1, \hbar_2$ are constants with action dimension. These constants play the same role here that $\hbar$ constant does on the quantum mechanic. These constants also need to be computed.

The form of the wanted operator is independent of the other component because of the independence of the action between the top and the bottom component of the extended operator. The action of the motion operator due to translation in the top position over the top component of a general ket for any operators located in the bottom position of the extended operator is written as
\begin{equation}
\extcolumn{\mathcal{T}^\dagger(dx')}{\mathbf{A}} \extket{\alpha}{\gamma} = 
\extcolumn{(\mathbf{1} + a_0^* \hbar_1^- \mathbf{p} \;d\mathbf{x}_n')^\dagger}{\mathbf{A}}
\extket{\alpha}{\gamma}. 
\end{equation}
We do not know the form of the action of the momentum over the ket component $\alpha$. However, we do know how the motion operator acts over the position ket.
Also, we assume that the action of the bottom component  of the operator, $\mathbf{A}$, over the bottom component of the ket, $\ket{\gamma}$, is always $\ket{\gamma'}$ like
\begin{equation}
\extcolumn{\mathcal{T}^\dagger(dx')}{\mathbf{A}} \extket{\alpha}{\gamma} = 
\extcolumn{\mathcal{T}^\dagger(dx')}{\mathbf{1}} \extket{\alpha}{\gamma'}. 
\end{equation}
We expand then the extended ket on terms of the position kets:
\begin{align}
\extcolumn{\mathcal{T^\dagger}(dx')}{\mathbf{A}} \extket{\alpha}{\gamma}
&= \sum_{x_n'} \extscalar{c_{x_n'}}{1} \extcolumn{\mathcal{T}^\dagger(dx')}{\mathbf{A}} \extket{x_n'}{\gamma}
\nonumber \\
&= \sum_{x_n'} \extscalar{c_{x_n'}}{1}  \extket{x_n' +d\mathbf{x}_n' }{\gamma'}
\nonumber \\
&= \sum_{x_n'} \extscalar{c_{(x_n' - d x_n')}}{1} \extket{x_n'}{\gamma'}
\end{align}
where we assume that the bottom component of the expansion coefficients is the unit because the expansion only occurs at the top component of the ket. On the last step, the translation $x_n' $ coordinate is replacing by $x_n' - d\mathbf{x}_n'$.

If the state are normalized then coefficients $c_{x_n'}$ can be expanded in series up to the second term:
\begin{align}
\extcolumn{\mathcal{T^\dagger}(dx')}{\mathbf{A}} \extket{\alpha}{\gamma}  &= 
\sum_{x_n'} \extscalarReal{c_{x_n'} -  dx_n'\frac{\partial c_{x_n'}}{\partial x_n'} }{1}   \extket{x_n'}{\gamma'}
\end{align}

On the other side, we have that
\begin{align}
\extcolumn{(\mathbf{1} + a_0 \hbar_1^- \mathbf{p} \;d\mathbf{x}_n')^\dagger}{\mathbf{A}}
\extket{\alpha}{\gamma} &= 
\extcolumn{(\mathbf{1} + a_0 \hbar_1^- \mathbf{p} \;d\mathbf{x}_n')^\dagger}{\mathbf{A}} \sum_{x_n'} \extscalar{c_{x_n'}}{1}\extket{x_n'}{\gamma'}
\nonumber \\
&= \sum_{x_n'} \extcolumn{1 + a_0^* \hbar_1^- p_u(x_n')\; dx_n'}{1} \extscalar{c_{x_n'}}{1}\extket{x_n'}{\gamma'}
\end{align}
Comparing both sides of the equation, we have
\begin{equation}
\extscalarReal{c_{x_n'} - dx_n' \frac{\partial c_{x_n'}}{\partial x_n'}}{1} =
\extcolumn{1 + a_0^* \hbar_1^- p_u(x_n')\; dx_n'}{1} \extscalar{c_{x_n'}}{1}\label{upperMomentumExp00}
\end{equation}
which we can rewrite using the standard and complex products:
\begin{align}
\big( c_{x_n'} -  dx_n' \frac{\partial c_{x_n'}}{\partial x_n'}  \big) \odot 1 &= 
\big[ \big( 1 + a_0^* \hbar_1^- p_u\; dx_n' \big) \odot 1 \big] \cdot
\big[c_{x_n'} \odot 1 \big]
\end{align}
or 
\begin{align}
\big( c_{x_n'} -  dx_n' \frac{\partial c_{x_n'}}{\partial x_n'}  \big)^*&=\big[ 1 + a_0^* \hbar_1^- p_u\; dx_n' \big]^* \cdot c_{x_n'}^* \label{upperMomentumExp0}
\end{align}
where $p_u(x_n') \equiv p_u$. 

It is possible to express the right member of the last equation as
\begin{equation}
\big[ 1 + a_0^* \hbar_1^- p_u\; dx_n' \big]^* \cdot c_{x_n'}^*=
\big[ (1 + a_0^* \hbar_1^- p_u\; dx_n' )c_{x_n'} + G'(x_n',c_{x_n'})dx_n'\big]^*\label{upperMomentumExp}.
\end{equation}
Indeed, using the extended algebra, with a few effort, we can obtain the equation
\begin{equation}
\big[ 1 + a_0^* \hbar_1^- p_u\; dx_n' \big]^* \cdot c_{x_n'}^*=
\big[ (1 + a_0^* \hbar_1^- p_u\; dx_n' )c_{x_n'} + G(x_n',c_{x_n'},dx_n')\big]^* 
\label{upperMomentumExp1}
\end{equation}
where the $ G(x_n',c_{x_n'},dx_n')$ is a function whose extended and imaginary parts have the expressions
\begin{align}
&G(x_n',c_{x_n'},dx_n')_E = \frac{1}{z_1^{(N)}}
\Big\{
(1 + a_0^* \hbar_1^- p_u\; dx_n')_E^* c_{{x_n'}_E}^* \big[ z_1^{(N_1)} z_1^{(N_2)}z_0 + w_1^{(N_2)}  +   w_1^{(N_1)} z_1^{(N_2)} + z_0^* z_1^{(N)})   \big]
\nonumber \\
&+ (1 + a_0^* \hbar_1^- p_u\; dx_n')_E^* c_{{x_n'}_I}^* \big[ z_1^{(N_1)} - z_1^{(N)}  \big]
+ (1 + a_0^* \hbar_1^- p_u\; dx_n')_I^* c_{{x_n'}_E}^* \big[ z_1^{(N_2)} - z_1^{(N)}  \big]
\Big\}
\end{align}
and
\begin{align}
&G(x_n',c_{x_n'},dx_n')_I = - w_1^{(N)} G(x_n',c_{x_n'},dx_n')_E^* + 
(1 + a_0^* \hbar_1^- p_u\; dx_n')_E^* c_{{x_n'}_E}^* \big[ z_1^{(N_1)} z_1^{(N_2)}w_0 + w_1^{(N_2)}  
\nonumber \\
&+   w_1^{(N_1)} w_1^{(N_2)} 
- z_0^* w_1^{(N)})  + w_0^* \big]
+ (1 + a_0^* \hbar_1^- p_u\; dx_n')_E^* c_{{x_n'}_I}^* \big[ w_1^{(N_1)} - w_1^{(N)}  \big]
\nonumber \\
&+ (1 + a_0^* \hbar_1^- p_u\; dx_n')_I^* c_{{x_n'}_E}^* \big[ w_1^{(N_2)} - w_1^{(N)}  \big],
\end{align}
respectively. The letters $N,N_1,N_2$ stand for the extended numbers 
\begin{equation}
N \equiv 1 + a_0^* \hbar_1^- p_u\; dx_n' + c_{x_n'},
\quad N_1 \equiv 1 + a_0^* \hbar_1^- p_u\; dx_n',  
\quad \text{and} \quad N_2 \equiv c_{x_n'}.
\end{equation}

On the other side, if the value $dx_n'=0$ is replaced in the definition of the function  $G(x_n',c_{x_n'},dx_n')$ on equation \ref{upperMomentumExp1}, we have that
\begin{equation}
c_{x_n'}^*= \big[c_{x_n'} + G(x_n',c_{x_n'},0)\big]^*,
\end{equation}
from we conclude that $G(x_n',c_{x_n'},0) = 0$. 

We can then, expand function $G(x_n',c_{x_n'},dx_n')$ around point $dx_n'=0$ to the first order like
\begin{align}
G(x_n',c_{x_n'},dx_n')
&\sim  G(x_n',c_{x_n'},0) + \frac{\partial G(x_n',c_{x_n'},dx_n')}{\partial (dx_n')} \bigg\rvert_{dx_n'=0} dx_n' 
\nonumber \\
&= \frac{\partial G(x_n',c_{x_n'},dx_n')}{\partial (dx_n')} \bigg\rvert_{dx_n'=0} dx_n' \equiv  G'(x_n',c_{x_n'})dx_n, \label{GFunctionOperTopComponent}
\end{align}
from we obtain the form of the function $ G'(x_n',c_{x_n'})$, that we name $G$-function. Replacing this result on equation \ref{upperMomentumExp1}, we arrive to the intended equation \ref{upperMomentumExp}. Replacing equation  \ref{upperMomentumExp} on equation \ref{upperMomentumExp0}, we have
\begin{equation}
\Big( c_{x_n'} -  dx_n'\frac{\partial c_{x_n'}}{\partial x_n'} \Big)^*= 
\Big[ (1 + a_0^* \hbar_1^- p_u(x_n')\; dx_n' )c_{x_n'} + G'(x_n',c_{x_n'})dx_n'\Big]^*
\end{equation}
or like in the representation of extended scalars
\begin{equation}
\extscalarReal{c_{x_n'} - dx_n' \frac{\partial c_{x_n'}}{\partial x_n'}}{1} =
\extcolumn{(1 + a_0^* \hbar_1^- p_u(x_n')\; dx_n' )c_{x_n'} + G'(x_n',c_{x_n'})dx_n'}{1}.
\end{equation}
Comparing the top components of the scalars, we finally obtain
\begin{equation}
p_u(x_n',c_{x_n'}) = -  a_0^{*-} \hbar_1 \Big[ \frac{\partial c_{x_n'}}{\partial x_n' } + G'(x_n',c_{x_n'}) \Big].
\end{equation}

The expression for the linear momentum on the bottom component of the operator can be founded using a similar method.
 
Expanding the state ket with eigenvectors whose bottom component is the position coordinate and replacing the expansion coefficients $c_{x_n'}$ with series up to the second term, we have:
\begin{align}
\extcolumn{A^\dagger}{\mathcal{T}(dx')} \extket{\gamma}{\alpha}  &= 
\sum_{x_n'} \extscalarReal{1}{c_{x_n'} -  dx_n'\frac{\partial c_{x_n'}}{\partial x_n'} } \extket{\gamma'}{x_n'}.
\end{align}
Also, the action of the translation operator results in the equation
\begin{equation}
\extcolumn{A^\dagger}{\mathbf{1} + b_0 \hbar_1^- \mathbf{p} \;d\mathbf{x}_n'}
\extket{\gamma}{\alpha} = \sum_{x_n'} \extcolumn{1}{1 + b_0 \hbar_1^- p_d(x_n')\; dx_n'} \extscalar{1}{c_{x_n'}}\extket{\gamma'}{x_n'}.
\end{equation}

Comparing both sides of the equation, we have
\begin{equation}
\extscalarReal{1}{c_{x_n'} - dx_n' \frac{\partial c_{x_n'}}{\partial x_n'}} =
\extcolumn{1}{1 + b_0 \hbar_1^- p_d(x_n')\; dx_n'} \extscalar{1}{c_{x_n'}}.
\end{equation}
Using the standard and complex products the equality is:
\begin{equation}
{1}\odot ( c_{x_n'} - dx_n' \frac{\partial c_{x_n'}}{\partial x_n'})  = \big[{1}\odot ( 1 + b_0 \hbar_1^- p_d(x_n')\; dx_n')\big] \cdot ({1}\odot{c_{x_n'}})
\end{equation}
or
\begin{equation}
c_{x_n'} - dx_n' \frac{\partial c_{x_n'}}{\partial x_n'} =  (1 + b_0 \hbar_1^- p_d(x_n')\; dx_n') {c_{x_n'}} = c_{x_n'} +  b_0 \hbar_1^-  dx_n' p_d(x_n') c_{x_n'}
\end{equation}
The comparison, in this case is more straightforward, and results in
\begin{equation}
p_d(x_n',c_{x_n'}) = -  b_0^{-} \hbar_1  \frac{\partial c_{x_n'}}{\partial x_n'}.
\end{equation}
which resembles the form of the linear momentum operator in the ordinary quantum mechanics.

We can also find the form of functions $l_u(\xi'_i)$ and $l_d(\xi'_i)$ in the scalars which result from:
\begin{equation}
\extoperator{ \mathbf{l} }{ A } \extket{\xi'_i}{\gamma'} = \extscalar{l_u(\xi'_i)}{\gamma'} \extket{\xi'_i}{\gamma'}
\qquad 
\extoperator{ A }{ \mathbf{l} } \extket{\gamma'}{\xi'_i} = \extscalar{\gamma'}{l_d(\xi'_i)} \extket{\gamma'}{\xi'_i}
\end{equation}

From the extended classical mechanic, the angular momentum operator is obtained as the generator for the rotation of the position of the particle. Functions $l_u(\xi'_i)$ and $l_d(\xi'_i)$ can be founded because of the known form of the action of the infinitesimal rotation operator over the state vector. The action of the infinitesimal rotation laying as the top or bottom component of an external operator over a state vector whose top or bottom component, respectively, is the position of the particle expressed in angular coordinates, regardless the operator of the other component, have the form
\begin{equation}
\extcolumn{\mathcal{R}^\dagger(d\xi_n')}{\mathbf{A}} \extket{\xi_{n}'}{\gamma} = \extcolumn{(\mathbf{1} + a_0^* \hbar_2^-\mathbf{l} \;d\mathbf{\xi}_n')^\dagger}{\mathbf{A}} \extket{\xi_{n}'}{\gamma} = \extket{\xi_{n}' + d\mathbf{\xi}_n'}{\gamma }
\end{equation}
and
\begin{equation}
\extcolumn{\mathbf{A}^\dagger}{\mathcal{R}(d\xi_n')} \extket{\gamma}{\xi_{n}'} = \extcolumn{\mathbf{A}^\dagger}{\mathbf{1} + b_0 \hbar_2^-\mathbf{l} \;d\mathbf{\xi}_n'\;} \extket{\gamma}{\xi_{n}'} = \extket{\gamma }{\xi_{n}' + d\mathbf{\xi}_n'}.
\end{equation}
Following the above method for the functions $p_u(x_n')$ and $p_d(x_n')$ with a few changes, we can find the desired functions, and their expressions are:
\begin{equation}
l_u(\xi_{n}',c_{\xi_{n}'}) = -  a_0^{*-} \hbar_2 \Big[ \frac{\partial c_{\xi_{n}'}}{\partial \xi_{n}' } + G'(\xi_{n}',c_{\xi_{n}'}) \Big]
\end{equation}
and
\begin{equation}
l_d(\xi_{n}',c_{\xi_{n}'}) = -  b_0^{-} \hbar_2  \frac{\partial c_{\xi_{n}'}}{\partial \xi_{n}' }, 
\end{equation}
where function $G'(\xi_{n}',c_{\xi_{n}'}) $ have the form of a $G$-function from equation \ref{GFunctionOperTopComponent} depending now in the angular coordinates and with letters $N,N_1,N_2$ standing for $N \equiv 1 + a_0^* \hbar_2^- l_u\; \xi_{n}' + c_{\xi_{n}'}$, $N_1 \equiv 1 + a_0^* \hbar_2^- l_u\; d\xi_{n}'$ and $N_2 \equiv c_{\xi_{n}'}$.

For example, the action of the extended operator $\extoperator{p}{l}$ over a general position ket is:
\begin{equation}
\extoperator{p}{l} \extket{\gamma}{\beta}= \sum_{x_n',\xi_n'} \extscalarReal{  - a_0^{*-}\hbar_1 \big[ \frac{\partial c_{x_n'}}{\partial x_n'}  +  G'(x_n',c_{x_n'}) \big]}{ -  b_0^{-} \hbar_2  \frac{\partial c_{\xi_{n}'}}{\partial \xi_{n}' } } \extket{x_n'}{\xi_n'}.
\end{equation}

If we express the bottom component the extended ket as function of rectangular coordinates, then we can replace the derivative $\frac{\partial c_{\xi_{n}'}}{\partial \xi_{n}' }$ using the transformation relations between the rectangular and the angular coordinates. In that case, the equation take the form
\begin{equation}
\extoperator{p}{l}= \sum_{x_n',x_n''} \extscalarReal{  a_0^{*-} \hbar_1\big[ \frac{\partial c_{x_n'}}{\partial x_n'}  +  G'(x_n',c_{x_n'}) \big]}{ b_0^- \hbar_2\big\{ \big[ D_{\;\xi}^{''\nu} \square_\nu'' + \frac{d }{d\tau} (D_{\;\xi}^{''\nu} ) \square_{\dot{\nu}}'' + \frac{d^2}{d\tau^2} (D_{\;\xi}^{''\nu}  ) \square_{\ddot{\nu}}''
\big]  c_{x_n''} \big \} } \extket{x_n'}{x_n''}.
\end{equation}

\subsubsection{Momentums \textbf{s} and \textbf{b} as generators of the displacements of the correlation function's pole in rectangular and angular coordinates.}

Another magnitude which appears on the extension of classical mechanics and plays a fundamental role in the description of $n$-VMVF systems is the second order momentum $s$. We define then a quantum operator related to this classical magnitude whose form of its action over a state ket, same as linear momentum, should also be determined. The extended operator of momentum $s$ in both coordinates can be defined as
\begin{equation*}
\extoperator{s}{b},
\end{equation*}
where $b$ is the second order momentum in angular coordinates. We follow in here the convention of putting the rectangular coordinates at the top component and the angular's at the bottom's.

We know from the study of the proposed classical solution for $n$-VMVF systems, that the momentum $s_n$ is the generator for an inverse displacement of the pole of the correlation function $\mathcal{F}_n(\bar{\dot{q}})$. For that reason, same as the motion state ket, and according to the extension of classical mechanics, we define the state:
\begin{equation}
\extket{f_{x_n}'}{f_{\xi_n}'}
\end{equation}
whose eigenvalues are the scalars containing the values of the pole of the correlation functions in the rectangular and the angular coordinates. They are related to the time derivative of the eigenvalue of the position of all particles as:
\begin{equation}
\mathcal{F}_{x_n'}(\bar{\dot{x}}_m') = {f_{x_n}'} \qquad \text{and} \qquad {\mathcal{F}_{\xi_n}(\bar{\dot{\xi}}_{m}')} = {f_{\xi_n}'}
\end{equation}
where the form of the set of functions $\mathcal{F}_{x_n}(\bar{\dot{x}}_m)$ and  $ \mathcal{F}_{\xi_n}(\bar{\dot{\xi}}_{m})$ are extracted from the extended classical mechanics, and we supposed already known.

Let us represent the infinitesimal operator that dislocate the values $f_{n}'$, in both set of coordinates, as $\mathcal{T}_{f}(df_{x_n}')$ and $\mathcal{R}_f(df_{\xi_n}')$. Each component of this operator should have the general form of the infinitesimal extended operators while the generator of each component, same as before, should be obtained from the classical theory. The action of such operator over a state vector characterized by the pole of the correlation functions has the form:
\begin{align}
\extcolumn{\mathcal{T}_{f}^\dagger(df_{x_n}')}{\mathcal{R}_f(df_{\xi_n}')} \extket{f_{x_1}',..f_{x_n}', .. f_{x_N}'}{f_{\xi_1}',..f_{\xi_n}',..f_{\xi_N}'} 
&= \extcolumn{(\mathbf{1} + a_0^* \hslash_1^- \mathbf{s} \;df_{x_n}')^\dagger}{\mathbf{1} + b_0 \hslash_2^- \mathbf{b} \;df_{\xi_n}'} 
\extket{f_{x_1}',..f_{x_n}', .. f_{x_N}'}{f_{\xi_1}',..f_{\xi_n}',..f_{\xi_N}'}
\nonumber \\
&= \extket{f_{x_1}',..f_{x_n}' - df_{x_n}', .. f_{x_N}'}{f_{\xi_1}',..f_{\xi_n}' - df_{\xi_n}',..f_{\xi_N}'},
\end{align}
where $s$ and $b$ are effectively included in the operator as the generator of the displacement of the pole of functions $F_{q_n}'$ for linear and angular coordinates. Also, $\hslash_1, \hslash_2$ (``h-slash'') are constants with action dimension and they play the role of adjusting the result to the classical measures.

Performing the same method as before, its possible to obtain the functions $s_u(f_{x_n}')$ and $s_d(f_{x_n}')$ of the referred operator which satisfy
\begin{equation}
\extoperator{ \mathbf{s} }{ A } \extket{f_{x_n}'}{\gamma} = \extscalar{s_u(f_{x_n}')}{1} \extket{f_{x_n}'}{\gamma} 
\qquad 
\extoperator{ A }{ \mathbf{s} } \extket{\gamma}{f_{x_n}'} = \extscalar{1}{s_d(f_{x_n}')} \extket{\gamma} {f_{x_n}'}.
\end{equation}
Also, the functions $b_u(f_{\xi_n}')$ and $b_d(f_{\xi_n}')$ if the operators linear and angular momentum, are on the bottom component of the extended operator like
\begin{equation}
\extoperator{ \mathbf{b} }{ A } \extket{f_{\xi_n}'}{\gamma} = \extscalar{b_u(f_{\xi_n}')}{1} \extket{f_{\xi_n}'}{\gamma}
\qquad 
\extoperator{ A }{ \mathbf{b} } \extket{\gamma}{f_{\xi_n}'} = \extscalar{1}{b_d(f_{\xi_n}')} \extket{\gamma}{f_{\xi_n}'}.
\end{equation}

Let us start with operator momentum $\mathbf{s}$ placed on the upper component. If the state is normalized then coefficients $c_{f_{x_n}'}$ can be expanded in series up to the second term:
\begin{align}
\extcolumn{\mathcal{T}_f^\dagger(df_{x_n}')}{A}& \extket{\alpha}{\gamma}  = 
\extcolumn{\mathcal{T}_f^\dagger(df_{x_n}')}{A} \sum_{f_{x_n}'}\extscalarReal{c_{f_{x_n}'}}{1}   \extket{f_{x_n}'}{\gamma'}
\nonumber \\
&= \sum_{f_{x_n}'}\extscalarReal{c_{f_{x_n}'}}{1}   \extket{f_{x_n}' - df_{x_n}'}{\gamma'}
=\sum_{f_{x_n}'} \extscalarReal{c_{f_{x_n}'} +  d f_{x_n}'\frac{\partial c_{f_{x_n}'}}{\partial f_{x_n}'} }{1}   \extket{f_{x_n}'}{\gamma'}
\end{align}
where $f_{x_n}'$ was substitute by $f_{x_n}' - df_{x_n}'$. On the other side, we have that
\begin{equation}
\extcolumn{(\mathbf{1} + a_0^* \hslash_1^- \mathbf{s} \;df_{x_n}')^\dagger}{A}
\extket{\alpha}{\gamma} = \sum_{f_{x_n}'} \extcolumn{1 + a_0^* \hbar_1^- s_u(f_{x_n}')\; df_{x_n}'}{1} \extscalar{c_{f_{x_n}'}}{1}\extket{f_{x_n}'}{\gamma'}
\end{equation}

Comparing both sides of the equation, we have
\begin{equation}
\extscalarReal{c_{f_{x_n}'} +  d f_{x_n}'\frac{\partial c_{f_{x_n}'}}{\partial f_{x_n}'} }{1}  =\extcolumn{1 + a_0^* \hslash_1^- s_u(f_{x_n}')\; df_{x_n}'}{1} \extscalar{c_{f_{x_n}'}}{1}\label{upperMomentumSExp00}.
\end{equation}
Applying the extended algebra, we can transform the right member like
\begin{equation}
\extscalarReal{c_{f_{x_n}'} +  d f_{x_n}'\frac{\partial c_{f_{x_n}'}}{\partial f_{x_n}'} }{1}  =
\extcolumn{(1 + a_0^* \hslash_1^- s_u(f_{x_n}')\; df_{x_n}')c_{f_{x_n}'} + G'(f_{x_n}',c_{f_{x_n}'})df_{x_n}'}{1},
\end{equation}
being function $G'(f_{x_n}',c_{f_{x_n}'})$ a $G$-function, from equation \ref{GFunctionOperTopComponent}, depending on the poles $f_{x_n}$. The equality of the top components take us to
\begin{equation}
s_u(f_{x_n}',c_{f_{x_n}'}) = a_0^{*-} \hslash_1 \Big[\frac{\partial c_{f_{x_n}'}}{\partial f_{x_n}'} - G'(f_{x_n}',c_{f_{x_n}'}) \Big].
\end{equation}

The same analysis is made if the second order momentum $\mathbf{s}$ are on the bottom component of the operator acting over the bottom component of the ket, and it results in a scalar with the function $s_d(f_{x_n}',c_{f_{x_n}'})$ being in the form:
\begin{equation}
s_d(f_{x_n}',c_{f_{x_n}'}) = b_0 \hslash_2 \Big[\frac{\partial c_{f_{x_n}'}}{\partial f_{x_n}'} \Big].
\end{equation}
Also, we can obtain the expressions
\begin{equation}
b_u(f_{\xi_n}',c_{f_{\xi_n}'}) = a_0^{*-} \hslash_1 \Big[\frac{\partial c_{f_{\xi_n}'}}{\partial f_{\xi_n}'} - G'(f_{\xi_n}',c_{f_{\xi_n}'}) \Big]
\end{equation}
and 
\begin{equation}
b_d(f_{\xi_n}',c_{f_{\xi_n}'}) = b_0 \hslash_2 \Big[\frac{\partial c_{f_{\xi_n}'}}{\partial f_{\xi_n}'} \Big].
\end{equation}

The action of the extended operator $\extoperator{s}{b}$ over a general extended ket then have the form:
\begin{equation}
\extoperator{s}{b}  \extket{\alpha}{\beta}= \sum_{f_{x_n}' f_{\xi_n'}} 
\extscalarReal{ a_0^{*-} \hslash_1 \Big[\frac{\partial c_{f_{x_n}'}}{\partial f_{x_n}'} - G'(f_{x_n}',c_{f_{x_n}'})\Big] } 
{b_0 \hslash_2 \frac{\partial c_{f_{x_n}'}}{\partial f_{x_n}'} } 
\extket{f_{x_n}'}{f_{\xi_n}'}
\end{equation}

%

The general expansion of a state vector is composed on the expansion over the position ket and functions $f_{\xi_n''}$ $e.i$, and it should look like
\begin{equation}
\extket{\alpha}{\beta} = \sum_{\substack{ x_n',\xi_{n}'\\ f_{x_n}' f_{\xi_n'}}} \extscalar{c_{(x_n',f_{x_n}')}}{c_{(\xi_{n}',f_{\xi_n'})}} \extket{x_n', f_{x_n}'}{\xi_{n}', f_{\xi_n'}} 
\end{equation}
alternatively, if we factorize the expansion coefficients, based on the independence of position coordinates and functions $f_{\xi_n''}$,
\begin{equation}
\extket{\alpha}{\beta} = \sum_{\substack{ x_n',\xi_{n}'\\ f_{x_n}' f_{\xi_n'}}} \extscalar{c_{x_n'} c_{f_{x_n}'}}{c_{\xi_{n}'} c_{f_{\xi_n'}}} \extket{x_n', f_{x_n}'}{\xi_{n}', f_{\xi_n'}} 
\end{equation}

In this case, an operator whose component is a product of momentum $p$ and $s$ acting over a general ket modify the expansion component as:
\begin{equation}
\extcolumn{ c_1 \mathbf{p} \cdot\mathbf{s}}{c_2\mathbf{l}\cdot\mathbf{b} }\extket{\alpha}{\beta} =  
\sum_{\substack{ x_n',\xi_{n}'\\ f_{x_n}' f_{\xi_n'}}} 
\extscalarReal{ - c_1 a_0^{*-2} \hbar_1 \hslash_1 \big[\frac{\partial c_{f_{x_n}'}}{\partial f_{x_n}'} - G'(f_{x_n}',c_{f_{x_n}'})\big] 
\big[  \frac{\partial c_{x_n'}}{\partial x_n'}  +  G'(x_n',c_{x_n'}) \big] } 
{-c_2 b_0^{-2} \hbar_2 \hslash_2 \big[\frac{\partial c_{f_{x_n}'}}{\partial f_{x_n}'} \big] \big[\frac{\partial c_{\xi_{n}'}}{\partial \xi_{n}'} \big]}
 \extket{x_n', f_{x_n}'}{\xi_{n}', f_{\xi_n'}} 
\end{equation}
where $c_1$ and $c_2$ are two constant settled for the dimensional fit.

\subsection{Energy of particle system}
The energy of the system is the eigenvalue of an operator whose top and bottom components have the form of the energy function from the classical theory, equation \ref{extEnergyFunction}:
\begin{equation*}
h= \sum_i p_i \dot{q}_i - \dot{s}_i\dot{q}_i + s_i\ddot{q}_i  - L 
= \sum_i H - \dot{s}_i\dot{q}_i, 
\end{equation*}
for both sets of coordinates. The components of the energy eigenvalue are related to the translation and rotation energy, respectively. The total energy extended operator whose top component is described by the rectangular coordinates $x^\nu$ and the bottom's by the angular's $\xi_i$ have the form
\begin{equation}
\extcolumn{\mathbf{H}_T - \sum_n \mathbf{\dot{s}}^\nu_n \mathbf{\dot{x}}_{n;\nu}}{\mathbf{H}_R - \sum_n \mathbf{\dot{b}}_{n;i} \mathbf{\dot{\xi}}_{n;i}}
\end{equation}
where we substitute the classical magnitudes with its extended operators and use the Einstein summation notation. According to the extension of classical mechanics, this eigenvalue must remain constant in time. Unfortunately, energy function depends on the time derivative of momentum $s$ and positions operators, and for that reason, we set a time evolution study before treating this problem.   

\subsection{Time Evolution}
In this section, we study the evolution of the $n$-VMVF quantum systems with time. The inclusion of the four component on particle position $x_n^\nu$ on the extended Quantum Mechanics imply that the evolution is related not to time, but to proper time $\tau$. On the Special Theory of Relativity, time is treated in the system as another degree of freedom which is related with the others by the Lorentz constraint which, following Dirac's ideas, we still consider it in here as a ``weak condition''. In this case, same as the quantum mechanic, proper time should be considered as a parameter and not an observable. From now,  we use the word ``time'' instead of ``proper time'' in the benefit of simplicity. When the reference of time as the four component of the particle position is needed, we will explicitly express it as such.

Along the development of this quantum theoretic proposal, we consider the action of the operator over the state vector divided in the action of each component of the operator over the corresponding component of the ket. We contemplate such actions independent one from the other. This statement is not entirely accurate. The action of each component have one thing in common: the same parameter describes their temporal evolution, the proper time of the inertial frame of reference from where the $n$-VMVF system is studied. The temporal evolution of all the magnitudes on both components occur simultaneously.

Following the above ideas and the quantum mechanic reasoning, we represent the space ket at a given observation time $\tau$, started to be observed at a given initial time $\tau_0$ as:
\begin{equation}
\extket{\alpha,\tau_0;\tau}{\beta,\tau_0;\tau} \qquad \qquad (\tau> \tau_0),
\end{equation}
being $\tau$ a continuous parameter. As the system was supposed to be in the state $\extket{\alpha}{\beta}$ at the reference time $\tau_0$, its expected then that:
\begin{equation}
\lim_{\tau \to \tau_0} \extket{\alpha,\tau_0;\tau}{\beta,\tau_0;\tau} = \extket{\alpha,\tau_0;\tau_0}{\beta,\tau_0;\tau_0} \equiv  \extket{\alpha}{\beta}.
\end{equation}

The time evolution of the system can be studied using an extended operator named the time evolution operator whose action over a state vector evolve the state from time $\tau_0$ to $\tau$ such as:
\begin{equation}
\extcolumn{\mathcal{U}_u^\dagger(\tau,\tau_0)}{\mathcal{U}_b(\tau,\tau_0)}  \extket{\alpha}{\beta} \equiv \extcolumn{\mathcal{U}_u^\dagger(\tau,\tau_0)}{\mathcal{U}_b(\tau,\tau_0)} \extket{\alpha,\tau_0;\tau_0}{\beta,\tau_0;\tau_0}= \extket{\alpha,\tau_0;\tau}{\beta,\tau_0;\tau}
\end{equation}
where subscripts $u$ and $b$ refers to the time operator components that evolve the up and bottom component of the state vector respectively. The time evolution operator act over both component of the extended ket with the same value of the parameter $\tau$. Time evolution operator must satisfy some properties:
\begin{enumerate}
\item Unitarity:
This property means that the resulting ket must satisfy the normalization relations. As shown, it is sufficient that the operator satisfies
\begin{equation}
\extcolumn{\mathcal{U}_u^\dagger(\tau,\tau_0)}{\mathcal{U}_b(\tau,\tau_0)}^\ddagger 
\extcolumn{\mathcal{U}_u^\dagger(\tau,\tau_0)}{\mathcal{U}_b(\tau,\tau_0)}
= \extcolumn{1}{1},
\end{equation}
\item Composition property: 

The time evolution operator must satisfy this property on its components since, as explained on the introduction of the infinitesimal extended operators, we cannot assure that the evolution of the hidden variables, which is replaced by the evolution of one component, satisfy the composition law. 
We propose then that the composition property is satisfied at every component like:
\begin{equation}
\mathcal{U}_{u}(\tau_2,\tau_0) =  
\mathcal{U}_u(\tau_2,\tau_1) \mathcal{U}_u(\tau_1,\tau_0) 
\quad \text{and} \quad 
\mathcal{U}_b(\tau_2,\tau_0) =  
\mathcal{U}_b(\tau_2,\tau_1) \mathcal{U}_b(\tau_1,\tau_0) \label{CompositionPro}
\end{equation}
alternatively, in the extended operator notation:
\begin{equation}
\extcolumn{\mathcal{U}_{u}^\dagger(\tau_2,\tau_0)}{\mathcal{U}_{b}(\tau_2,\tau_0)}  = \extcolumn{\big[ \mathcal{U}_u(\tau_2,\tau_1) \mathcal{U}_u(\tau_1,\tau_0) \big]^\dagger}
{\big[ \mathcal{U}_b(\tau_2,\tau_1) \mathcal{U}_b(\tau_1,\tau_0)\big] \;}.
\end{equation}
\item An infinitesimal time evolution operator is a particular case of the time evolution operators such that
\begin{equation}
\extcolumn{\mathcal{U}_u^\dagger(\tau_0 + d\tau,\tau_0)}{\mathcal{U}_b(\tau_0 + d\tau,\tau_0)} \extket{\alpha,\tau_0;\tau_0}{\beta,\tau_0;\tau_0}= 
\extket{\alpha,\tau_0;\tau + d\tau}{\beta,\tau_0;\tau + d\tau}.
\end{equation}
\item Because of the continuity property, the infinitesimal time evolution extended operator must reduce to identity operator when $d\tau$ goes to zero:
\begin{equation}
\lim_{d\tau \to 0} \extcolumn{\mathcal{U}_u^\dagger(\tau_0 + d\tau,\tau_0)}{\mathcal{U}_b(\tau_0 + d\tau,\tau_0)} =  \extcolumn{1}{1}.
\end{equation}
\end{enumerate}

Same as motion operator, the infinitesimal time evolution operator can be written as
\begin{equation}
\extcolumn{\mathcal{U}_u^\dagger(\tau_0 + d\tau,\tau_0)}{\mathcal{U}_b(\tau_0 + d\tau,\tau_0)} = \extcolumn{\mathbf{1} + a_0^* \mathbf{\Omega}_u^\dagger \;d\tau}{\mathbf{1} + b_0 \mathbf{\Omega}_d \;d\tau}
\end{equation}
where each component of the operator $\extcolumn{\mathbf{\Omega}_u^\dagger}{\mathbf{\Omega}_b}$ satisfy
\begin{equation}
{\mathbf{\Omega}_u^\dagger} = \mathbf{\Omega}_u^{\dagger \ddagger} \qquad   
{\mathbf{\Omega}_b} = {\mathbf{\Omega}_b}^\ddagger  \qquad
{\mathbf{\Omega}_b^\dagger} = \mathbf{\Omega}_d^{\dagger \ddagger} \qquad   
{\mathbf{\Omega}_u} = {\mathbf{\Omega}_u}^\ddagger  .
\end{equation}

The composition property can also be verified on each component:
\begin{align}
\extcolumn{\big[ \big( \mathbf{1} + a_0^* \mathbf{\Omega}_u \;d\tau_1 \big) \big(
\mathbf{1} + a_0^* \mathbf{\Omega}_u \;d\tau_2 \big)\big]^\dagger}
{\big(\mathbf{1} + b_0 \mathbf{\Omega}_d \;d\tau_1 \big) \big( \mathbf{1} + b_0 \mathbf{\Omega}_d \;d\tau_2\big)} 
= 
\extcolumn{\big[ \mathbf{1} + a_0^* \mathbf{\Omega}_u \;(d\tau_1 + d\tau_2) + \mathcal{O}(d\tau^2)\big]^\dagger}{\mathbf{1} + b_0 \mathbf{\Omega}_d \;(d\tau_1 + d\tau_2)  + \mathcal{O}(d\tau^2)}
\end{align}

Now we can take borrowed from the extended classical mechanic, the form of the function \ref{timeGenerator1}
\begin{equation*}
\mathcal{G} = H+\sum_i \dot{s}_i\delta \dot{q}_i - \ddot{s}_i q_i.
\end{equation*}
or its approximate expression from equation \ref{timeGeneratorApx1}
\begin{equation*}
\mathcal{G} \sim H + \sum_i  \dot{s}_i \dot{f}_i(\bar{\dot{q}}) t - \ddot{s}_i q_i,
\end{equation*}
as the generator of time evolution of the system. If the top ket component is described by the rectangular coordinates $x^\nu$ and the bottom's are described using the angular coordinates $\xi_i$, the generators $\mathbf{\Omega}_u$ and $\mathbf{\Omega}_d$ can be replaced by the operators
\begin{equation}
\extcolumn{\mathcal{U}_T^\dagger(\tau_0 + d\tau,\tau_0)}
{\mathcal{U}_R(\tau_0 + d\tau,\tau_0)} 
= \extcolumn{\big[ \mathbf{1} + a_0^* \hbar_3^- (\mathbf{H}_T - \mathbf{\ddot{s}}^\nu \mathbf{x}_\nu - \mathbf{\dot{s}}^\nu \mathbf{\dot{f}}_{T_n;\nu} \tau)
\;d\tau \big]^\dagger}
{\mathbf{1} + b_0 \hbar_4^- (\mathbf{H}_R - \sum_i \mathbf{\ddot{b}}_i \mathbf{\xi}_i - \mathbf{\dot{b}}_i \mathbf{\dot{f}}_{R_{n;i}} \tau) \;d\tau}\label{extTimeQMOPerator}
\end{equation}
where $\mathbf{H}_T$ and $\mathbf{H}_R$ are the translational and rotational Hamiltonian operators, respectively and $\hbar_3$ and $\hbar_4$ are two constant introduced for the dimensional fitting. The values of these constants must be established in future works.

\subsection{The extended Schr\"odinger equation}
We derive the fundamental differential equation for the time evolution operator of each component, known as the Schr\"odinger equation. We apply the composition property on the equation \ref{CompositionPro} of the extended time evolution operator for each component letting $\tau_1 \to \tau$ and  $\tau_2 \to \tau +d \tau$. Using the top component as an example, we have
\begin{equation}
\mathcal{U}_{T}( \tau +d \tau,\tau_0)= \mathcal{U}_{T}( \tau +d \tau, \tau) \mathcal{U}_{T}( \tau,\tau_0)=  (\mathbf{1} + a_0^* \hbar_3^-\mathbf{\Omega}_T d\tau) \mathcal{U}_{T}( \tau,\tau_0)
\end{equation}
where the difference $\tau - \tau_0$ do not need to be infinitesimal. We have then
\begin{equation}
\mathcal{U}_{T}( \tau +d \tau,\tau_0) - \mathcal{U}_{T}( \tau,\tau_0) =  a_0^* \hbar_3^-\mathbf{\Omega}_T \mathcal{U}_{T}( \tau,\tau_0) d\tau
\end{equation}
which in differential form is
\begin{equation}
\frac{\partial}{\partial \tau} \mathcal{U}_{T}( \tau,\tau_0) =  a_0^* \hbar_3^-\mathbf{\Omega}_T \mathcal{U}_{T}( \tau,\tau_0) d\tau
\end{equation}

We find then the  Schr\"odinger equation for each component. The Schr\"odinger equation for the extended operator have the form:
\begin{equation}
 \extcolumn{ \frac{\partial}{\partial \tau} \mathcal{U}_{T}^\dagger( \tau,\tau_0)}
 {\frac{\partial}{\partial \tau}  \mathcal{U}_{R}( \tau,\tau_0)} =  
 \extcolumn{ \big [ a_0^* \hbar_3^-\mathbf{\Omega}_T \mathcal{U}_{T}( \tau,\tau_0) \big]^\dagger   }{\big [ b_0\hbar_4^- \mathbf{\Omega}_R \mathcal{U}_{R}( \tau,\tau_0)\big ]}. \label{extSchEq}
\end{equation}
So far we treat the evolution of the system through the evolution of its component, and there is nothing new despite the fact the theory is developed on a new space. The difference comes once we look at the evolution of the extended ket
\begin{equation}
\frac{\partial}{\partial \tau}  \extket{\alpha,\tau_0;\tau}{\beta,\tau_0;\tau}
= \frac{\partial}{\partial \tau}  \extcolumn{ \mathcal{U}_{T}^\dagger( \tau,\tau_0)}{ \mathcal{U}_{R}( \tau,\tau_0)}\extket{\alpha,\tau_0;\tau_0}{\beta,\tau_0;\tau_0}.
\end{equation}
That is because
\begin{equation}
\frac{\partial}{\partial \tau}  \extcolumn{ \mathcal{U}_{T}^\dagger( \tau,\tau_0)}{ \mathcal{U}_{R}( \tau,\tau_0)}
\neq
\extcolumn{ \frac{\partial}{\partial \tau} \mathcal{U}_{T}^\dagger( \tau,\tau_0)}
 {\frac{\partial}{\partial \tau}  \mathcal{U}_{R}( \tau,\tau_0)}
\end{equation}
The derivative of a complex product $a\odot b$ with respect to time is
\begin{align}
\frac{d}{d \tau} (a\odot b) &= \frac{d}{d \tau} \big[ 
(a_E^* b_I z_1^{(a)} + a_I^*b_E )\extk + \imagi a_E^* b_E + a_E^* b_I z_1^{(a)} + a_I^* b_I
\big]
\nonumber \\
&= ( \frac{d a_E^*}{d \tau} b_I z_1^{(a)}  
+  a_E^*\frac{d b_I}{d \tau}  z_1^{(a)} 
+  a_E^* b_I \frac{d z_1^{(a)}}{d \tau}  
+ \frac{d a_I^*}{d \tau} b_E 
+  a_I^*\frac{d b_E}{d \tau}  )\extk 
+ \imagi \frac{d a_E^*}{d \tau} b_E 
+ \imagi a_E^*\frac{d b_E}{d \tau} 
\nonumber \\
&+ \frac{d a_E^*}{d \tau} b_I w_1^{(a)}  
+  a_E^*\frac{d b_I}{d \tau}  w_1^{(a)} 
+  a_E^* b_I \frac{d w_1^{(a)}}{d \tau}
+ \frac{d a_I^*}{d \tau} b_I 
+  a_I^*\frac{d b_I}{d \tau}.  
\end{align}
Also, we have
\begin{equation}
\frac{d a}{d \tau} \odot b =  
( \frac{d a_E^*}{d \tau} b_I z_1^{(\dot{a})} + \frac{d a_I^*}{d \tau} b_E )\extk 
+ \imagi \frac{d a_E^*}{d \tau} b_E + \frac{d a_E^*}{d \tau} b_I w_1^{(\dot{a})} 
+ \frac{d a_I^*}{d \tau} b_I
\end{equation}
and 
\begin{equation}
a \odot \frac{d b}{d \tau} =  
( a_E^*\frac{d b_I}{d \tau}  z_1^{(a)} + a_I^*\frac{d b_E}{d \tau}  )\extk 
+ \imagi  a_E^*\frac{d b_E}{d \tau} + a_E^*\frac{d  b_I}{d \tau} w_1^{(a)} 
+ a_I^*\frac{d b_I}{d \tau}. 
\end{equation}
Putting all together we have the relation:
\begin{equation}
\frac{d}{d \tau} (a\odot b)  = \frac{d a}{d \tau} \odot b + a \odot \frac{d b}{d \tau}
+ \mathcal{G}(a,b)
\end{equation}
where
\begin{equation}
\mathcal{G}(a,b) \equiv [ \frac{d a_E^*}{d \tau} b_I( z_1^{(a)} - z_1^{(\dot{a})} )
+  a_E^* b_I \frac{d z_1^{(a)}}{d \tau}
]\extk
+ \frac{d a_E^*}{d \tau} b_I( w_1^{(a)} - w_1^{(\dot{a})} )
+  a_E^* b_I \frac{d w_1^{(a)}}{d \tau}.
\end{equation}

An extended operators must satisfy the same derivative properties:
\begin{equation}
\frac{d}{d \tau} \extoperator{A}{B}  = \extoperator{\frac{d A}{d \tau}}{B} + \extoperator{A}{\frac{d B}{d \tau}} +  
+  \extcolumn{\sqrt[*]{\mathcal{G}(\mathbf{A},\mathbf{B})}}{\sqrt[*]{\mathcal{G}(\mathbf{A},\mathbf{B})}}  \label{extOperTimeDeriv}
\end{equation}
Deriving twice with respect to time, we obtain
\begin{align}
&\frac{d^2}{d \tau^2} \extoperator{A}{B}  = \extoperator{\frac{d^2 A}{d \tau^2}}{B} 
+ \extoperator{A}{\frac{d^2 B}{d \tau^2}} + 
 2  \extoperator{\frac{d A}{d \tau}}{\frac{d B}{d \tau}}
+  \frac{d}{d \tau} \extcolumn{\sqrt[*]{\mathcal{G}(\mathbf{A},\mathbf{B})}}{\sqrt[*]{\mathcal{G}(\mathbf{A},\mathbf{B})}}  
\nonumber \\
&+  \extcolumn{\sqrt[*]{\mathcal{G}(\mathbf{\dot{A}},\mathbf{B})}}{\sqrt[*]{\mathcal{G}(\mathbf{\dot{A}},\mathbf{B})}}  
+  \extcolumn{\sqrt[*]{\mathcal{G}(\mathbf{A},\mathbf{\dot{B}})}}{\sqrt[*]{\mathcal{G}(\mathbf{A},\mathbf{\dot{B}})}}  \label{extOperTimeDeriv2}
\end{align}

The derivative of a general ket with respect to time which was at the initial time $\tau_0$ is
\begin{align}
&\frac{\partial}{\partial \tau}  \extket{\alpha,\tau_0;\tau}{\beta,\tau_0;\tau}
= \frac{\partial}{\partial \tau}  \extcolumn{ \mathcal{U}_{T}^\dagger( \tau,\tau_0)}{ \mathcal{U}_{R}( \tau,\tau_0)}
\extket{\alpha,\tau_0;\tau_0}{\beta,\tau_0;\tau_0} 
\nonumber \\
&=\left[ \extcolumn{ \frac{\partial\mathcal{U}_{T}^\dagger( \tau,\tau_0) }{\partial \tau}}
{ \mathcal{U}_{R}( \tau,\tau_0)}
+\extcolumn{ \mathcal{U}_{T}^\dagger( \tau,\tau_0)}{ \frac{\partial\mathcal{U}_{R}( \tau,\tau_0)}{\partial \tau}}
+  \extcolumn{\sqrt[*]{\mathcal{G}(\mathcal{U}_{T},\mathcal{U}_{R})}} {\sqrt[*]{\mathcal{G}(\mathcal{U}_{T},\mathcal{U}_{R})}}\right] 
\extket{\alpha,\tau_0;\tau_0}{\beta,\tau_0;\tau_0} \label{extQMTimeOprInfEq}
\end{align}
Each component $\frac{\partial\mathcal{U}_{T}^\dagger( \tau,\tau_0) }{\partial \tau}$ and $\frac{\partial\mathcal{U}_{R}( \tau,\tau_0)}{\partial \tau}$ can be replaced using the Schr\"odinger equation finally obtaining:
\begin{align}
&\frac{\partial}{\partial \tau}  \extket{\alpha,\tau_0;\tau}{\beta,\tau_0;\tau}
= \frac{\partial}{\partial \tau}  \extcolumn{ \mathcal{U}_{T}^\dagger( \tau,\tau_0)}{ \mathcal{U}_{R}( \tau,\tau_0)}
\extket{\alpha,\tau_0;\tau_0}{\beta,\tau_0;\tau_0} 
\nonumber \\
&=\left[ \extcolumn{ \big [ a_0^* \hbar_3^-\mathbf{\Omega}_T \mathcal{U}_{T}( \tau,\tau_0) \big]^\dagger}
{ \mathcal{U}_{R}( \tau,\tau_0)}
+\extcolumn{ \mathcal{U}_{T}^\dagger( \tau,\tau_0)}{\big [ b_0\hbar_4^- \mathbf{\Omega}_R \mathcal{U}_{R}( \tau,\tau_0)\big ]}
+  \extcolumn{\sqrt[*]{\mathcal{G}(\mathcal{U}_{T},\mathcal{U}_{R})}} {\sqrt[*]{\mathcal{G}(\mathcal{U}_{T},\mathcal{U}_{R})}}\right] 
\extket{\alpha,\tau_0;\tau_0}{\beta,\tau_0;\tau_0} \label{extQMTimeOprInfEq1}
\end{align}

The methodology to obtain the extended operator of the time evolution is to find the time evolution operator for each component using the Schr\"odinger equation. There are two well-defined cases to derive the time evolution operator for every component: when the generator is time independent and when it depends on time. In the ordinary quantum mechanic, the Hamiltonian for an isolated particle, as the time generator operator, is time independent, from where it is obtained the well-known exponential form for this operator. In the present case, however, things are not that easy, and the components of the extended operator $\extcolumn{ \mathcal{U}_{T}^\dagger( \tau,\tau_0)}{ \mathcal{U}_{R}( \tau,\tau_0)}$ explicitly depend on time eq. \ref{extTimeQMOPerator}: 
\begin{equation*}
\extcolumn{\mathcal{U}_T^\dagger(\tau_0 + d\tau,\tau_0)}
{\mathcal{U}_R(\tau_0 + d\tau,\tau_0)} 
= \extcolumn{\big[ \mathbf{1} + a_0^* \hbar_3^- (\mathbf{H}_T - \mathbf{\ddot{s}}^\nu \mathbf{x}_\nu - \mathbf{\dot{s}}^\nu \mathbf{\ddot{x}}_\nu \tau)
\;d\tau \big]^\dagger}
{\mathbf{1} + b_0 \hbar_4^- (\mathbf{H}_R - \sum_i \mathbf{\ddot{b}}_i \mathbf{\xi}_i - \mathbf{\dot{b}}_i \mathbf{\ddot{\xi}}_i \tau) \;d\tau}
\end{equation*}
In that case, the solution of the above equation \ref{extQMTimeOprInfEq1} will probably be in some Dyson1's type series, but it is out of the scope of this work and should be studied in future studies. 

\subsection{The extended Heisenberg equation}
The time evolution of a general ket demands the solution of a Schr\"odinger equations for each component. On the top of this, the time evolution operator of both components \ref{extTimeQMOPerator} explicitly depends on the derivative with respect to time of physical magnitudes like the position coordinates $ \mathbf{\ddot{x}}_\nu, \bm{\xi}_i $ and the second order momentums $\mathbf{\dot{s}}^\nu, \mathbf{\ddot{b}}_i $. In that case, we need to find the analytical expressions for the time derivative of the component of an extended operators.

To find the form of the derivative of the extended quantum operators with respect to time, we take advantage of the Heisenberg approach from the ordinary Quantum mechanic. In our case, because of the associative property of the extended vectors, it is possible to approach the extended quantum dynamics from a picture where operators are constant, and the state vectors evolve with time into a new formulation where the operators are the ones whose change with time while physical states remain constant. These approaches are known as the Schr\"odinger and the Heisenberg pictures respectively. Indeed, the associative axiom of the inner product allows switching between both pictures. The unitary property of time evolution operator guarantee the invariability of the inner product with the temporal evolution of the system since
\begin{equation}
\extinnerprod{\alpha}{\beta}{\alpha}{\beta} \to \extbra{\alpha}{\beta}
\extcolumn{ \mathcal{U}_{T}^\dagger\; }{ \mathcal{U}_{R}  }^\ddagger
\extcolumn{ \mathcal{U}_{T}^\dagger\; }{ \mathcal{U}_{R}\; } 
\extket{\alpha}{\beta} =\extinnerprod{\alpha}{\beta}{\alpha}{\beta}.
\end{equation}
If states evolve with time ket while the operator remain unchanged, the quantity
\begin{equation}
\extaverage{\alpha}{\beta}{A}{B}{\alpha}{\beta}
\end{equation}
evolve with time like:
\begin{align}
\extaverage{\alpha}{\beta}{A}{B}{\alpha}{\beta} \to 
\left[
\extbra{\alpha}{\beta} 
\extcolumn{ \mathcal{U}_{T}^\dagger\; }{ \mathcal{U}_{R}  }^\ddagger
\right]
\extoperator{A}{B}
\left[
\extcolumn{ \mathcal{U}_{T}^\dagger\; }{ \mathcal{U}_{R}\; } 
\extket{\alpha}{\beta} 
\right].
\end{align}
From the associative axiom we have
\begin{align}
\extaverage{\alpha}{\beta}{A}{B}{\alpha}{\beta} \to 
= \extbra{\alpha}{\beta} 
\left[
\extcolumn{ \mathcal{U}_{T}^\dagger\; }{ \mathcal{U}_{R}^{\dagger}}
\extoperator{A}{B}
\extcolumn{ \mathcal{U}_{T}^\dagger\; }{ \mathcal{U}_{R}\; } 
\right]
\extket{\alpha}{\beta} \label{extHeisSchrEq}
\end{align}
that suggest that the operators can be the ones evolving with time while the state vector remains constant. This point of view is closer to the reality, where quantities like position and momentum, represented in here in the component of the operators, are one that changes with time in classical mechanics.

Last equation can be written as:
\begin{equation}
\extaverage{\alpha}{\beta}{A}{B}{\alpha}{\beta} \to \extaveragearg{\alpha}{\beta}{A}{{}^{(H)}(\tau)}{B}{{}^{(H)}(\tau)}{\alpha}{\beta}
\end{equation}
being
\begin{equation}
\extcolumn{\mathbf{A}^{\dagger(H)}(\tau)}{\mathbf{B}^{(H)}\;(\tau)} = 
\extcolumn{ \mathcal{U}_{T}^\dagger\; }{ \mathcal{U}_{R}  }^\ddagger
\extcolumn{\mathbf{A}^{\dagger(S)}}{\mathbf{B}^{(S)}\;}
\extcolumn{ \mathcal{U}_{T}^\dagger\; }{ \mathcal{U}_{R}\; } \label{extHeisSchrRel}
\end{equation}
where superscripts $S$ and $H$ stands for Schr\"odinger and Heisenberg picture, respectively. At $\tau=0$, operators and physical kets coincide at both pictures.

We derive now the time derivative for an extended operator which not explicitly depend on time. Differentiating equation \ref{extHeisSchrEq}, we obtain:
\begin{align}
\frac{d}{d\tau}
\extcolumn{\mathbf{A}^{\dagger(H)}(\tau)}{\mathbf{B}^{(H)}\;(\tau)}= &
\frac{d}{d\tau}
\left[
\extcolumn{ \mathcal{U}_{T}^\dagger\; }{ \mathcal{U}_{R}}^\ddagger
\right]
\extcolumn{\mathbf{A}^{\dagger(S)}}{\mathbf{B}^{(S)}}
\extcolumn{ \mathcal{U}_{T}^\dagger\; }{ \mathcal{U}_{R}\; }
+ 
\nonumber \\
&\extcolumn{ \mathcal{U}_{T}^\dagger\; }{ \mathcal{U}_{R}}^\ddagger
\extcolumn{\mathbf{A}^{\dagger(S)}}{\mathbf{B}^{(S)}}
\frac{d}{d\tau}
\left[
\extcolumn{ \mathcal{U}_{T}^\dagger\; }{ \mathcal{U}_{R}\; }
\right].
\end{align}
or applying the time derivative properties for extended operators:
\begin{align}
\frac{d}{d\tau}
\extcolumn{\mathbf{A}^{\dagger(H)}(\tau)}{\mathbf{B}^{(H)}\;(\tau)}= &
\extcolumn{ \frac{d\mathcal{U}_{T}^\dagger}{d\tau}\; }{ \mathcal{U}_{R}}^\ddagger
\extcolumn{\mathbf{A}^{\dagger(S)}}{\mathbf{B}^{(S)}}
\extcolumn{ \mathcal{U}_{T}^\dagger\; }{ \mathcal{U}_{R}\; }
+
\extcolumn{ \mathcal{U}_{T}^\dagger\; }{ \frac{d\mathcal{U}_{R}}{d\tau}}^\ddagger
\extcolumn{\mathbf{A}^{\dagger(S)}}{\mathbf{B}^{(S)}}
\extcolumn{ \mathcal{U}_{T}^\dagger\; }{ \mathcal{U}_{R}\; }
\nonumber \\
+& 
\extcolumn{ \mathcal{U}_{T}^\dagger\; }{ \mathcal{U}_{R}}^\ddagger
\extcolumn{\mathbf{A}^{\dagger(S)}}{\mathbf{B}^{(S)}}
\extcolumn{ \frac{d\mathcal{U}_{T}^\dagger}{d\tau} }{ \mathcal{U}_{R}\; }
+
\extcolumn{ \mathcal{U}_{T}^\dagger\; }{ \mathcal{U}_{R}}^\ddagger
\extcolumn{\mathbf{A}^{\dagger(S)}}{\mathbf{B}^{(S)}}
\extcolumn{ \mathcal{U}_{T}^\dagger\; }{ \frac{d\mathcal{U}_{R}}{d\tau}}
\nonumber \\
+
\extcolumn{\sqrt[*]{\mathcal{G}(\mathcal{U}_{T}^{\dagger \ddagger},\mathcal{U}_{R}^\ddagger)}}
{\sqrt[*]{\mathcal{G}(\mathcal{U}_{T}^{\dagger \ddagger},\mathcal{U}_{R}^\ddagger)}}
&\extcolumn{\mathbf{A}^{\dagger(S)}}{\mathbf{B}^{(S)}}
\extcolumn{ \mathcal{U}_{T}^\dagger\; }{ \mathcal{U}_{R}\; }
+ 
\extcolumn{ \mathcal{U}_{T}^\dagger\; }{ \mathcal{U}_{R}}^\ddagger
\extcolumn{\mathbf{A}^{\dagger(S)}}{\mathbf{B}^{(S)}}
\extcolumn{\sqrt[*]{\mathcal{G}(\mathcal{U}_{T}^{\dagger},\mathcal{U}_{R})}}
{\sqrt[*]{\mathcal{G}(\mathcal{U}_{T}^{\dagger},\mathcal{U}_{R})}}
\end{align}
where the operators $\frac{d\mathcal{U}_{T}^\dagger}{d\tau}$ and $\frac{d\mathcal{U}_{R}}{d\tau}$ can be replaced using the equation of  Schr\"odinger \ref{extOperTimeDeriv}.

The previous equation is insufficient to find the derivative of one component of the operator with respect to time. The reason is the existence in the derivative of the operator with time of the residual function $\mathcal{G}(\mathcal{U}_{T}^{\dagger},\mathcal{U}_{R})$ which depends on both components of the operator. However, taking advantage of the connection between both components given by the simultaneous time evolution, the derivative of the components with respect to time can be found by computing the previous equation whit the same operator at both components of the extended operator:
\begin{align}
\frac{d}{d\tau}
\extcolumn{\mathbf{A}^{\dagger(H)}(\tau)}{\mathbf{A}^{(H)}\;(\tau)}= &
\extcolumn{ \frac{d\mathcal{U}_{T}^\dagger}{d\tau}\; }{ \mathcal{U}_{R}}^\ddagger
\extcolumn{\mathbf{A}^{\dagger(S)}}{\mathbf{A}^{(S)}}
\extcolumn{ \mathcal{U}_{T}^\dagger\; }{ \mathcal{U}_{R}\; }
+
\extcolumn{ \mathcal{U}_{T}^\dagger\; }{ \frac{d\mathcal{U}_{R}}{d\tau}}^\ddagger
\extcolumn{\mathbf{A}^{\dagger(S)}}{\mathbf{A}^{(S)}}
\extcolumn{ \mathcal{U}_{T}^\dagger\; }{ \mathcal{U}_{R}\; }
\nonumber \\
+& 
\extcolumn{ \mathcal{U}_{T}^\dagger\; }{ \mathcal{U}_{R}}^\ddagger
\extcolumn{\mathbf{A}^{\dagger(S)}}{\mathbf{A}^{(S)}}
\extcolumn{ \frac{d\mathcal{U}_{T}^\dagger}{d\tau} }{ \mathcal{U}_{R}\; }
+
\extcolumn{ \mathcal{U}_{T}^\dagger\; }{ \mathcal{U}_{R}}^\ddagger
\extcolumn{\mathbf{A}^{\dagger(S)}}{\mathbf{A}^{(S)}}
\extcolumn{ \mathcal{U}_{T}^\dagger\; }{ \frac{d\mathcal{U}_{R}}{d\tau}}
\nonumber \\
+
\extcolumn{\sqrt[*]{\mathcal{G}(\mathcal{U}_{T}^{\dagger \ddagger},\mathcal{U}_{R}^\ddagger)}}
{\sqrt[*]{\mathcal{G}(\mathcal{U}_{T}^{\dagger \ddagger},\mathcal{U}_{R}^\ddagger)}}
&\extcolumn{\mathbf{A}^{\dagger(S)}}{\mathbf{A}^{(S)}}
\extcolumn{ \mathcal{U}_{T}^\dagger\; }{ \mathcal{U}_{R}\; }
+ 
\extcolumn{ \mathcal{U}_{T}^\dagger\; }{ \mathcal{U}_{R}}^\ddagger
\extcolumn{\mathbf{A}^{\dagger(S)}}{\mathbf{A}^{(S)}}
\extcolumn{\sqrt[*]{\mathcal{G}(\mathcal{U}_{T}^{\dagger},\mathcal{U}_{R})}}
{\sqrt[*]{\mathcal{G}(\mathcal{U}_{T}^{\dagger},\mathcal{U}_{R})}}.
\end{align}
Both components are related now since one is the Hermitian operator of the other. The form of the Hermitian operator still needs to be also computed with the study of the extended numbers. In that case, we can use the Heisenberg equation to find the expressions for the derivative of operators of each component with respect to time.

\subsection{Time dependent eigenvalue equation}
The eigenvalue equation is also modified with the time evolution. The stationary equation
\begin{equation}
\extoperator{A}{B} \extket{a'}{b'} = \extscalar{a'}{b'} \extket{a'}{b'}.
\end{equation}
under the Schr\"odinger picture does not change with the evolution of the system with time. The operator does not change with time, so its eigenkets and eigenvalues at any time moment. Under the Heisenberg scope, we have a different behavior since the operator is time-dependent. Multiplying both sides of the eigenvalue equation by 
$
\extcolumn{ \mathcal{U}_{T}^\dagger\; }{ \mathcal{U}_{R}  }^\ddagger
$, 
and using the unitary property of time evolution operator, we can have
\begin{equation}
\extcolumn{ \mathcal{U}_{T}^\dagger\; }{ \mathcal{U}_{R}  }^\ddagger
\extoperator{A}{B} 
\extcolumn{ \mathcal{U}_{T}^\dagger\; }{ \mathcal{U}_{R}\; }
\extcolumn{ \mathcal{U}_{T}^{\dagger}\; }{ \mathcal{U}_{R}  }^\ddagger
\extket{a'}{b'} 
 = 
\extcolumn{ \mathcal{U}_{T}^{\dagger}\; }{ \mathcal{U}_{R}  }^\ddagger
\extscalar{a'}{b'} \extket{a'}{b'}.
\end{equation}
Using Heisenberg picture, last equation can be rewritten as:
\begin{equation}
\extcolumn{\mathbf{A}^{\dagger(H)}(\tau)}{\mathbf{B}^{(H)}(\tau)} 
\left[
\extcolumn{ \mathcal{U}_{T}^{\dagger}\; }{ \mathcal{U}_{R}  }^\ddagger
\extket{a'}{b'} 
\right]
 =
 \left[
\extcolumn{ \mathcal{U}_{T}^{\dagger} \; }{ \mathcal{U}_{R} }^\ddagger
\extscalar{a'}{b'} \extket{a'}{b'}
\right],
\end{equation}
where the is shown that the eigenvalues remain constants while the eigenkets evolve with time but under the operator $\extcolumn{ \mathcal{U}_{T}^{\dagger} \; }{ \mathcal{U}_{R} }^\ddagger$.

\subsection{Single particle energy eigenvalue equations}	
According to our approach, the energy for $n$-VMVF systems is divided between the energy of the translation and the rotation of the system. Each component of the energy operator can be extracted from the classical energy function, equation\ref{extEnergyFunction} 
\begin{equation*}
h= \sum_i p_i \dot{q}_i - \dot{s}_i\dot{q}_i + s_i\ddot{q}_i  - L 
=  \sum_i p_i H - \dot{s}_i\dot{q}_i.
\end{equation*}
The energy function has terms involving the derivative with time of some of the canonical variables, and also it differs from the generator of the motion of the system with time.

From the classical theory, we know that the energy function of the single particle on equation \ref{extPartEnergyFunction} 
\begin{equation*}
h_n = H_n - \dot{s}_{n}\dot{q}_{n}
\end{equation*}
is not a constant of motion so, the eigenvalue equation is written like
\begin{equation}
\extcolumn{(\mathbf{H}_{n_T} - \mathbf{\dot{s}}^\nu_n \mathbf{\dot{x}}_{n;\nu} + \sum_a \lambda_a \Omega_{T_a} )^\dagger}
{\mathbf{H}_{n_R} -  \mathbf{\dot{b}}_{n;i} \mathbf{\dot{\xi}}_{n;i} + \sum_a \lambda_b \Omega_{R_b}} 
\left[
\extcolumn{ \mathcal{U}_{T}^{\dagger}\; }{ \mathcal{U}_{R}  }^\ddagger
\extket{E_{n_T}}{E_{n_R}}
\right]
 =
 \left[
\extcolumn{ \mathcal{U}_{T}^{\dagger}\; }{ \mathcal{U}_{R}  }^\ddagger
\extscalar{E_{n_T}}{E_{n_R}} \extket{E_{n_T}}{E_{n_R}}
\right].
\end{equation}
On the other side, the total energy of the system, as the sum of all the single particle energies should remain constant along the evolution of the system with time. In this case, the eigenvalues equation is stationary and should have the form:
\begin{equation}
\extcolumn{(\sum_n \mathbf{H}_{n_T} - \mathbf{\dot{s}}^\nu_n \mathbf{\dot{x}}_{n;\nu} + \sum_a \lambda_a \Omega_{T_a} )^\dagger}
{\sum_n  \mathbf{H}_{n_R} -  \mathbf{\dot{b}}_{n;i} \mathbf{\dot{\xi}}_{n;i} + \sum_a \lambda_b \Omega_{R_b}} 
\extket{E_{T}}{E_{R}}
 =
\extscalar{E_{T}}{E_{R}} \extket{E_{T}}{E_{R}}.
\end{equation}
\newpage

\section{Conclusions}
In this work, we propose a quantum theory for particle systems with variable masses connected by a field with no predefined form ($n$-VMVF systems). We define $n$ as the maximum number of particle involved in the evolution of the system during a time interval. The masses and the field are considered variables of the system and they depend on the positions and velocities of the particles. We follow the methodology for constructing the modern quantum theory by de definition of a vector space and quantum operators borrowed from the canonical transformations of the corresponding classical theory.

The primary motivation to develop this approach, once we assume the mass as a variable quantity, is to search for the canonical transformation for the mass variation, from where can outcome a quantum operator. It was not possible, however, to reach this goal. Instead, we were able to propose a solvable set of classical equations from where the mass of the particle can be computed. The solution manifests a two set of equation structure which is the base for the vector space and the quantum theory. Our proposal is not yet complete. However, we think that the text embrace the main aspects and ideas that should be included, added or modified for a consistent quantum mechanics theory for $n$-VMVF systems. We review and discuss the main results of the work. 

\subsection*{Revisiting the classical mechanics}

In the construction of the classical theory, we found that if the mass of the particle is considered variable, then the application of the second law of Newton to an isolated particle cause a violation of the relativity principle under a Galilean transformation. However, the second Newton law is satisfied not by one particle but to whole the system of particles. We introduce a new type of constraints that relates all the variables of the system, including the masses and the field functions different from the geometrical constraint that set geometric restrictions only to the position of all the particles. For that reason we conclude that the well-known D'Alembert's principle can be no longer applied to the $n$-VMVF systems. We proposed then, a modification of the principle of D'Alembert for those kinds of systems such as
\begin{quote}
``The total virtual work of the sum of the impressed, the intrinsic constraint and the inertial forces vanishes for the reversible displacements of any particle of the system''.
\end{quote}
We obtain the inertial forces by assuming that 
\begin{quote}
``The net applied and inertial forces acting on every particle of the system are the same forces acting on each particle measured by an external observer on an inertial frame. In that case, each particle of the system is considered as isolated and the external forces acting over it, as the action of the other particles through the field.''
\end{quote} 
From the point of view of an external observer on an inertial frame, each particle of the system is considered as isolated and under the action of the external field, which depends on the others particle's position and velocities as parameters.

The second law of Newton is satisfied by the $n$-VMVF system, even when it is not satisfied by each particle by its own, only if the amount of momentum which causes the violation because of the mass variation is suppressed by the variation of the variables of the others particles of the system. Such harmonic variations could only exist through an internal field. The internal field has no defined form and has the function of connecting the particles of the system. Because the variable mass is the cause of its existence, the field is then connected to the variations of mass of the particles, which is unknown. We assume that the field is also unknown with an undetermined nature, and is a function of the positions and velocities of all the particles. It is the final solution that will reveal the nature and its fundamental interaction type.

We conclude in this part that a system of particle with variables masses must have two or more particles and they must be connected by an internal field. Because of that, the axiom of single particle can not be applied for this type of systems. Also that the system can be studied assuming the masses and the field as functions of the particle positions and velocities, excluding the existences of hidden variables. The modified D'Alembert principle also show a degeneracy which shows that we are in presence of a constrained system.

\subsection*{The construction of the Lagrangian theory for $n$-VMVF systems}
The solution for the $n$-VMVF systems is proposed by forcing the isolated system to satisfy the equilibrium condition. Combining the Lagrange equations for a starting Lagrangian, the modified principle of D'Alembert and the principle for obtaining the inertial and applied forces, we can identify the constraints for the isolated $n$-VMVF system remain at equilibrium. The chosen Lagrangian must agree with a known Lagrangian when the variation of masses is neglected. We proposed the Lagrange function $L_{{sp}_n} = \frac{1}{2}m_n\dot{\textbf{r}}^2_n -\phi + \mathbf{A}\cdot \dot{\mathbf{r}}_n$ as the known Lagrange function an also as the starting Lagrangian, being the form of the potential function equal to the potential function for electromagnetic field.

The first set of obtained equations resulting from the above method were the Lagrange equations in the rectangular coordinates for the starting Lagrangian. The equilibrium condition using the rectangular coordinates is referred to the conservation of the linear momentum in agreement with the homogeneity property of space. By combining the D'Alembert extended principle and the already defined inertial and applied forces on each particle, we can obtain a new group of constrained equations for the variables of the system. 

However, the number of variables still surpass the number of the identified equations. To find another set of equations, we proposed to describe the system with a set of coordinates independent from the $3D$-rectangular coordinates: the $3D$-angular coordinates. The equilibrium condition using the angular coordinates is referred now to the conservation of the angular momentum, in agreement this time with the isotropy property of space. Both sets, for particle systems with more of one particle, correspond to independent momentum conservation laws and symmetries, completing the equilibrium condition for any physical systems. 

The representation of the particle position on the $3D$-angular coordinates, $\vec{r} = \vec{r}(\theta,\phi,\chi)$, is carried out by expand the three-dimensional space to a four dimensions space and choosing a constraint with a new parameter $R$, such that Pythagorean theorem holds in the new $4-D$ space $x^2 + y^2 +z^2 +w^2 = R^2$. The method introduces $R$ quantity as a parameter, which should depend on the energy of the particle system and need to be defined on future works. On the other hand, the method introduces the four four-dimensional space-time to the problem. Because of that we increase the dimension of all the rest of the variables of the problem and assume the compliance of the first postulate of the Theory of Relativity. The rectangular coordinates are now expanded to the Lorentzian coordinates.

The second postulate of the Theory of Relativity, however, is based on the invariance of Maxwell's equations, which cannot be settled under this approach because the field is precisely one of the variables to be found on the system. Because of that, we do not take the compliance of the second postulate for granted in detriment of imposing additional constraints the motion of the particles and a limit for the rotation energy. This means that under the present approach $\dot{r}^\nu_n \dot{r}_{n,\nu} \neq c^2$ and also that it should be theme of futures discussions.

The combination of the modified D'Alembert principle and the definition of the inertial and applied forces on each particle using now the angular coordinates, permit us to obtain a new group of constrained equations for the variables of the system and double the existing equations for solving the problem. The number of equations, nevertheless, remains shorter than the number of variables if we suppose masses and field are functions depending on all particles positions and velocities. In that case, some approximations must be made like set the mass of one particle depending on its particle position. This issue shows that in this approach, the form of the functions of field and masses can be proposed in different ways as long as the number of equations and degree of freedom are the same.

The constraints show the $\ddot{x}_n^\nu$ dependency of the constraint equations for both sets of coordinates, which means that the Lagrangians of the system must also have the same dependency. That forced us to use the second order Lagrange equations and expand the Multiplier Lagrange's method also to the second order. The constraints are added to the starting Lagrangian using this method an obtaining two independent Lagrangians.

From this part, we can conclude that it is possible to solve the classical problem of a system with the mass of the particle and the field connecting those particles as unknown quantities depending on the positions and velocities of the particles without introducing any hidden variable. The study reveals that for the solution we need not solve one set of second order Lagrange equation but two. The solution also treats the masses and the field derivatives as variables of the whole system; however they are not part of the configuration space of any of the two variational problems. A solution with this kind of treatment and with more than a Lagrange function, is new as a solution of a classical problem, at least to the best of our knowledge.

Also, because of the need of expressing the position of the particle as a function of angular coordinates, the four-dimensional space-time is naturally included in the problem when the 3-D space of the angular coordinates is found to be the stereographic projection of the 4-D sphere defined by the Lorentz condition in the space-time. That means that the classical solution for $n$-VMVF systems, is necessarily a relativistic problem.

The obtained classical equations for $n$-VMVF systems have the double of variables and equations and include approximations for the particle masses and the field functions. The proposed solution corresponds to a constrained second order set of Lagrange equations with a notable complexity even for the simplest case involving only two particles and a linear series expansion for the particle masses and the field functions. However, we think the proposed solution can be considered as a universal solution, since it has the masses of the particles and the internal field as a solution to the problem, Also the equations are designed to satisfy the most verified conservation laws in physics for particle systems: the conservation of the linear and angular momentum, which corresponds to the two most verified properties of space: the homogeneity and the isotropy. The solution then, once all approximations are settled, and for a defined number of particles can be used for all the system with such features. Once the Lagrange functions are entirely determined, then externals field can be added in the standard way as an external field with a defined form, and the geometric constraints can be included as they commonly do.

\subsection*{The construction of the Hamiltonian theory for $n$-VMVF systems}
Because of the $\ddot{x}_n^\nu$ dependency of the Lagrangians, the Hamiltonian theory for each Lagrangian is extended up to the second order. We define the second order momentum as $s_i=\frac{\partial L}{\partial \ddot{q}_i}$, which is different from the Ostrogradsky's definition. We define the extended Hamilton function, following the Legendre procedure, as
$
H = \sum_i p_i \dot{q}_i + s_i\ddot{q}_i  - L
$
and obtain a set of second order Hamilton equations different from Ostrogradsky. Nevertheless, the equations report the same instability as the author, which is the appearance of $n$ more degrees of freedom than on the Lagrange approach. 

It is the Identity canonical transformation that reveals the existence of $n$ constraints which add a new set of equations in the form $\mathcal{F}_{2_i}=0$, where the correlation functions $\mathcal{F}_{2_i}$ depends only on the generalized velocities. The form of the correlation functions is fixed and they should be chosen according the studied system, which in our case, embrace the number of particles and the approximations made for the mass and the field functions. This topic should go into a deeper investigation on futures works. This set of constraints removes the Ostrogradsky instability matching the number of equations with the number of variables of the Lagrange approach.

The canonical transformations of second order Hamiltonians show that the generalized linear momentum remains as the generator of the displacement of the generalized coordinate. They also show that the new momentum $s$ is the generator of a negative displacement of the pole of the correlation functions, $f_i$, causing, due to the definition of the correlation functions, the collective motion of the system. That is a consistent result since in our problem, the new momentum $s_i$ appears on the Lagrangian because of the mass variation, which cause that a single particle cannot satisfy Newton's second law. Instead, the collective motion, expressed in the constraint $\mathcal{F}_{2_i}=f_i$, allows the system to satisfy the second Newton's law as a single physical object.

The addition of new constraints to the Hamilton extended equations, force us to expand the study of the constrained system up to the second order. In the text, we make the first proposal for this problem. However, it depends on the form of the Lagrangian and the Hamiltonian, which turn out to be extensive. A more accurate solution should be developed on futures works.

The extension of the Hamilton classical theory for the $n$-VMVF systems, maintains the set of two equations structure for solving the problem: one Hamiltonian related to the translation motion and the other related to the rotation. Each one includes the position of the particle, the pole of the correlation function, and the linear and second-order momentum expressed in the Lorentzian and angular coordinates, respectively. Also, for each one, a set of canonical transformations and their generators are defined that act over the canonical variables. The two set of Hamilton equations needed to solve the problem of $n$-VMVF systems point out the two-component canonical transformations that is one of the bases for constructing the new quantum mechanics for $n$-VMVF systems.

\subsection*{The extension of complex numbers}
The structure of two set of second order Hamilton equations in the classical theory and with it the need of two simultaneous canonical transformations for the modification of the system, points out the need for a new vector space. We propose that extension by defining the unit of the new domain from an unsolvable equation in the complex number domain: $|x|^2 = x^*x = \imagi$. We represent it as $\extk = \sqrt[*]{\imagi}$, where the complex root $\sqrt[*]{a}=b$ if $a^*a = b$. The mentioned unsolvable equation point out the existence of two new operations within the complex numbers which is the complex sum and multiplication which stands as the sum and multiplication of a number by the complex conjugate of the other, respectively. This methods shows as another conclusion of this work, that it is possible to expand the complex numbers without violating the fundamental theorem of algebra by defining the already mentioned new operations.  

The new operations are named the complex product and complex sum, and they are represented as $a \odot b \equiv a^* b$ and $a\oplus b \equiv a^* + b$, respectively. The standard and the complex multiplication of of two extended numbers is based on the standard and the complex multiplication of the extended unit. The first type introduces two complex parameters $z_0, w_0$ whose values must be computed on future works.

From the extended complex unit definition, we also show that the inner product must have at least four terms. We define a new isomorphism represented by $z^\bullet$, and $z^*$, which guarantee that any defined inner product of the new space can satisfy the Positive-definiteness axiom. The properties of the new operations are studied using their abstract expressions, and it shows that the standard sum and multiplication as defined, is commutative, distributive and have an additive identity and inverse element different from the conjugated sum and multiplication.

According to the properties of the space operations, we define the inner product of four extended numbers grouping by pairs such as they can satisfy the commutativity and distribution properties and with that, a future superposition law. We use the Positive-definiteness property to obtained equations for finding the new isomorphisms, which depend on the numbers  $z_1,w_1,  z_2,w_2 \in \mathbb{C}$. The obtained equations are extensive and proven to depend on the number itself. 

Using the abstract expression of both isomorphisms, we defined the inverse of a complex extended number and with it, the division of two extended numbers. We also start the study the linearity and conjugated symmetry properties among others topics. Different from the pure complex space, the properties are satisfied only under certain conditions that should be checked. Remains only to see what is the meaning of these conditions in future works. 

We define a new map $J \equiv ()^\cbullet$ for an extended number $\lambda$, so the linearity and the conjugate symmetry axiom of the inner product are consistent. We set the condition of satisfying conjugate symmetry as a requirement of an inner product. Future works need to verify this property.

We conclude in this part that is possible to define a two component vector space according to our classical needs. The space as defined from the expansion of the complex numbers, without violating the fundamental theorem of algebra. We proposed new properties, among we can find the inner commutation, which studies the switching between terms the same component on the multiplication of two inner products. We expect that those new properties are connected to the physical phenomena of any theory developed over the extended complex space.

The new space must be studied in depth. Among futures studies must include with top priority the solution for the isomorphisms, for which it must be defined the parameters $z_0,w_0$ and even if that is the case, redefine the extended unit from a different unsolvable equation in the complex domain. The solution also must satisfy the conjugate symmetry axiom. A general review will cause for sure the modification or correction of all the results obtained in here. We are aware that futures changes on the extended space concepts and results will directly reflect into the concepts and obtained results of the quantum theory for the $n$-VMVF systems.

\subsection*{Proposing a new quantum approach for $n$-VMVF systems}
The two equation structure from the extension of the classic mechanics applied to $n$-VMVF systems suggests the extension of the complex space and also the reason of defining an inner product as a such. Specifically, the two canonical transformations structure needed for a complete evolution of the system point out a new structure for the quantum equation. Because of that, we propose a quantum state as a two-component object like $\extket{\alpha}{\beta}$, the operators: $\extoperator{A}{B}$ and the scalars: $\extscalar{\alpha}{\beta}$, resembling the extended complex quantity $\alpha \odot \beta$. In general, the bi-dimensional structure of the action of an operator acting over a state vector is
\begin{equation*}
\extoperator{A}{B} \extket{\alpha}{\beta} = \extket{\gamma}{\zeta}.
\end{equation*}

The bi-dimensional operators act over bi-dimensional vectors component by component, in agreement with the independence of the two Lagrangian set of equations in the classical theory. We also introduce the extended bar space, whose vectors are represented as $\extbra{\alpha}{\beta}$, which allows the definition of the inner product between an extended bra and an extended ket like $\extinnerprod{\alpha}{\beta}{\gamma}{\lambda}$. We set the Positive-definiteness and linearity axioms for the kets and bra spaces, based on the analysis of extended complex vectors. 

In this work we analyse the normalization condition which, from our point of view, can seen as a relationship more than an axiom if it is written such as
\begin{quote}
``The expectation value of the operator identity defined on a vector space \textbf{1} must be 1 for any physical vector of such space'', 
\end{quote}
which means that the normalization relations and the expectation value must be consistently defined. We introduce two normalization relations for the extended quantum mechanic for an arbitrary state vector, according to the geometrical reinterpretation for the extended quantum mechanics.

We study the eigenvalues and eigenvectors of an extended operator and we show that, because of the extended properties of the inner product, the eigenvalue of operators which represent physical magnitudes, are pure complex numbers. The series expansion of a state vector  are proposed in a similar form as the ordinary quantum mechanics based on the linear property of the extended kets, whose coefficients can be determined using the normalization relations. 

This result shows that the expansion coefficients, constrained by the new normalization relations and following the measurement principle for quantum observables, are connected to probabilistic quantities. The new Hilbert space over the extended complex numbers permits the existence of negative or imaginary probabilities. The physical meaning of the previous concepts demands a more in-depth study of this issues and the verification of the solution with experimental data.

In here, is also assumed that the measurements principle follows the Dirac's idea that ''A measurement always cause the system to jump to an eigenstate of the dynamical variable that is being measured''.

We show that in general, each component of any physical infinitesimal extended operator can be represented in the same form than the classical canonical transformations, as the sum of the Identity operator and its generator, times the value of the magnitude of the transformation. 

Taking borrowed the linear momentum as the generator of the displacement of the generalized coordinate, we find the functions $p_u(x_n')$ and $l_u(\xi'_i)$ that appears as result of the action of an operator which have the linear and the angular momentum, respectively, at the top component. We also find the functions $p_b(x_n')$ and $l_b(\xi'_i)$ that appears when such operators are at the bottom component. Following the same method, and being the second order momentum $s_i$ the generator of the inverse displacement of the pole of the correlation function, $f_i(\{\bar{q}\})$, we find the form of the functions  $s_u(f_{x_n}')$, $s_d(f_{x_n}')$, $b_u(f_{x_n}')$ and $b_d(f_{x_n}')$, where $b$ is the second order momentum in the angular coordinates.

The obtained form for the functions when the corresponding operator, as the generator of the displacement of variable $q_n$, is at the bottom component of the operator is similar to the form obtained on the ordinary quantum mechanics $e.i.$ $\sim \frac{\partial c_{q_n'}}{\partial q_n'}$. However, the form of the function from the top component adds an extra function, $\sim \frac{\partial c_{q_n'}}{\partial q_n'} + G(q_n')$, as a direct consequence of the theory being developed on a Hilbert space over the extended numbers.

We study the dynamics and time evolution of $n$-VMVF quantum system. As the quantum theory for $n$-VMVF systems is intrinsically a relativistic theory, we consider the evolution of such systems not with time but with the proper time $\tau$, which is considered as a parameter and not an observable. We propose each component of the extended time evolution operator like the operators in the ordinary quantum mechanics, $e.i$ considering the composition property. Each component ``borrowed'' the classical generator of the time evolution canonical transformation as the generator of the time evolution of the system. The action of the operator of the time evolution of the system is fundamentally different from the action of the others quantum operator because the action of each of its component is related by the proper time since all magnitudes on both components simultaneously evolve with time. We study the evolution of the state vectors and the operators, using the  Schr\"odinger and Heisenberg frames, respectively.

\subsection*{General considerations}
The solution for solving the quantum problem of $n$-VMVF systems is extensive and include some approximations. Starting from the obtained Lagrangian functions with their constraints, continuing to the inclusion of new constraints related to the correlation functions from the extension of the Hamilton theory and arriving to the complexity of the equations of the extended complex vector space, whose inner product, isomorphisms and most of the properties are incomplete, this work is far from finished. However, from the above development, we can conclude that it is possible to develop the basis for a quantum theory for particle systems with variable masses connected by an undefined field without hidden variables or an extra degree of freedom, using only the Lorentzian coordinates and the particle masses and field derivatives.

The classical solution for the $n$-VMVF systems shows that the phenomena related to those systems have a relativistic nature. The choice of setting the field as a degree of freedom while space is flat and remains unchanged establish a different point of view from theories like General Relativity, where the form of the field is fixed, and it is the space which is modified. As the field is proposed as the unique physical object that connects particle, represented by a function with no defined form, it includes all possible fundamental interactions. The well-known modification of space-time because of the presence of a field or a massive particle is treated in here as the modification of form of the field leaving the metric of the space constant.

The final solution of a quantum $n$-VMVF system should be the quantification of the energy of the particle $E_n(\tau)$, the probability of finding any of the particle of the system in any specific region from the inertial frame $x^\nu_n(\tau)$, the momentums $p^\nu_n(\tau)$ and $s^\nu_n(\tau)$. Also we should obtain the particle masses's derivatives $ \frac{\partial m_n}{\partial x^\nu_n}(\tau)$ and the field derivatives $ \frac{\partial A^\mu_n}{\partial x^\nu_n}(\tau)$ and $ \frac{\partial A^\mu_n}{\partial \dot{x}^\nu_n}(\tau)$. The derivatives functions, according the particle masses and field functions approaches will lead to the $ m_n(\tau)$ and $A^\mu_n(\tau)$ final expressions.

The present approach does not contradict the existing theoretical theories or models in Particle Physics. Indeed, it assumes the variable character of the mass and the field but set no specific form of these variations and are, instead, found as a solution to the problem. The variations of the mass and field are not related on any nature, model or effects like the relativistic's, only in its inertial properties. The last means that concepts like the rest mass or fundamental interactions can be no longer applied. The masses and the field are functions constrained only by the conservation of the linear and angular momentum and the principle of least action. Also, the Theory of relativity is included only up to its spacial theory related only to the Lorentz conditions and the first principle.

From classical mechanics, we show that the isolated particle whose mass varies with no restriction can no longer exist without violate the relativity principle under a Galilean transformation. If the mass of a particle as a quantum state of the mass operator, then an isolated particle can be seen as a quantum system of two or more particles. One of them is the particle we see and the others, particles of vacuum. That is consistent since, from our point of view, matter cannot exist without the vacuum. The something cannot exist without anything. One is the non-existence of the other. One is the reference to the other so it can exist. 

In the light of these new proposals, we can re-examine some physical concepts.  For example, the vacuum state can be defined as the lowest quantum state of matter which is referred to a quantum state where all the new quantum numbers of the system are the lowest. Also, some of the already known properties for single particles can be reviewed as a particle system. For example, the spin as a property of an ``isolated'' particle now it can be related to a quantized 2-particle system, so it can be studied as the quantized angular momentum of a 2-particle system being one particle a massive particle and the another particle are a vacuum particle.

\newpage
\appendix

\section{Conjugate symmetry} \label{ApxConjSymm}
\begin{align}
&\big [ (\gamma^\bullet \odot \gamma^\bullet) (\delta \odot \delta)\big ]_E = 
		\gamma^{\bullet^*}_{E} \gamma^{\bullet}_{I} \delta^*_{E} \delta_{I} z_1^{(\gamma^{\bullet})} z_1^{(\delta)} z_0 
+ 		\gamma^{\bullet^*}_{E} \gamma^{\bullet}_{I} \delta^*_{I} \delta_{E} z_1^{(\gamma^{\bullet})} z_0
+ 		\gamma^{\bullet^*}_{I} \gamma^{\bullet}_{E} \delta^*_{E} \delta_{I} z_1^{(\delta)} z_0
\nonumber \\
& \qquad 
+ 		\gamma^{\bullet^*}_{I} \gamma^{\bullet}_{E} \delta^*_{I} \delta_{E} z_0
+ 		\gamma^{\bullet^*}_{E} \gamma^{\bullet}_{I} \delta^*_{E} \delta_{I} z_1^{(\gamma^{\bullet})} w_1^{(\delta)}
+\imagi \gamma^{\bullet^*}_{E} \gamma^{\bullet}_{I} \delta^*_{E} \delta_{E} z_1^{(\gamma^{\bullet})}
+ 		\gamma^{\bullet^*}_{E} \gamma^{\bullet}_{I} \delta^*_{I} \delta_{I} z_1^{(\gamma^{\bullet})}
\nonumber \\
& \qquad 
+ 		\gamma^{\bullet^*}_{I} \gamma^{\bullet}_{E} \delta^*_{E} \delta_{I} w_1^{(\delta)}
+\imagi \gamma^{\bullet^*}_{I} \gamma^{\bullet}_{E} \delta^*_{E} \delta_{E}
+		\gamma^{\bullet^*}_{I} \gamma^{\bullet}_{E} \delta^*_{I} \delta_{I}
+ 		\gamma^{\bullet^*}_{E} \gamma^{\bullet}_{I} \delta^*_{E} \delta_{I} w_1^{(\gamma^{\bullet})} z_1^{(\delta)}
\nonumber \\
& \qquad 
+		\gamma^{\bullet^*}_{E} \gamma^{\bullet}_{I} \delta^*_{I} \delta_{E} w_1^{(\gamma^{\bullet})}
+\imagi \gamma^{\bullet^*}_{E} \gamma^{\bullet}_{E} \delta^*_{E} \delta_{I} z_1^{(\delta)}
+\imagi \gamma^{\bullet^*}_{E} \gamma^{\bullet}_{E} \delta^*_{I} \delta_{E}
+		\gamma^{\bullet^*}_{I} \gamma^{\bullet}_{I} \delta^*_{E} \delta_{I} z_1^{(\delta)}
\nonumber \\
& \qquad 
+		\gamma^{\bullet^*}_{I} \gamma^{\bullet}_{I} \delta^*_{I} \delta_{E}. 
\end{align}

\begin{align}
&\big [ (\gamma^\bullet \odot \gamma^\bullet) (\delta \odot \delta)\big ]_I = 
		\gamma^{\bullet^*}_{E} \gamma^{\bullet}_{I} \delta^*_{E} \delta_{I} z_1^{(\gamma^{\bullet})} z_1^{(\delta)} w_0 
+ 		\gamma^{\bullet^*}_{E} \gamma^{\bullet}_{I} \delta^*_{I} \delta_{E} z_1^{(\gamma^{\bullet})} w_0
+ 		\gamma^{\bullet^*}_{I} \gamma^{\bullet}_{E} \delta^*_{E} \delta_{I} z_1^{(\delta)} w_0
\nonumber \\
& \qquad 
+ 		\gamma^{\bullet^*}_{I} \gamma^{\bullet}_{E} \delta^*_{I} \delta_{E} w_0
+ 		\gamma^{\bullet^*}_{E} \gamma^{\bullet}_{I} \delta^*_{E} \delta_{I} w_1^{(\gamma^{\bullet})} w_1^{(\delta)}
+\imagi \gamma^{\bullet^*}_{E} \gamma^{\bullet}_{I} \delta^*_{E} \delta_{E} w_1^{(\gamma^{\bullet})}
+		\gamma^{\bullet^*}_{E} \gamma^{\bullet}_{I} \delta^*_{I} \delta_{I} w_1^{(\gamma^{\bullet})}
\nonumber \\
& \qquad 
+\imagi \gamma^{\bullet^*}_{E} \gamma^{\bullet}_{E} \delta^*_{E} \delta_{I} w_1^{(\delta)}
- 		\gamma^{\bullet^*}_{E} \gamma^{\bullet}_{E} \delta^*_{E} \delta_{E}
+\imagi \gamma^{\bullet^*}_{E} \gamma^{\bullet}_{E} \delta^*_{I} \delta_{I}
+ 		\gamma^{\bullet^*}_{E} \gamma^{\bullet}_{I} \delta^*_{E} \delta_{I} w_1^{(\delta)}
\nonumber \\
& \qquad 
+\imagi \gamma^{\bullet^*}_{I} \gamma^{\bullet}_{I} \delta^*_{E} \delta_{E}
+		\gamma^{\bullet^*}_{I} \gamma^{\bullet}_{I} \delta^*_{I} \delta_{I} 
\end{align}
Substituting the extended numbers
\begin{align}
&\big [ (\gamma^\bullet \odot \gamma^\bullet) (\delta \odot \delta)\big ]_E = 
		\gamma^{\bullet^*}_{E} (\gamma_{E} w_2^{(\gamma)} + \gamma_{I}) \delta^*_{E} \delta_{I} z_1^{(\gamma^{\bullet})} z_1^{(\delta)} z_0 
+ 		\gamma_{E}^* z_2^{(\gamma)^*} (\gamma_{E} w_2^{(\gamma)} + \gamma_{I}) \delta^*_{I} \delta_{E} z_1^{(\gamma^{\bullet})} z_0
\nonumber \\
&
+ 		(\gamma_{E}^* w_2^{(\gamma)^*} + \gamma_{I}^*) \gamma_{E} z_2^{(\gamma)} \delta^*_{E} \delta_{I} z_1^{(\delta)} z_0
+ 		(\gamma_{E}^* w_2^{(\gamma)^*} + \gamma_{I}^*) \gamma_{E} z_2^{(\gamma)} \delta^*_{I} \delta_{E} z_0
+ 		\gamma_{E}^* z_2^{(\gamma)^*} (\gamma_{E} w_2^{(\gamma)} + \gamma_{I}) \delta^*_{E} \delta_{I} z_1^{(\gamma^{\bullet})} w_1^{(\delta)}
\nonumber \\
&
+\imagi \gamma_{E}^* z_2^{(\gamma)^*} (\gamma_{E} w_2^{(\gamma)} + \gamma_{I}) \delta^*_{E} \delta_{E} z_1^{(\gamma^{\bullet})}
+ 		\gamma_{E}^* z_2^{(\gamma)^*} (\gamma_{E} w_2^{(\gamma)} + \gamma_{I}) \delta^*_{I} \delta_{I} z_1^{(\gamma^{\bullet})}
+ 		(\gamma_{E}^* w_2^{(\gamma)^*} + \gamma_{I}^*) \gamma_{E} z_2^{(\gamma)} \delta^*_{E} \delta_{I} w_1^{(\delta)}
\nonumber \\
&
+\imagi (\gamma_{E}^* w_2^{(\gamma)^*} + \gamma_{I}^*) \gamma_{E} z_2^{(\gamma)} \delta^*_{E} \delta_{E}
+		(\gamma_{E}^* w_2^{(\gamma)^*} + \gamma_{I}^*) \gamma_{E} z_2^{(\gamma)} \delta^*_{I} \delta_{I}
+ 		\gamma_{E}^* z_2^{(\gamma)^*} (\gamma_{E} w_2^{(\gamma)} + \gamma_{I}) \delta^*_{E} \delta_{I} w_1^{(\gamma^{\bullet})} z_1^{(\delta)}
\nonumber \\
&
+		\gamma_{E}^* z_2^{(\gamma)^*} (\gamma_{E} w_2^{(\gamma)} + \gamma_{I}) \delta^*_{I} \delta_{E} w_1^{(\gamma^{\bullet})}
+\imagi \gamma_{E}^* z_2^{(\gamma)^*} \gamma_{E} z_2^{(\gamma)} \delta^*_{E} \delta_{I} z_1^{(\delta)}
+\imagi \gamma_{E}^* z_2^{(\gamma)^*} \gamma_{E} z_2^{(\gamma)} \delta^*_{I} \delta_{E}
\nonumber \\
&
+		(\gamma_{E}^* w_2^{(\gamma)^*} + \gamma_{I}^*) (\gamma_{E} w_2^{(\gamma)} + \gamma_{I}) \delta^*_{E} \delta_{I} z_1^{(\delta)}
+		(\gamma_{E}^* w_2^{(\gamma)^*} + \gamma_{I}^*) (\gamma_{E} w_2^{(\gamma)} + \gamma_{I}) \delta^*_{I} \delta_{E}. 
\end{align}

\begin{align}
&\big [ (\gamma^\bullet \odot \gamma^\bullet) (\delta \odot \delta)\big ]_I = 
		\gamma_{E}^* z_2^{(\gamma)^*} (\gamma_{E} w_2^{(\gamma)} + \gamma_{I}) \delta^*_{E} \delta_{I} z_1^{(\gamma^{\bullet})} z_1^{(\delta)} w_0 
+ 		\gamma_{E}^* z_2^{(\gamma)^*} (\gamma_{E} w_2^{(\gamma)} + \gamma_{I}) \delta^*_{I} \delta_{E} z_1^{(\gamma^{\bullet})} w_0
\nonumber \\
&
+ 		(\gamma_{E}^* w_2^{(\gamma)^*} + \gamma_{I}^*) \gamma_{E} z_2^{(\gamma)} \delta^*_{E} \delta_{I} z_1^{(\delta)} w_0
+ 		(\gamma_{E}^* w_2^{(\gamma)^*} + \gamma_{I}^*) \gamma_{E} z_2^{(\gamma)} \delta^*_{I} \delta_{E} w_0
+ 		\gamma_{E}^* z_2^{(\gamma)^*} (\gamma_{E} w_2^{(\gamma)} + \gamma_{I}) \delta^*_{E} \delta_{I} w_1^{(\gamma^{\bullet})} w_1^{(\delta)}
\nonumber \\
&
+\imagi \gamma_{E}^* z_2^{(\gamma)^*} (\gamma_{E} w_2^{(\gamma)} + \gamma_{I}) \delta^*_{E} \delta_{E} w_1^{(\gamma^{\bullet})}
+		\gamma_{E}^* z_2^{(\gamma)^*} (\gamma_{E} w_2^{(\gamma)} + \gamma_{I}) \delta^*_{I} \delta_{I} w_1^{(\gamma^{\bullet})}
+\imagi \gamma_{E}^* z_2^{(\gamma)^*} \gamma_{E} z_2^{(\gamma)} \delta^*_{E} \delta_{I} w_1^{(\delta)}
\nonumber \\
&
- 		\gamma_{E}^* z_2^{(\gamma)^*} \gamma_{E} z_2^{(\gamma)} \delta^*_{E} \delta_{E}
+\imagi \gamma_{E}^* z_2^{(\gamma)^*} \gamma_{E} z_2^{(\gamma)} \delta^*_{I} \delta_{I}
+ 		\gamma_{E}^* z_2^{(\gamma)^*} (\gamma_{E} w_2^{(\gamma)} + \gamma_{I}) \delta^*_{E} \delta_{I} w_1^{(\delta)}
\nonumber \\
&
+\imagi (\gamma_{E}^* w_2^{(\gamma)^*} + \gamma_{I}^*) (\gamma_{E} w_2^{(\gamma)} + \gamma_{I}) \delta^*_{E} \delta_{E}
+		(\gamma_{E}^* w_2^{(\gamma)^*} + \gamma_{I}^*) (\gamma_{E} w_2^{(\gamma)} + \gamma_{I}) \delta^*_{I} \delta_{I} 
\end{align}
Expanding terms

\begin{align}
&\big [ (\gamma^\bullet \odot \gamma^\bullet) (\delta \odot \delta)\big ]_E = 
		\gamma^{\bullet^*}_{E} (\gamma_{E} w_2^{(\gamma)}) \delta^*_{E} \delta_{I} z_1^{(\gamma^{\bullet})} z_1^{(\delta)} z_0 
+		\gamma^{\bullet^*}_{E} (\gamma_{I}) \delta^*_{E} \delta_{I} z_1^{(\gamma^{\bullet})} z_1^{(\delta)} z_0 
+ 		\gamma_{E}^* z_2^{(\gamma)^*} (\gamma_{E} w_2^{(\gamma)}) \delta^*_{I} \delta_{E} z_1^{(\gamma^{\bullet})} z_0
\nonumber \\
&
+ 		\gamma_{E}^* z_2^{(\gamma)^*} (\gamma_{I}) \delta^*_{I} \delta_{E} z_1^{(\gamma^{\bullet})} z_0
+ 		(\gamma_{E}^* w_2^{(\gamma)^*}) \gamma_{E} z_2^{(\gamma)} \delta^*_{E} \delta_{I} z_1^{(\delta)} z_0
+ 		(\gamma_{I}^*) \gamma_{E} z_2^{(\gamma)} \delta^*_{E} \delta_{I} z_1^{(\delta)} z_0
+ 		(\gamma_{E}^* w_2^{(\gamma)^*}) \gamma_{E} z_2^{(\gamma)} \delta^*_{I} \delta_{E} z_0
\nonumber \\
&
+ 		(\gamma_{I}^*) \gamma_{E} z_2^{(\gamma)} \delta^*_{I} \delta_{E} z_0
+ 		\gamma_{E}^* z_2^{(\gamma)^*} (\gamma_{E} w_2^{(\gamma)}) \delta^*_{E} \delta_{I} z_1^{(\gamma^{\bullet})} w_1^{(\delta)}
+ 		\gamma_{E}^* z_2^{(\gamma)^*} (\gamma_{I}) \delta^*_{E} \delta_{I} z_1^{(\gamma^{\bullet})} w_1^{(\delta)}
\nonumber \\
&
+\imagi \gamma_{E}^* z_2^{(\gamma)^*} (\gamma_{E} w_2^{(\gamma)}) \delta^*_{E} \delta_{E} z_1^{(\gamma^{\bullet})}
+\imagi \gamma_{E}^* z_2^{(\gamma)^*} (\gamma_{I}) \delta^*_{E} \delta_{E} z_1^{(\gamma^{\bullet})}
+ 		\gamma_{E}^* z_2^{(\gamma)^*} (\gamma_{E} w_2^{(\gamma)}) \delta^*_{I} \delta_{I} z_1^{(\gamma^{\bullet})}
+ 		\gamma_{E}^* z_2^{(\gamma)^*} (\gamma_{I}) \delta^*_{I} \delta_{I} z_1^{(\gamma^{\bullet})}
\nonumber \\
&
+ 		(\gamma_{E}^* w_2^{(\gamma)^*}) \gamma_{E} z_2^{(\gamma)} \delta^*_{E} \delta_{I} w_1^{(\delta)}
+ 		(\gamma_{I}^*) \gamma_{E} z_2^{(\gamma)} \delta^*_{E} \delta_{I} w_1^{(\delta)}
+\imagi (\gamma_{E}^* w_2^{(\gamma)^*}) \gamma_{E} z_2^{(\gamma)} \delta^*_{E} \delta_{E}
+\imagi (\gamma_{I}^*) \gamma_{E} z_2^{(\gamma)} \delta^*_{E} \delta_{E}
\nonumber \\
&
+		(\gamma_{E}^* w_2^{(\gamma)^*}) \gamma_{E} z_2^{(\gamma)} \delta^*_{I} \delta_{I}
+		(\gamma_{I}^*) \gamma_{E} z_2^{(\gamma)} \delta^*_{I} \delta_{I}
+ 		\gamma_{E}^* z_2^{(\gamma)^*} (\gamma_{E} w_2^{(\gamma)}) \delta^*_{E} \delta_{I} w_1^{(\gamma^{\bullet})} z_1^{(\delta)}
+ 		\gamma_{E}^* z_2^{(\gamma)^*} (\gamma_{I}) \delta^*_{E} \delta_{I} w_1^{(\gamma^{\bullet})} z_1^{(\delta)}
\nonumber \\
&
+		\gamma_{E}^* z_2^{(\gamma)^*} (\gamma_{E} w_2^{(\gamma)}) \delta^*_{I} \delta_{E} w_1^{(\gamma^{\bullet})}
+		\gamma_{E}^* z_2^{(\gamma)^*} (\gamma_{I}) \delta^*_{I} \delta_{E} w_1^{(\gamma^{\bullet})}
+\imagi \gamma_{E}^* z_2^{(\gamma)^*} \gamma_{E} z_2^{(\gamma)} \delta^*_{E} \delta_{I} z_1^{(\delta)}
+\imagi \gamma_{E}^* z_2^{(\gamma)^*} \gamma_{E} z_2^{(\gamma)} \delta^*_{I} \delta_{E}
\nonumber \\
&
+		(\gamma_{E}^* w_2^{(\gamma)^*}) (\gamma_{E} w_2^{(\gamma)} + \gamma_{I}) \delta^*_{E} \delta_{I} z_1^{(\delta)}
+		(\gamma_{E}^* w_2^{(\gamma)^*}) (\gamma_{E} w_2^{(\gamma)}) \delta^*_{E} \delta_{I} z_1^{(\delta)}
+		(\gamma_{I}^*) (\gamma_{E} w_2^{(\gamma)}) \delta^*_{E} \delta_{I} z_1^{(\delta)}
\nonumber \\
&
+		(\gamma_{I}^*) (\gamma_{I}) \delta^*_{E} \delta_{I} z_1^{(\delta)}
+		(\gamma_{E}^* w_2^{(\gamma)^*}) (\gamma_{E} w_2^{(\gamma)}) \delta^*_{I} \delta_{E}
+		(\gamma_{E}^* w_2^{(\gamma)^*}) (\gamma_{I}) \delta^*_{I} \delta_{E}
+		(\gamma_{I}^*) (\gamma_{E} w_2^{(\gamma)}) \delta^*_{I} \delta_{E}
+		(\gamma_{I}^*) (\gamma_{I}) \delta^*_{I} \delta_{E}. 
\end{align}

\begin{align}
&\big [ (\gamma^\bullet \odot \gamma^\bullet) (\delta \odot \delta)\big ]_I = 
		\gamma_{E}^* z_2^{(\gamma)^*} (\gamma_{E} w_2^{(\gamma)}) \delta^*_{E} \delta_{I} z_1^{(\gamma^{\bullet})} z_1^{(\delta)} w_0 
+		\gamma_{E}^* z_2^{(\gamma)^*} (\gamma_{I}) \delta^*_{E} \delta_{I} z_1^{(\gamma^{\bullet})} z_1^{(\delta)} w_0 
\nonumber \\
&
+ 		\gamma_{E}^* z_2^{(\gamma)^*} (\gamma_{E} w_2^{(\gamma)}) \delta^*_{I} \delta_{E} z_1^{(\gamma^{\bullet})} w_0
+ 		\gamma_{E}^* z_2^{(\gamma)^*} (\gamma_{I}) \delta^*_{I} \delta_{E} z_1^{(\gamma^{\bullet})} w_0
+ 		(\gamma_{E}^* w_2^{(\gamma)^*}) \gamma_{E} z_2^{(\gamma)} \delta^*_{E} \delta_{I} z_1^{(\delta)} w_0
\nonumber \\
&
+ 		(\gamma_{I}^*) \gamma_{E} z_2^{(\gamma)} \delta^*_{E} \delta_{I} z_1^{(\delta)} w_0
+ 		(\gamma_{E}^* w_2^{(\gamma)^*}) \gamma_{E} z_2^{(\gamma)} \delta^*_{I} \delta_{E} w_0
+ 		(\gamma_{I}^*) \gamma_{E} z_2^{(\gamma)} \delta^*_{I} \delta_{E} w_0
+ 		\gamma_{E}^* z_2^{(\gamma)^*} (\gamma_{E} w_2^{(\gamma)}) \delta^*_{E} \delta_{I} w_1^{(\gamma^{\bullet})} w_1^{(\delta)}
\nonumber \\
&
+ 		\gamma_{E}^* z_2^{(\gamma)^*} (\gamma_{I}) \delta^*_{E} \delta_{I} w_1^{(\gamma^{\bullet})} w_1^{(\delta)}
+\imagi \gamma_{E}^* z_2^{(\gamma)^*} (\gamma_{E} w_2^{(\gamma)}) \delta^*_{E} \delta_{E} w_1^{(\gamma^{\bullet})}
+\imagi \gamma_{E}^* z_2^{(\gamma)^*} (\gamma_{I}) \delta^*_{E} \delta_{E} w_1^{(\gamma^{\bullet})}
\nonumber \\
&
+		\gamma_{E}^* z_2^{(\gamma)^*} (\gamma_{E} w_2^{(\gamma)}) \delta^*_{I} \delta_{I} w_1^{(\gamma^{\bullet})}
+		\gamma_{E}^* z_2^{(\gamma)^*} (\gamma_{I}) \delta^*_{I} \delta_{I} w_1^{(\gamma^{\bullet})}
+\imagi \gamma_{E}^* z_2^{(\gamma)^*} \gamma_{E} z_2^{(\gamma)} \delta^*_{E} \delta_{I} w_1^{(\delta)}
- 		\gamma_{E}^* z_2^{(\gamma)^*} \gamma_{E} z_2^{(\gamma)} \delta^*_{E} \delta_{E}
\nonumber \\
&
+\imagi \gamma_{E}^* z_2^{(\gamma)^*} \gamma_{E} z_2^{(\gamma)} \delta^*_{I} \delta_{I}
+ 		\gamma_{E}^* z_2^{(\gamma)^*} (\gamma_{E} w_2^{(\gamma)}) \delta^*_{E} \delta_{I} w_1^{(\delta)}
+ 		\gamma_{E}^* z_2^{(\gamma)^*} (\gamma_{I}) \delta^*_{E} \delta_{I} w_1^{(\delta)}
+\imagi (\gamma_{E}^* w_2^{(\gamma)^*}) (\gamma_{E} w_2^{(\gamma)}) \delta^*_{E} \delta_{E}
\nonumber \\
&
+\imagi (\gamma_{E}^* w_2^{(\gamma)^*}) (\gamma_{I}) \delta^*_{E} \delta_{E}
+\imagi (\gamma_{I}^*) (\gamma_{E} w_2^{(\gamma)}) \delta^*_{E} \delta_{E}
+\imagi (\gamma_{I}^*) (\gamma_{I}) \delta^*_{E} \delta_{E}
+		(\gamma_{E}^* w_2^{(\gamma)^*}) (\gamma_{I}) \delta^*_{I} \delta_{I} 
\nonumber \\
&
+		(\gamma_{E}^* w_2^{(\gamma)^*}) (\gamma_{E} w_2^{(\gamma)}) \delta^*_{I} \delta_{I} 
+		(\gamma_{I}^*) (\gamma_{E} w_2^{(\gamma)}) \delta^*_{I} \delta_{I} 
+		(\gamma_{I}^*) (\gamma_{I}) \delta^*_{I} \delta_{I} 
\end{align}

\begin{align}
&\big [ (\gamma \odot \gamma) (\delta^\bullet \odot \delta^\bullet)\big ]_E = 
		\gamma^*_{E} \gamma_{I} \delta^{\bullet^*}_{E} \delta^{\bullet}_{I} z_1^{(\gamma)} z_1^{(\delta^\bullet)} z_0 
+ 		\gamma^*_{E} \gamma_{I} \delta^{\bullet^*}_{I} \delta^{\bullet}_{E} z_1^{(\gamma)} z_0
+ 		\gamma^*_{I} \gamma_{E} \delta^{\bullet^*}_{E} \delta^{\bullet}_{I} z_1^{(\delta^\bullet)} z_0
\nonumber \\
& \qquad 
+ 		\gamma^*_{I} \gamma_{E} \delta^{\bullet^*}_{I} \delta^{\bullet}_{E} z_0
+ 		\gamma^*_{E} \gamma_{I} \delta^{\bullet^*}_{E} \delta^{\bullet}_{I} z_1^{(\gamma)} w_1^{(\gamma)}
+\imagi \gamma^*_{E} \gamma_{I} \delta^{\bullet^*}_{E} \delta^{\bullet}_{E} z_1^{(\gamma)}
+ 		\gamma^*_{E} \gamma_{I} \delta^{\bullet^*}_{I} \delta^{\bullet}_{I} z_1^{(\gamma)}
\nonumber \\
& \qquad 
+ 		\gamma^*_{I} \gamma_{E} \delta^{\bullet^*}_{E} \delta^{\bullet}_{I} w_1^{(\delta^\bullet)}
+\imagi \gamma^*_{I} \gamma_{E} \delta^{\bullet^*}_{E} \delta^{\bullet}_{E}
+		\gamma^*_{I} \gamma_{E} \delta^{\bullet^*}_{I} \delta^{\bullet}_{I}
+ 		\gamma^*_{E} \gamma_{I} \delta^{\bullet^*}_{E} \delta^{\bullet}_{I} w_1^{(\gamma)} z_1^{(\delta^\bullet)}
\nonumber \\
& \qquad 
+		\gamma^*_{E} \gamma_{I} \delta^{\bullet^*}_{I} \delta^{\bullet}_{E} w_1^{(\gamma)}
+\imagi \gamma^*_{E} \gamma_{E} \delta^{\bullet^*}_{E} \delta^{\bullet}_{I} z_1^{(\delta^\bullet)}
+\imagi \gamma^*_{E} \gamma_{E} \delta^{\bullet^*}_{I} \delta^{\bullet}_{E}
+		\gamma^*_{I} \gamma_{I} \delta^{\bullet^*}_{E} \delta^{\bullet}_{I} z_1^{(\delta^\bullet)}
\nonumber \\
& \qquad 
+		\gamma^*_{I} \gamma_{I} \delta^{\bullet^*}_{I} \delta^{\bullet}_{E}. 
\end{align}

\begin{align}
&\big [ (\gamma \odot \gamma) (\delta^\bullet \odot \delta^\bullet)\big ]_I = 
		\gamma^*_{E} \gamma_{I} \delta^{\bullet^*}_{E} \delta^{\bullet}_{I} z_1^{(\gamma)} z_1^{(\delta^\bullet)} w_0 
+ 		\gamma^*_{E} \gamma_{I} \delta^{\bullet^*}_{I} \delta^{\bullet}_{E} z_1^{(\gamma)} w_0
+ 		\gamma^*_{I} \gamma_{E} \delta^{\bullet^*}_{E} \delta^{\bullet}_{I} z_1^{(\delta^\bullet)} w_0
\nonumber \\
& \qquad 
+ 		\gamma^*_{I} \gamma_{E} \delta^{\bullet^*}_{I} \delta^{\bullet}_{E} w_0
+ 		\gamma^*_{E} \gamma_{I} \delta^{\bullet^*}_{E} \delta^{\bullet}_{I} w_1^{(\gamma)} w_1^{(\delta^\bullet)}
+\imagi \gamma^*_{E} \gamma_{I} \delta^{\bullet^*}_{E} \delta^{\bullet}_{E} w_1^{(\gamma)}
+		\gamma^*_{E} \gamma_{I} \delta^{\bullet^*}_{I} \delta^{\bullet}_{I} w_1^{(\gamma)}
\nonumber \\
& \qquad 
+\imagi \gamma^*_{E} \gamma_{E} \delta^{\bullet^*}_{E} \delta^{\bullet}_{I} w_1^{(\delta^\bullet)}
- 		\gamma^*_{E} \gamma_{E} \delta^{\bullet^*}_{E} \delta^{\bullet}_{E}
+\imagi \gamma^*_{E} \gamma_{E} \delta^{\bullet^*}_{I} \delta^{\bullet}_{I}
+ 		\gamma^*_{E} \gamma_{I} \delta^{\bullet^*}_{E} \delta^{\bullet}_{I} w_1^{(\delta^\bullet)}
\nonumber \\
& \qquad 
+\imagi \gamma^*_{I} \gamma_{I} \delta^{\bullet^*}_{E} \delta^{\bullet}_{E}
+		\gamma^*_{I} \gamma_{I} \delta^{\bullet^*}_{I} \delta^{\bullet}_{I} 
\end{align}

\begin{align}
&\big [ (\gamma \odot \gamma) (\delta^\bullet \odot \delta^\bullet)\big ]_E = 
		\gamma^*_{E} \gamma_{I} \delta_{E}^* z_2^{(\delta)^*} (\delta_{E} w_2^{(\delta)} + \delta_{I}) z_1^{(\gamma)} z_1^{(\delta^\bullet)} z_0 
+ 		\gamma^*_{E} \gamma_{I} (\delta_{E}^* w_2^{(\delta)^*} + \delta_{I}^*) \delta_{E} z_2^{(\delta)} z_1^{(\gamma)} z_0
\nonumber \\
& \qquad 
+ 		\gamma^*_{I} \gamma_{E} \delta_{E}^* z_2^{(\delta)^*} (\delta_{E} w_2^{(\delta)} + \delta_{I}) z_1^{(\delta^\bullet)} z_0
+ 		\gamma^*_{I} \gamma_{E} (\delta_{E}^* w_2^{(\delta)^*} + \delta_{I}^*) \delta_{E} z_2^{(\delta)} z_0
+ 		\gamma^*_{E} \gamma_{I} \delta_{E}^* z_2^{(\delta)^*} (\delta_{E} w_2^{(\delta)} + \delta_{I}) z_1^{(\gamma)} w_1^{(\gamma)}
\nonumber \\
& \qquad 
+\imagi \gamma^*_{E} \gamma_{I} \delta_{E}^* z_2^{(\delta)^*} \delta_{E} z_2^{(\delta)} z_1^{(\gamma)}
+ 		\gamma^*_{E} \gamma_{I} (\delta_{E}^* w_2^{(\delta)^*} + \delta_{I}^*) (\delta_{E} w_2^{(\delta)} + \delta_{I}) z_1^{(\gamma)}
+ 		\gamma^*_{I} \gamma_{E} \delta_{E}^* z_2^{(\delta)^*} (\delta_{E} w_2^{(\delta)} + \delta_{I}) w_1^{(\delta^\bullet)}
\nonumber \\
& \qquad 
+\imagi \gamma^*_{I} \gamma_{E} \delta_{E}^* z_2^{(\delta)^*} \delta_{E} z_2^{(\delta)}
+		\gamma^*_{I} \gamma_{E} (\delta_{E}^* w_2^{(\delta)^*} + \delta_{I}^*) (\delta_{E} w_2^{(\delta)} + \delta_{I})
+ 		\gamma^*_{E} \gamma_{I} \delta_{E}^* z_2^{(\delta)^*} (\delta_{E} w_2^{(\delta)} + \delta_{I}) w_1^{(\gamma)} z_1^{(\delta^\bullet)}
\nonumber \\
& \qquad 
+		\gamma^*_{E} \gamma_{I} (\delta_{E}^* w_2^{(\delta)^*} + \delta_{I}^*) \delta_{E} z_2^{(\delta)} w_1^{(\gamma)}
+\imagi \gamma^*_{E} \gamma_{E} \delta_{E}^* z_2^{(\delta)^*} (\delta_{E} w_2^{(\delta)} + \delta_{I}) z_1^{(\delta^\bullet)}
+\imagi \gamma^*_{E} \gamma_{E} (\delta_{E}^* w_2^{(\delta)^*} + \delta_{I}^*) \delta_{E} z_2^{(\delta)}
\nonumber \\
& \qquad 
+		\gamma^*_{I} \gamma_{I} \delta_{E}^* z_2^{(\delta)^*} (\delta_{E} w_2^{(\delta)} + \delta_{I}) z_1^{(\delta^\bullet)}
+		\gamma^*_{I} \gamma_{I} (\delta_{E}^* w_2^{(\delta)^*} + \delta_{I}^*) \delta_{E} z_2^{(\delta)}. 
\end{align}

\begin{align}
&\big [ (\gamma \odot \gamma) (\delta^\bullet \odot \delta^\bullet)\big ]_I = 
		\gamma^*_{E} \gamma_{I} \delta_{E}^* z_2^{(\delta)^*} (\delta_{E} w_2^{(\delta)} + \delta_{I}) z_1^{(\gamma)} z_1^{(\delta^\bullet)} w_0 
+ 		\gamma^*_{E} \gamma_{I} (\delta_{E}^* w_2^{(\delta)^*} + \delta_{I}^*) \delta_{E} z_2^{(\delta)} z_1^{(\gamma)} w_0
\nonumber \\
& \qquad 
+ 		\gamma^*_{I} \gamma_{E} \delta_{E}^* z_2^{(\delta)^*} (\delta_{E} w_2^{(\delta)} + \delta_{I}) z_1^{(\delta^\bullet)} w_0
+ 		\gamma^*_{I} \gamma_{E} (\delta_{E}^* w_2^{(\delta)^*} + \delta_{I}^*) \delta_{E} z_2^{(\delta)} w_0
+ 		\gamma^*_{E} \gamma_{I} \delta_{E}^* z_2^{(\delta)^*} (\delta_{E} w_2^{(\delta)} + \delta_{I}) w_1^{(\gamma)} w_1^{(\delta^\bullet)}
\nonumber \\
& \qquad 
+\imagi \gamma^*_{E} \gamma_{I} \delta_{E}^* z_2^{(\delta)^*} \delta_{E} z_2^{(\delta)} w_1^{(\gamma)}
+		\gamma^*_{E} \gamma_{I} (\delta_{E}^* w_2^{(\delta)^*} + \delta_{I}^*) (\delta_{E} w_2^{(\delta)} + \delta_{I}) w_1^{(\gamma)}
+\imagi \gamma^*_{E} \gamma_{E} \delta_{E}^* z_2^{(\delta)^*} (\delta_{E} w_2^{(\delta)} + \delta_{I}) w_1^{(\delta^\bullet)}
\nonumber \\
& \qquad 
- 		\gamma^*_{E} \gamma_{E} \delta_{E}^* z_2^{(\delta)^*} \delta_{E} z_2^{(\delta)}
+\imagi \gamma^*_{E} \gamma_{E} (\delta_{E}^* w_2^{(\delta)^*} + \delta_{I}^*) (\delta_{E} w_2^{(\delta)} + \delta_{I})
+ 		\gamma^*_{E} \gamma_{I} \delta_{E}^* z_2^{(\delta)^*} (\delta_{E} w_2^{(\delta)} + \delta_{I}) w_1^{(\delta^\bullet)}
\nonumber \\
& \qquad 
+\imagi \gamma^*_{I} \gamma_{I} \delta_{E}^* z_2^{(\delta)^*} \delta_{E} z_2^{(\delta)}
+		\gamma^*_{I} \gamma_{I} (\delta_{E}^* w_2^{(\delta)^*} + \delta_{I}^*) (\delta_{E} w_2^{(\delta)} + \delta_{I}) 
\end{align}

Expanding
\begin{align}
&\big [ (\gamma \odot \gamma) (\delta^\bullet \odot \delta^\bullet)\big ]_E = 
		\gamma^*_{E} \gamma_{I} \delta_{E}^* z_2^{(\delta)^*} (\delta_{E} w_2^{(\delta)} ) z_1^{(\gamma)} z_1^{(\delta^\bullet)} z_0 
+		\gamma^*_{E} \gamma_{I} \delta_{E}^* z_2^{(\delta)^*} ( \delta_{I}) z_1^{(\gamma)} z_1^{(\delta^\bullet)} z_0 
\nonumber \\
&
+ 		\gamma^*_{E} \gamma_{I} (\delta_{E}^* w_2^{(\delta)^*}) \delta_{E} z_2^{(\delta)} z_1^{(\gamma)} z_0
+ 		\gamma^*_{E} \gamma_{I} (\delta_{I}^*) \delta_{E} z_2^{(\delta)} z_1^{(\gamma)} z_0
+ 		\gamma^*_{I} \gamma_{E} \delta_{E}^* z_2^{(\delta)^*} (\delta_{E} w_2^{(\delta)}) z_1^{(\delta^\bullet)} z_0
+ 		\gamma^*_{I} \gamma_{E} \delta_{E}^* z_2^{(\delta)^*} (\delta_{I}) z_1^{(\delta^\bullet)} z_0
\nonumber \\
&
+ 		\gamma^*_{I} \gamma_{E} (\delta_{E}^* w_2^{(\delta)^*}) \delta_{E} z_2^{(\delta)} z_0
+ 		\gamma^*_{I} \gamma_{E} (\delta_{I})\delta_{E} z_2^{(\delta)} z_0
+ 		\gamma^*_{E} \gamma_{I} \delta_{E}^* z_2^{(\delta)^*} (\delta_{E} w_2^{(\delta)}) z_1^{(\gamma)} w_1^{(\gamma)}
+ 		\gamma^*_{E} \gamma_{I} \delta_{E}^* z_2^{(\delta)^*} ( \delta_{I}) z_1^{(\gamma)} w_1^{(\gamma)}
\nonumber \\
&
+\imagi \gamma^*_{E} \gamma_{I} \delta_{E}^* z_2^{(\delta)^*} \delta_{E} z_2^{(\delta)} z_1^{(\gamma)}
+ 		\gamma^*_{E} \gamma_{I} (\delta_{E}^* w_2^{(\delta)^*}) (\delta_{E} w_2^{(\delta)}) z_1^{(\gamma)}
+ 		\gamma^*_{E} \gamma_{I} (\delta_{I}^*) (\delta_{E} w_2^{(\delta)}) z_1^{(\gamma)}
+ 		\gamma^*_{E} \gamma_{I} (\delta_{E}^* w_2^{(\delta)^*}) (\delta_{I}) z_1^{(\gamma)}
\nonumber \\
&
+ 		\gamma^*_{E} \gamma_{I} (\delta_{I}^*) (\delta_{I}) z_1^{(\gamma)}
+ 		\gamma^*_{I} \gamma_{E} \delta_{E}^* z_2^{(\delta)^*} (\delta_{E} w_2^{(\delta)}) w_1^{(\delta^\bullet)}
+ 		\gamma^*_{I} \gamma_{E} \delta_{E}^* z_2^{(\delta)^*} (\delta_{I}) w_1^{(\delta^\bullet)}
+\imagi \gamma^*_{I} \gamma_{E} \delta_{E}^* z_2^{(\delta)^*} \delta_{E} z_2^{(\delta)}
\nonumber \\
&
+		\gamma^*_{I} \gamma_{E} (\delta_{E}^* w_2^{(\delta)^*}) (\delta_{E} w_2^{(\delta)})
+		\gamma^*_{I} \gamma_{E} (\delta_{E}^* w_2^{(\delta)^*}) (\delta_{I})
+		\gamma^*_{I} \gamma_{E} (\delta_{I}^*) (\delta_{E} w_2^{(\delta)})
+		\gamma^*_{I} \gamma_{E} (\delta_{I}^*) (\delta_{I})
\nonumber \\
&
+ 		\gamma^*_{E} \gamma_{I} \delta_{E}^* z_2^{(\delta)^*} (\delta_{E} w_2^{(\delta)}) w_1^{(\gamma)} z_1^{(\delta^\bullet)}
+ 		\gamma^*_{E} \gamma_{I} \delta_{E}^* z_2^{(\delta)^*} (\delta_{I}) w_1^{(\gamma)} z_1^{(\delta^\bullet)}
+		\gamma^*_{E} \gamma_{I} (\delta_{E}^* w_2^{(\delta)^*}) \delta_{E} z_2^{(\delta)} w_1^{(\gamma)}
\nonumber \\
&
+		\gamma^*_{E} \gamma_{I} (\delta_{I}^*) \delta_{E} z_2^{(\delta)} w_1^{(\gamma)}
+\imagi \gamma^*_{E} \gamma_{E} \delta_{E}^* z_2^{(\delta)^*} (\delta_{E} w_2^{(\delta)}) z_1^{(\delta^\bullet)}
+\imagi \gamma^*_{E} \gamma_{E} \delta_{E}^* z_2^{(\delta)^*} (\delta_{I}) z_1^{(\delta^\bullet)}
+\imagi \gamma^*_{E} \gamma_{E} (\delta_{E}^* w_2^{(\delta)^*}) \delta_{E} z_2^{(\delta)}
\nonumber \\
&
+\imagi \gamma^*_{E} \gamma_{E}  \delta_{I}^*) \delta_{E} z_2^{(\delta)}
+		\gamma^*_{I} \gamma_{I} \delta_{E}^* z_2^{(\delta)^*} (\delta_{E} w_2^{(\delta)}) z_1^{(\delta^\bullet)}
+		\gamma^*_{I} \gamma_{I} \delta_{E}^* z_2^{(\delta)^*} (\delta_{I}) z_1^{(\delta^\bullet)}
+		\gamma^*_{I} \gamma_{I} (\delta_{E}^* w_2^{(\delta)^*}) \delta_{E} z_2^{(\delta)}
\nonumber \\
&
+		\gamma^*_{I} \gamma_{I} (\delta_{I}^*) \delta_{E} z_2^{(\delta)}. 
\end{align}

\begin{align}
&\big [ (\gamma \odot \gamma) (\delta^\bullet \odot \delta^\bullet)\big ]_I = 
		\gamma^*_{E} \gamma_{I} \delta_{E}^* z_2^{(\delta)^*} (\delta_{E} w_2^{(\delta)}) z_1^{(\gamma)} z_1^{(\delta^\bullet)} w_0 
+		\gamma^*_{E} \gamma_{I} \delta_{E}^* z_2^{(\delta)^*} ( \delta_{I}) z_1^{(\gamma)} z_1^{(\delta^\bullet)} w_0 
\nonumber \\
&
+ 		\gamma^*_{E} \gamma_{I} (\delta_{E}^* w_2^{(\delta)^*}) \delta_{E} z_2^{(\delta)} z_1^{(\gamma)} w_0
+ 		\gamma^*_{E} \gamma_{I} (\delta_{I}^*) \delta_{E} z_2^{(\delta)} z_1^{(\gamma)} w_0
+ 		\gamma^*_{I} \gamma_{E} \delta_{E}^* z_2^{(\delta)^*} (\delta_{E} w_2^{(\delta)}) z_1^{(\delta^\bullet)} w_0
\nonumber \\
&
+ 		\gamma^*_{I} \gamma_{E} \delta_{E}^* z_2^{(\delta)^*} (\delta_{I}) z_1^{(\delta^\bullet)} w_0
+ 		\gamma^*_{I} \gamma_{E} (\delta_{E}^* w_2^{(\delta)^*}) \delta_{E} z_2^{(\delta)} w_0
+ 		\gamma^*_{I} \gamma_{E} (\delta_{I}^*) \delta_{E} z_2^{(\delta)} w_0
+ 		\gamma^*_{E} \gamma_{I} \delta_{E}^* z_2^{(\delta)^*} (\delta_{E} w_2^{(\delta)}) w_1^{(\gamma)} w_1^{(\delta^\bullet)}
\nonumber \\
&
+ 		\gamma^*_{E} \gamma_{I} \delta_{E}^* z_2^{(\delta)^*} (\delta_{I}) w_1^{(\gamma)} w_1^{(\delta^\bullet)}
+\imagi \gamma^*_{E} \gamma_{I} \delta_{E}^* z_2^{(\delta)^*} \delta_{E} z_2^{(\delta)} w_1^{(\gamma)}
+		\gamma^*_{E} \gamma_{I} (\delta_{E}^* w_2^{(\delta)^*}) (\delta_{E} w_2^{(\delta)}) w_1^{(\gamma)}
+		\gamma^*_{E} \gamma_{I} (\delta_{E}^* w_2^{(\delta)^*}) (\delta_{I}) w_1^{(\gamma)}
\nonumber \\
&
+		\gamma^*_{E} \gamma_{I} (\delta_{I}^*) (\delta_{E} w_2^{(\delta)}) w_1^{(\gamma)}
+		\gamma^*_{E} \gamma_{I} (\delta_{I}^*) (\delta_{I}) w_1^{(\gamma)}
+\imagi \gamma^*_{E} \gamma_{E} \delta_{E}^* z_2^{(\delta)^*} (\delta_{E} w_2^{(\delta)}) w_1^{(\delta^\bullet)}
+\imagi \gamma^*_{E} \gamma_{E} \delta_{E}^* z_2^{(\delta)^*} (\delta_{I}) w_1^{(\delta^\bullet)}
\nonumber \\
&
- 		\gamma^*_{E} \gamma_{E} \delta_{E}^* z_2^{(\delta)^*} \delta_{E} z_2^{(\delta)}
+\imagi \gamma^*_{E} \gamma_{E} (\delta_{E}^* w_2^{(\delta)^*}) (\delta_{E} w_2^{(\delta)})
+\imagi \gamma^*_{E} \gamma_{E} (\delta_{E}^* w_2^{(\delta)^*}) (\delta_{I})
+\imagi \gamma^*_{E} \gamma_{E} (\delta_{I}^*) (\delta_{E} w_2^{(\delta)})
\nonumber \\
&
+\imagi \gamma^*_{E} \gamma_{E} (\delta_{I}^*) (\delta_{I})
+ 		\gamma^*_{E} \gamma_{I} \delta_{E}^* z_2^{(\delta)^*} (\delta_{E} w_2^{(\delta)}) w_1^{(\delta^\bullet)}
+ 		\gamma^*_{E} \gamma_{I} \delta_{E}^* z_2^{(\delta)^*} (\delta_{I}) w_1^{(\delta^\bullet)}
+\imagi \gamma^*_{I} \gamma_{I} \delta_{E}^* z_2^{(\delta)^*} \delta_{E} z_2^{(\delta)}
\nonumber \\
&
+		\gamma^*_{I} \gamma_{I} (\delta_{E}^* w_2^{(\delta)^*}) (\delta_{E} w_2^{(\delta)}) 
+		\gamma^*_{I} \gamma_{I} (\delta_{E}^* w_2^{(\delta)^*}) (\delta_{I}) 
+		\gamma^*_{I} \gamma_{I} (\delta_{I}^*) (\delta_{E} w_2^{(\delta)}) 
+		\gamma^*_{I} \gamma_{I} (\delta_{I}^*) (\delta_{I}) 
\end{align}

\newpage
\bibliography{ArticleVMVF}

\end{document}